\documentclass[aps,tightenlines,superscriptaddress,titlepage]{revtex4-1}
\usepackage[dvips]{graphicx}
\usepackage{dcolumn}
\usepackage{epsfig}
\usepackage{color}
\usepackage{footnote} 
\usepackage{subfigure}
\usepackage{amsmath}
\usepackage{amssymb}
\usepackage{amsfonts}
\usepackage{amsthm}
\usepackage[T1]{fontenc}
\usepackage{slashed}
\usepackage{times}
\usepackage[table]{xcolor}

\usepackage{xcolor}
\usepackage[percent]{overpic}

\usepackage{textcomp}

\usepackage{graphicx}

\makeatletter
\renewcommand*{\p@subsection}{}
\renewcommand*{\p@subsubsection}{}
\makeatother

  \makeatletter
  \let\ps@plain=\ps@empty
  \makeatother

\begin{document}

\title{Studies of Nucleon Resonance Structure in Exclusive Meson Electroproduction}

\newcommand*{\Kroll}{Fachbereich Physik, Universit\"at Wuppertal, 42097 Wuppertal, Germany}
\newcommand*{\CSSM}{CSSM and CoEPP, School of Chemistry and Physics, 
University of Adelaide, Adelaide SA 5005, Australia}
\newcommand*{\Idaho}{Idaho State University, Department of Physics, Pocatello, Idaho, 83209, USA}
\newcommand*{\Argonne}{Physics Division, Argonne National Laboratory, Argonne Illinois 60439, USA\\Department of Physics, Illinois Institute of Technology, Chicago, Illinois 60616, USA}

\newcommand*{\William}{College of William and Mary, Williamsburg, Virginia 23187, USA}

\newcommand*{\Brazil}{International Institute of Physics,
Federal University of Rio Grande do Norte
Natal, RN 59078-400, Brazil}

\newcommand*{\Yerevan}{Yerevan Physics Institute, Yerevan, Armenia}

\newcommand*{\JLAB}{Thomas Jefferson National Accelerator Facility,  Newport News, Virginia 23606, USA}

\newcommand*{\Stanford}{Stanford National Accelerator Laboratory,
Stanford University, Stanford, California 94025, USA}

\newcommand*{\USC}{University of South Carolina,  Columbia, South Carolina 29208, USA}

\newcommand*{\CP}{CP3-Origins, Southern Denmark University,  Odense, Denmark}

\newcommand*{\CostaRica}{Universidad de Costa Rica,  San Jos\'e, Costa Rica }

\newcommand*{\UNGenova}{Dipartimento di Fisica, Universit\`a 
di Genova, Italy}

\newcommand*{\INFNGenova}{Istituto Nazionale di Fisica Nucleare,  Sezione di Genova, Italy}

\newcommand*{\SKOBLO}{Skobeltsyn Institute Nuclear Physics at Moscow State University, 119899 Moscow, Russia}

\newcommand*{\CFTP}{CFTP, IST, Universidade T\'ecnica de Lisboa,  UTL, Portugal}
\newcommand*{\Lisboa}{Departamento de F\'isica, IST, Universidade T\'ecnica de Lisboa,
       UTL, Portugal}

\newcommand*{\Washington}{Department of Physics, University of Washington, Seattle, Washington 98195, USA} 

\newcommand*{\Regensburg}{Institut f\"ur Theoretische Physik, Universit\"at Regensburg, 93040 Regensburg, Germany}

\newcommand*{\Morelia}{Instituto de F\'{i}sica y Matem\'aticas, Universidad Michoacana de San Nicol\'as de Hidalgo, Edificio C-3, Ciudad Universitaria, Morelia, Michoac\'an 58040, M\'exico}

\newcommand*{\Julich}{Forschungszentrum J\"ulich, D-52425 J\"ulich, Germany}

\newcommand*{\Hefei}{Institute for Theoretical Physics and Department of Modern Physics, University of Science and Technology of China, Hefei 230026, P. R. China}

\newcommand*{\Illinois}{Department of Physics, Illinois Institute of Technology, Chicago, Illinois 60616, USA}

\newcommand*{\Cruzeiro}{Universidade Cruzeiro do Sul, Rua Galv\~ao Bueno, 868,  01506-000 S\~ao Paulo, SP, Brazil}

\newcommand*{\Paulista}{Instituto de F\'isica Te\'orica, Universidade Estadual Paulista,  Rua Dr.\,Bento Teobaldo Ferraz, 271, 01140-070 S\~ao Paulo, SP, Brazil}

\newcommand*{\ODU}{Department of Physics, Old Dominion University, Norfolk, Virginia 23529, USA}

\author{I.~G.~Aznauryan}
\affiliation{\Yerevan}
\affiliation{\JLAB}

\author{A.~Bashir}
\affiliation{\Morelia}

\author{V.~M.~Braun}
\affiliation{\Regensburg}

\author{S.~J.~Brodsky}
\affiliation{\Stanford}
\affiliation{\CP}

\author{V.~D.~Burkert}
\affiliation{\JLAB}

\author{L.~Chang}
\affiliation{\Argonne}
\affiliation{\Julich}

\author{Ch.~Chen}
\affiliation{\Argonne}
\affiliation{\Hefei}
\affiliation{\Illinois}

\author{B.~El-Bennich}
\affiliation{\Cruzeiro}
\affiliation{\Paulista}

\author{I.~C.~Clo\"et}
\affiliation{\Argonne}
\affiliation{\CSSM}

\author{P.~L.~Cole}
\affiliation{\Idaho}

\author{R.~G.~Edwards}
\affiliation{\JLAB}

\author{G.~V.~Fedotov}
\affiliation{\USC}
\affiliation{\SKOBLO}

\author{M.~M.~Giannini}
\affiliation{\UNGenova}
\affiliation{\INFNGenova}

\author{R.~W.~Gothe}
\affiliation{\USC}

\author{F.~Gross}
\affiliation{\JLAB}
\affiliation{\William}

\author{Huey-Wen~Lin}
\affiliation{\Washington}

\author{P.~Kroll}
\affiliation{\Kroll}
\affiliation{\Regensburg}

\author{T.-S.~H.~Lee}
\affiliation{\Argonne}

\author{W.~Melnitchouk}
\affiliation{\JLAB}

\author{V.~I.~Mokeev}
\affiliation{\JLAB}
\affiliation{\SKOBLO}

\author{M.~T.~Pe\~na}
\affiliation{\CFTP}
\affiliation{\Lisboa}

\author{G.~Ramalho}
\affiliation{\CFTP}

\author{C.~D.~Roberts}
\affiliation{\Argonne}
\affiliation{\Illinois}

\author{E.~Santopinto}
\affiliation{\INFNGenova}

\author{G.~F.~de~Teramond}
\affiliation{\CostaRica}

\author{K.~Tsushima}
\affiliation{\CSSM}
\affiliation{\Brazil}

\author{D.~J.~Wilson}
\affiliation{\Argonne}
\affiliation{\ODU}

\thispagestyle{empty}

\begin{abstract}
\vspace*{0.5cm}
\section*{Abstract}
Studies of the structure of excited baryons are key to the $N^*$ program
at Jefferson Lab.  Within the first year of data taking with the Hall~B
CLAS12 detector following the 12~GeV upgrade, a dedicated experiment will
aim to extract the $N^*$ electrocouplings at high photon virtualities
$Q^2$.  This experiment will allow exploration of the structure of $N^*$
resonances at the highest photon virtualities ever yet achieved, with a
kinematic reach up to $Q^2 = 12$~GeV$^2$.  This high-$Q^2$ reach will
make it possible to probe the excited nucleon structures at distance scales ranging
from where effective degrees of freedom, such as constituent quarks,
are dominant through the transition to where nearly massless bare-quark
degrees of freedom are relevant.  In this document, we present a detailed
description of the physics that can be addressed through $N^*$ structure
studies in exclusive meson electroproduction.  The discussion includes
recent advances in reaction theory for extracting $N^*$ electrocouplings
from meson electroproduction off protons, along with QCD-based approaches
to the theoretical interpretation of these fundamental quantities.
This program will afford access to the dynamics of the nonperturbative
strong interaction responsible for resonance formation, and will be crucial in understanding the nature of confinement and dynamical chiral
symmetry breaking in baryons, and how excited nucleons emerge from QCD.
\end{abstract}

\maketitle

\clearpage
\tableofcontents

\newpage


\section{The Case for Nucleon Resonance Structure Studies at High Photon Virtualities  \label{Introduction and Motivation}}

\subsection{Background}
\begin{figure}[Ht]
\begin{center}
\includegraphics[clip,height=.55\textheight]{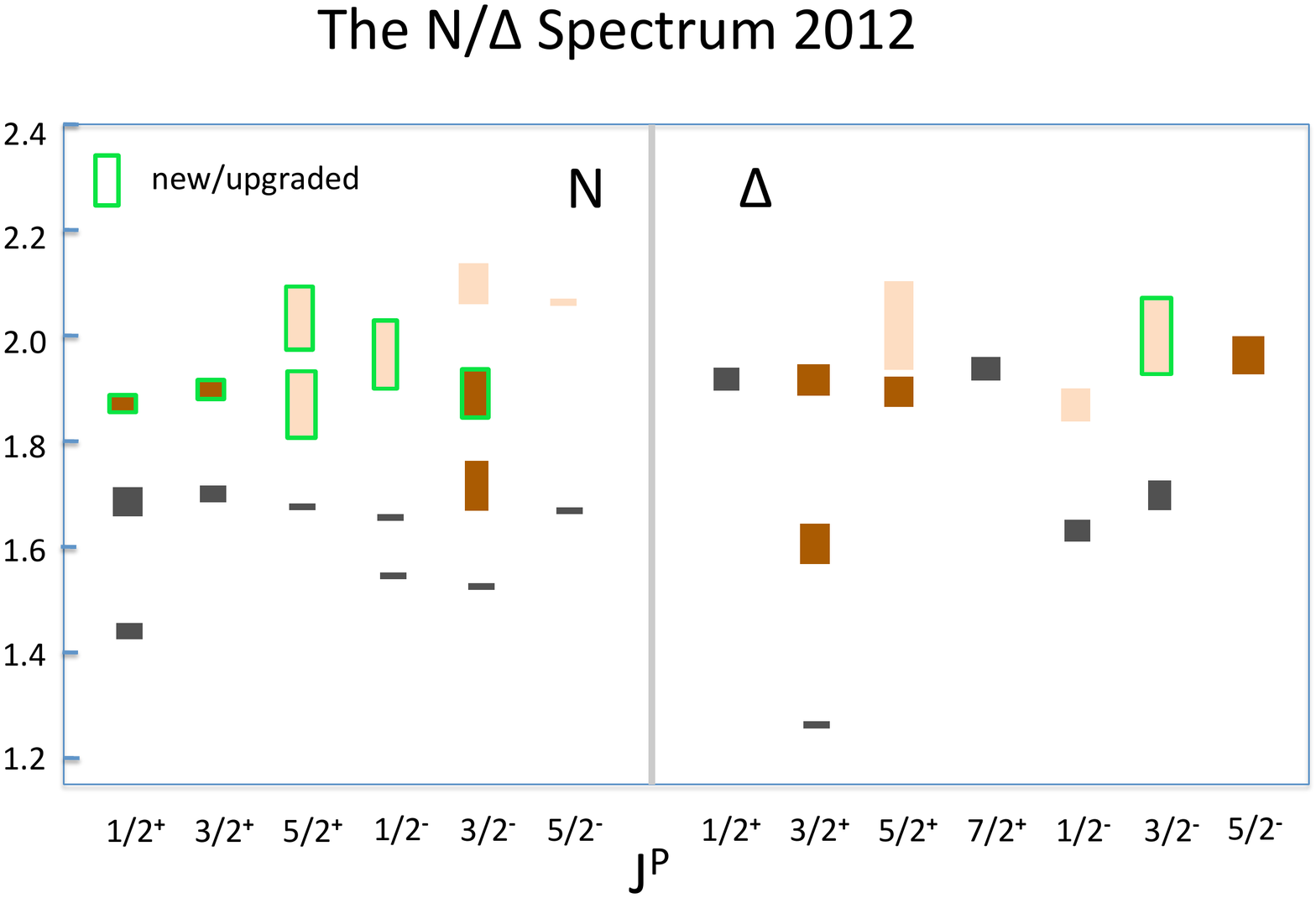}
\caption{(color online) Nucleon resonance spectrum as known in 2012 \cite{Nakamura:2010zzi,Klempt:2012gu,Anisovich:2011fc}. $N^*$ states of spin-parity $J^{P}$ are shown by boxes centred at the Breit-Wigner masses (vertical axis), while the box heights correspond to the $N^*$ mass uncertainties. Resonance state ratings: black, dark brown, and light brown colors stand for four, three, and two star RPP (Review of Particle Physics) status, respectively. Recent
updates from  the Bonn-Gatchina coupled-channel analysis \cite{Anisovich:2011fc} that included the new high precision meson photoproduction data including strangeness channels \cite{Burkert:2012ee} are shown by the boxes framed in green.}
\label{fig:spectrum-1a}
\end{center}
\end{figure}

  It has been known since the 1970s that the nucleon is a bound state
of three valence quarks surrounded by an infinite sea of gluons and
quark-antiquark pairs.  In the underlying theory that describes the
quark-gluon interactions, Quantum Chromodynamics (QCD), the quarks exist
in three states (``colors'') that transform into each other under a
local SU(3) gauge transformation which is an {\it exact\/} local gauge symmetry of
nature.   Only colorless  states are observed as free particles in
nature; they are composed of either (i) quark-antiquark pairs (with
canceling colors) in a sea of gluons (known as mesons), or (ii)  three
valence quarks (each with a different color combining to produce a
colorless state) in a sea of gluons and quark-antiquark pairs (known as
baryons, including the $N^*$ states discussed in this review).   An
immediate implication of this understanding is that when energy is
dumped into the nucleon ground states, they are excited, and can
lose their energies only by emitting colorless states, mostly mesons.  Many
excited states or nucleon resonances, referred to generically as $N^*$s,
have been observed.  The spectrum of these excited states is summarized
in Fig.~\ref{fig:spectrum-1a}. The studies of this resonance spectrum
represent a rapidly evolving area and recent updates can be found in
\cite{Burkert:2012ee,Klempt:2012gu,Anisovich:2011fc}.

QCD has the feature that the elementary couplings between quarks and
gluons and the couplings of gluons to themselves become very small as
the momentum carried by the quarks or gluons becomes very large. This
feature, known as {\it asymptotic freedom\/}, makes it possible to use
perturbation theory to describe precisely the interactions of quarks and
gluons at the distances which are much smaller than the typical hadron
size.  Perturbative QCD (or pQCD for short) has been very successful in
explaining many high energy and/or short distance phenomena, and it is
because of this success that QCD is believed to be the fundamental
theory of the strong interactions. However, at the energies and distance
scales found within the nucleon resonances, the quark-gluon running
coupling becomes large.  One consequence of this large coupling is that
the quarks and gluons are {\it confined\/}; no matter how much energy is
pumped into a nucleon, it is impossible to liberate either a colored
quark or a colored gluon. Another feature of QCD in this region is that
mass is spontaneously generated, and this effect is so large that it
accounts for most (about 98\%) of the mass of the nucleon and 
its excited states.  Neither the mechanism of mass generation nor quark-gluon
confinement can be understood by employing pQCD; a completely new
understanding is needed.  The need to understand QCD in this new,
nonperturbative region is a fundamental problem that the study of $N^*$
states seeks to address.  This review examines various approaches that
have been used to study this problem, describing current successes and
the outlook for future progress.

\begin{figure}[Hb]
\includegraphics[clip,height=.2\textheight]{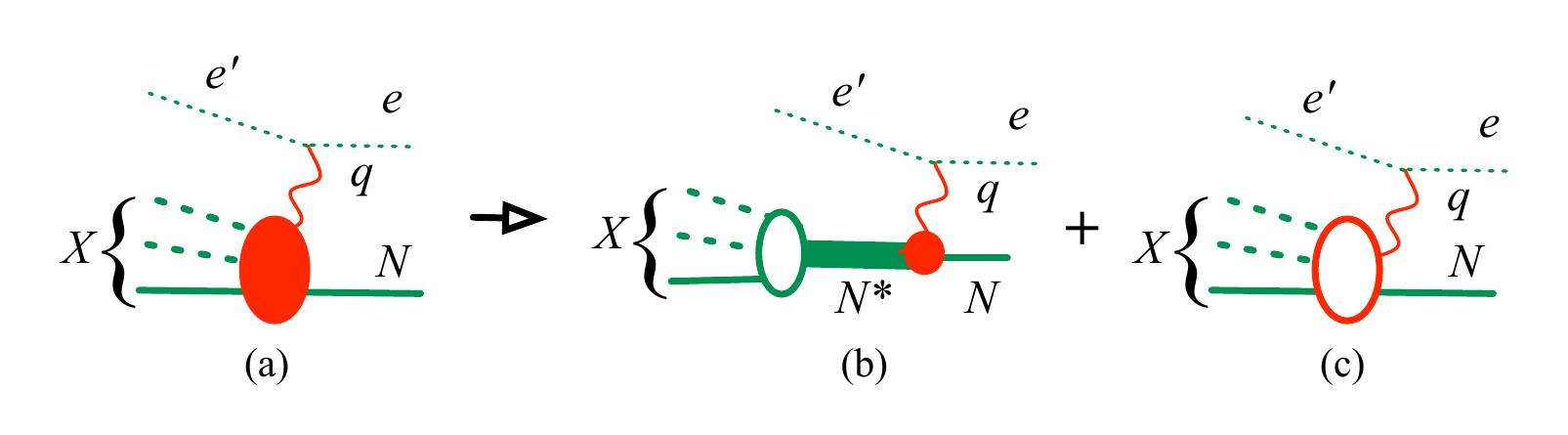}
\caption{Diagram (a) represents the scattering of an electron from a nucleon with the creation of a final state $X$.  It is built from two contributions: (b) the excitation of several $N^*$ intermediate states (only one shown in the figure), and (c) background contributions. Note that time flows from right to left in the diagrams.
}
\label{fig:scattering}
\end{figure}

\subsection{Excitation of nucleon resonances by photons and electrons}

 The states in the excited nucleon spectrum are studied at Jefferson Lab by scattering a beam of electrons or tagged photons from  a nucleon target and observing the hadrons emitted in the final state, as illustrated in Fig.~\ref{fig:scattering}.  Electrons interact with the target by exchanging a virtual photon, with a four-momentum squared $q^2=-Q^2$, and for electron--nucleon scattering one has $Q^2 \ge 0$.  As $Q^2$ increases, the virtual photon becomes capable of resolving the structures at smaller distances   and the scattering is sensitive to the structure at shorter distance scales.  The study of electroproduction of a final state $X$ at high photon virtualities (high $Q^2$)  reveals the inner structure of the resonance.

In May of 2012 the 6-GeV program with the CEBAF Large Acceptance Spectrometer (CLAS) in Hall B at Jefferson Lab was successfully completed.  Among the many data runs with photons and electrons were several dedicated experiments focusing on  the spectroscopy and  structure of $N^*$ states.   CLAS was a unique instrument formed from a set of detectors and designed for the comprehensive exploration of exclusive meson electroproduction off nucleons.  The CLAS detector afforded excellent opportunities to study the electroexcitation of nucleon resonances with great precision, and has contributed the lion's share of the world's data on meson photo- and electroproduction in the resonance excitation region~\cite{Aznauryan:2011ub,Aznauryan:2011qj,Ripani:2002ss,Fedotov:2008aa,Denizli:2007tq,Carman:2009fi,Ambrozewicz:2006zj}.  For the first time detailed information from sets of unpolarized cross sections and different single- and double-polarization asymmetries have become available for many  different meson electroproduction channels off protons and neutrons.

In order to extract resonance information from the electroproduction data it must be analyzed and separated into the two contributions shown in Fig.~\ref{fig:scattering}. The contribution in Fig.~\ref{fig:scattering}(b)  gives information about the  $\gamma_{v}NN^*$ electrocouplings as they evolve with photon virtuality $Q^2$~\cite{Burkert:2012zza,Roberts:2011rr,Dudek:2012ag,Gothe:2011up,Aznauryan:2009da,deTeramond:2012rt,Gothe:clas12}, and how the $N^*$ decays into any final state $X$, $N^*\to X$.    Experimental studies of the structure of all prominent $N^*$ states, in terms of the $\gamma_{v}NN^*$ electrocoupling evolution with $Q^2$, offer promising means of delineating the nature of the strong interaction  in the nonperturbative regime of large quark-gluon running couplings.   These studies are the necessary first steps in understanding how QCD  generates most of the matter or mass in the real world, namely,  mesons, baryons, and atomic nuclei. 

To extract this information from the data, the process in Fig.~\ref{fig:scattering}(b) must be separated from the background contribution in Fig.~\ref{fig:scattering}(c).  This is a major theoretical effort that has been approached in several different ways (see Chapter \ref{Analysis Approaches}).
 The electroexcitation amplitudes for the low-lying resonances $P_{33}(1232)$, $P_{11}(1440)$, $D_{13}(1520)$, and $S_{11}(1535)$ were determined over a wide range of $Q^2$ in independent analyses of  the $\pi^+n$, $\pi^0 p$, $\eta p$, and $\pi^+\pi^-p$ electroproduction channels \cite{Aznauryan:2009mx,Denizli:2007tq,2012sha}. Two of them, the $P_{11}(1440)$ and $D_{13}(1520)$ electrocouplings, have become available through independent analyses of single- and charged-double-pion electroproduction channels. The successful description of the measured observables in these exclusive channels resulting in the same $\gamma_{v}NN^*$ electrocoupling values confirms their reliable extraction from the experimental data. These results have recently been complemented by still preliminary electrocouplings of high-lying resonances with masses above 1.6~GeV~\cite{Aznauryan:2011td} in the $\pi^+\pi^-p$ electroproduction channel, which is particularly sensitive to high-mass resonances.
An alternative resonance electrocoupling extraction in a combined multi-channel analysis  of the $N\pi$, $N\eta$, and $KY$ channels  within the framework of the advanced Excited Baryon Analysis Center (EBAC) Dynamic Coupled Channel (DCC)  approach is in progress \cite{Kamano:2012it,Suzuki:2010yn} (see  Chapter~\ref{EBAC status}).

\subsection{How can we isolate the quark degrees of freedom?}

The nonperturbative strong interaction is enormously challenging. Owing to the quark-gluon interaction, the elementary quarks and gauge-field gluons employed in the QCD Lagrangian are dressed by a cloud of gluons and quark-anti-quark pairs coupled to gluon fields. In the regime of large quark-gluon running coupling this dressing generates effective objects -- dressed quarks and dressed gluons. The dressing of the quarks  gives rise to a momentum-dependent dynamical mass that reflects the structure of the  dressed quarks, and  these dressed {\it constituent\/} quarks become the effective degrees of freedom of normal nuclear matter. In the process of being dressed, the mass of the lightest quarks changes from a few MeV to an effective mass in the range of 250-400~MeV, and the theoretical {\it constituent quark models\/} treat these as real physical particles.  In the evolution of the strong interaction from the pQCD regime of almost point-like and weakly coupled quarks and gluons (distances $< 10^{-17}$~m) to the nonperturbative regime, where dressed quarks and gluons acquire dynamical mass and structure (distances $\approx 10^{-15}$~m), two major nonperturbative phenomena emerge: a) quark-gluon confinement and b) Dynamical Chiral Symmetry Breaking (DCSB). Quark-gluon confinement locks colored quarks and gluons inside the hadron interior. It should be understood that confinement is \textit{not} merely the statement that QCD supports only color singlet asymptotic states. Instead, it is the statement that the QCD spectrum contains only color singlet states of small spatial extent, such that the individual colored constituents cannot be isolated empirically. The dynamical generation of a length scale that characterizes these bound states is a crucial piece in the puzzles posed by confinement and DCSB. It is believed that the DCSB mechanism is responsible for the generation of the large effective quark masses, and both of these new phenomena are completely outside of the  scope of pQCD.   More than 98\% of the hadron mass is generated nonperturbatively through the DCSB processes, while the Higgs mechanism accounts for less than 2\%  of the light-quark baryon masses.  The nonperturbative strong interaction is responsible for the formation of all individual $N^*$ states as  bound systems of quarks and gluons.   The DCSB mechanism will be discussed in more detail in Sec.~\ref{sec:DSE} and in Chapter~\ref{DSE} below.

\begin{figure}[Hb]
\includegraphics[clip,height=.2\textheight]{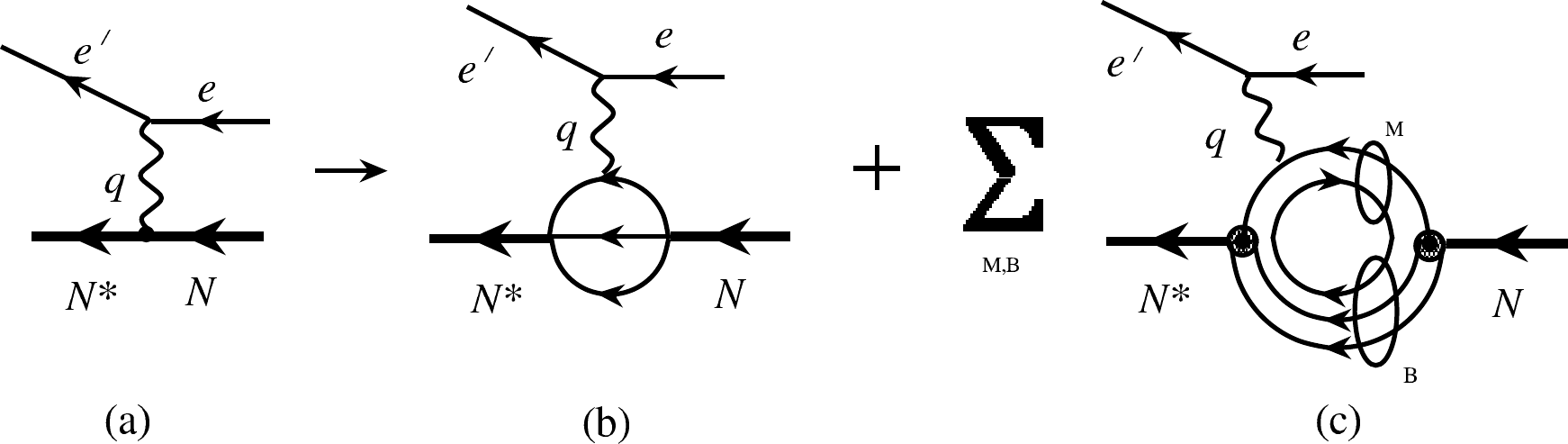}
\caption{ Diagram (a) represents the dressed $\gamma_vNN^*$ electrocoupling that determines the $N^*$-state contribution to the resonant part of meson electroproduction amplitudes. Diagram (b) is the contribution of the three-quark core (with quarks denoted by thin lines). Diagram (c) shows a contribution from the meson-baryon cloud. Here the baryons (B) are represented by their three-quark cores, the mesons (M) by their valence quark-antiquark structure, and the sum is over all meson and baryon states.}
\label{fig:scatteringN*}
\end{figure}

Extraction of the  $\gamma_{v}NN^*$ electrocouplings \cite{Burkert:2012rh,Aznauryan:2011ub,Aznauryan:2011qj,Tiator:2011pw}  gives information on the dressed-quark mass, structure, and nonperturbative interaction, which is critical in exploring the nature of quark/gluon confinement and DCSB in baryons.   Figure~\ref{fig:scatteringN*} illustrates  two of the contributions to these electrocouplings. In Fig.~\ref{fig:scatteringN*}(b) the virtual photon interacts directly with the constituent quark, an interaction which is sensitive to the {\it quark current\/} and depends on the quark-mass function.  This contribution is the quark ``core'' interaction. The general unitarity condition for full meson electroproduction amplitudes 
requires contributions from
non-resonant meson electroproduction and hadronic scattering amplitudes 
to the $\gamma_{v}pN^*$ vertex, as depicted in Fig.~\ref{fig:scatteringN*}(c). This contribution incorporates
all possible intermediate meson-baryon states and subsequent meson-baryon 
scattering processes
that eventually result in the $N^*$ formation.   This contribution is referred to as the meson-baryon cloud (or sometimes simply as the {\it pion\/} cloud since the pion usually makes the largest contribution to
the meson-baryon loops).  These two contributions can be separated from one another using, for example, coupled channel analyses.

\begin{figure}[Ht]
\begin{center}
\includegraphics[width=7.5cm]{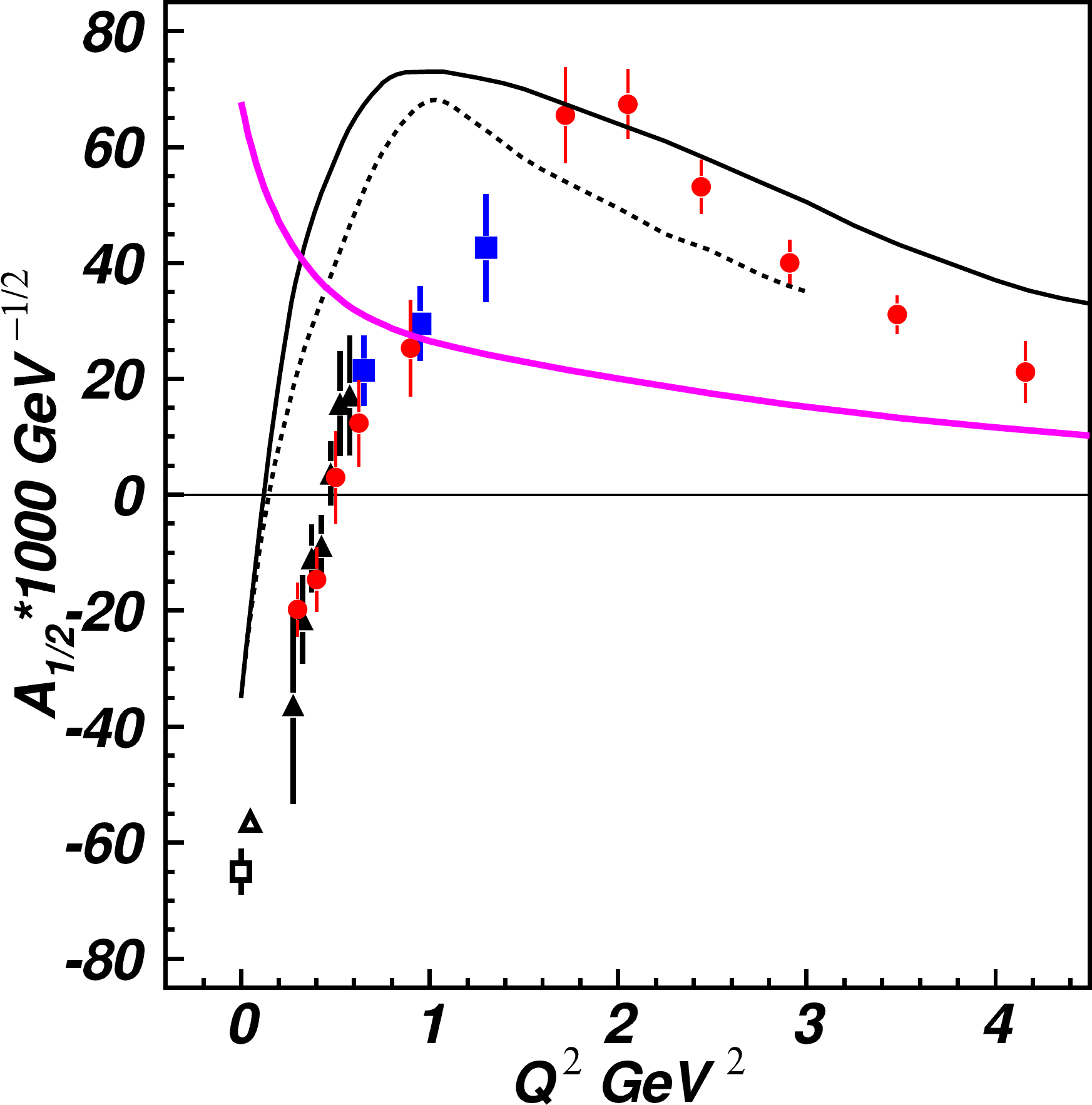}
\includegraphics[width=7.5cm]{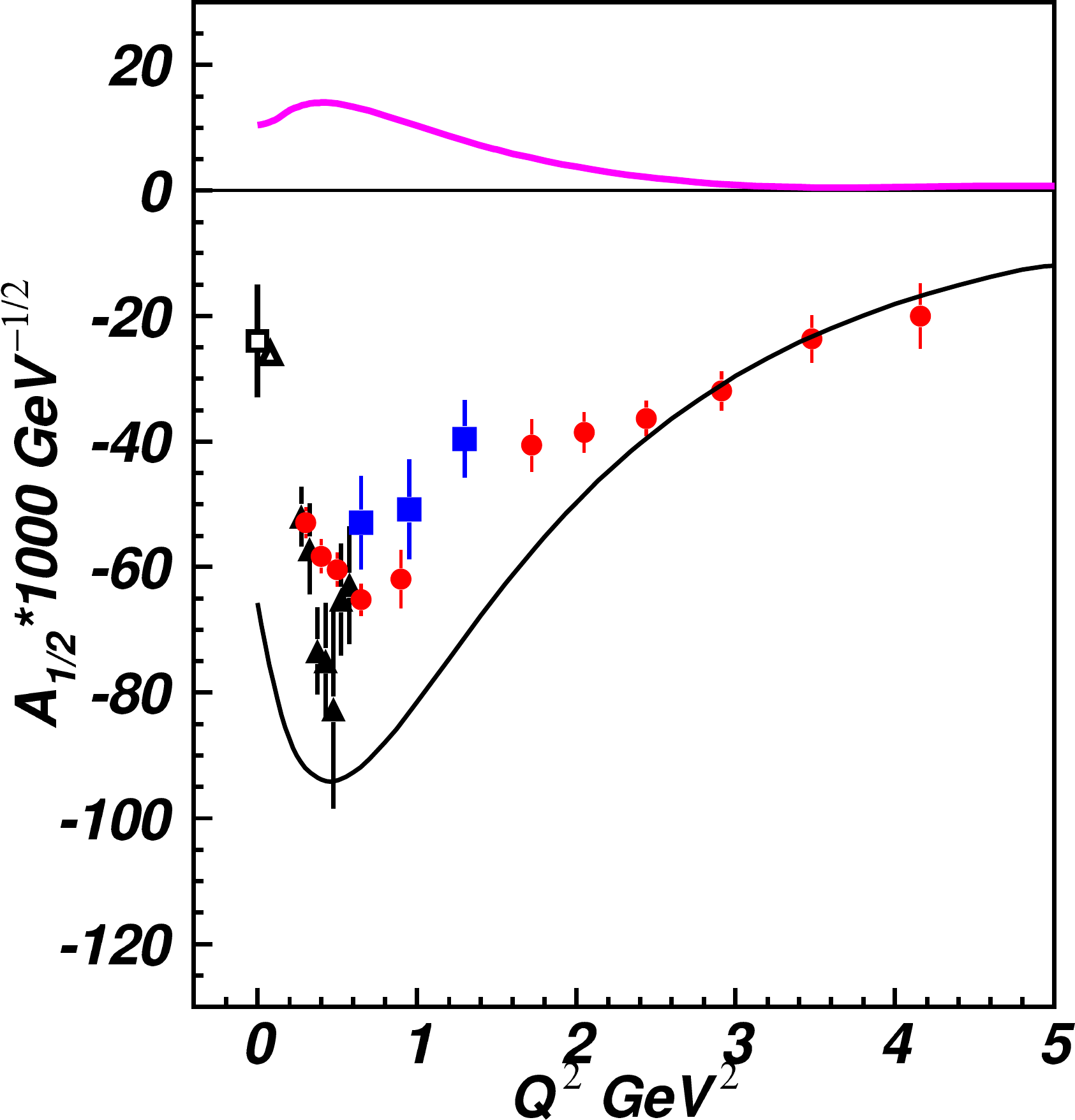}
\caption{\small (Left) The $A_{1/2}$ electrocoupling of the $P_{11}(1440)$ excited state from the analyses of the $N\pi$ electroproduction data \cite{Aznauryan:2009mx} (circles), $\pi^+\pi^-p$ electroproduction data \cite{Fedotov:2008aa} (triangles), and preliminary results from the $\pi^+\pi^-p$ electroproduction data at $Q^2$ from 0.5 to 1.5 GeV$^2$ \cite{Ripani:2002ss} (squares). The photocouplings are taken from RPP \cite{Nakamura:2010zzi} (open square) and the CLAS data analysis \cite{Dugger:2009pn} (open triangle). Predictions from relativistic light-front quark models \cite{Aznauryan:2007ja,Capstick:1994ne} are shown by black solid and dashed lines, respectively. The absolute value of the meson-baryon cloud contribution as determined by the EBAC-DCC coupled-channel analysis \cite{JuliaDiaz:2007fa} is shown by the magenta thick solid line. (Right) The $A_{1/2}$ electrocoupling of the $D_{13}(1520)$ state. The data symbols are the same as in the left panel. The results of the hypercentral constituent quark model \cite{Aiello:1998xq} and the absolute value of meson-baryon dressing amplitude \cite{JuliaDiaz:2007fa} are presented by the black thin and magenta thick solid lines, respectively.}  
\label{p11d13}
\end{center}
\vspace{-3in}\hspace{-1in} ${\bf P_{11}(1440)}$\hspace{3in} ${\bf D_{13}(1520)}$\hspace{-2in}\vspace{3in}
\end{figure}  

It was found that the structure of nucleon resonances in the $Q^2$ range below about 1.5~GeV$^2$ is determined by contributions from both diagrams in Fig.~\ref{fig:scatteringN*}.
As an example, these contributions to the structure of the $P_{11}(1440)$ and $D_{13}(1520)$ states are shown in Fig.~\ref{p11d13}. The absolute values of the meson-baryon dressing amplitudes are maximal for $Q^2 < 1$~GeV$^2$, and decrease with $Q^2$.  In the region of $Q^2 > 1$~GeV$^2$, there is a gradual transition to where the quark degrees of freedom dominate, as is demonstrated by a better description of the $P_{11}(1440)$ and $D_{13}(1520)$ electrocouplings obtained within the framework of quark models.

\begin{figure}[htp]
\begin{center}
\includegraphics[width=9cm]{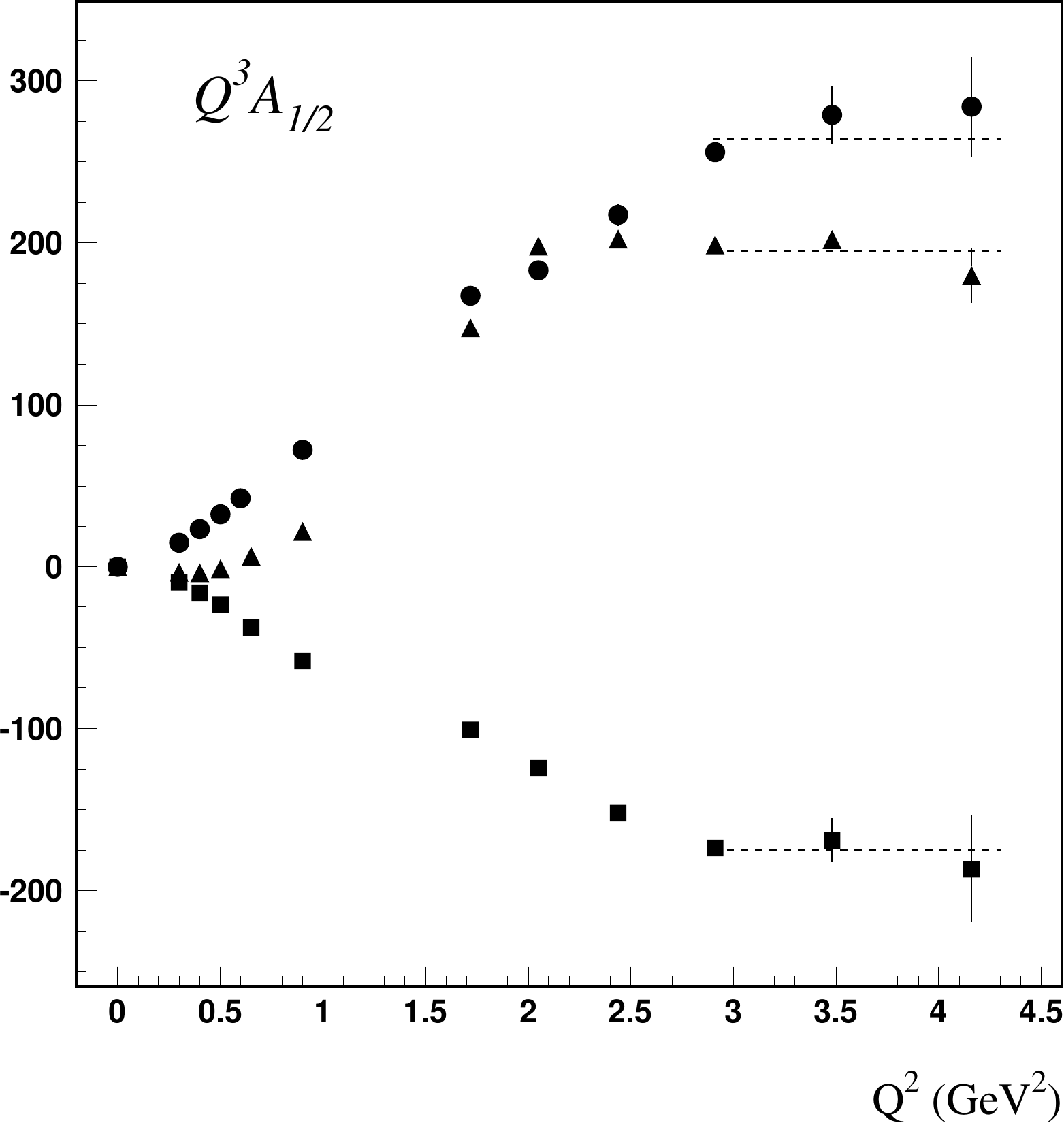}
\caption{\small The $A_{1/2}$ electrocouplings of $P_{11}(1440)$ (triangles), $D_{13}(1520)$ (squares) and $S_{11}(1535)$ (circles)  scaled  with $Q^3$ from the CLAS data analysis \cite{Aznauryan:2009mx}. }  
\label{scaling}
\end{center}
\end{figure} 

\begin{figure}[htp]
\begin{center}
\includegraphics[width=10cm]{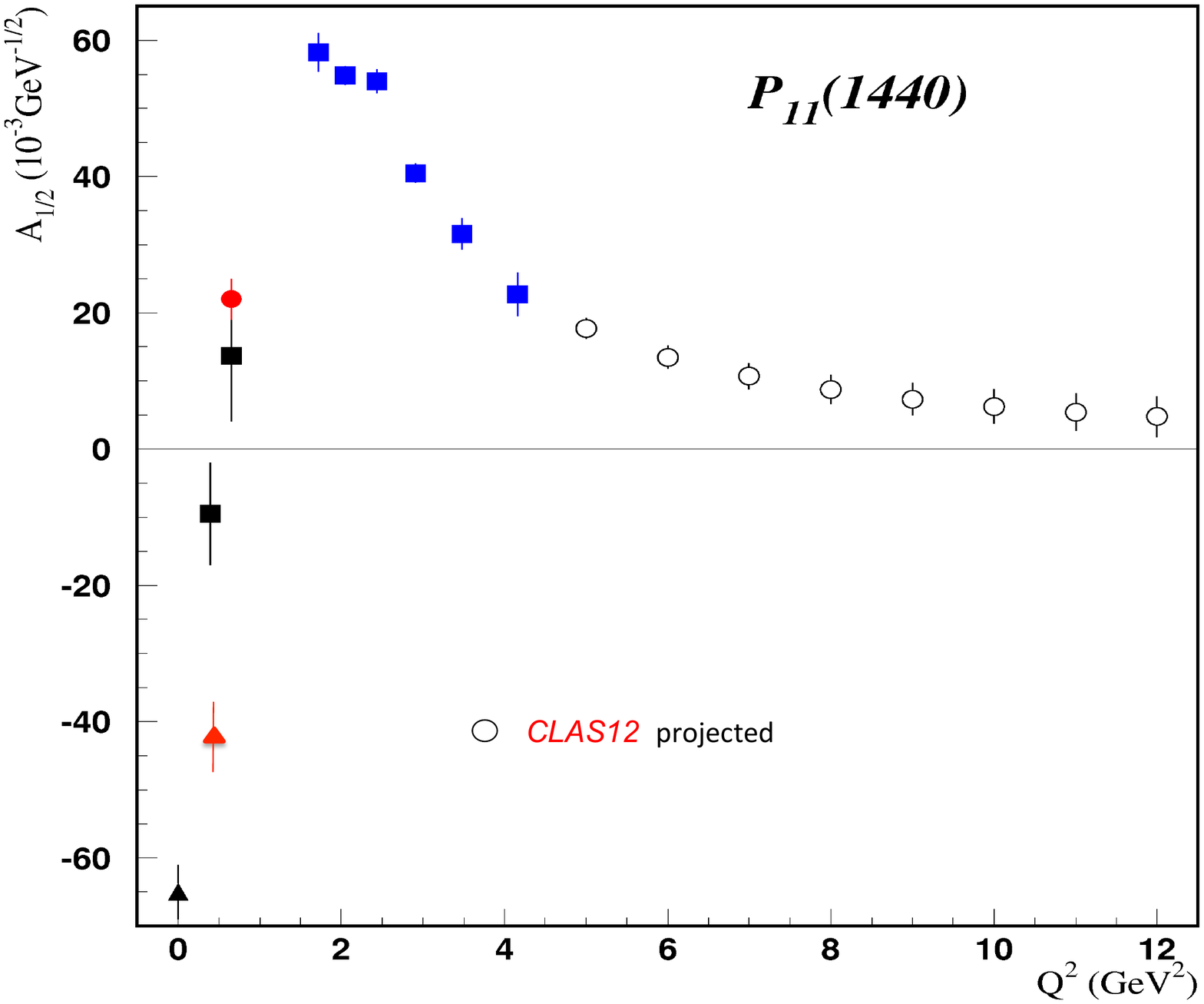}
\caption{\small Available (filled symbols) and projected CLAS12 \cite{Gothe:clas12} (open symbols) $A_{1/2}$ electrocouplings of the $P_{11}(1440)$ excited state.}  
\label{p11_projected}
\end{center}
\end{figure} 

At photon virtualities  of $Q^2 > 5$~GeV$^2$, the quark degrees of freedom are expected to dominate $N^*$ structures \cite{Aznauryan:2009da}.  This expectation is supported by the present analysis of the high-$Q^2$ behavior of the $\gamma_{v}pN^*$ electrocouplings ~\cite{Aznauryan:2009mx} shown in Fig.~\ref{scaling}, where the electrocoupling values scaled with $Q^3$ are plotted.  The $Q^3$ scaling is predicted from constituent counting rules derived from pQCD, and this result implies that the meson cloud contributions are small.  The indicated onset of scaling seen for  $Q^2 > 3$~GeV$^2$ is likely related to the  preferential interaction of the photon with dressed quarks, while interactions with the meson-baryon cloud results in strong deviations from this scaling behavior at smaller photon virtualities (greater distances).

Therefore, in the $\gamma_{v}pN^*$ electrocoupling studies for $Q^2 > 5$~GeV$^2$, the quark degrees of freedom, as revealed by the $N^*$ structure, will be directly accessible from experiment with only small contributions from the meson-baryon cloud.  This will mark the first time that this new and unexplored region in the electroexcitation of nucleon resonances will be investigated.  A dedicated experiment on the $N^*$ studies in exclusive meson electroproduction off protons with the CLAS12 detector (E12-09-003)~\cite{Gothe:clas12} is scheduled  to take place within the first year after the completion of the Jefferson Lab 12-GeV Upgrade Project. By measuring the differential cross sections off protons in the exclusive single-meson and double-pion electroproduction channels, complemented by single- and double-polarization asymmetries in single-meson electroproduction, this experiment seeks to obtain the world's only foreseen data on the electrocouplings of all prominent $N^*$ states in the still unexplored domain of photon virtualities up to 12~GeV$^2$. As an example, the projected $A_{1/2}$ electrocoupling values of $P_{11}(1440)$ at photon virtualities from 5 to 12~GeV$^2$ are shown in Fig.~\ref{p11_projected}. A similar quality of results is expected for the electrocouplings of all other prominent $N^*$ states. The available reaction models for the extraction of the resonance electrocouplings have to be extended toward these high photon virtualities with the goal to reliably extract the $\gamma_{v}pN^*$ electrocouplings from the anticipated data on meson electroproduction off protons. In particular, the new reaction models must account for a gradual transition from meson-baryon to quark degrees of freedom in the non-resonant reaction mechanisms. The current status and prospects of the reaction model developments are discussed in Chapter~\ref{Analysis Approaches}. 

\vspace{0.2in}

\subsection{What can we learn?}

The analysis of the electroproduction data leads to the extraction of the amplitudes for the transition between the initial virtual photon-nucleon and the final $N^*$ state (so-called $\gamma_{v}NN^*$ electrocoupling) that describe the physics, and that can be calculated theoretically.  Among these amplitudes are $A_{1/2}(Q^2)$ and $A_{3/2}(Q^2)$, which describe the resonance electroexcitation for two different helicities of the initial transverse photon and nucleon, as well as $S_{1/2}(Q^2)$ for the description of the resonance electroexcitation by longitudinal photons of zero helicity.  In a case of an  $N^*$ state of 1/2-spin, there is an alternative and equivalent  way of describing resonance electroexcitation in terms of the form factors $F_1^{N^*}(Q^2)$ and $F_2^{N^*}(Q^2)$ that correspond to the helicity-conserving and helicity-flipping parts of the scattering in the infinite momentum frame. The confrontation of the theoretical predictions for these amplitudes with their experimental measurements is the basis for further understanding of nonperturbative QCD.  

A strong collaboration between experimentalists and theorists has been established to achieve the  challenging objectives in pursuing $N^*$ studies at high photon virtualities. Three topical Workshops \cite{EmNN,NStruct,EmNN12} have been organized by Hall B, the Theory Center at Jefferson Lab, and the University of South Carolina to foster these efforts and create opportunities to facilitate and stimulate further growth in this field. This overview is prepared based on the  presentations and discussions at  these dedicated workshops with the goal to develop:
\begin{enumerate}
\item reaction models for the extraction of the $\gamma_{v}pN^*$ electrocouplings from the data on single-meson and double-pion electroproduction off protons at photon virtualities from 5 to 12~GeV$^2$, by incorporating the transition from meson-baryon to quark degrees of freedom into the reaction mechanisms;
\item approaches for the theoretical interpretation of $\gamma_{v}pN^*$ electrocouplings, which are capable of exploring how $N^*$ states are generated nonperturbatively by strong interactions, and how these processes emerge from QCD.
\end{enumerate}

Current theoretical approaches fall into two broad categories: (i) those that tackle the QCD equations of motion directly, and (ii) those that use models inspired by our knowledge of QCD.  It is important to realize that even those approaches that attempt to solve QCD directly can only do so approximately, with further improvements in these calculations expected over time. Furthermore, the demanding theoretical effort that is required might not be undertaken without the data obtained from electroproduction experiments.  It is the interplay between theory and experiment that leads to progress.  Until exact calculations exist, approaches that model QCD have an important role to play.  They can give insight into different aspects of the problem and show us what to expect.

In this review six approaches are discussed.  Those that attempt to solve QCD  directly are (i) Lattice QCD (LQCD) (Chapter~\ref{lqcd1}), (ii)  QCD applications of the Dyson-Schwinger equation (DSEQCD) (Chapter~\ref{DSE}), and (iii) light-cone sum rules (Chapter~\ref{Distribution Amplitudes}).  Those that model various aspects of QCD or employ phenomenological constraints are (i) studies of quark-hadron duality (Chapter~\ref{duality}), (ii) light-front holographic QCD (AdS/QCD) (Chapter~\ref{Light-Front Holographic}), and (iii) constituent quark models (Chapter~\ref{qm}).  Conclusions are given in Chapter~\ref{Conclusion}.

\subsubsection{Lattice QCD (LQCD)}


QCD can be solved numerically on a four-dimensional lattice of space-time points.   In order for the numerical solutions to converge, the physical Minkowski space is mapped into a Euclidean space in which physical time is replaced by imaginary time ($t\to i\tau$).  The accuracy of the solutions obtained from LQCD are limited by the statistical accuracy of the numerical methods (so that, like experimental data, lattice ``data'' comes with error bars), and also by other approximations that derive from the fact that the lattice approximates  continuous and infinite  space-time by a discrete and finite lattice of space-time points.   In addition, when finding  $N^*$ states on the lattice, operators that project an initial ''white'' source onto the quantum numbers of the  $N^*$ are constructed, and as the number of such independent operators is increased, the predictions for the states tends to become more reliable. Finally, current lattice methods require very long times to calculate results for the very small masses of the {\it  up\/} and {\it down\/} quarks from which normal nuclear states are constructed (the masses are usually expressed in terms of the mass of the pion on the lattice, which is typically much larger than the physical mass of about 140 MeV).    In looking at lattice calculations, one must be careful to note the effective pion mass associated with the calculation, and to observe whether or not the calculation   is {\it quenched\/} (approximate calculations which omit quark-antiquark loops) or {\it unquenched\/} (calculations that include such loops and hence correspond to real QCD).  All of these approximations are continually being improved.

Recent advances have shown  the promising potential of LQCD in describing the resonance $\gamma_{v}NN^*$ electrocouplings from QCD.
Proof-of-principle results on the  $Q^2$-evolution of the $F^{P_{11}}_{1}(Q^2)$ and $F^{P_{11}}_{2}(Q^2)$ form factors for the transition from the ground state proton to the excited $P_{11}(1440)$ state have recently become available, employing unquenched LQCD calculations  \cite{Lin:2011da}, see Fig.~\ref{fig:F12pR-dyn} from Chapter~\ref{lqcd1}. The corresponding experimental values of the $F^{P_{11}}_{1}(Q^2)$ and $F^{P_{11}}_{2}(Q^2)$ form factors are computed from the CLAS results \cite{Aznauryan:2009mx} on the $\gamma_{v}pP_{11}(1440)$ electrocouplings. 

Despite the simplified basis of projection operators used in these computations, along with relatively large pion masses of $\approx 400$~MeV, a reasonable description of the experimental data from CLAS \cite{Aznauryan:2009mx,2012sha} was still achieved. In the future, when the LQCD evaluation of $\gamma_{v}NN^*$ electrocouplings will become available, employing a realistic basis of projection operators in a volume of relevant size with the physical pion mass, we will be able to compare with the experimental electrocoupling results for all prominent $N^*$ states. Such comparisons will allow us to test our understanding of how the full complexity of strongly interacting phenomena responsible for generating the ground and excited state hadrons can arise from QCD.


LQCD results \cite{Dudek:2012ag} further predict the contributions for particular configurations in the resonance structure that should couple strongly to   gluons.  These $N^*$ states with very large contributions coming from confined gluons are referred to as {\it baryon hybrids\/} and it is of great interest to find them.  The observation of such hybrids,  with {\it massive constituent\/} ``valence'' gluons, would eliminate the distinction between gluon fields as the sole carriers of the strong interaction and quarks as the sole sources of massive matter.  The proposed search for hybrid $N^*$s \cite{Burkert:2012rh} opens up yet another door  in the $N^*$ program with the CLAS and CLAS12 detectors. Based on the LQCD results \cite{Dudek:2012ag}, the hybrid $N^*$ masses are expected to be heavier than 1.9~GeV. 
Since hybrid $N^*$s have the same quantum numbers as regular $N^*$s, a measurement of the $Q^2$ evolution of the $\gamma_{v}pN^*$  electrocouplings (sensitive to the characteristic $Q^2$ behavior of constituent gluons) will be essential in order to distinguish hybrids from regular states. The high-$Q^2$  regime is of particular interest, since it is in this region where the contribution from quark and gluon degrees of freedom to the $N^*$ structure is expected to dominate.

\subsubsection{Dyson-Schwinger Equation of QCD} \label{sec:DSE}
The Dyson-Schwinger equations of QCD (DSEQCD) represent the tower of coupled integral equations which infers the dressed quark and gluon propagators as well as full vertex for interaction of dressed quarks and dressed gluons from the QCD Lagrangian.  Having these basic ingredients of hadron structure derived from QCD, the hadron spectra, elastic  meson and nucleon form factors and transition $\gamma_{v}NN^*$ electrocouplings  can be evaluated employing Poincare covariant kernel for dressed quark and gluon scattering amplitudes which is responsible for generation of mesons and baryons.  

\begin{figure}[t]
\begin{center}
\includegraphics[width=10cm]{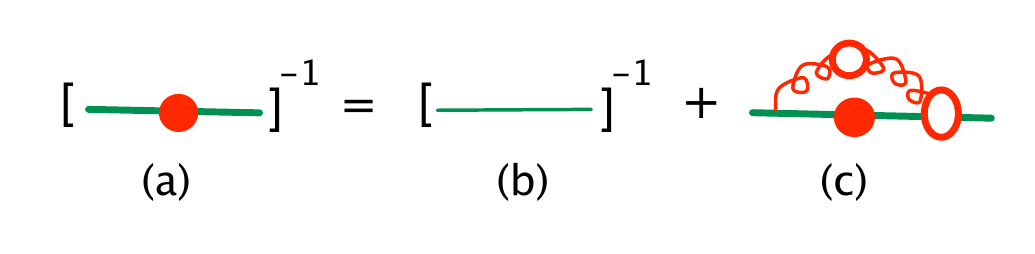}
\caption{\small  Graphical representation of the Dyson-Schwinger equation for the dressed quark propagator.  Diagram (a) is the (inverse) of the dressed quark propagator, (b) is the (inverse) of the bare quark propagator, and (c) is the self energy, with the curly line representing the gluon, and  the open circles representing the dressing of the gluon propagator and the full $gqq$ vertex function.}  
\label{fig:dse}
\end{center}
\end{figure}

The DSEQCD, illustrated in Fig.~\ref{fig:dse}, displays the quark propagator (from which the mass function is extracted) as the sum of the undressed propagator (or the propagator for a {\it current\/} quark) plus the self energy of the quark, as generated by QCD from the interaction of a quark with its own gluon field, which is described by the dressed gluon propagator and dressed quark-gluon vertex.  If nonperturbative QCD is the infinite sum of all quark and gluon interactions, then the Dyson-Schwinger Equation gives the dressed propagator exactly, provided the {\it exact\/} dressed gluon propagator and the {\it exact\/} quark-gluon vertex are inserted into the equation.  In practice, it is necessary make an {\it Ansatz\/}  for quark-gluon vertex employing certain truncation of full $qqg$-vertex. Furthermore, frequently the simplifying parametrizations are also employed in the treatment of dressed gluon propagators.  Agreement of the results obtained from the  two conceptually different frameworks of DSEQCD and LQCD with the experimental measurements of the $\gamma_{v}NN^*$ electrocouplings will offer compelling evidence for both (i) the validity of QCD in the nonperturbative regime, and (ii) the use of the Dyson-Schwinger equations as a practical way to approximate QCD in the strong-coupling region solving the QCD equation of motion in continuous Euclidean space-time.

\begin{figure}[b]
\begin{center}
\begin{overpic}[width=9cm]{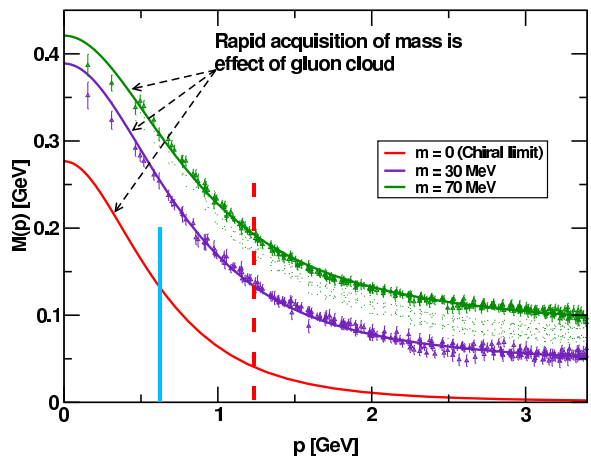}
\put(39.3,42){\color{red}\rule{1.5pt}{5pt}}
\put(39.3,37){\color{red}\rule{1.5pt}{5pt}}
\put(39.3,32){\color{red}\rule{1.5pt}{5pt}}
\put(39.3,27){\color{red}\rule{1.5pt}{5pt}}
\put(39.3,22){\color{red}\rule{1.5pt}{5pt}}
\put(39.3,17){\color{red}\rule{1.5pt}{5pt}}
\put(39.3,12){\color{red}\rule{1.5pt}{5pt}}
\definecolor{lightblue}{rgb}{0.0,0.75,1.0}
\put(25.5,11.7){\color{lightblue}\rule{1.5pt}{66pt}}
\end{overpic}
\caption{\small 
The mass function for a dressed-quark evaluated within the framework of LQCD \cite{Bowman:2005vx} (points with error bars) and DSEQCD \cite{Bhagwat:2006tu,Roberts:2007ji} (solid lines) for two values of bare masses, 70 MeV and 30 MeV, are shown in green and magenta, respectively. The chiral limit of zero bare quark mass, which is close to the bare masses of $u$ and $d$ quarks, is shown in red. Momenta $p < 0.4$~GeV correspond to confinement, while those at $p > $ 2.~GeV correspond to the regime which is close to pQCD. The areas that are accessible for mapping of the dressed quark mass function by the $\gamma_{v}NN^*$ electrocoupling studies with 6~GeV and 11~GeV electron beams are shown to the left of the blue solid and red dashed lines, respectively. 
}  
\label{mass_funct_p11}
\end{center}
\end{figure}

As an example, Fig.~\ref{mass_funct_p11}  shows the dressed-quark masses as a function of momentum running over the quark propagator as calculated by DSEQCD \cite{Bhagwat:2006tu,Roberts:2007ji} and LQCD \cite{Bowman:2005vx} for different values of the bare-quark mass. The sharp increase from the mass of almost undressed current quarks ($p >  2$~GeV) to dressed constituent quarks ($p < 0.4$~GeV) clearly demonstrates that the dominant part of the dressed quark and consequently the  hadron mass {\it in toto} is generated nonperturbatively by the strong interaction. The bulk of the dressed quark mass arises from a cloud of low-momentum gluons attaching themselves to the current-quark in the regime where the running quark-gluon coupling is large, and which is completely inaccessible by pQCD. The region where the dressed quark mass displays the fastest increase represents the transition domain from pQCD ($p > 2$~GeV)  to confinement ($p < 0.4$~GeV). The solution of the DSEQCD gap equation  (illustrated in Fig.~\ref{fig:dse}) \cite{Chen:2012qr} shows that the propagator pole in the  confinement regime leaves the real-momentum axis and  the momentum squared $p^2$ of the dressed quark becomes substantially different from that for the dynamical mass squared $[M(p)]^2$. This means that the dressed quark in the confinement regime will never be on-shell, as is required for a free particle when it propagates through space-time. Dressed quarks have therefore to be strongly bound, locked inside the nucleon, and confinement becomes a property of the dressed quark and gluon propagators. 

These dressing mechanisms are also responsible for  the phenomenon of DCSB.  If the bare quark masses are zero, then QCD is chirally symmetric, meaning that the helicity of a quark (the projection of its spin in the direction of its motion, either $+1/2$ or $-1/2$) is unchanged by the scattering.  This symmetry is dynamically broken by the QCD interactions, so that even if the quark masses were zero the dressed quark mass would be nonzero.  Such a phenomena cannot be generated in pQCD: it is completely nonperturbative.  It can be explained by the DSE  shown in Fig.~\ref{fig:dse}.  This equation can generate a dynamical mass even if the bare quark mass is zero (see also Fig.~\ref{mass_funct_p11} for $m=0$, the chiral limit).   In the physical world, the bare quark masses are only a few MeV, while the dressed quark masses are several hundreds of MeV, showing that  more than 98\% of dressed quark and hadron masses are generated nonperturbatively through DCSB processes.

\begin{figure}
\begin{center}
\includegraphics[width=11cm]{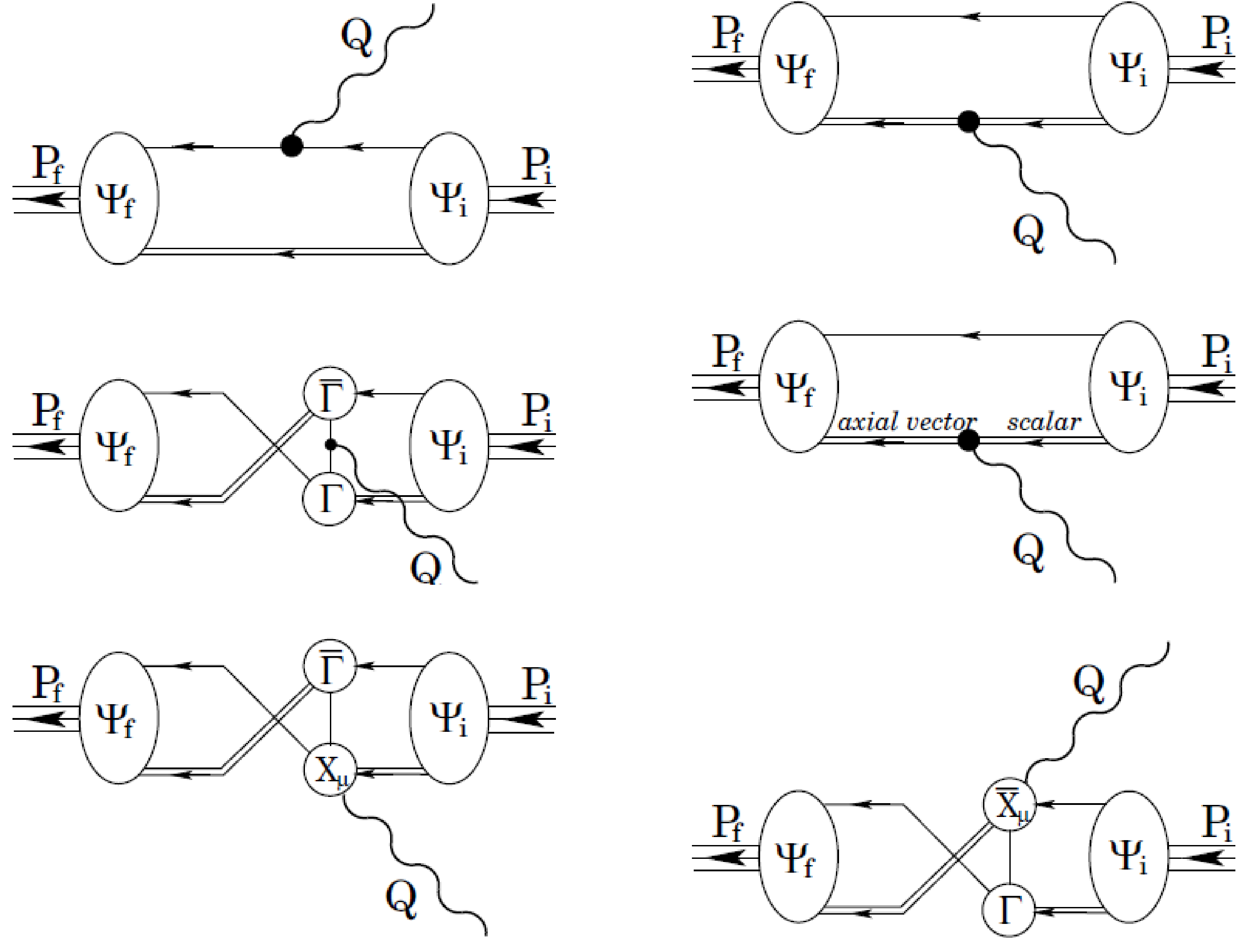}
\caption{\small  The dressed quark interactions for the quark-core contribution to the electromagnetic transition amplitudes ($\gamma_{v}NN^*$ electrocouplings) from the ground nucleon state of four-momentum $P_{i}$ to excited nucleon states of four-momentum $P_{f}$ in the DSEQCD approach \cite{Chen:2012qr}. Solid and double-solid lines stand for dressed quark and the superposition of scalar and axial-vector diquark propagators, respectively. The $\Gamma$ vertices  describe the transition amplitudes between two-quark and diquark states, while the $X$-vertices represent the Schwinger interaction of the virtual photon with the transition current between the diquark and two-quark states. The $\psi_{i}$ and $\psi_{f}$ amplitudes describe the transitions between the intermediate diquark quark states and the initial nucleon and final $N^*$ states, respectively.}  
\label{dse_nst_ex}
\end{center}
\end{figure}

The DSEQCD studies \cite{Roberts:2011rr,Chen:2012qr} suggest  that the quark-core contribution  [illustrated in Fig.~\ref{fig:scatteringN*}(b)] to the electromagnetic transition amplitudes  from the ground to excited nucleon states are determined primarily by the processes as depicted in Fig.~\ref{dse_nst_ex}. The momentum-dependent dressed-quark mass affects all quark propagators and the virtual photon interaction with the dressed-quark electromagnetic currents  affords access to the dynamical quark structure. The 
 interaction of the virtual photon with the transition  currents between diquark and two-quark states elucidates the very details of the strong interaction between and among dressed quarks. The value of momentum  that  is carried away by a single quark can roughly be estimated by assuming equal sharing of the virtual photon momentum among all three dressed quarks. Under this assumption it is reasonably straightforward to see that the measurements of $\gamma_{v}NN^*$ electrocouplings at 5~GeV$^2 < Q^2 < 12$~GeV$^2$ will be able to span nearly the entire range of quark momenta where the transition from confinement to pQCD occurs, as seen Fig.~\ref{mass_funct_p11}. Therefore, the DSEQCD analyses of the $\gamma_{v}NN^*$ electrocouplings of all prominent $N^*$ states expected from CLAS12 will offer a unique way to explore how quark-gluon confinement emerges from QCD. Strong constraints on the dressed quark mass function can be derived from the data on elastic nucleon form factors and from $\gamma_{v}NN^*$  electrocouplings of different excited nucleon states. Consistent results obtained from the studies of elastic form factors and  $\gamma_{v}NN^*$  electrocouplings of all prominent excited proton states are of particular importance for reliable evaluation of this fundamental ingredient of nonperturbative strong interaction from the experimental data.

DSEQCD studies have revealed the sensitivity of $\gamma_{v}NN^*$  electrocouplings to diquark correlations in baryons. This $qq$-pair correlation is generated by a nonperturbative strong interaction, which is responsible for meson formation, and  can be described by the finite sizes of quasi-particles formed of paired quarks.   In the DSEQCD approach, depending on the parity of the $N^*$ state, the two-quark assembly is described by either a superposition of scalar and axial-vector diquarks, or a superposition of pseudoscalar and vector diquarks.   It turns out that the relative contributions of the possible diquark components strongly depend on the quantum numbers of the $N^*$ state. Furthermore, the amplitudes shown in Fig.~\ref{dse_nst_ex}, which describe the transitions between the intermediate diquark-quark state and  the initial ground $\psi_{i}$ or the final $N^*$ state $\psi_{f}$, are strongly dependent on the quantum numbers of both the initial nucleon and the final $N^*$ states.  The information on the electrocouplings of as many $N^*$ states as possible is needed for fully separating  and identifying the mechanisms in  the electroexcitation of nucleon resonances.  

\subsubsection{Light-cone sum rules (LCSR)}

A particularly successful way to analyze the structure of a bound system of quarks is to look at the system in the {\it infinite momentum frame\/}, where the momentum of the system, $P_z$  (assumed to be in the $\hat z$ direction), is very large.  Since any successful theory must  be  Poincare\'e invariant, the physics cannot depend on the frame in which the interaction is observed, but {\it approximations\/} that may be valid in one frame may not be easy to implement or even understand in another frame.   For example, in the  infinite-momentum frame it is natural to assume that the $\hat z$ component of the  momenta of the constituents particles, $(k_i)_z$, are also large (because they must be carried along with the bound state), and to ignore (at least as a first approximation) the transverse components $({\bf k}_i)_\perp$.  In this case the $\hat z$ component of the momenta of each constituent quark is written in terms of its momentum fraction $x_i$, so that $(k_i)_z=x_iP_z$.  As $P_z\to\infty$, the mass of the particle becomes  less and less important, and  the particle moves closer and closer to the {\it light-cone\/} swept out by a light beam.  For this reason, analyses in the infinite momentum  frame are described in terms of light-cone coordinates, and it turns out that this becomes a completely general, frame independent formalism which has many theoretical advantages, as well as some non-intuitive properties.

As reviewed briefly in Chapter~\ref{Distribution Amplitudes}, the LCSR technique provides a way to relate the $\gamma_{v}NN^*$  electrocoupling amplitudes to integrals over matrix elements involving {\it distribution amplitudes\/} (DA) that depend only on the $x_i$ (and are related to the bound state wave function integrated over the transverse momentum components).  The DA can be expanded in a series of terms which can be calculated with the help of LQCD.  This series has the nice feature that the first few few terms dominate the physics at $Q^2\sim 2-10$~GeV$^2$, and that at very high $Q^2$ the results of pQCD emerge automatically.  So far this method has been used to make predictions about the nucleon elastic form factor ($\gamma_{v}NN$) and $N^*(1535)$ electroproduction ($\gamma_{v}NN^*$).

\subsubsection{Duality}

All of the methods discussed so far have used quarks and gluons (either
bare or dressed) to describe the physics.  This is a natural choice in
that these are the degrees of freedom that appear in the QCD Lagrangian.
However, the physics can also be discussed directly in terms of the
colorless degrees of freedom one observes in nature: nucleons, $N^*$s, and mesons.  Many years ago it was observed that the total cross section
for electrodisintegration of a nucleon,
		$\gamma_{v} + N \to$~{\it anything\/},
which depends on two variables, $Q^2$ and $\nu = P\cdot q/M$
(where $P$ is the four-momentum of the nucleon and $M$ its mass),
shows non-trivial resonance structure, which appears as ridges at
the values of the resonance masses $M^{*2}=(P+q)^2=M^2+2M\nu-Q^2$.
As $Q^2$ increases, these ridges were found to follow a smooth function
which describes the high-energy behavior of the cross section,
extrapolated into the resonance region.  This function is sometimes
referred to as the {\it scaling\/} function, since it depends only
on the variable $x_B=Q^2/(2M\nu)$, and is approximately independent
of the scale $Q^2$.  It is related to the distribution amplitudes
mentioned above, and hence can be understood as a consequence of
the underlying quark degrees of freedom in the target nucleon.
At asymptotically large $Q^2$, in a given interval of $x_B$ 
the cross section is dominated by an
ever increasing number of overlapping resonances, whose collective
behavior is given by the scaling function.

At modest $Q^2$, however, where the resonances appear as distinct
peaks in the cross section (when plotted as a function of $x_B$ for
fixed $Q^2$), it has been discovered that the {\it average\/} of the
cross sections over an interval of $x_B$ agrees with the average of
the scaling function over the same interval.  This feature, know as
quark-hadron {\it duality\/}, suggests that the average behavior of
the resonances is the same as the average behavior of the quarks,
establishing a connection between hadronic degrees of freedom and
quark degrees of freedom.  As discussed in Chapter~\ref{duality},
there are indications that duality also works for exclusive processes,
$\gamma_v+N\to X_i$ (where $X_i$ is a particular final state, such as
$X_i=N+\pi$, for example).  These connections can provide constraints
on the exclusive $N^*$ production channels that can relate them to
the underlying quark degrees of freedom.

\subsubsection{Light-front holographic QCD (AdS/QCD)}

A new method for the study of hadron physics is  derived from the idea of mapping QCD, defined using light-front coordinates in four-dimensional Minkowski space-time, onto  a five-dimensional anti-de Sitter space (AdS) familiar from studies of gravity.  This is {\it holographic\/} in the sense that it defines a correspondence between theories in a different number of space-time dimensions, much as a two-dimensional film can be made to produce a three-dimensional image.   
AdS/QCD has led to new insights into the color-confining, nonperturbative dynamics and the internal structure of relativistic light-hadron bound states.  The formalism is relativistic and frame-independent. Hadrons are described as eigenstates of a light-front Hamiltonian with a specific color-confining potential. A single parameter $\kappa$ sets the mass scale of the hadrons.  The hadronic spectroscopy of the light-front holographic model gives a good description of the masses of  the observed light-quark mesons and baryons. 
Elastic and transition form factors are computed from the overlap of the light-front wave functions. Many predictions of light-front holography for baryon resonances can be tested at the 12-GeV Jefferson Lab facility.   
An outline of this new method is given in Chapter~\ref{Light-Front Holographic}.

\subsubsection{Constituent quark models}

{\it Constituent quark models\/} offer an alternative to the approaches discussed above.  These models postulate that constituent (dressed) quarks are physical particles with a definite mass, and usually replace the difficult nonperturbative interactions between dressed quarks by  a potential  that  includes a confining term that will not allow two quarks to be separated.
The simplicity of these models makes the calculation of the resonance electrocouplings a
comparatively straightforward problem. They are currently the only available phenomenological tool
to investigate the electrocouplings over the full range of the baryon spectrum and the physical insight they provide make them very attractive.  While some aspects of constituent quark models can be justified
from QCD in the heavy quark limit, the challenge
is to employ the information available from observable analyses within the quark models in order to understand the underlying physics of light quarks
  described by QCD.  

Different approaches to the physical analyses of $\gamma_{v}NN^*$ electrocouplings at high photon virtualities within the framework of constituent quark models is discussed in Chapter~\ref{qm}.  In some of these the covariant wave functions of an $N^*$, which normally would be calculated from a covariant wave equation using a QCD inspired kernel (the covariant version of a potential), are modeled directly.  While this method simplifies the problem by bypassing the need to model the kernel and solve the equation, the consequence is that the masses of the resonances cannot be predicted and the wave functions of different resonances are not derived from a common dynamics.  Nevertheless, these calculations do provide a simple prediction for the quark core interactions, depicted in Fig.~\ref{fig:scatteringN*}(b), and provide useful insights into the $Q^2$ behavior of these core diagrams.   Approaches of this type are described in Secs.~\ref{sec:CST} and \ref{sec:LFquark}.   One of these uses the covariant spectator theory, the other light-front dynamics.

Section \ref{sec:cqm2} summarizes a number of dynamical models that calculate the three-quark wave functions from model potentials.  Most of these models assume a linear confining interaction and a short range one-gluon-exchange interaction with a Coulomb-like structure.  The three-quark equations are solved using variational methods, and the parameters in the potentials fixed to the observed spectrum of $N^*$ states.  Electromagnetic transitions between the states are then calculated. 


\subsection{Reflections}
The studies of the electromagnetic transition amplitudes between the nucleon ground and its excited states in a wide range of photon virtualities elucidate relevant degrees of freedom in the $N^*$ structure at different distances and eventually will allow us to access the complexity of the nonperturbative strong interaction, which is responsible for the formation of various resonance states with different quantum numbers.
With the 12-GeV upgrade of the Jefferson Lab accelerator and the new CLAS12 detector, we have a unique opportunity to study the structure of the nucleon resonances at high $Q^2$ beginning the first year of running after completion of the Jefferson Lab 12-GeV Upgrade Project. These studies will address some of the most fundamental issues of present-day hadron physics:

\begin{enumerate}
\item How does nature achieve confinement?
\item How is confinement tied into dynamical chiral symmetry breaking, which describes the origin of  most of the 
visible mass in the universe?
\item Can the fundamental QCD Lagrangian successfully describe the complex structure of all the $N^*$ states?
\end{enumerate}
These questions are very difficult to answer.  As discussed above, study of the the behavior of the resonances at high $Q^2$ provides a decisive advantage; in this region the meson cloud contributions become small and the quark core contributions we seek to study are exposed.   With the massive $N^*$ states that can be excited at high $Q^2$ it is also possible that we will see the widely anticipated, but not yet established, hybrid baryons in which massive confined gluons play a role equal to that of valence quarks.   All of these efforts will be greatly enhanced by the new CLAS12 detector which will provide simultaneous data for a wide range of $N^*$ states from major exclusive meson electroproduction channels, which theorists can use to untangle the competing contributions  that enter each state in a different way.

A coordinated effort between experimental and theoretical physicists is required.  This review summarizes  where we are today and can be used, we hope, to bring other interested scientists up-to-date  so that they may join this exciting effort.   For further conclusions, see Chapter \ref{Conclusion}.

\section{Analysis Approaches for Evaluation of Nucleon Resonance
Electrocouplings from the CLAS Data: Status and Prospects \label{Analysis Approaches}}

\subsection{Introduction}
 The CLAS detector at  Jefferson Lab is a unique instrument, which has provided the lion's share of the world's data on meson photo- and electroproduction in the resonance excitation region. Cross sections and polarization asymmetries collected with the CLAS detector have made it  possible for the first time to determine electrocouplings of all prominent $N^*$ states over a wide range of photon virtualities (\mbox{$Q^2 < 5.0$~GeV$^2$}) allowing for a comprehensive analysis of exclusive single-meson ( $\pi^+n$, $\pi^0p$, $\eta p$, and $KY$) reactions in electroproduction off protons.  Furthermore, CLAS was able to precisely measure $\pi^+\pi^-p$ electroproduction differential cross sections owing to the nearly full kinematic coverage of the detector for charged particles.

With the advent of the future CLAS12 detector,   the $Q^{2}$ reach  will be considerably extended  for exploring the nature of confinement and Dynamical Chiral Symmetry Breaking in baryons for our  $N^*$ structure studies.   Indeed, the CLAS12 detector will be the sole instrument worldwide that will allow for performing experiments to determine the  $\gamma_{v}NN^*$ electrocouplings of prominent excited proton states as listed in the Table~\ref{bf1pi2pi}.  These will be the highest photon virtualities yet achieved for $N^*$ studies with photon virtualities in the range between 5.0  and 10.0  to 12.0~GeV$^2$, where the upper $Q^{2}$ boundary  depends on the mass of excited proton state.  The primary objective of the dedicated experiment E12-09-003, {\it Nucleon Resonance Studies with CLAS12}~\cite{Gothe:clas12},  is to determine the  $\gamma_{v}NN^*$ electro\-couplings from the  exclusive-meson electroproduction reactions,  $\pi^+n$, $\pi^0p$, and $\pi^+\pi^-p$,  off protons.  The CLAS12 experiment E12-09-003~\cite{Gothe:clas12} was approved for 40 days of running time and it is scheduled to start in the first year of running with the CLAS12 detector, that is right after the Hall~B 12-GeV upgrade.  This experiment represents the next step towards extending our current $N^*$ Program with the CLAS detector~\cite{Aznauryan:2011ub,Aznauryan:2011qj}; it will make use of the 11-GeV continuous electron beam that will be delivered to Hall~B of Jefferson Lab.   We remark that the maximum energy of the electron beam to Hall B will be 11~GeV for the 12~GeV upgrade to JLab.

The data from the $\pi^+n$,
$\pi^0p$, and $\pi^+\pi^-p$ electroproduction channels will play a key role in the
evaluation of $\gamma_{v}NN^*$ electrocouplings. The primary $N\pi$ and $\pi^+\pi^-p$ exclusive channels, combined, account for approximately 90\%  of the total cross section for meson electroproduction in the resonance excitation region of $W < 2.0$~GeV. Both single- and
charged-double-pion electroproduction channels are sensitive to $N^*$ contributions, as can be seen in
Table~\ref{bf1pi2pi}.  Further, these two channels offer complementary  information  for cross checking the $N^*$ parameters derived in the fits to the observables in the exclusive channel.

\begin{table}
\begin{center}
\begin{tabular}{|c|c|c|c|c|}
\hline
$N^{*},\Delta^{*}$ & Branching fraction & Branching fraction &
Prominent\, in  & Prominent \, in the \\
 states &  $N\pi$ \, [\%]  &  $N\pi\pi$ \, [\%]  &
$N\pi$ \, exclusive   & $\pi^+\pi^-p$ \, exclusive\,   \\
 & & & channels & channel \\
\hline
$P_{33}(1232)$ & 100 & 0 & * & \, \, \\
$P_{11}(1440)$ & 60 & 40 & * & * \,\\
$D_{13}(1520)$ & 60 & 40 & * & * \, \\
$S_{11}(1535)$ & 45 & $<$ 10 & * & \, \, \\
$S_{31}(1620)$ & $<$ 25 & 75 & \, \, & * \\
$S_{11}(1650)$ & 75 & $<$ 15 & * & \, \, \\
$F_{15}(1680)$ & 65 & 35 & * & * \, \\
$D_{33}(1700)$ & $<$ 15 & 85 & \, \, & * \, \\
$P_{13}(1720)$ & $<$ 15 & $>$70 & \, \, & * \, \\
$F_{35}(1905)$ & $<$ 10 & 90 & \, \, & * \, \\
$F_{37}(1950)$ &  40 & $>$25 & * & * \, \\
\hline
\end{tabular}
\caption{\label{bf1pi2pi}  $N\pi$ and $N\pi\pi$ branching fractions for decays of excited
proton states that have prominent contributions to
the exclusive single- and/or charged-double-pion electroproduction channels. The values are taken from \cite{Nakamura:2010zzi} or from the CLAS data
analyses \cite{Aznauryan:2009mx,Aznauryan:2011td}. The asterisks * mark most suitable exclusive channel(s) for the studies of a particular 
$N^*$ state.}
\end{center}
\end{table}

 A necessary first step towards extracting the $\gamma_{v}NN^*$ electrocouplings, in a  robust way, requires that we employ  independent analyses of  the single- and charged-double-pion electroproduction data within the framework of several different phenomenological  reaction models. A reliable separation of resonant and non-resonant contributions, moreover, is crucial 
for evaluating the $\gamma_{v}NN^*$ electrocouplings within each of the frameworks provided by these approaches.  Independent analyses of major meson electroproduction channels offer a sensitive test as well as a  quality check in separating the contributions of resonant and non-resonant mechanisms.   Consistent extractions of  the $N^*$ electrocoupings among different channels are imperative, as well as $Q^2$-independent values of their hadronic decays and masses.
The $\pi^{+}n$, $\pi^0p$, and $\pi^+\pi^-p$ exclusive electroproduction channels, for example,  
have entirely different non-resonant mechanisms.  Clearly, the value of the $\gamma_{v}NN^*$ electrocouplings must be analysis independent and must remain the same in all exclusive channels, since resonance electroexcitation amplitudes do not depend on the final states that will populate the different exclusive reaction channels.  Therefore, consistency in ascertaining the values of the $\gamma_{v}NN^*$ electrocouplings from a large body of observables, as measured in  $\pi^{+}n$, $\pi^0p$, and $\pi^+\pi^-p$ electroproduction reactions, will give good measure of the reliability of  the extraction of these fundamental quantities.
In the next section we shall review the current status and prospects for developing several phenomenological reaction models
with the primary objective of determining $\gamma_{v}NN^*$ electrocouplings from independent analyses of 
the single- and charge-double-pion electroproduction data ($\pi^{+}n$, $\pi^0p$, and $\pi^+\pi^-p$).

\subsection{Approaches for independent analyses of the CLAS data on single- and charged-double-pion
electroproduction off protons  \label{tools}}

Several phenomenological reaction models \cite{Aznauryan:2009mx,Aznauryan:2002gd,Aznauryan:2004jd,Aznauryan:2011td,Mokeev:2008iw,2012sha} were developed by the CLAS Collaboration for evaluating the $\gamma_{v}NN^*$ electrocouplings in
independent analyses of the data on $\pi^+n$, $\pi^0p$, and $\pi^+\pi^-p$ electroproduction off protons. These models were 
successfully applied to single-pion electroproduction for  $Q^2  < 5.0$~GeV$^2$ and $W < 1.7$~GeV.  The $\gamma_{v}NN^*$ electrocouplings were extracted from the CLAS $\pi^+\pi^-p$ electroproduction data for the kinematical ranges of $Q^2 < 1.5$~GeV$^2$ and 
$W  < 1.8$~GeV  \cite{Burkert:2012rh,Aznauryan:2011ub,Aznauryan:2011td}. The CLAS data on single-pion exclusive electroproduction were also analyzed within the framework of the MAID \cite{Tiator:2011pw} and the SAID \cite{Arndt:1985vj,Arndt:2006bf} approaches. These reaction models have allowed us to access resonant amplitudes by fitting all available
observables in each channel independently and within the framework of different reaction models. 
Consequently, the $\gamma_{v}NN^*$ electrocouplings, along with the respective $N\pi$ and $N\pi\pi$ hadronic decay widths,
were determined by employing a Breit-Wigner parameterization of the resonant amplitudes. 

\subsubsection{CLAS collaboration approaches for resonance electrocoupling extraction from the data on single-pion electroproduction off protons \label{1pisec}}

Analyses of the rich CLAS data samples  have extended our knowledge considerably of single-meson electroproduction reactions off protons, i.e.~the $\pi^{+}n$ and $\pi^0p$ channels. 
A total of nearly 120,000 data points have been collected on  reactions arising from unpolarized differential cross sections, longitudinally-polarized beam asymmetries as well as from
longitudinal-target and beam-target asymmetries with a nearly complete coverage in the phase space for exclusive reactions~\cite{Aznauryan:2009mx}.  The data were analyzed within the framework of two conceptually
different approaches, namely:  a) the unitary isobar model (UIM) and b) a model employing dispersion relations
\cite{Aznauryan:2002gd,Aznauryan:2004jd}. All well-established $N^*$ states in the mass range of $M_{N^*} <1.8$~GeV were 
incorporated into the $N\pi$ channel analyses.

The UIM follows the approach detailed in Ref.~\cite{Tiator:2011pw}. The $N\pi$ electroproduction amplitudes are described as a superposition of $N^*$ electroexcitations in the $s$-channel and non-resonant Born terms. A Breit-Wigner ansatz, with  energy-dependent hadronic decay widths \cite{Walker:1968xu}, is employed  for the resonant amplitudes. Non-resonant amplitudes are described by a gauge-invariant superposition of nucleon $s$- and $u$-channel exchanges and in the $t$-channel by  $\pi$, $\rho$, and $\omega$ exchanges. The latter are reggeized; this allows for a better description of the data in the second- and the third-resonance regions, whereas for $W < 1.4$~GeV, the role of Regge-trajectory exchanges becomes insignificant. The Regge-pole amplitudes were constructed using the prescription delineated in Refs.~\cite{Guidal:1997hy,Guidal:1997by} allowing us to preserve gauge invariance of the non-resonant amplitudes.
 The final-state interactions are treated as $\pi N$ rescattering in the K-matrix approximation~\cite{Aznauryan:2002gd}.

In another approach, dispersion relations relate the real and imaginary parts of the invariant amplitudes, which describe  $N\pi$ electroproduction  in a model-independent way~\cite{Aznauryan:2002gd}.  For the 18 independent invariant amplitudes describing the
$\gamma_{v} N \rightarrow N \pi$ transition electromagnetic current, dispersion relations (17 unsubtracted and 1 subtracted) at fixed $t$ are employed.  This analysis has shown that the imaginary parts of amplitudes are dominated by resonant contributions  for $W > 1.3$~GeV.  That is to say,  in this kinematical region, they are described solely by resonant contributions.  For smaller $W$ values, both the resonant and non-resonant contributions to the imaginary part of the amplitudes are taken into  account based on an analysis of  $\pi N$ elastic scattering and by making use of the Watson theorem and the requisite dispersion relations.

For either of these approaches, the $Q^2$ evolution of the non-resonant amplitudes is determined by the behavior of the hadron electromagnetic form factors, which are probed at different photon virtualities.  The $s$- and $u$-channel nucleon exchange amplitudes depend on the proton and neutron electromagnetic form factors, respectively.  The $t$-channel  $\pi$, $\rho$, $\omega$ exchanges  depend on pion electromagnetic form factors and  $\rho (\omega)$ $\rightarrow$ $\pi\gamma$ transition form factors. The exact parameterization of these electromagnetic form factors as a function of $Q^2$  that are employed in the analyses of the CLAS single-pion electroproduction data can be found in Ref.~\cite{Aznauryan:2009mx}. These analyses have demonstrated that for photon virtualities  $Q^2 > 0.9$~GeV$^2$, the reggeization of the Born amplitudes becomes insignificant in the resonance region for  $W < 1.9$~GeV. Consequently, at these photon virtualities,
the background of UIM was constructed  from the nucleon exchanges in the $s$- and $u$-channels and in the $t$-channel
through $\pi$, $\rho$ and $\omega$ exchanges. In the approach based on dispersion relations, we additionally take into account the
$Q^2$ evolution of the subtraction function $f_{\rm sub}(Q^2,t)$.  The subtraction function was determined using a linear parameterization over the
Mandelstam variable $t$  and through fitting two parameters to the data in each bin of $Q^2$ \cite{Aznauryan:2009mx}. Employing information on the $Q^2$ evolution of hadron electromagnetic form factors from other experiments or from our fits to the CLAS data, we are able to predict the $Q^2$ evolution of the non-resonant contributions to single-pion electroproduction in the region of $Q^2$, where  the meson-baryon degrees of freedom remain relevant.

\begin{figure*}[htp]
\begin{center}
\includegraphics[width=12.0cm]{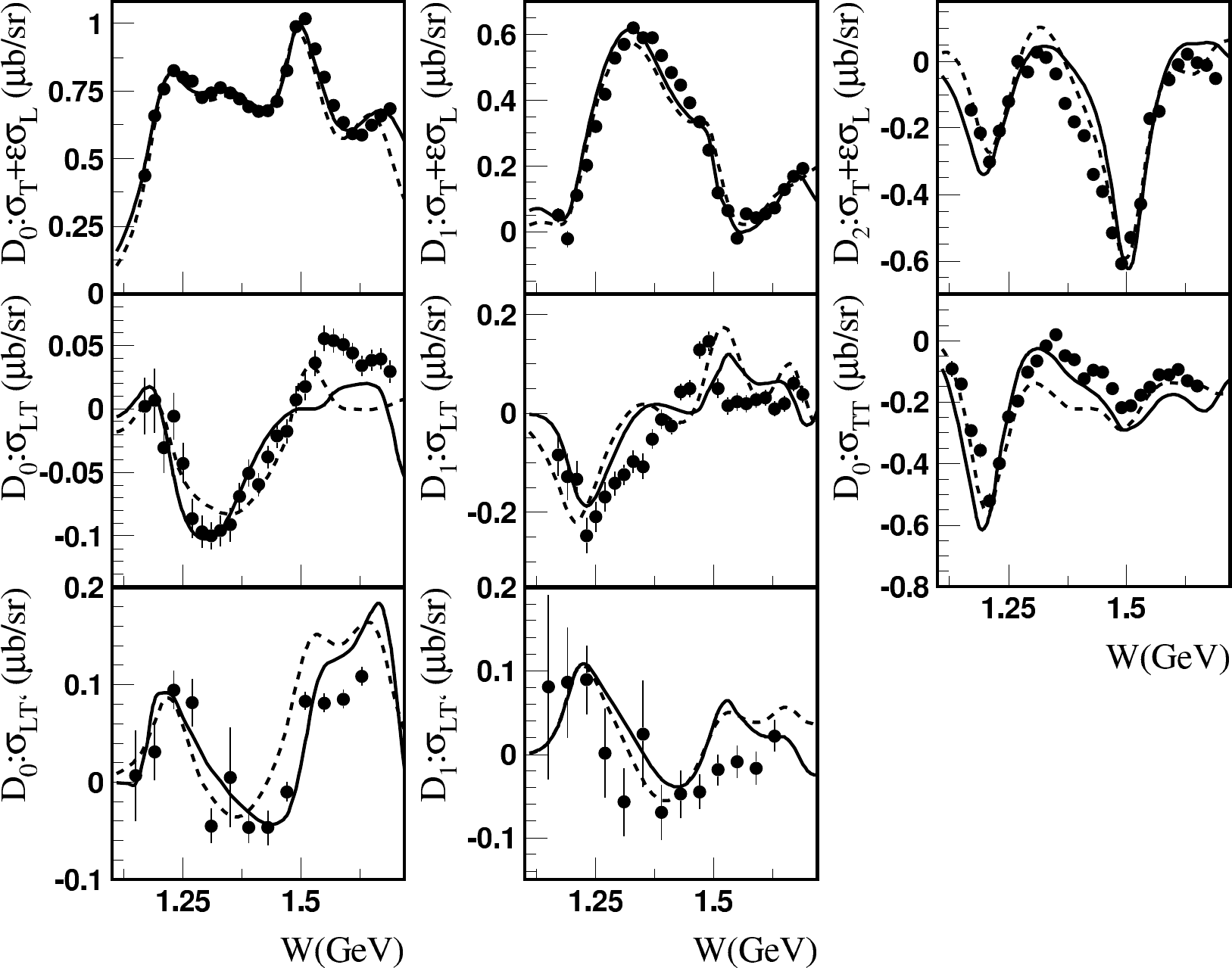}
\vspace{-0.1cm}
\caption{\small 
Results for the Legendre moments of the
$\vec{e}p\rightarrow en\pi^+$ structure functions
in comparison with experimental data \cite{Park:2007tn} for $Q^2=2.44~$GeV$^2$.
The solid (dashed) curves
correspond to the results obtained using DR (UIM) approach.
\label{str_1pi}}
\end{center}
\end{figure*}

These two approaches provide a good  description of the $N\pi$ exclusive channel observables in the entire range  covered by the CLAS
measurements: $W < 1.7$~GeV and $Q^2 < 5.0$~GeV$^2$, resulting in $\chi^2{\rm /d.p}. < 2.0$ for $Q^2 < 1.0$~GeV$^2$ and 
$\chi^2{\rm /d.p}. < 3.0$ at $Q^2$ from 1.5 to 4.5 GeV$^2$ \cite{Aznauryan:2009mx}. Exclusive structure functions $\sigma_{T} + \varepsilon \sigma_{L}$, $\sigma_{TT}$, $\sigma_{LT}$, and
$\sigma_{LT'}$ were derived from the measured CLAS cross sections and polarization asymmetries.
An example of the description of the structure function moments is shown in Fig.~\ref{str_1pi}. The results from these two approaches further provide information for setting the  systematical uncertainties associated with the models.  And  a consistent description  of a large body of observables in the $N\pi$ exclusive channels, obtained within the respective frameworks of 
two conceptually different approaches, lends credibility to a correct evaluation of the resonance contributions.

\subsubsection{Evaluation of $\gamma_{v}NN^*$ resonance electrocouplings from the data on charged-double-pion 
electroproduction off protons \label{2pisec}}

\begin{figure*}[htp]
\begin{center}
\includegraphics[width=14.0cm]{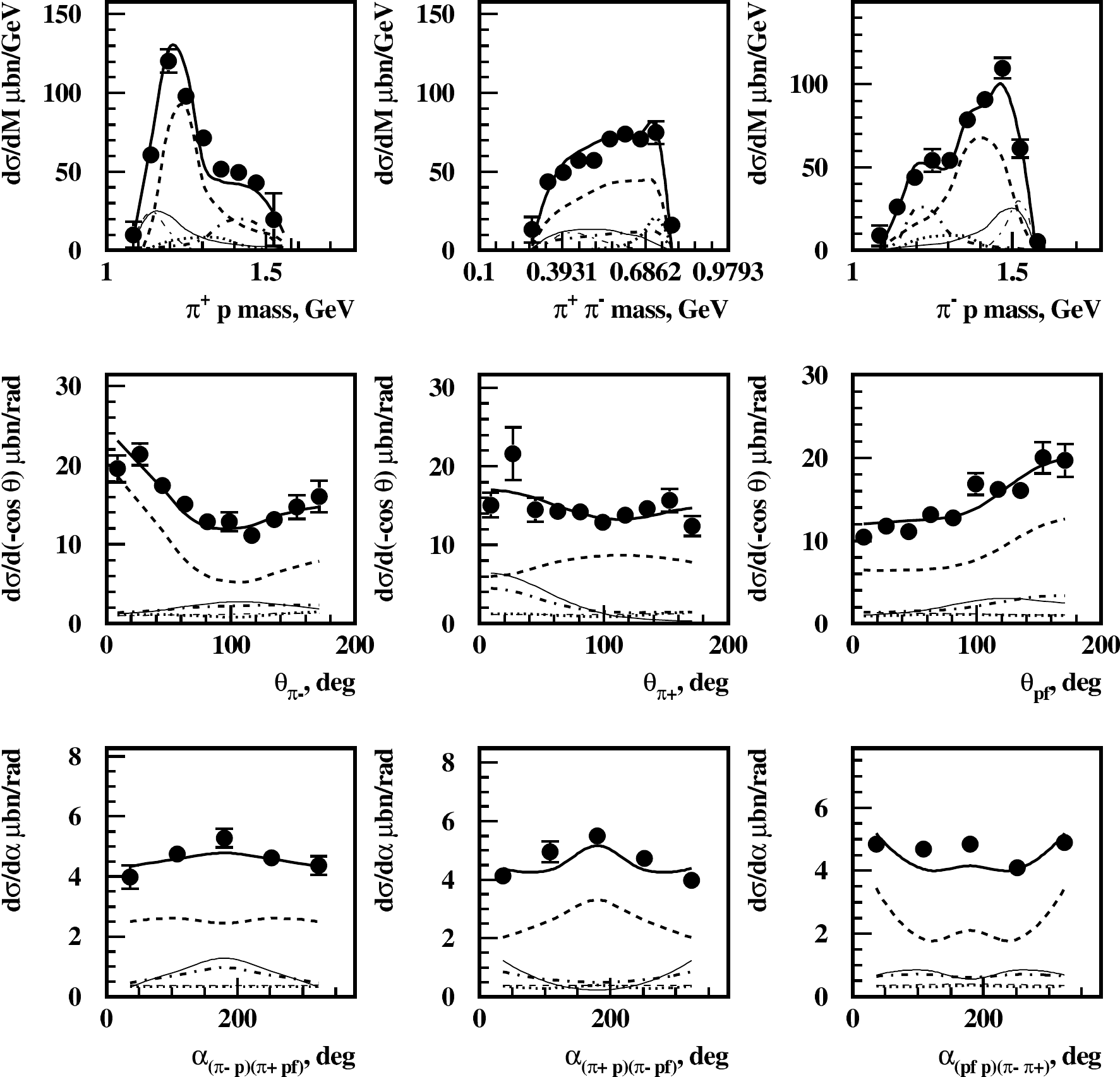}
\vspace{-0.1cm}
\caption{\small 
Fits to  the CLAS
$ep\rightarrow e'\pi^+\pi^- p$ data \cite{Ripani:2002ss} within the framework of JM model
\cite{Aznauryan:2011td,Aznauryan:2011td,Mokeev:2008iw} at $W$ = 1.71 GeV and $Q^2$=0.65 GeV$^2$. Full model results are shown by thick solid lines
together with the contributions from $\pi^-\Delta^{++}$ (dashed thick lines), $\rho p$ (dotted thick lines), 
$\pi^+ \Delta^0$ (dash-dotted thick lines), 
$\pi^+ D^{0}_{13}(1520)$ (thin solid lines), and $\pi^+F^{0}_{15}(1685)$ (dash-dotted thin lines) isobar channels. 
The contributions from other mechanisms described in 
the Section~\ref{2pisec} are comparable with the data error bars and they are not shown in the plot. 
\label{isochan}}
\end{center}
\end{figure*}

\begin{figure*}[btp]
\begin{center}
\includegraphics[width=14cm]{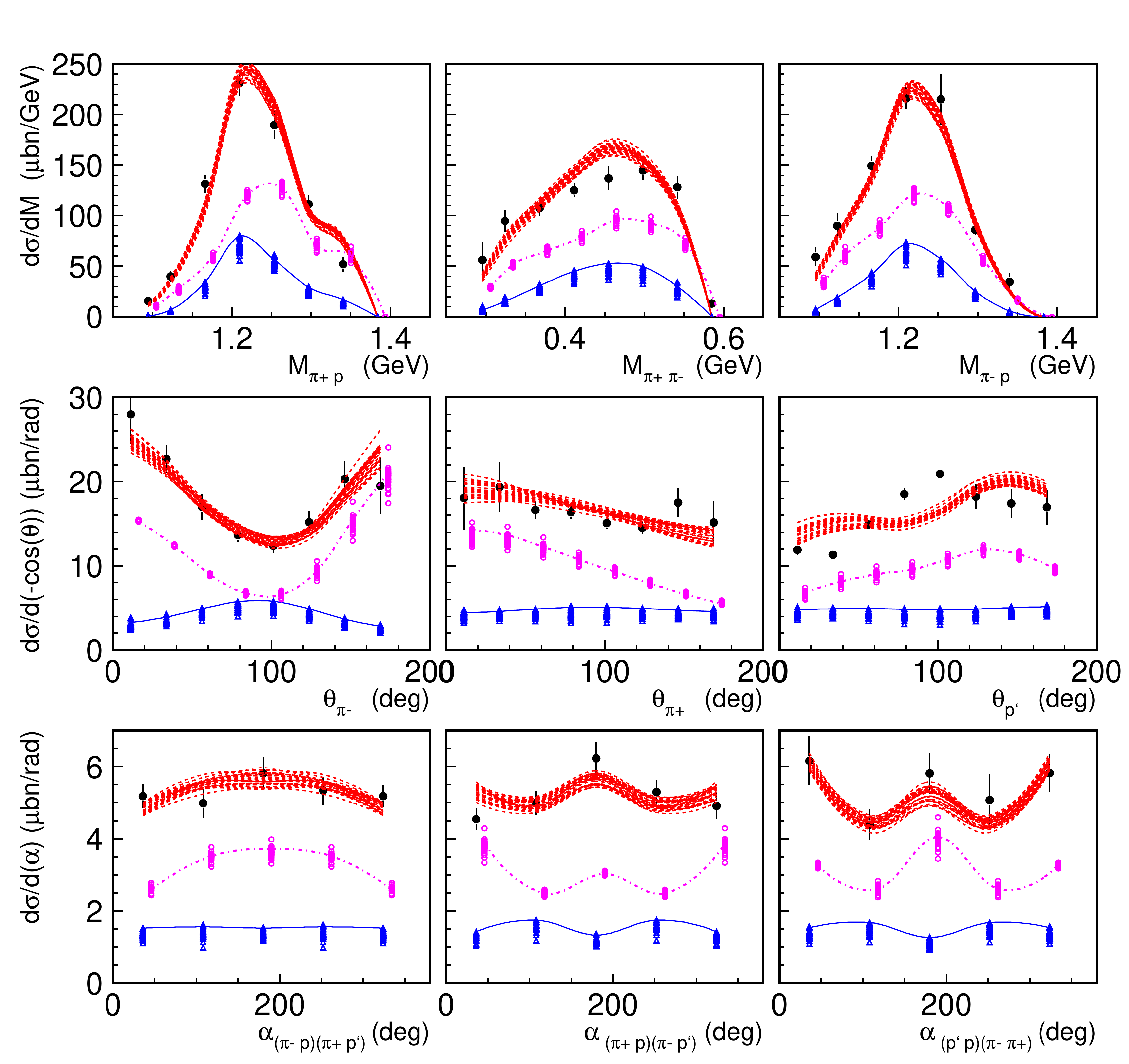}
\caption{\small (color online) Resonant (blue triangles) and non-resonant (green open circles) contributions
to the differential cross sections (red lines) obtained from the CLAS data \cite{Fedotov:2008aa}, fit within the framework of
the JM model at
$W$= 1.51 GeV, $Q^2$=0.38 GeV$^2$. The solid blue and dotted-dashed green lines stand for the resonant and non-resonant contributions, respectively, which were computed  for minimal $\chi^2/d.p.$ achieved in the data fit.  The points for resonant and non-resonant contributions are shifted on each panel for better visibility. Dashed lines show selected
fits.}
\label{nsdtback}
\end{center}
\end{figure*}

The $\pi^+\pi^- p$ electroproduction data measured with the CLAS detector \cite{Fedotov:2008aa,Ripani:2002ss}  provide information on 
nine independent one-fold differential and fully-integrated cross sections in a mass range of
$W < 2.0$~GeV and for photon virtualities  in the range of $0.25  < Q^2 < 1.5$~GeV$^2$.  Examples of  the available $\pi^+\pi^-p$ one-fold
differential cross-section
data for specific bins in $W$ and $Q^2$ are shown in Figs.~\ref{isochan} and \ref{nsdtback}. 
Analysis of these data have
allowed us to establish  which mechanisms contribute to the 
measured cross sections.  The peaks in the invariant mass distributions provide evidence for the presence 
of the channels  arising from $\gamma_{v}p \rightarrow Meson+Baryon \rightarrow \pi^+\pi^- p$ having an unstable baryon or meson in the
intermediate state. Pronounced dependences
in angular distributions further allow us to establish the relevant $t$-, $u$-, and $s$-channel exchanges. 
The mechanisms without 
pronounced kinematical dependences are identified through examination in various 
differential cross sections, with their presence emerging from correlation patterns.  
The phenomenological reaction model JM \cite{Mokeev:2008iw,Aznauryan:2011td,2012sha,Mokeev:NSTAR2007}  was developed in collaboration between Hall~B at
Jefferson Lab and the Skobeltsyn Nuclear Physics Institute in Moscow State University.  The primary  objective  of this work is to  determine the   $\gamma_{v}NN^*$ electrocouplings, and corresponding  $\pi \Delta$ and $\rho p$  partial hadronic decay widths from fitting all measured observables in the $\pi^+\pi^- p$ electroproduction channel.

The amplitudes of  the $\gamma_{v} p \rightarrow \pi^+\pi^- p$ reaction are described in the JM model as a superposition of the
$\pi^-\Delta^{++}$, $\pi^+\Delta^{0}$, $\rho p$, $\pi^{+} D_{13}^{0}(1520)$, $\pi^{+} F_{15}^{0}(1685)$, and
$\pi^{-} P_{33}^{++}(1600)$ sub-channels with subsequent decays of unstable hadrons to the final $\pi^+\pi^-p$ state, 
and additional direct 2$\pi$-production mechanisms, 
where the
final  $\pi^+\pi^- p$ state comes about without going through the intermediate process  of forming unstable hadron states.

The JM model
incorporates contributions from all well-established $N^*$ states  listed in the Table~\ref{bf1pi2pi} to the $\pi \Delta$ and $\rho p$ sub-channels only. We also have
included the $3/2^{+}(1720)$ candidate state, whose existence is suggested in the analysis \cite{Ripani:2002ss} of the CLAS $\pi^+\pi^- p$ electroproduction data.
In the current version of the JM model version (2012)~\cite{2012sha}, the resonant amplitudes are described by a unitarized 
Breit-Wigner ansatz as
proposed in Ref.~\cite{Aitchison:1972ay}; it was modified to make it fully consistent with the parameterization of individual $N^*$ state contributions by a relativistic 
Breit-Wigner ansatz with energy-dependent hadronic decay widths \cite{Ripani:2000va} employed in the JM model. 
unitarity~\cite{2012sha}.
Quantum number conservation in strong interactions allows the transitions between $D_{13}(1520)/D_{13}(1700)$, $S_{11}(1535)/S_{11}(1650)$, and $3/2^+(1720)/P_{13}(1720)$ pairs of $N^*$ states incorporated into the JM model. 
We found that use of the unitarized Breit-Wigner ansatz has a minor influence on 
the $\gamma_{v}NN^*$ electrocouplings, but  it may substantially affect the $N^*$
hadronic decay widths determined from fits to the CLAS data. 
    
 Non-resonant contributions 
to the $\pi \Delta$ sub-channels incorporate a minimal set of current-conserving Born terms \cite{Mokeev:2008iw,Ripani:2000va}. They consist of $t$-channel pion exchange, $s$-channel nucleon exchange, $u$-channel $\Delta$ exchange, and contact terms.  Non-resonant Born terms were reggeized and current conservation was preserved as proposed in Refs.~\cite{Guidal:1997hy,Guidal:1997by}. The initial- and
final-state interactions in $\pi \Delta$ electroproduction are treated in an absorptive approximation, with the absorptive
coefficients estimated from the data from $\pi N$ scattering \cite{Ripani:2000va}. Non-resonant contributions to 
the $\pi \Delta$ sub-channels further include additional contact terms that
have different Lorentz-invariant structures with respect to the contact terms in the sets of Born terms. These
extra contact terms  effectively account for non-resonant processes in the $\pi \Delta$ sub-channels beyond the Born terms,
as well as for the final-state interaction effects that are beyond those taken into account by absorptive
approximation. Parameterizations of the extra contact terms in the $\pi \Delta$ sub-channels are given in  Ref.~\cite{Mokeev:2008iw}. 

Non-resonant amplitudes in  the $\rho p$ sub-channel are described within the framework of a diffractive approximation, which also takes into account the effects caused by $\rho$-line shrinkage \cite{Burkert:2007kn}. The latter effects play a significant role in the $N^*$ excitation region, and in particular, in near-threshold and sub-threshold $\rho$-meson production for $W < 1.8$~GeV. Even in this kinematical regime the $\rho p$ sub-channel may affect the one-fold differential cross sections due to the
contributions from nucleon resonances that decay into the $\rho p$ final states. Therefore, a reliable and credible treatment of non-resonant contributions in the $\rho p$ sub-channel becomes important for ascertaining the electrocouplings and corresponding hadronic parameters of these resonances. The analysis of 
the CLAS data  \cite{Fedotov:2008aa,Ripani:2002ss} has revealed the  presence of the $\rho p$ sub-channel contributions for $W > 1.5$~GeV.

The $\pi^{+} D_{13}^{0}(1520)$, $\pi^{+} F_{15}^{0}(1685)$,  and
$\pi^{-} P_{33}^{++}(1600)$ sub-channels are described in the JM model by non-resonant contributions only. The amplitudes of the $\pi^{+} D_{13}^{0}(1520)$ sub-channel  were derived from the non-resonant Born terms in the $\pi \Delta$ sub-channels  by implementing an additional $\gamma_{5}$-matrix 
 that accounts for the opposite
parities of  the $\Delta$ and $ D_{13}(1520)$ \cite{Mokeev:2005re}. The magnitudes of the
$\pi^{+} D_{13}^{0}(1520)$ production
amplitudes were independently fit to the data for each bin in $W$ and $Q^2$. The contributions from  the
$\pi^{+} D_{13}^{0}(1520)$ sub-channel should be taken into account for $W > 1.5$~GeV.

The $\pi^{+} F_{15}^{0}(1685)$ and
$\pi^{-} P_{33}^{++}(1600)$ sub-channel contributions are seen in the data \cite{Ripani:2002ss} at $W > 1.6$~GeV. These
contributions are almost negligible at smaller $W$.
The effective contact terms were employed in the JM model for parameterization of these sub-channel 
amplitudes~\cite{Mokeev:2005re,Mokeev:NSTAR2007}.
Magnitudes of the $\pi^{+} F_{15}^{0}(1685)$ and
$\pi^{-} P_{33}^{++}(1600)$ sub-channel amplitudes were fit to the data for each bin in $W$ and $Q^2$.

A general unitarity condition for  $\pi^+\pi^- p$ electroproduction amplitudes requires the presence of so-called 
direct-$2\pi$-production mechanisms, where the final $\pi^+\pi^- p$  state is created without going through the intermediate step of forming  
unstable hadron states \cite{Aitchison:1978pw}. These  intermediate-stage processes are
beyond those aforementioned contributions from two-body sub-channels. Direct 2$\pi$ production amplitudes were established for the first time in the analysis of the CLAS 
$\pi^+\pi^-p$ electroproduction data \cite{Mokeev:2008iw}. They are described in the JM model by a sequence of two  exchanges in $t$ and/or $u$ channels
by unspecified particles. The amplitudes of  the $2\pi$-production mechanisms are parameterized by an
Lorentz-invariant
contraction between spin-tensors of the initial and final-state particles, while two exponential propagators describe the above-mentioned exchanges by unspecified particles. The magnitudes of these amplitudes are fit to the data for each bin
in $W$ and $Q^2$. Recent
studies of the correlations between the final-hadron angular distributions have allowed us to establish the phases of
the $2\pi$ direct-production amplitudes \cite{Mokeev:Washington2011}. The contributions from  the $2\pi$ direct-production mechanisms are
maximal and substantial ($\approx$ 30\% ) for $W < 1.5$~GeV and they decrease with increasing $W$, contributing less than 10\% for $W > 1.7$~GeV. However, even in this kinematical regime,  $2\pi$ direct-production mechanisms  can be seen in the  $\pi^+\pi^-p$ electroproduction cross sections due to an interference of the amplitudes from two-body sub-channels.

The JM model provides a reasonable description of the $\pi^+\pi^- p$ differential cross sections for
$W  < 1.8$~GeV and $Q^2 < 1.5$~GeV$^2$ 
with a $\chi^2{\rm /d.p.} < 3.0$, accounting only for statistical uncertainties in the experimental data. As a typical example, the  nine
one-fold differential 
cross sections for $W = 1.71$~GeV and $Q^{2}$ = 0.65~GeV$^{2}$, with fits, are shown in
Fig.~\ref{isochan}, together with the contributions from each of the
individual mechanisms incorporated into the JM description. Each contributing mechanism has a distinctive
shape for the cross section as is depicted by the observables in Fig.~\ref{isochan}. Furthermore, any contributing  
mechanism will be manifested by
substantially different shapes in the cross sections for the observables,  all of which are highly correlated through the 
underlying-reaction dynamics. 
The fit  takes into account all of the nine one-fold differential cross sections simultaneously and allows  for identifying the essential mechanisms
contributing to $\pi^+\pi^-p$ electroproduction off protons.  Such a global fit serves towards understanding the underlying mechanisms and thereby affording access to the dynamics.

This successful fit to the CLAS $\pi^+\pi^-p$ electroproduction data has further allowed us to
determine the resonant parts of cross sections. An example is shown in Fig.~\ref{nsdtback}. The uncertainties associated with the resonant part are comparable with those of the experimental
data. It therefore provides strong evidence for an unambiguous separation of resonant/non-resonant contributions.
A credible means for separating resonances from background was achieved by fitting CLAS data  within the framework of the JM model and it is of particular importance in the extraction of the
$\gamma_{v}NN^*$  
electrocouplings, as well as for evaluating each of the excited states decay widths into the $\pi \Delta$ and $\rho p$ channels. 
A special fitting procedure for the extraction of resonance electrocouplings with the
full and partial $\pi \Delta$ and $\rho p$ hadronic decay widths was developed, thereby allowing us to obtain  uncertainties of resonance parameters and which account for both experimental data uncertainties and for the systematical uncertainties from the JM reaction model~\cite{2012sha}.

\subsection{Resonance electrocouplings from the CLAS pion electroproduction data}

\begin{figure}
  \includegraphics[height=.29\textheight]{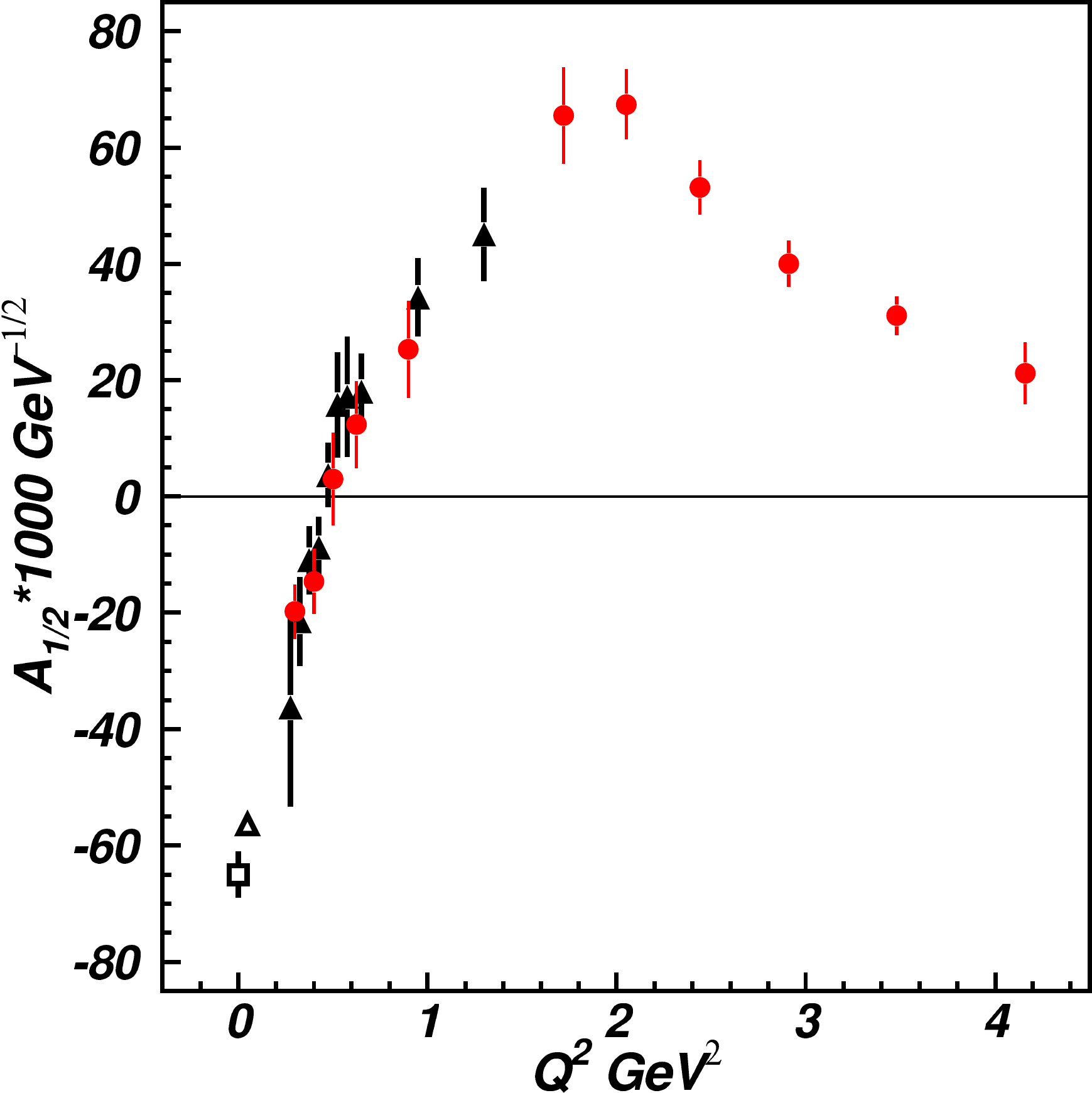}
  \includegraphics[height=.29\textheight]{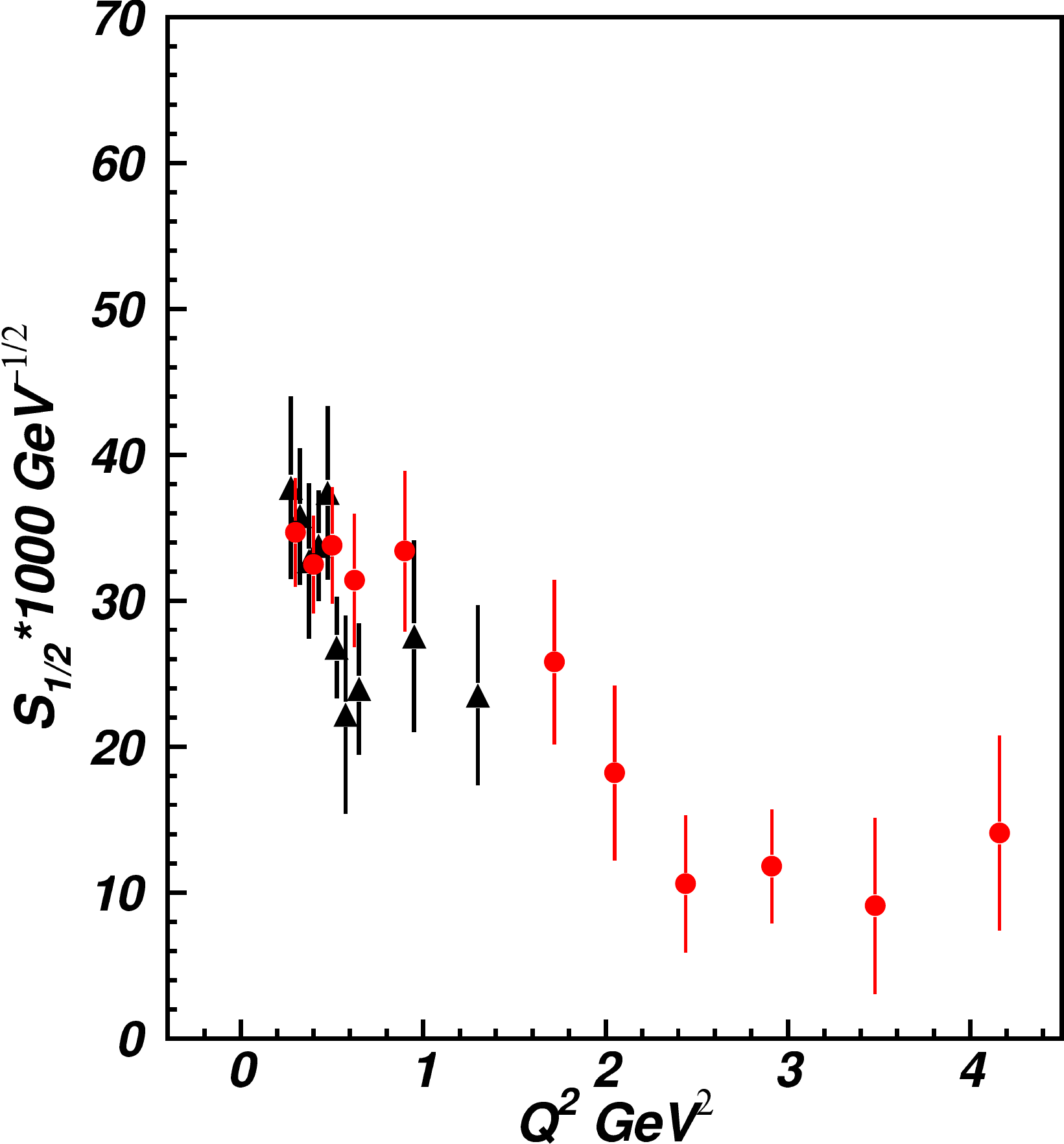}
  \caption{$A_{1/2}$ (left) and $S_{1/2}$ (right) electrocouplings of the $P_{11}(1440)$ resonance determined in independent analyses 
  of the CLAS data on $N\pi$ (circles) \cite{Aznauryan:2009mx}, and $\pi^+\pi^- p$ (triangles) \cite{Aznauryan:2011td}
  electroproduction off protons. Squares and triangles at $Q^2$=0 GeV$^2$ correspond to  \cite{Nakamura:2010zzi} and the CLAS $N\pi$  
  \cite{Dugger:2009pn} photoproduction
  results, respectively.}
  \label{p11}
\end{figure}

\begin{figure}
  \includegraphics[height=.22\textheight]{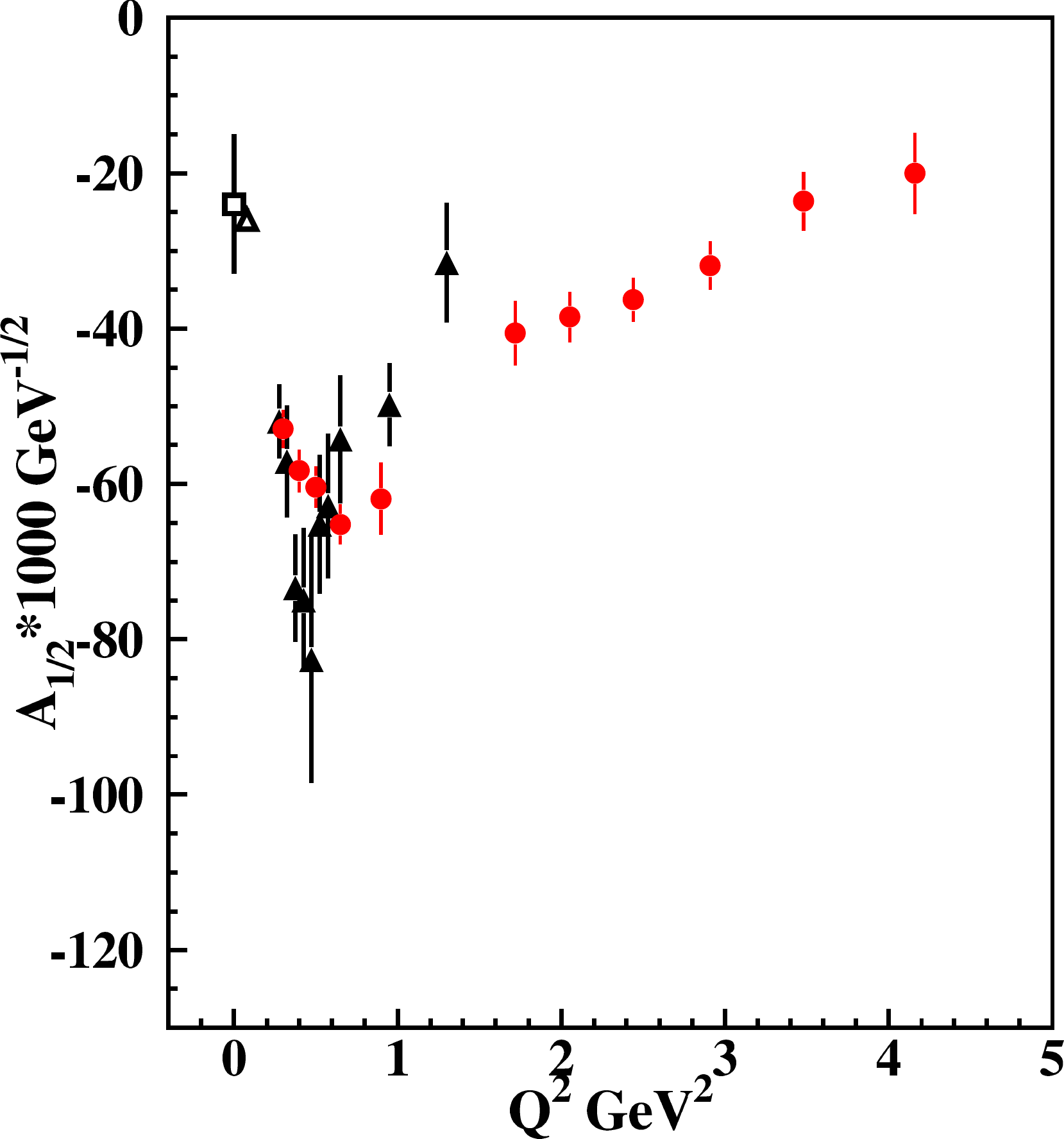}
  \includegraphics[height=.22\textheight]{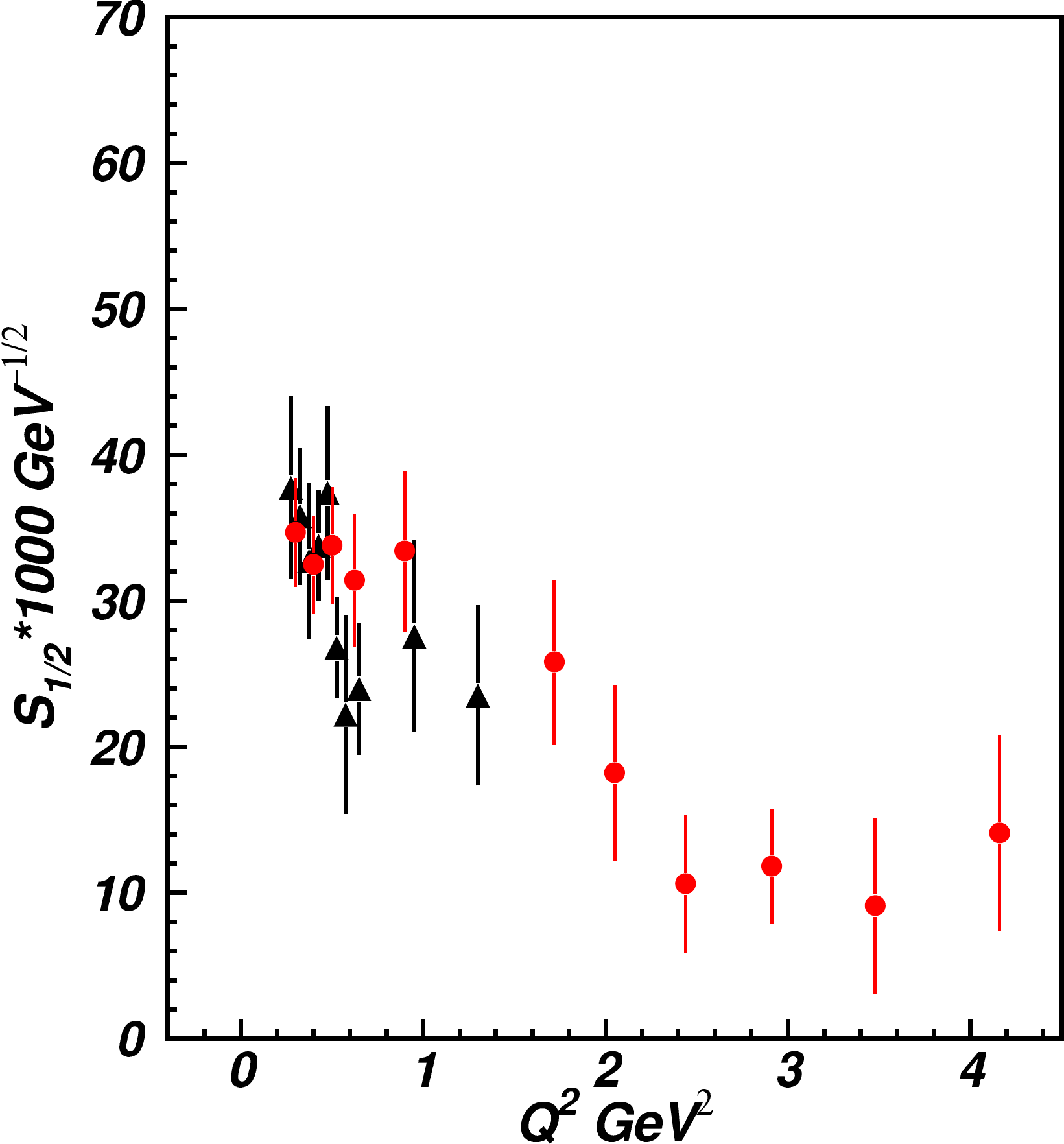}
  \includegraphics[height=.22\textheight]{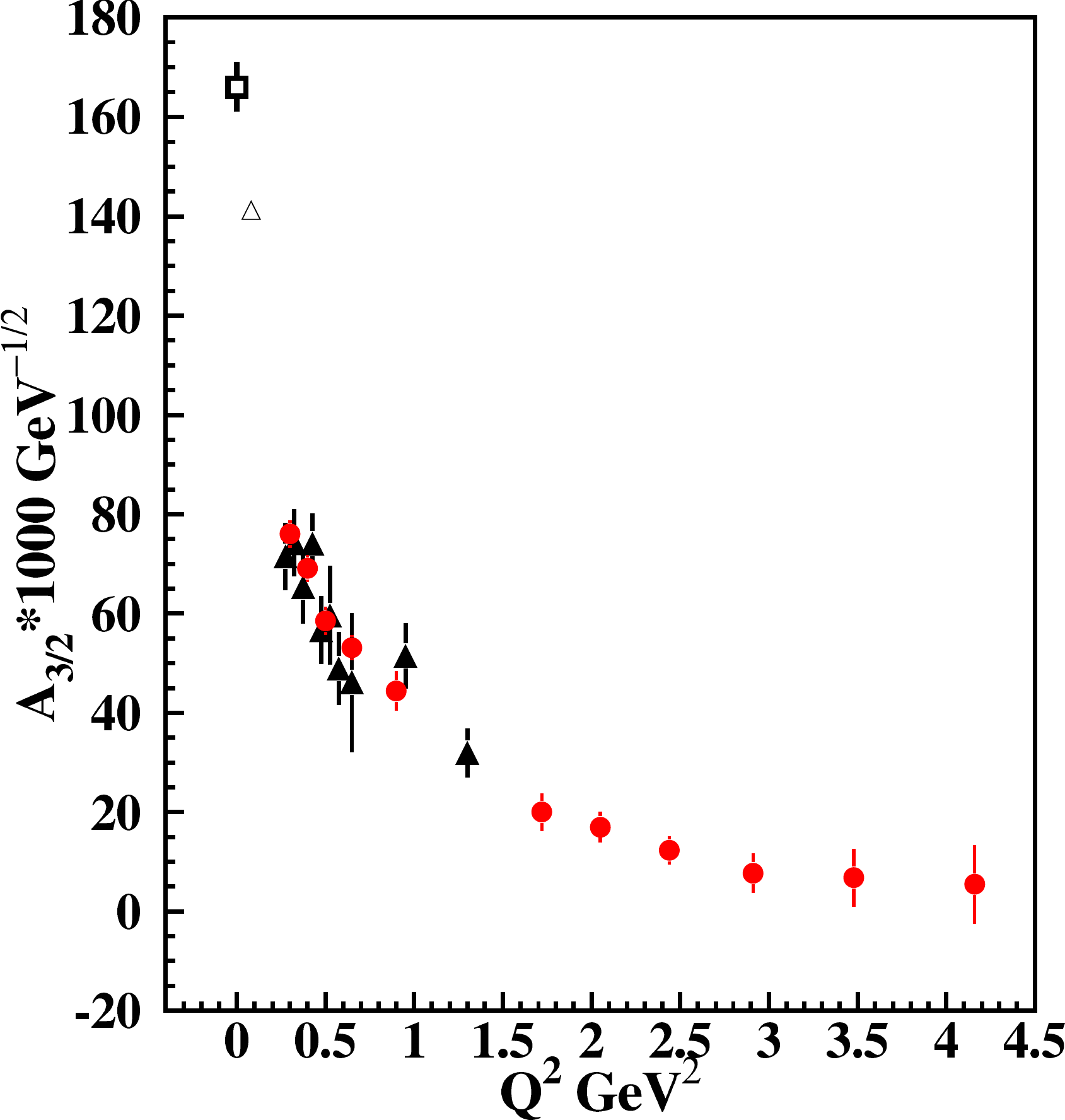}
  \caption{$A_{1/2}$ (left), $S_{1/2}$ (middle), and $A_{3/2}$ (right) electrocouplings of the $D_{13}(1520)$ resonance 
  determined in independent analyses 
  of the CLAS data on $N\pi$ (circles) \cite{Aznauryan:2009mx}, and $\pi^+\pi^- p$ (triangles) \cite{Aznauryan:2011td}
  electroproduction off protons. Squares and triangles at $Q^2$=0 GeV$^2$ correspond to \cite{Nakamura:2010zzi} and the CLAS $N\pi$  
  \cite{Dugger:2009pn} photoproduction
  results, respectively.}
  \label{d13}
\end{figure}

Several analyses of CLAS data were carried out on single- and charged-double-pion electroproduction 
off protons within the framework of
fixed-$t$ dispersion relations, the UIM model, and the JM model 
as described in Sections~\ref{1pisec} and \ref{2pisec}, which have provided, for the first time, information on electrocouplings of the
$P_{11}(1440)$, $D_{13}(1520)$, and $F_{15}(1685)$ resonances from independent analyses of $\pi^+n$, $\pi^0p$, and
$\pi^+\pi^-p$ electroproduction channels \cite{Aznauryan:2009mx,Aznauryan:2011td,2012sha}.  The electrocouplings of the $P_{11}(1440)$ and $D_{13}(1520)$ 
resonances determined
from these channels are shown in Figs.~\ref{p11} and~\ref{d13}. They are consistent within uncertainties. 
The longitudinal $S_{1/2}$ electrocouplings of the
$D_{13}(1520)$, $S_{11}(1535)$, $S_{31}(1620)$, $S_{11}(1650)$, $F_{15}(1685)$, $D_{33}(1700)$,
and $P_{13}(1720)$ excited proton states have become available from the CLAS data 
for the first time as well~\cite{Aznauryan:2009mx,Aznauryan:2011td}.

Consistent results on $\gamma_{v}NN^*$ electrocouplings 
from the  $P_{11}(1440)$, $D_{13}(1520)$, and $F_{15}(1685)$ resonances that were determined from
independent analyses of the major meson electroproduction  channels, $\pi^+n$, $\pi^0p$, and $\pi^+\pi^-p$, demonstrate that the extraction of these fundamental quantities are reliable as these different electroproduction channels have quite different 
backgrounds.
Furthermore, this consistency also strongly suggests that 
the reaction models described 
in  sections~\ref{1pisec} and \ref{2pisec} will provide  a reliable 
evaluation of the $\gamma_{v}NN^*$ electro\-couplings for analyzing either single- or charged-double-pion 
electroproduction data.  It therefore makes it possible to determine 
electrocouplings  for all resonances that decay preferentially 
to the  $N\pi$ and$/$or $N\pi\pi$ final states. 

The studies of $N\pi$ exclusive channels 
are the primary source of information on electrocouplings of the $N^*$ states with masses below 
1.6~GeV \cite{Aznauryan:2009mx}.  The reaction kinematics restrict the $P_{33}(1232)$ state to only the $N\pi$ exclusive channels. 
The $P_{11}(1440)$ and $D_{13}(1520)$ resonances have contributions to both single- and double-pion electroproduction channels, which are sufficient for the extraction of their respective electrocouplings.
Analysis of the $\pi^+\pi^-p$ electroproduction off protons allows us to check
the results of $N\pi$ exclusive channels for the resonances that have substantial decays to both the  $N\pi$
and $N\pi\pi$ channels. 

\begin{figure*}[tb]
\begin{center}
\includegraphics[height=.26\textheight]{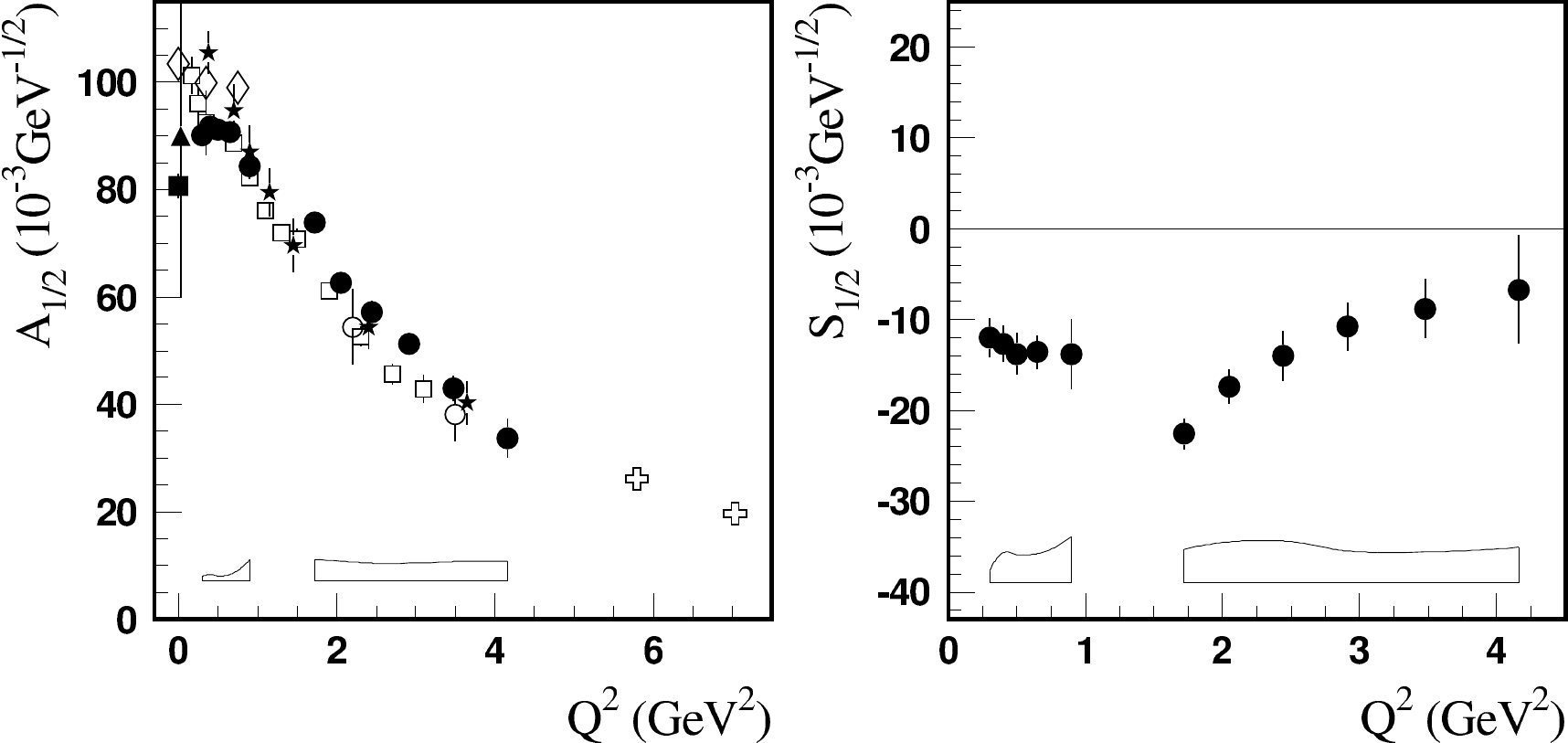}
\end{center}
\caption{
Transverse electrocoupling $A_{1/2}$
of the $\gamma^* p \rightarrow~S_{11}(1535)$ transition.
The full circles are 
the electrocouplings extracted from $N\pi$ electroproduction
data \cite{Aznauryan:2009mx}.
The electrocouplings extracted from $N\eta$ electroproduction
data are:
the stars \cite{Thompson:2000by},
the open boxes \cite{Denizli:2007tq},
the open circles \cite{Armstrong:1998wg},
the crosses \cite{Dalton:2008aa}, and
the rhombuses \cite{Aznauryan:2004jd,Aznauryan:2002gd}.
The full box and triangle at $Q^2=0$ correspond
to \cite{Nakamura:2010zzi} and the CLAS $N\pi$ \cite{Dugger:2009pn}
photoproduction
results, respectively.} 
\label{s11}
\end{figure*}

For the 
$S_{11}(1535)$ resonance, the hadronic decays to the $N\pi\pi$ final state is unlikely (see Table
~\ref{bf1pi2pi}). Therefore,  the studies of this very pronounced  $N\pi$-electroproduction
resonance become problematic in the charged-double-pion electroproduction off protons. 
On the other hand, the $S_{11}(1535)$ resonance has a  large branching ratio
to the $\pi N$ and $\eta N$ channels.   And since 1999, this resonance has been extensively
studied  at JLab over a wide range of $Q^2$ up to $4.5$ and $7$~GeV$^2$ 
for the channels $N\pi$ and $N\eta$, respectively in the 
electroproduction off protons 
(see Fig.~\ref{s11}).
For $N\eta$ electroproduction,
the $S_{11}(1535)$ strongly dominates the cross section
for $W<1.6$~GeV and is extracted from the data in a nearly
model-independent way using a Breit-Wigner form
for the resonance contribution \cite{Thompson:2000by,Denizli:2007tq,Armstrong:1998wg,Dalton:2008aa}. 
These analyses assume that the longitudinal contribution 
is small enough to have a negligible effect on the
extraction of the transverse amplitude.  This assumption
is confirmed by the analyses of the CLAS
$N\pi$ electroproduction data \cite{Aznauryan:2009mx}.
Accurate results were obtained in both reactions
for the transverse electrocoupling $A_{1/2}$;
they show a consistent $Q^2$ slope
and allowed for the determination of the 
branching ratios
to the $N\pi$ and $N\eta $ channels \cite{Aznauryan:2009mx}.
Transverse $A_{1/2}$ electrocouplings of the $S_{11}(1535)$ extracted in independent 
analyses of $N\pi$ and $N\eta $ electro\-production channels are in a reasonable agreement,
after taking  the systematical uncertainties of the analysis \cite{Aznauryan:2009mx} into consideration. Expanding the proposal, Nucleon Resonance Studies with CLAS12
\cite{Gothe:clas12}, by further incorporating $N\eta$ electroproduction at high $Q^2$ would considerably enhance
 our capabilities for extracting self-consistent and reliable results for the $S_{11}(1535)$ 
electrocouplings in independent
analyses of the $N\pi$ and $N\eta$ electroproduction channels.

\begin{figure}
  \includegraphics[height=.22\textheight]{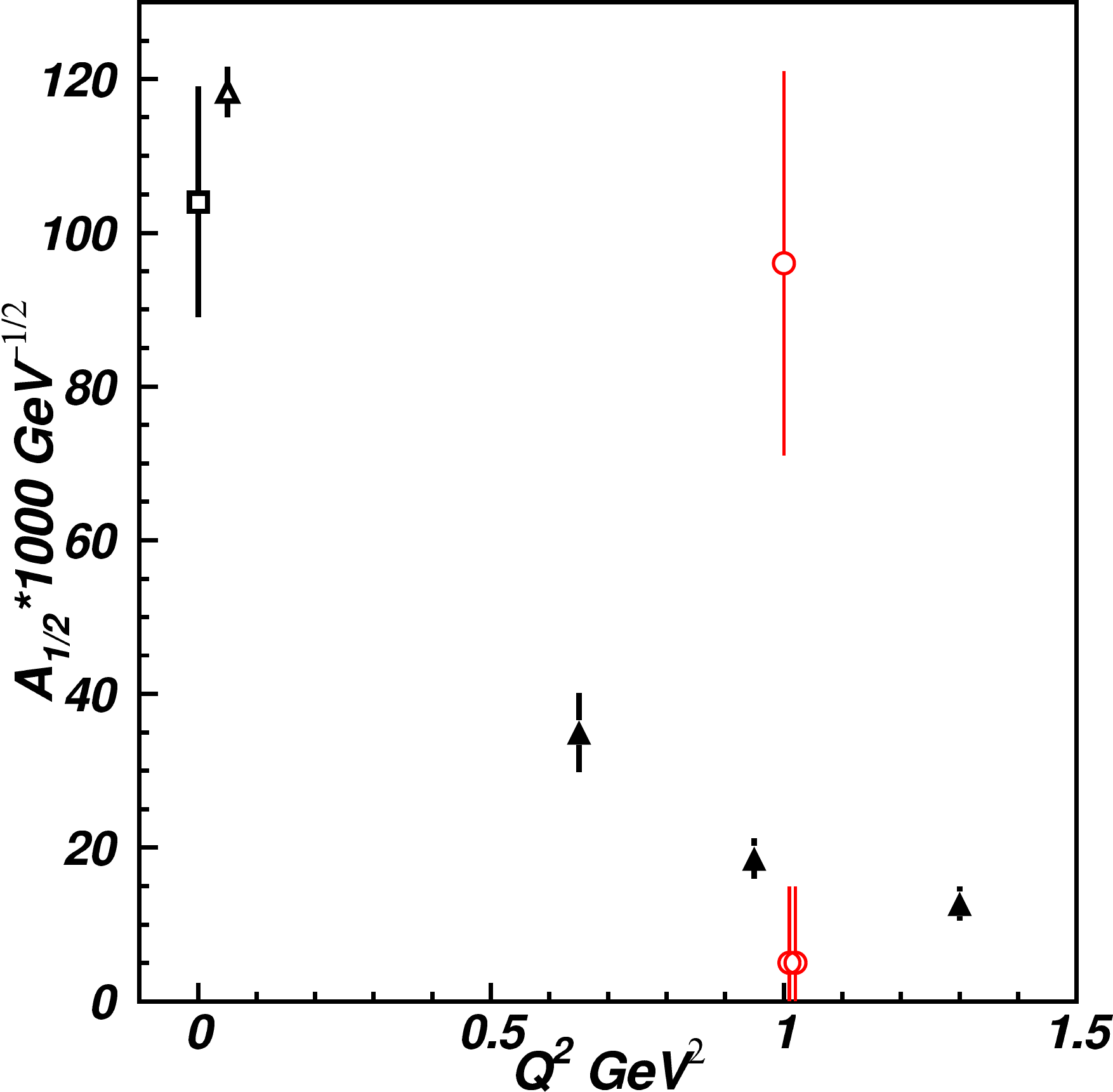}
  \includegraphics[height=.22\textheight]{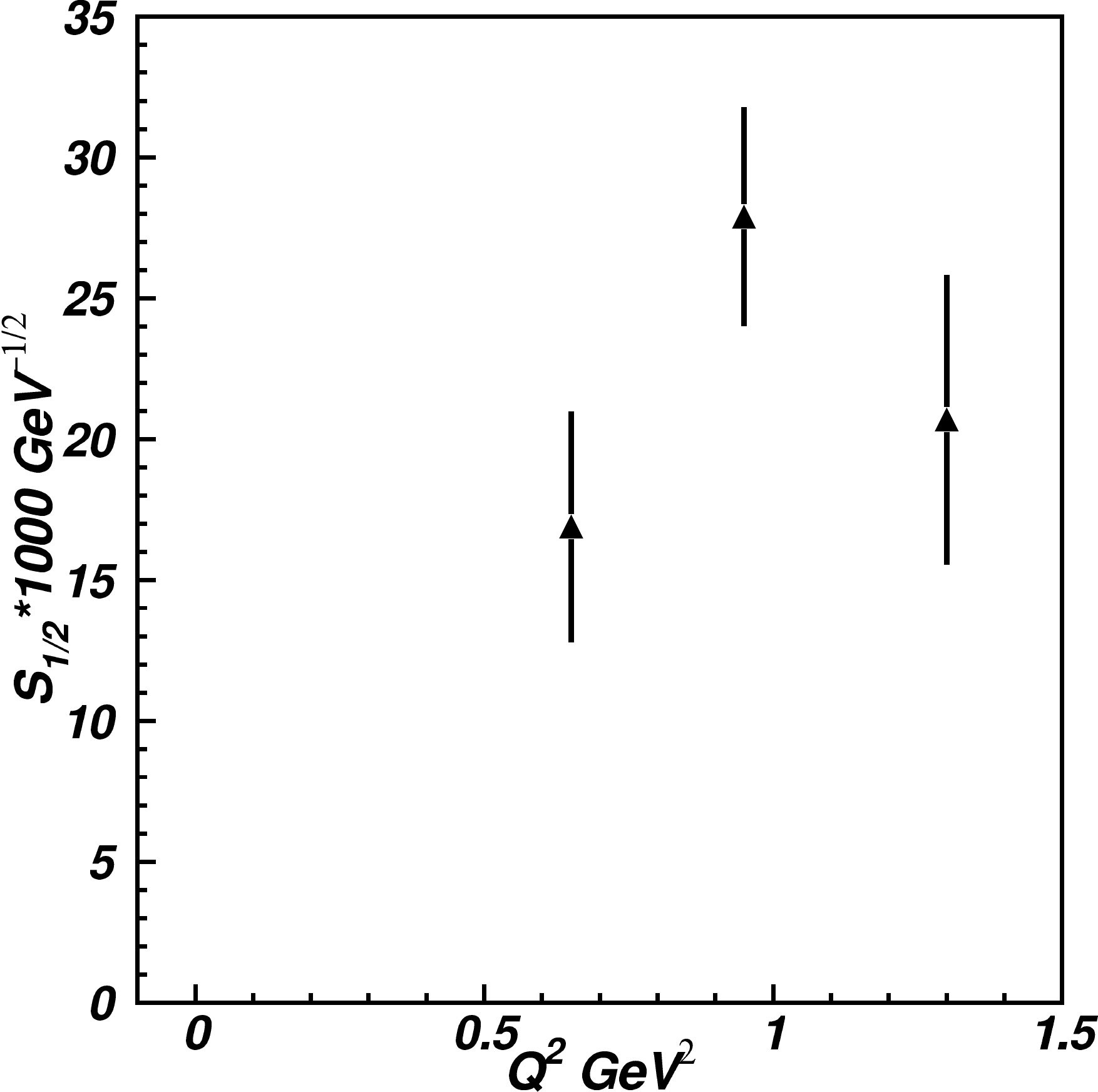}
  \includegraphics[height=.22\textheight]{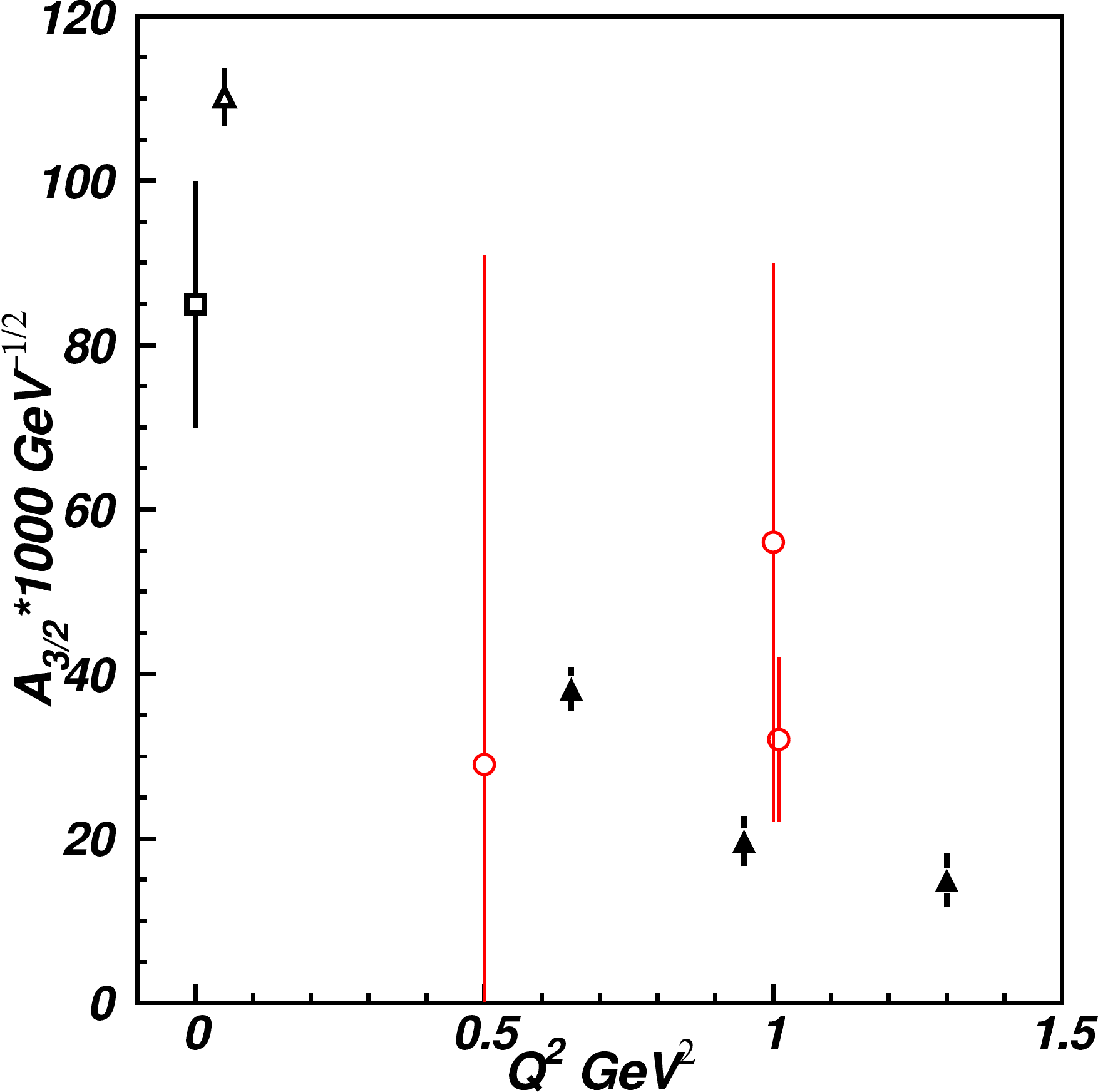}
  \caption{Electrocouplings of $D_{33}(1700)$ resonance $A_{1/2}$ (left), $S_{1/2}$ (middle) and $A_{3/2}$ (right)  
  determined in analyses the CLAS  $\pi^+\pi^- p$ electroproduction data \cite{Mokeev:2008iw} and world data on $N\pi$
  electroproduction off protons \cite{Burkert:2002zz}.}
  \label{d33}
\end{figure}

The charged-double-pion electroproduction channel
is of particular 
importance for evaluation of  high-lying resonance electrocouplings, 
 since most $N^*$ states with masses above 1.6~GeV  decay
 preferentially by two~pion emission (Table~\ref{bf1pi2pi}). 
Preliminary results on the electrocouplings of the $S_{31}(1620)$, $S_{11}(1650)$, $F_{15}(1685)$, $D_{33}(1700)$, and $P_{13}(1720)$ 
 resonances were
 obtained from an analysis of the CLAS $\pi^+\pi^-p$ electroproduction data \cite{Ripani:2002ss} within the framework of the JM model \cite{Aznauryan:2011td}. As an example, electrocouplings 
 of the $D_{33}(1700)$ resonance were determined from analysis of the CLAS  $\pi^+\pi^-p$  electroproduction data and
 are shown in Fig.~\ref{d33} in comparison with the previous world's data taken from Ref.~\cite{Burkert:2002zz}. 
 The $D_{33}(1700)$ resonance decays preferentially to $N\pi\pi$ final states with the branching 
 fraction exceeding 80\%. Consequently, 
 electrocouplings of this resonance determined from the $N\pi$ electroproduction channels have large
 uncertainties due to insufficient sensitivity of these exclusive channels contributing to 
 the $D_{33}(1700)$ resonance.
 The CLAS results have considerably improved our knowledge on electrocouplings of the
 $S_{31}(1620)$, $S_{11}(1650)$, $F_{15}(1685)$, $D_{33}(1700)$, and $P_{13}(1720)$ resonances. 
 They have provided accurate information on the $Q^2$ evolution of the transverse 
 electrocouplings, while longitudinal
 electrocouplings of these states  were determined, again, for the first time.

Most of the $N^*$ states with masses above 1.6~GeV decay preferentially through channels with two pions in the final state, thus making it difficult to explore these states in single-pion electroproduction channels. The CLAS $KY$ electroproduction data \cite{Carman:2009fi,Ambrozewicz:2006zj} may potentially provide independent information on the electrocouplings of these states. At the time of this writing, however, reliable information on $KY$ hadronic decays from $N^*$s are not yet available. The $N^*$ hyperon decays can be obtained from fits to the CLAS $KY$ electroproduction data \cite{Carman:2009fi,Ambrozewicz:2006zj}, which should be carried out independently in different bins of $Q^2$ by utilizing the $Q^2$-independent behavior of resonance hadronic decays. The development of reaction models for the extraction of $\gamma_{v}NN^*$ electrocouplings from the $KY$ electroproduction channels is urgently needed. Furthermore, complementary studies of the  $KY$ decay mode can  be carried out with future data from the Japan Proton Accelerator Research Complex (J-PARC) and through $J/\Psi$ decays to various $\bar{N}N^{*}$ channels at the Beijing Electron Positron Collider (BEPC).

Most of the well-established resonances have substantial 
 decays to either the $N\pi$ or $N\pi\pi$ 
final states. Therefore, 
 studies of $N\pi$ and $\pi^+\pi^-p$ electroproduction off protons will allow us
to determine the electrocouplings of all prominent excited proton states and  such studies will mark the first step in the evaluation of resonance electrocouplings in the unexplored regime of photon virtualities ranging from 5 to 12 GeV$^2$.

 \subsection{Status and prospect of Excited Baryon Analysis Center (EBAC) \label{EBAC status}}

\subsubsection{The case for a multi-channel global analyses}

Interactions among different hadronic final states are termed final-state interactions (FSI). In exclusive-meson electroproduction, for example, FSI represent a key issue both in terms of the extraction as well as in the physical interpretation of  the nucleon resonance parameters.  In the reaction models for analyses of different exclusive-meson electroproduction channels,  as detailed above,  FSI  are treated phenomenologically for each specific reaction.  Analyses of exclusive hadroproduction have allowed us to explicitly establish the relevant mechanisms for hadron interactions among the various final states for different exclusive photo- and electroproduction channels in terms of the meson-baryon degrees of freedom. The information on meson electroproduction amplitudes comes mostly from CLAS experiments. These results on meson-baryon hadron interaction amplitudes open up additional opportunities for the extraction of resonances, their photo- and electrocouplings, as well as their associated hadronic decay parameters.  These parameters can be constrained through a  global analysis of all exclusive-meson electroproduction data from different photo- and electroproduction channels as analyzed within the framework of coupled-channel approaches \cite{Kamano:2011ut,Krewald:2012zz,Sato:2012zza}.

These approaches have allowed us to explicitly take into account the hadronic final-state interactions among the exclusive meson electroproduction channels and to build up reaction amplitudes consistent with the restrictions imposed by a general unitarity condition. Another profound consequence of unitarity is reflected by the relations among the non-resonant meson production mechanisms and the contributions 
from meson-baryon dressing amplitudes (i.e.~the meson-baryon cloud) to the resonance electrocouplings along with their hadronic decay parameters. Use of coupled-channel approaches have allowed us to determine such contributions in the fitting to the experimental data.
Therefore, global analyses of all exclusive meson photo- and electroproduction data  within the framework of coupled-channel approaches will reveal information on the resonance structure in terms of quark-core and meson-baryon cloud contributions at different distance scales.   

The $N\pi$ and $\pi^+\pi^-p$ electroproduction channels are strongly 
coupled through  final-state interactions. The data from experiments with hadronic probes  have shown that the 
$\pi N \rightarrow \pi\pi N$ reactions are the second biggest exclusive contributors to inclusive $\pi N$ interactions. Therefore, data on the mechanisms contributing to single- and charged-double-pion electroproduction off protons are needed for the development of global multi-channel analyses for the extraction of $\gamma_{v}NN^*$ electrocouplings within the framework of 
coupled-channel approaches.  A consistent description of hadronic interactions between the $\pi N$ and $\pi \pi N$ asymptotic states is critical for the reliable extraction of  $\gamma_{v}NN^*$ electrocouplings within the framework of 
coupled-channel approaches.

\subsection{Dynamical Coupled Channel model}

In this section, we report on the development and results of the EBAC-DCC approach spanning the period from January  2006  through March 2012. This  analysis project has three primary components:
\begin{enumerate}
\item perform a  dynamical coupled-channels analysis 
on the world data on meson production reactions from the nucleon
to determine the meson-baryon partial-wave amplitudes, 
\item extract the $N^*$ parameters from the determined partial-wave amplitudes,
\item  investigate the interpretations of the extracted $N^*$ properties in terms of the available hadron models and Lattice QCD. 
\end{enumerate}

The  Excited Baryon Analysis Center (EBAC)  is conducting dynamical coupled-channel (DCC) analyses of Jefferson Lab data and other relevant data in order to extract $N^{*}$ parameters and to investigate the reaction mechanisms for mapping out the important components of the $N^{*}$ structure as a function of distance or $Q^{2}$.  This work is predicated upon the dynamical model for the $\Delta$(1232) resonance~\cite{Matsuyama:2006rp}, which was developed  by the Argonne National Laboratory-Osaka University (ANL-Osaka) collaboration~\cite{Sato:1996gk}.
 In the EBAC extension to the ANL-Osaka formulation~\cite{Matsuyama:2006rp}, 
the reaction amplitudes $T_{\alpha,\beta}(p,p^{\prime};E)$ for each partial wave are calculated from
the following coupled-channels integral equations,
\begin{eqnarray}
T_{\alpha,\beta}(p,p';E)&=& V_{\alpha,\beta}(p,p') + \sum_{\gamma}
 \int_{0}^{\infty} q^2 d q  V_{\alpha,\gamma}(p, q )
G_{\gamma}(q ,E)
T_{\gamma,\beta}( q  ,p',E) \,, \label{eq:cct}\\
V_{\alpha,\beta}&=& v_{\alpha,\beta}+
\sum_{N^*}\frac{\Gamma^{\dagger}_{N^*,\alpha} \Gamma_{N^*,\beta}}
{E-M^*} \,,
\label{eq:ccv}
\end{eqnarray}
where $\alpha,\beta,\gamma = \gamma N, \pi N, \eta N, KY, \omega N$, and
$\pi\pi N$ which has  $\pi \Delta, \rho N, \sigma N$ resonant components,
$v_{\alpha,\beta}$ are meson-exchange interactions deduced from
the phenomenological Lagrangian, $\Gamma_{N^*,\beta}$  describes
the excitation of the nucleon to a bare $N^*$ state with a mass
$M^*$, and $G_{\gamma}(q ,E)$ is a meson-baryon propagator. 
The DCC model, defined by Eqs.~(\ref{eq:cct}) and~(\ref{eq:ccv}), respects unitarity for
both  two- and three-body reactions.


This dynamical coupled-channel model was used initially in fitting  $\pi N$ reactions from  elastic scattering to extract parameters associated with the strong-interaction parts of  $V_{\alpha,\beta}$ in Eq.~(\ref{eq:ccv})  and corresponding  electromagnetic components of  $V_{\alpha,\beta}$ came from fits to the  $\gamma p \rightarrow \pi^0p, \pi^+n$, and $p(e,e'\pi^{0,+})N$  data in the invariant mass range of  $W \le 2$~GeV.
To simplify the analysis during the developmental stage (2006-2010), the $KY$ and $\omega N$ channels were not included in these fits.

The resulting six-channel model was then tested by comparing the 
the predicted $\pi N, \gamma N \rightarrow \pi\pi N$ production cross sections with data.
In parallel to analyzing the data, a procedure
to analytically continue Eqs.~(\ref{eq:cct}) and~(\ref{eq:ccv}) 
to the complex-energy plane was developed to allow for extracting the positions and residues of several nucleon resonances.
In the following, we present a sample of some of our results in these efforts.

\subsubsection{Results for single-pion production reactions}

In fitting the  $\pi N$ elastic-scattering channel, we found that one or two bare $N^*$ states 
were needed for each partial wave. 
The coupling strengths of the $N^*\rightarrow MB$ vertex interactions
$\Gamma_{N^*,MB}$ with $MB=\pi N, \eta N, \pi\Delta, \rho N, \sigma N$
were then determined in the $\chi^2$-fits to the data and these results can be found in Ref.~\cite{JuliaDiaz:2007kz}.

\begin{figure}[!ht]
\includegraphics[clip,height=.35\textheight]{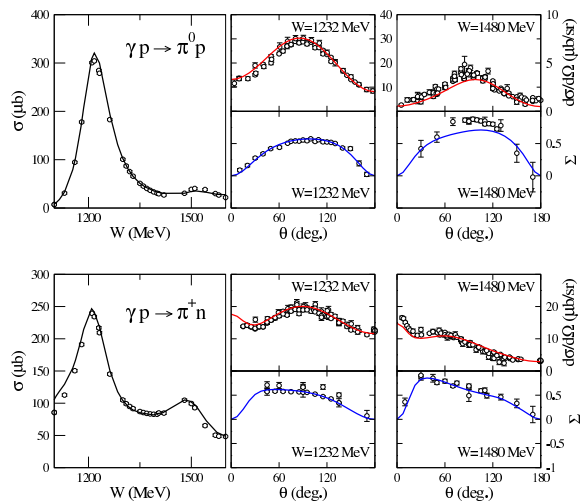}
\caption{The DCC results~\cite{JuliaDiaz:2007fa} of total cross sections ($\sigma$), 
differential cross sections ($d\sigma/d\Omega)$, and photon
asymmetry ($\Sigma$) of $\gamma p\rightarrow \pi^0p$ (upper parts), 
$\gamma p \rightarrow \pi^+n$ (lower parts).}
\label{fig:gnpin}
\end{figure}

Our next step was to determine the bare $\gamma N \rightarrow N^*$
interaction $\Gamma_{N^*,\gamma N}$ by fitting the data from $\gamma p \rightarrow
\pi^0p$ and $\gamma p \rightarrow \pi^+n$ reactions.

Because we did not adjust  any parameter which had already been fixed in earlier fits to the $\pi N$ elastic scattering, we found~\cite{JuliaDiaz:2007fa} that our fits to the data were sound only up to invariant masses not exceeding $W = 1.6$ GeV.   
In Fig.~\ref{fig:gnpin} are shown our results for total cross sections ($\sigma$),  
differential cross sections ($d\sigma/d\Omega)$, and the photon 
asymmetry ($\Sigma$).   The $Q^2$ dependence of the $\Gamma_{N^*,\gamma N}$ vertex functions were then
determined~\cite{JuliaDiaz:2009ww} by fitting the $p(e,e^{\prime}\pi^0)p$ and $p(e,e^{\prime}\pi^+)n$ data 
up to $W = 1.6$ GeV and $Q^2 = 1.5 $ (GeV/c)$^2$. 

\subsubsection{Results for two-pion production reactions}

\begin{figure}[!ht]
\includegraphics[clip,height=.35\textheight]{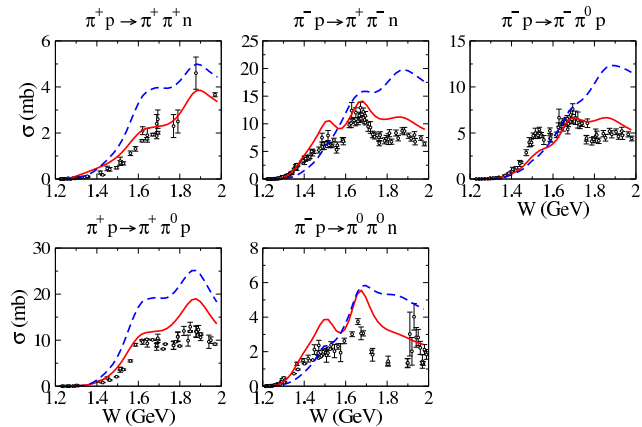}
\caption{The predicted~\cite{Kamano:2008gr} total cross sections of the
$\pi N \rightarrow \pi\pi N$ are compared with data.
The dashed curves come from switching off  the coupled-channel effects in the DCC model of Ref.~\cite{Matsuyama:2006rp}.}
\label{fig:pnppn}
\end{figure}

As delineated above, the dynamical coupled-channel model was  constructed from fitting single-pion data.  We then tested the efficacy of this model by examining to what extent the model could describe the  $\pi N \rightarrow \pi\pi N$
and the $\gamma N \rightarrow \pi\pi N$ data.  Interestingly, at the near--threshold 
region of $W \leq 1.4$~GeV,  we found~\cite{Kamano:2008gr,Kamano:2009im} that the predicted total cross sections
are in excellent agreement with the data.   At higher $W$, the predicted $\pi N \rightarrow \pi\pi N$ cross sections describe the major features of  the available data reasonably well, as is shown in Fig.~\ref{fig:pnppn}. 
Here, we further see the important role that the effects from coupled channels play.
The predicted $\gamma p \rightarrow \pi^+\pi^- p, \pi^0\pi^0p$ cross sections, however, exceed the data by about a factor of two,
while the fits describe, more or less, the overall shapes of the two-particle invariant-mass distributions.

\begin{figure}[Ht]
\includegraphics[clip,height=.25\textheight]{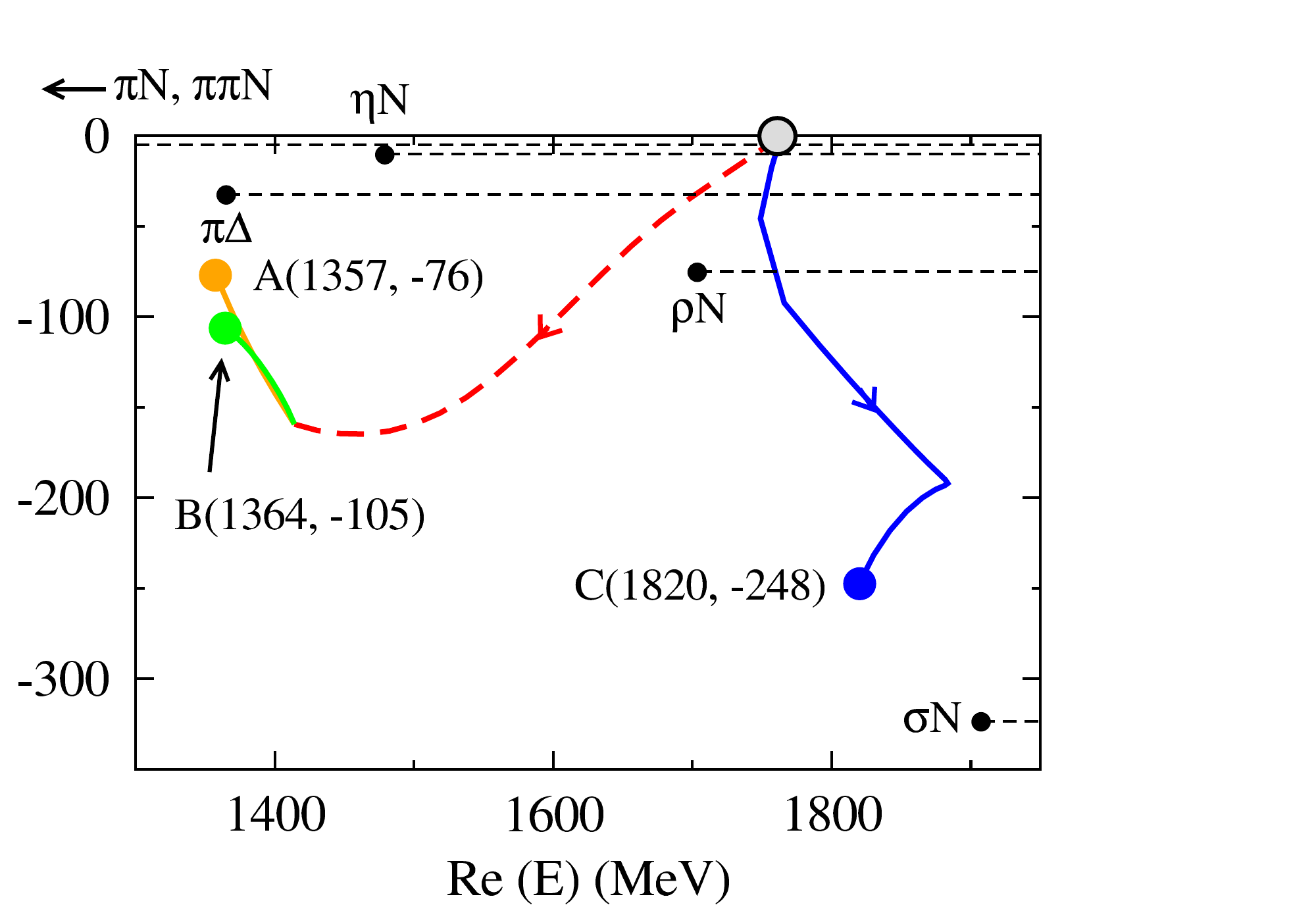}
\caption{The trajectories of the evolution of
three nucleon resonances in $P_{11}$ from the same bare $N^*$ state.
The results are from Ref.~\cite{Suzuki:2009nj}.}
\label{fig:p11-traj}
\end{figure}


\subsubsection{Resonance extractions}

We define resonances as the eigenstates of the Hamiltonian with the outgoing waves being the respective decay channels as is described in Refs.~\cite{Suzuki:2008rp,Suzuki:2010yn}. 
One can then show that the nucleon resonance positions are
the poles $M_R$ of meson-baryon scattering amplitudes as calculated
from Eqs.~(\ref{eq:cct}) and~(\ref{eq:ccv}) on 
the Riemann surface in the complex-$E$ plane.
The coupling of meson-baryon states with the resonances can be
determined by the residues $R_{N^*,MB}$ at the pole positions.
Our procedures for determining $M_R$ and $R_{N^*,MB}$ are further explained in our recent work (see: Refs.~\cite{Suzuki:2008rp,Suzuki:2010yn,Suzuki:2009nj,Kamano:2010ud}).

With our method of analytic continuation into the complex plane~\cite{Suzuki:2008rp,Suzuki:2010yn},
we are able to analyze the dynamical origins of the nucleon resonances within the framework of the
EBAC dynamical coupled-channel model~\cite{Matsuyama:2006rp}.
This was done by examining how the resonance positions move as each of the
coupled-channel  couplings are  systematically switched off.   For example, 
as illustrated in Fig.~\ref{fig:p11-traj} for the $P_{11}$ states,
this exercise revealed that two poles in the Roper region and the next-higher
pole are associated with the same bare state on the Riemann surface.

\subsubsection{Prospects and path forward}

\begin{figure}[Ht]
\includegraphics[clip,height=.3\textheight]{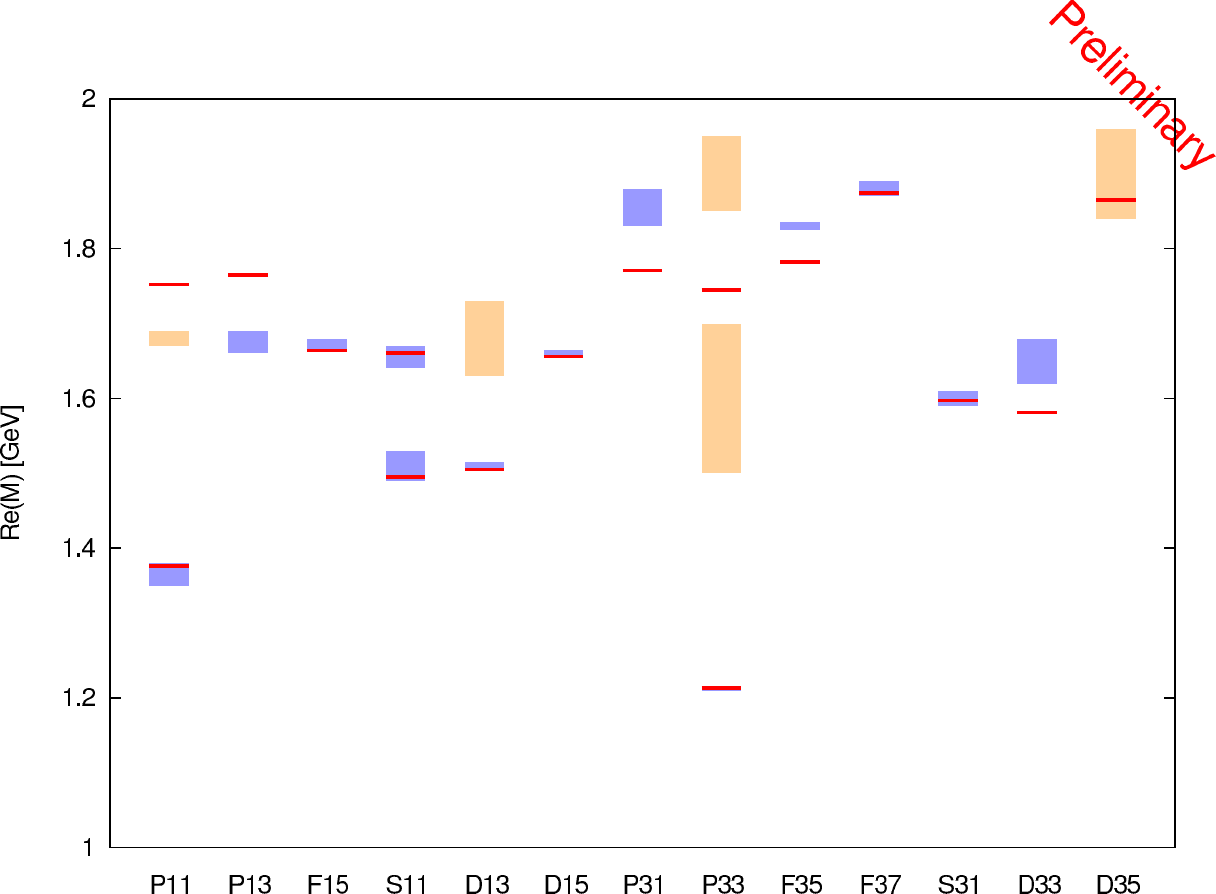}
\caption{Preliminary results (red bars) of the determined $N^*$ spectrum
are compared with 4-star (blue bands) and 3-star (brown bands) states
listed by the Particle Data Group.}
\label{fig:spectrum-1}
\end{figure}

During the developmental stage of the DCC analysis  by the EBAC collaboration
 in 2006-2010,  
the DCC model parameters were determined by separately analyzing 
 the following data sets: $\pi N\rightarrow \pi N$~\cite{JuliaDiaz:2007kz}, 
$\gamma N \rightarrow \pi N$~\cite{JuliaDiaz:2007fa}, $N(e,e'\pi)N$~\cite{JuliaDiaz:2009ww},
$\pi N\rightarrow \pi\pi N$~\cite{Kamano:2008gr}, 
and $\gamma N \rightarrow \pi\pi N$~\cite{Kamano:2009im}. 
The very extensive data on $K\Lambda$ and $K\Sigma$ production, however, were not included in this analysis.
To  afford the highest-precision extraction of nucleon resonances, it behooves us to
perform a combined and simultaneous coupled-channels analysis with all meson production reactions included.

In the summer of 2010 we initiated a comprehensive eight-channel combined analysis of  the world's data that include strange mesons in the final state, 
i.e.~$\pi N, \gamma N \rightarrow \pi N, \eta N, K\Lambda, K\Sigma$.   The EBAC collaboration came to an end on March 31, 2012 and, in its place,  the ANL-Osaka collaboration  has taken over this DCC 
analysis task.
Preliminary results of the full eight-channel combined  analysis of the  excited nucleon spectrum are shown in Fig.~\ref{fig:spectrum-1}. 
We expect to have completed the analysis by early 2013.  The ANL-Osaka analysis will then proceed to
 extract the  $\gamma N \rightarrow N^*$ form factors from the anticipated JLab data on meson
electroproduction, extending the momentum reach to much higher $Q^2$.   Further, we will explore the interpretations of the extracted resonance parameters 
in terms of the available hadron models, such as the Dyson-Schwinger-Equation model, constituent quark model,  and Lattice QCD.
Making these connections with these hadron models is needed to complete the DCC project with conclusive results, 
as is discussed in Refs.\cite{Matsuyama:2006rp,Sato:1996gk,Suzuki:2009nj,Wilson:2011aa}.

\subsection{Future developments \label{Future developments}}

The CLAS collaboration has provided a wealth of data, much of which is still being analyzed.  These rich data sets have impacted and expanded the $N^*$ program, through which reaction models can now be tuned to extract the $\gamma_{v}NN^*$ electrocouplings  for CLAS12 experiments with $Q^2 > 5.0$~GeV$^2$, thereby enabling deeper $N^*$ studies~\cite{Gothe:2011up}.
Preliminary CLAS data on charged-double-pion electroproduction for photon virtualities in the range of  $2.0 < Q^2 < 5.0$~GeV$^2$ have recently become available \cite{Isupov:Hadron}. 
They span the entire $N^*$ excitation region for $W < 2.0$~GeV and  the statistics allow for 115 bins in $W$ and $Q^2$. The data consist of nine one-fold differential cross sections 
 as is shown in Figs~\ref{isochan} and \ref{nsdtback}. The extension of the JM approach to higher $Q^2$ values up to 5.0 GeV$^2$ 
 covering the entire $N^*$ excitation region is in progress and will be completed within two years after the publication of this document.

After the completion of this data analysis, electrocouplings 
of the  $P_{11}(1440)$ and $D_{13}(1520)$ resonances will become available from both the $N\pi$ and $\pi^+\pi^-p$ 
electroproduction channels for $0.2 < Q^2 <  5.0$~GeV$^2$. We will then have
reliable information on the electrocouplings for these two states over a full range of distances that correspond to transitioning, wherein the 
quark degrees of freedom in the resonance structure dominate. 
The studies of  the $N^*$ meson-baryon dressing as described in Refs.~\cite{JuliaDiaz:2006xt,JuliaDiaz:2007fa} strongly suggest a nearly negligible contribution 
from the meson-baryon cloud to the $A_{1/2}$ electrocouplings of the $D_{13}(1520)$ resonance for $Q^2 > 1.5$~GeV$^2$. 
Therefore, theoretical interpretations of already available 
and future CLAS results on $A_{1/2}$ electrocouplings of the $D_{13}(1520)$ resonance are of
particular interest for approaches that are capable of describing the quark content of resonances based on  QCD.

Analysis of the CLAS $\pi^+\pi^-p$ electroproduction data \cite{Isupov:Hadron} within the framework 
of the JM approach will deliver the first information on electrocouplings of most of the high-lying excited proton states 
($M > 1.6$~GeV) for  $2.0 < Q^2 < 5.0$~GeV$^2$. This information will allow us to considerably extend our 
knowledge on how strong
interactions generate excited proton states having different quantum numbers.
 
There will also be analyses of the available and future CLAS results on electrocouplings of all prominent $N^*$ states for  
$Q^2 > 2.0$~GeV$^2$ within the framework
of the Light Cone Sum Rule approach as outlined in  Chapter~\ref{Distribution Amplitudes} that  will constrain the quark-distribution amplitudes of the various
$N^*$ states. Access to the quark-distribution amplitudes in the $N^*$ structure is of particular importance,
since these amplitudes can be evaluated
from QCD employing lattice calculations.

\begin{figure}[h]
\includegraphics[width=.41\textwidth]{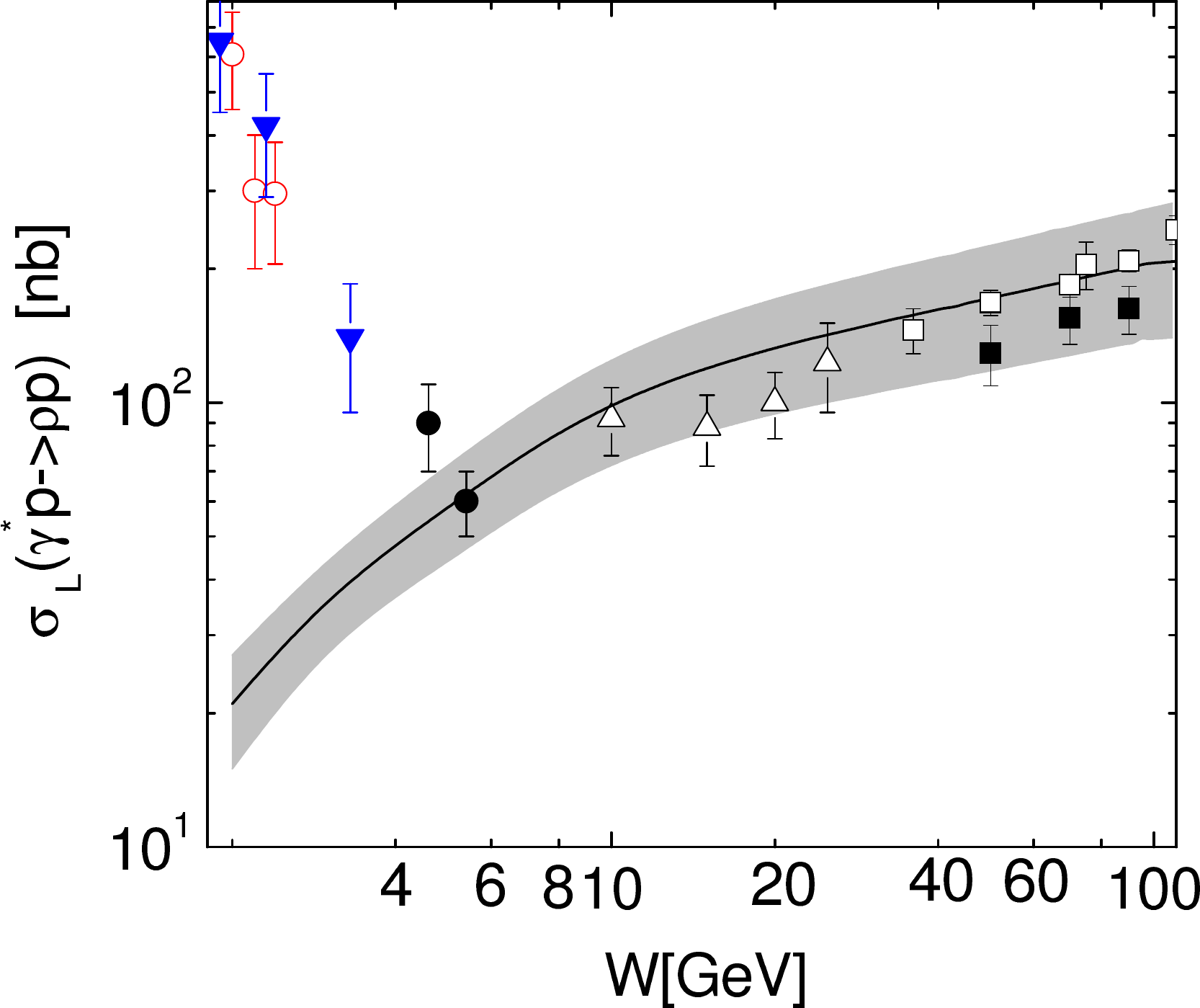}
\includegraphics[width=.41\textwidth]{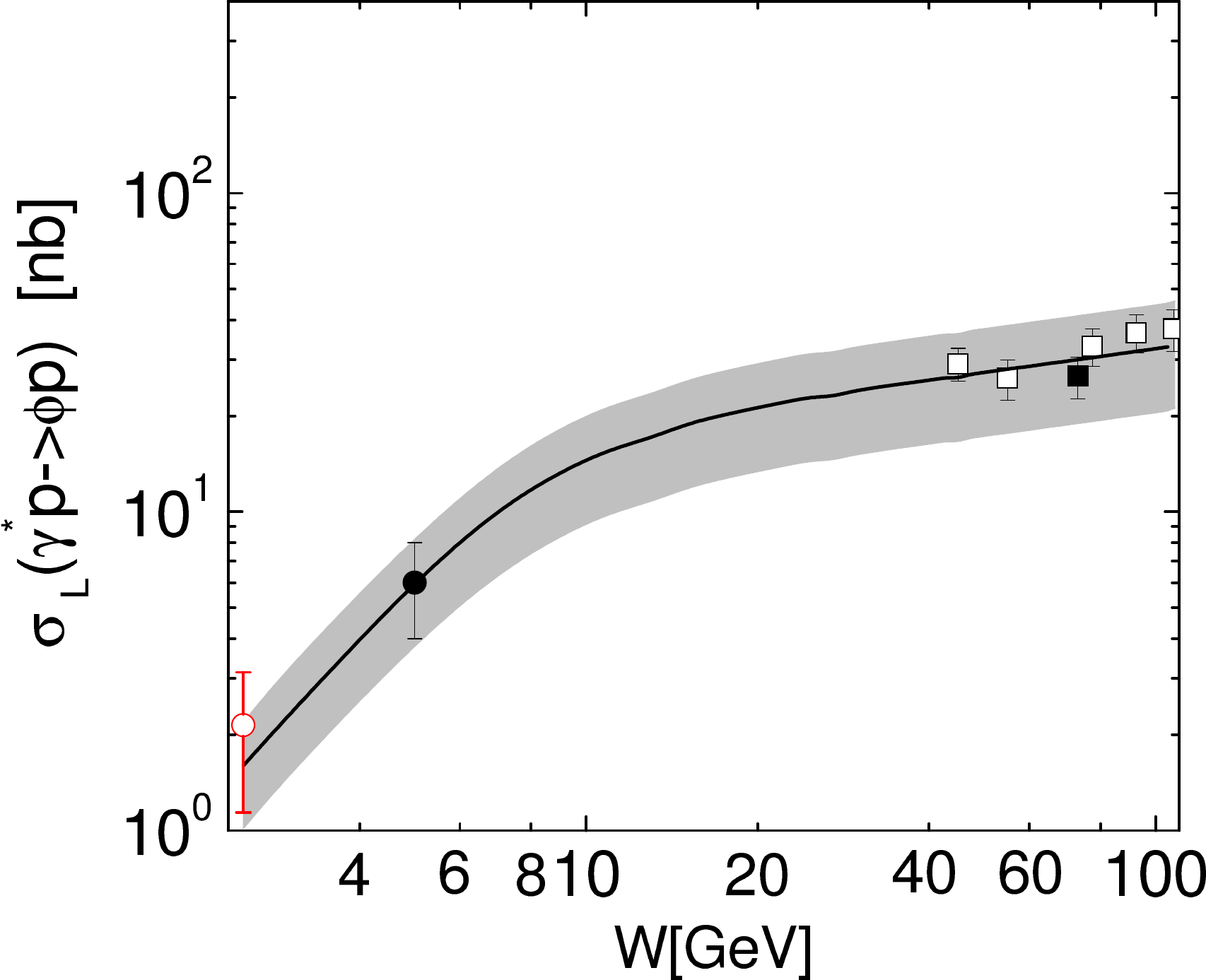}
\caption{Predictions of the longitudinal cross section 
  of $\rho^0$ (left) and $\phi$ (right) production  versus $W$ at $Q^2=4\,{\rm
    GeV}^2$. For references to data see \cite{Goloskokov:2007nt,Goloskokov:2009ia,Goloskokov:2011rd} and references therein.}
\label{rhophi}    
\end{figure}

 Information on the $Q^2$ evolution of non-resonant mechanisms  as obtained from analyses of the CLAS data on single- 
 and charged-double-pion electroproduction at $Q^2 < 5.0$~GeV$^2$  will serve as the starting point for 
 the development of reaction models that will make it possible 
 to determine the  $\gamma_{v}NN^*$ electrocouplings from fitting the anticipated CLAS12 data for $Q^2$ from 5.0 to 12.0 GeV$^2$.

A consistent description 
of a large body of observables in the $N\pi$ exclusive channels achieved within the framework of 
two conceptually different approaches as outlined in Section~\ref{1pisec} and with the success of the JM model in describing of $\pi^+\pi^-p$ electroproduction off
protons all serve to demonstrate that the meson-baryon
degrees of freedom play a significant role  
at photon virtualities of $Q^2 < 5.0$~GeV$^2$. 
Further development to the reaction models
is needed in analyzing these exclusive channels for the anticipated CLAS12 data, where the quark degrees of freedom 
are expected to dominate.  The reaction models for the description of $\pi^+n$ , $\pi^0p$, and $\pi^+\pi^-p$ electroproduction off protons
for $Q^2 > 5.0$~GeV$^2$ should explicitly account for contributions from these quark degrees of freedom.  At present, however,
there is no overarching theory of hadron interactions that will offer any "off-the-shelf" approach at these particular distance scales, 
where the quark degrees of freedom dominate, but are still well inside the regime of the nonperturbative strong
interaction. Given the state of hadronic theory, we are pursuing  a phenomenological way for evaluating the non-resonant mechanisms for the higher-$Q^2$ regime. 
We will explore the possibilities of implementing
different models  that employ quark degrees of freedom  by explicitly  comparing the predictions from these models directly to the data. 
First we will start from  models
that employ handbag diagrams for parameterizing non-resonant single-pion electroproduction and we will then
extend this work for a proper description of $\pi^+\pi^-p$ electroproduction off protons

For the kinematics accessible at the  JLab  energy upgrade, one reaches the 
region where a description of the processes of interest in terms of 
quark degrees of freedom applies. In this case, the calculation of 
cross sections and other observables can be performed within the 
handbag approach, which is based on QCD factorization of the scattering 
amplitudes in hard subprocesses, pion electroproduction off quarks, 
and the generalized parton distributions (GPDs) for $p\to p$ or
$p \to \Delta$ transitions.

In recent years, the data on the electroproduction of vector and pseudoscalar
mesons have been analyzed extensively. In particular, as described in Refs.~\cite{Goloskokov:2007nt,Goloskokov:2009ia,Goloskokov:2011rd},
a systematic analysis of these processes in the kinematical region of large
$Q^2$ ($>3$~GeV$^2$) and $W$ larger than about 4~GeV but having small Bjorken-$x$
(i.e.\ small skewness)  has led to a set of GPDs 
($H, E, \widetilde{H}, H_T, \cdots$) that respect all theoretical
constraints -- polynomiality, positivity, parton distributions ,
and nucleon form factors. These GPDs are also in reasonable agreement with 
moments calculated within lattice QCD \cite{Hagler:2007xi} and with the data on deeply 
virtual Compton scattering in the aforementioned kinematical region \cite{Kroll:2012sm}.
On the other hand, applications in the kinematical region accessible
presently from the 6-GeV JLab data as characterized by rather large values of
Bjorken-$x$ and small $W$, in general, do not lead to agreement with experiment. 
Predictions for $\rho^0$ electroproduction, for instance, fails
by an order of magnitude, whereas for $\phi$ electroproduction it works quite well, as can be 
seen in Fig.~\ref{rhophi}.
For the JLab energy upgrade, one can expect fair agreement between experiment
and predictions for meson electroproduction evaluated from these
set of GPDs~\footnotemark[1]. 

\footnotetext[1]{Tables of predictions for electroproduction of various mesons in this 
kinematical region can be obtained from the authors of \cite{Goloskokov:2011rd}
upon request} 

To describe the electroproduction of nucleon resonances in meson-baryon intermediate states in the $\pi^+\pi^-p$ exclusive electroproduction, 
one needs the $p\to N^*$
transition GPDs. In principle, these GPDs are new unknown functions. 
Therefore, straightforward predictions for reactions like 
$\gamma^* p\to \pi N^*$ are not possible at present. In the large 
$N_c$ limit, however, one can at least relate the $p\to\Delta^+$ GPDs 
to the flavor diagonal $p\to p$ ones  since the nucleon and the $\Delta$ 
are eigenstates of the same object, the chiral soliton \cite{Goeke:2001tz,Belitsky:2005qn}. 
The proton-proton GPDs always occur in the isovector combination 
$F^{(3)}=F^u-F^d$ where $F$ is a proton-proton GPD. With the help of flavor 
symmetry one can further relate the $p\to \Delta^+$ GPDs to all other 
octet-decuplet transitions. Using these theoretical considerations, the 
observables for $\gamma^*p\to\pi N^*$ can be estimated. One should be aware,
however, that the quality of the large $N_c$ and $SU(3)_F$ relations
are unknown; corrections of the order of $20$ to $30\%$ are to be expected.
One also should bear in mind that pions electroproduced by
transversely-polarized virtual photons must further be taken into account as has 
been shown in Refs.~\cite{Goloskokov:2009ia,Goloskokov:2011rd}. Within the handbag approach,  the contributions
from such photons are related to the transversity (helicity-flip) GPDs.
Despite this complication, an estimate of hard exclusive resonance production
seems feasible.

A well-developed program on resonance studies at high photon virtualities~\cite{Gothe:clas12} will allow us to determine 
electrocouplings of several high-lying $N^*$ states with dominant $N\pi\pi$ decays 
(see the Table~\ref{bf1pi2pi}) from the data on charged-double-pion electroproduction channel. However, reliable extraction of these 
electrocouplings for these states should be supported by independent analyses of other exclusive electroproduction channels 
having different non-resonant mechanisms.
 The $\eta p$ and $K\Lambda$ electro\-production channels may well improve our knowledge 
 on electro\-couplings of  the isospin 1/2 $P_{13}(1720)$ state due to  isospin filtering in these exclusive channels. 
 The studies of $K\Sigma$ and $\eta\pi N$ electroproduction may further offer 
 access to the electrocouplings of the
$D_{33}(1700)$ and  $F_{35}(1905)$ resonances. More detailed studies on the feasibility of incorporating these 
additional exclusive channels for evaluating the electrocouplings of high-lying resonances are, in any case, a clear and present need.

\section{$N^\ast$ Physics from Lattice QCD \label{lqcd1}}

\subsection{Introduction}

 Quantum ChromoDynamics (QCD), when combined with the electroweak interactions, underlies all of nuclear physics, from the spectrum and structure of hadrons to the most complex nuclear reactions. The underlying symmetries that are the basis of QCD were established long ago. Under very modest assumptions, these symmetries predict a rich and exotic spectrum of QCD bound states, few of which have been observed experimentally. While QCD predicts that quarks and gluons are the basic building blocks of nuclear matter, the rich structure that is exhibited by matter suggests there are underlying collective degrees of freedom. Experiments at nuclear and high-energy physics laboratories around the world measure the properties of matter with the aim to determine its underlying structure. Several such new experiments worldwide are under construction, such as the 12-GeV upgrade at Jefferson Lab's electron accelerator, its existing the experimental halls, as well as the new Hall~D.

To provide a theoretical determination and interpretation of the spectrum, \textit{ab initio} computations within lattice QCD have been used. Historically, the calculation of the masses of the lowest-lying states, for both baryons and mesons, has been a benchmark calculation of this discretized, finite-volume computational approach, where the aim is well-understood control over the various systematic errors that enter into a calculation; for a recent review, see~\cite{Lin:2011ti}. However, there is now increasing effort aimed at calculating the excited states of the theory, with several groups presenting investigations of the low-lying excited baryon spectrum, using a variety of discretizations, numbers of quark flavors, interpolating operators, and fitting methodologies (Refs.~\cite{Mahbub:2010me,Mahbub:2010jz,Engel:2010my,Mathur:2003zf}). Some aspects of these calculations remain unresolved and are the subject of intense effort, notably the ordering of the Roper resonance in the low-lying nucleon spectrum.

The Hadron Spectrum Collaboration, involving the Lattice Group at Jefferson Lab, Carnegie Mellon University, University of Maryland, University of Washington, and Trinity College (Dublin), is now several years into its program to compute the high-lying excited- state spectrum of QCD, as well as their (excited-state) electromagnetic transition form factors up to $Q^2 \sim 10\mbox{ GeV}^2$. This program has been utilizing ``anisotropic'' lattices, with finer temporal than spatial resolution, enabling the hadron correlation functions to be observed at short temporal distances and hence many energy levels to be extracted~\cite{Edwards:2008ja,Lin:2008pr}. Recent advances suggest that there is a rich spectrum of mesons and baryons, beyond what is seen experimentally. In fact, the HSC's calculation of excited spectra, as well as recent successes with GPUs, were featured in {\em Selected FY10 Accomplishments in Nuclear Theory in the FY12 Congressional Budget Request}.

\begin{figure}[t]
\begin{center}
\includegraphics[width=0.95\textwidth,angle=0]{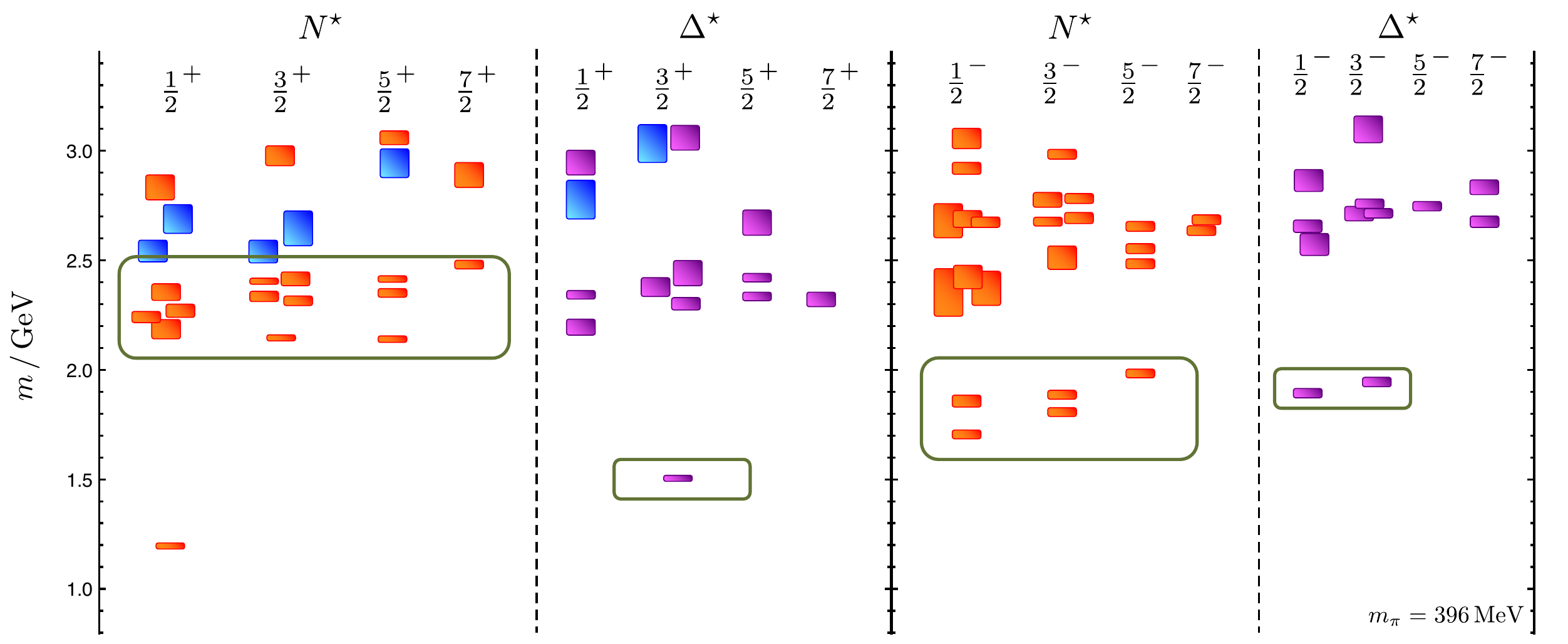} 
\end{center}
\caption{Results from Ref.~\cite{Edwards:2011jj,Dudek:2012ag} showing the spin-identified spectrum of Nucleons and Deltas from the
  lattices at $m_\pi = 396~{\rm MeV}$. The spectrum of states within the rectangles are identifiable as admixtures of  $SU(6)\otimes O(3)$ representations and a counting of levels, along with the spectral overlaps of interpolating fields onto the states, that is consistent with the nonrelativistic $qqq$ constituent quark model. In addition, there are additional states of hybrid character, shown in blue, that have relatively large overlap onto operators which sample gluonic excitations. The pattern of these positive-parity ``hybrid'' baryons is compatible with a color-octet gluonic excitation having $J^P=1^+$.
\label{fig:spin_840}
}
\end{figure}

\begin{figure*}[t]
 \centering
\includegraphics[width=0.99\textwidth,angle=0]{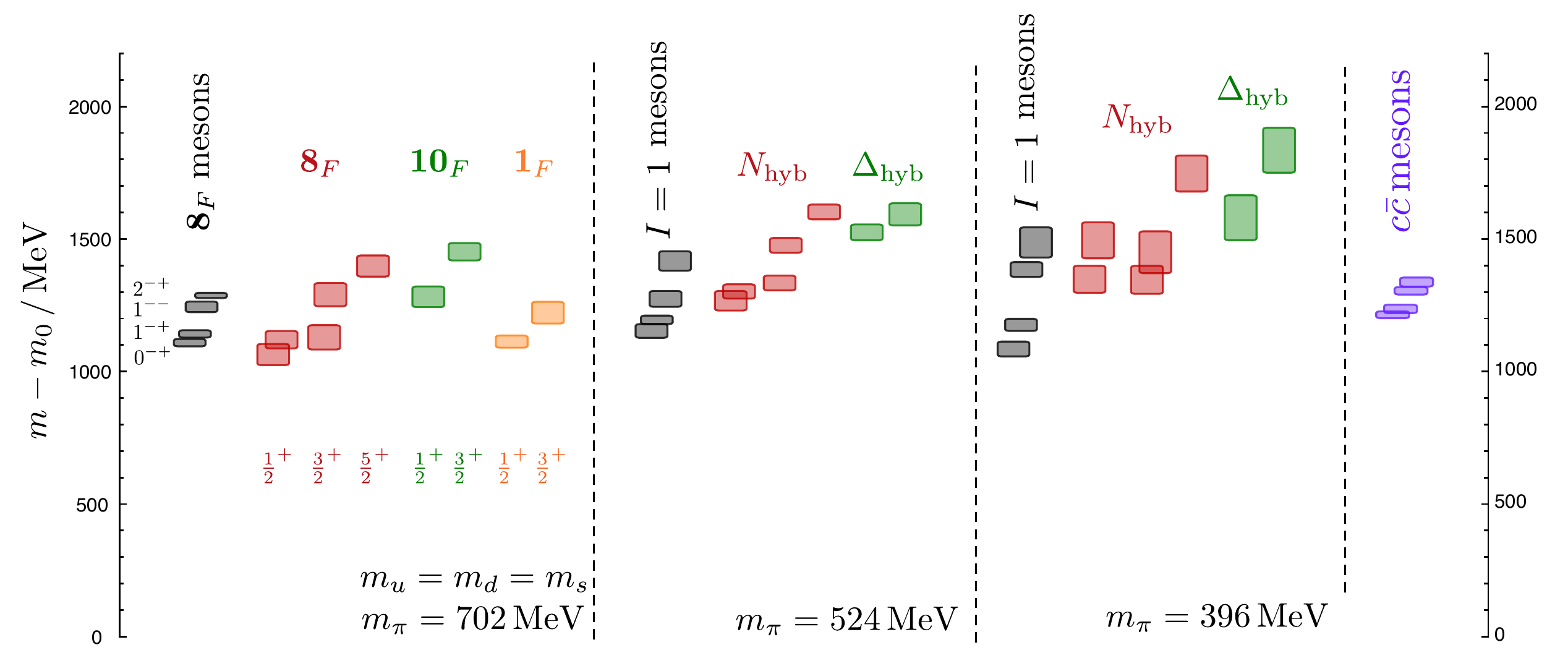}
\caption{Results from Ref.~\cite{Edwards:2011jj,Dudek:2012ag,Liu:2012ze} showing the spectrum of hybrid mesons and baryons for three light quark masses. Mass scale is $m-m_\rho$ for light quark mesons, $m-m_{J/\Psi}$, and $m-m_N$ for baryons to approximately subtract the effect of differing numbers of quarks. \label{fig:mesons}}
\end{figure*}

\subsection{Spectrum} 

The development of new operator constructions that follow from continuum symmetry constructions has allowed, for the first time, the reliable identification of the spin and masses of the single-particle spectrum at a statistical precision at or below about 1\%. In particular, the excited spectrum of isovector as well as isoscalar mesons (Refs.~\cite{Dudek:2009qf,Dudek:2010wm,Dudek:2011tt}) shows a pattern of states, some of which are familiar from the $q\bar{q}$ constituent quark model, with up to total spin $J=4$ and arranged into corresponding multiplets. In addition, there are indications of a rich spectrum of exotic $J^{PC}$ states, as well as a pattern of states interpretable as non-exotic hybrids~\cite{Dudek:2011bn}. The pattern of these multiplets of states, as well as their relative separation in energy, suggest a phenomenology of constituent quarks coupled with effective gluonic degrees of freedom. In particular, the pattern of these exotic and non-exotic hybrid states appears to be consistent with a bag-model description and inconsistent with a flux-tube model~\cite{Dudek:2011bn}.

Recently, this lattice program has been extended into the baryon spectrum, revealing for the first time, the excited-state single-particle spectrum of nucleons, Deltas, Lambdas, Sigmas, Xis, and Omegas along with their total spin up to $J=\tfrac{7}{2}$ in both positive and negative parity~\cite{Edwards:2011jj,Dudek:2012ag,Edwards:2012fx}. 
The results for the nucleon and Delta channels at the lightest pion-mass ensemble are shown in Fig.~\ref{fig:spin_840}. There was found a high multiplicity of levels spanning across $J^P$, which is consistent with $SU(6) \otimes O(3)$ multiplet counting, and hence with that of the non-relativistic $qqq$ constituent quark model. In particular, the counting of levels in the low-lying negative-parity sectors are consistent with the non-relativistic quark model and with the observed experimental states~\cite{Nakamura:2010zzi}. 
The spectrum observed in the first-excited positive-parity sector is also consistent in counting with the quark model, but the comparison with experiment is less clear, with the quark model predicting more states than are observed experimentally, spurring phenomenological investigations to explain the discrepancies (e.g., see Refs.~\cite{Nakamura:2010zzi,Isgur:1978wd,Capstick:1986bm,Anselmino:1992vg,Glozman:1995fu,Capstick:2000qj,Goity:2003ab}).

It was found that there is significant mixing among each of the allowed multiplets, including the $\mathbf{20}$-plet that is present in the non-relativistic $qqq$ quark model but does not appear in quark-diquark models~\cite{Anselmino:1992vg} (see in particular Ref.~\cite{Lichtenberg:1967zz}). These results lend credence to the assertion that there is no ``freezing'' of degrees of freedom with respect to those of the non-relativistic quark model.  These qualitative features of the calculated spectrum extend across all three of the quark-mass ensembles studied.  Furthermore, no evidence was found for the emergence of parity-doubling in the spectrum~\cite{Glozman:1999tk}.

In addition, it was found that there are states, above the first positive parity band of states, which have large overlap onto interpolating fields which transform like chromomagnetic fields (color octet, $J^{PC}=1^{+-}$). The form of the operators suggest that within a hybrid baryon the three quarks are arranged in a color octet with the chromomagnetic gluonic excitation making the state an overall color singlet. The low-lying hybrid states overlap strongly onto the ``non-relativistic" subset of the hybrid interpolators, those constructed using only upper components of Dirac spinors. We can interpret this as suggesting that the quarks within the lightest hybrid baryons are dominantly in $S$-waves. In light of this it seems unlikely that the Roper is dominantly of hybrid character as has been speculated in the past. 

In Fig.~\ref{fig:mesons}, we present the spectrum of light and charm quark mesons and light quark baryons identified as ``hybrid'' in origin. To isolate the energy scale associated with the chromomagnetic gluonic excitation, the energies splittings from a constituent quark mass scale are shown.
The structure of the operators that interpolate the hybrid states from the vacuum, along with the observation that the energy scale of gluonic excitations appears to be common for mesons and baryons, provides evidence that the gluonic excitation sector in QCD may turn out to be relatively simple. We suggest that a chromomagnetic excitation ($J^{PC}=1^{+-}$) is lightest with an energy scale in the region of 1.3 GeV.

It was argued that the extracted $N$ and $\Delta$ spectrum can be interpreted in terms of single-hadron states, and based on investigations in the meson sector~\cite{Dudek:2010wm} and initial investigations of the baryon sector at a larger volume~\cite{Edwards:2011jj}, little evidence was found for multi-hadron states.  To study multi-particle states, and hence the resonant nature of excited states, operator constructions with a larger number of fermion fields are needed. Such constructions are in progress~\cite{Thomas:2011rh,Dudek:2012gj}, and it is believed that the addition of these operators will lead to a denser spectrum of states. With suitable understanding of the discrete energy spectrum of the system, the L\"uscher formalism~\cite{Luscher:1991cf} and its inelastic extensions (for example, see Ref.~\cite{Lage:2009zv}) can be used to extract the energy dependent phase shift for a resonant system, such as has been performed for the $I=1$ $\rho$ system~\cite{Dudek:2012xn}. The energy of the resonant state is determined from the energy dependence of the phase shift. It is this resonant energy that is suitable for chiral extrapolations.
Suitably large lattice volumes and smaller pion masses are needed to adequately control the systematic uncertainties in these calculations.

\begin{figure}
\includegraphics[height=.22\textheight]{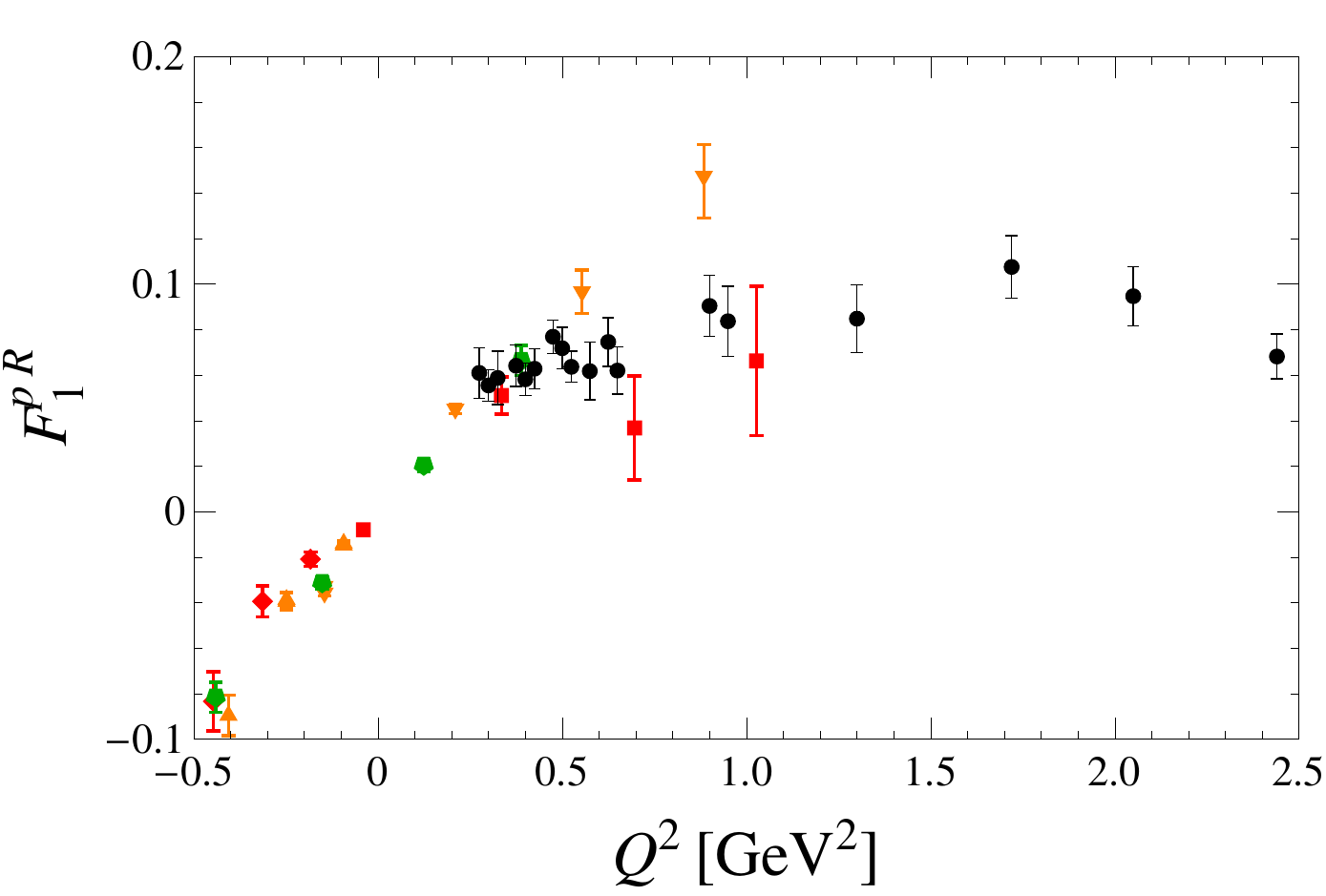}
\includegraphics[height=.22\textheight]{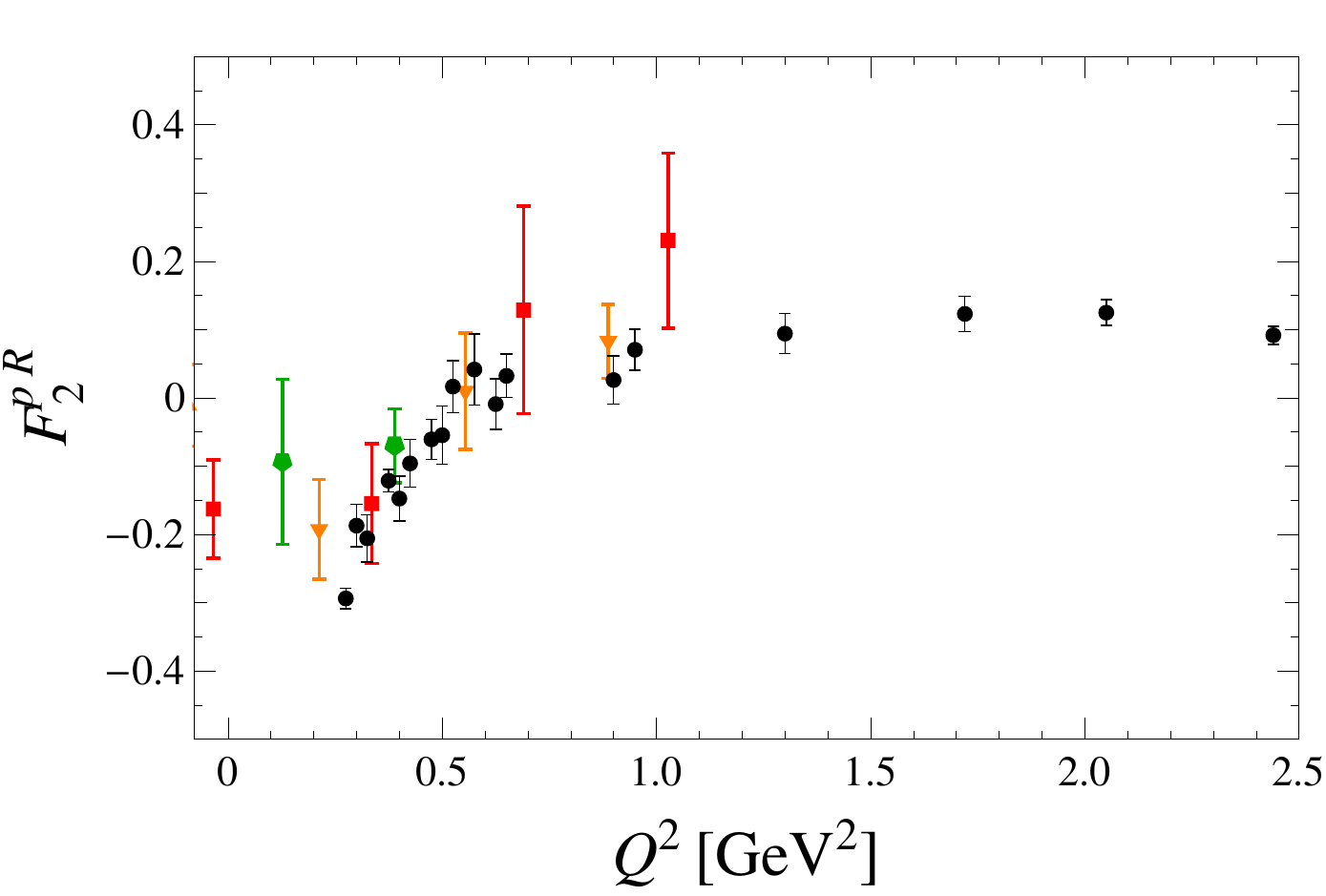}
\caption{\label{fig:F12pR-dyn}
Proton-Roper transition form factor $F_1^{pR}(Q^2)$ (left) and $F_2^{pR}(Q^2)$ (right) on the $N_f=2+1$ anisotropic lattices with $M_\pi\approx 390$, 450, 875~MeV whose volumes are 3, 2.5, 2.5~fm, respectively.
}
\end{figure}

\subsection{Electromagnetic transition form factors}

 The measurement of the excited-to-ground state radiative transition form factors in the baryon sector provides a probe into the internal structure of hadrons. Analytically, these transition form factors can be expressed in terms of matrix elements between states $\langle N(p_f)|V_\mu(q)|N^\ast(p_i)\rangle$ where $V_\mu$ is a vector (or possibly axial-vector) current with some four-momentum $q = p_f - p_i$ between the final ($p_f$) and initial ($p_i$) states. This matrix element can be related to the usual form factors $F^\ast_1(q^2)$ and $F^\ast_2(q^2)$. However, the exact meaning as to the initial state $|N^\ast\rangle$ is the source of some ambiguity since in general it is a resonance. In particular, how is the electromagnetic decay disentangled from that of some $N\pi$ hadronic contribution?

Finite-volume lattice-QCD calculations are formulated in Euclidean space, and as such, one does not directly observe the imaginary part of the pole of a resonant state. However, the information is encoded in the volume and energy dependence of excited levels in the spectrum. L\"uscher's formalism~\cite{Luscher:1991cf} and its many generalizations show how to relate the infinite-volume energy-dependent phase shifts in resonant scattering to the energy dependence of levels determined in a continuous but finite-volume box in Euclidean space. In addition, infinite-volume matrix elements can be related to those in finite-volume~\cite{Kim:2005gf} up to a factor which can be determined from the derivative of the phase shift.

For the determination of transition form factors, what all this means in practice is that one must determine the excited-state transition matrix element from each excited level in the resonant region of a state, down to the ground state. The excited levels and the ground state might each have some non-zero momentum, arising in some $Q^2$ dependence. In finite volume, the transition form factors are both $Q^2$ and energy dependent, the latter coming from the discrete energies of the states within the resonant region. The infinite-volume form factors are related to these finite-volume form factors via the derivative of the phase shift as well as another kinematic function. Sitting close to the resonant energy, in the large volume limit the form factors become independent of the energy as expected.

The determination of transition form factors for highly excited states was first done in the charmonium sector with quenched QCD~\cite{Dudek:2007wv,Dudek:2009kk}. Crucial to these calculations was the use of a large basis of non-local operators to form the optimal projection onto each excited level. In a quenched theory, the excited charmonium states are stable and have no hadronic decays, thus there is no correction factor.

The determination of the electromagnetic transitions in light-quark baryons will eventually require the determination of the transition matrix elements from multiple excited levels in the resonance regime, the latter determined through the spectrum calculations in the previous section. However, as a first step, the $Q^2$ dependence of transition form factors between the ground and first-excited state can be investigated within a limited basis. These first calculations of the $F^{pR}_{1,2}(Q^2)$ excited transition levels, in Refs.~\cite{Lin:2008qv,Lin:2011da} already have shown many interesting features.

The first calculations of the $P_{11}\rightarrow \gamma N$ transition form factors were performed a few years ago using the quenched approximation~\cite{Lin:2008qv}. Since then, these calculations have been extended to full QCD with two light quarks and one strange quark ($N_f=2+1$) using the same anisotropic lattice ensembles as for the spectrum calculations. Preliminary results~\cite{Lin:2011da} of the $Q^2$ dependence of the first-excited nucleon (the Roper) to the ground-state proton, $F_{1,2}^{pR}$, are shown in Fig.~\ref{fig:F12pR-dyn}. These results focus on the low-$Q^2$ region. At the unphysical pion masses used, some points are in the time-like region. What is significant in these calculations with full-QCD lattice ensembles is that the sign of $F_2$ at low $Q^2$ has flipped compared to the quenched result, which had relatively mild $Q^2$ dependence at similar pion masses. These results suggest that at low $Q^2$ the pion-cloud dynamics are significant in full QCD.

\begin{figure}
\includegraphics[width=0.6\textwidth]{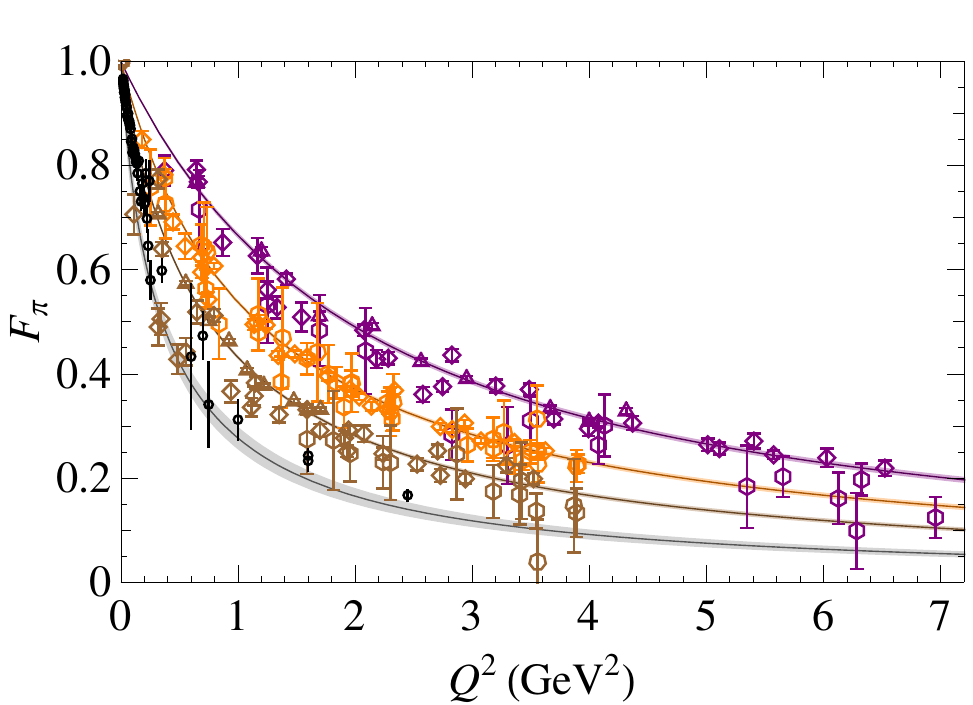}
\caption{\label{fig:Fpi}
Pion form factor utilizing an extended basis of smearing functions to increase the range of $Q^2$ with multiple pion masses at 580, 875, 1350~MeV. The experimental points are shown as (black) circles while the lowest gray band is the extrapolation to the physical pion mass using lattice points from Ref.~\cite{Lin:2011sa}.
}
\end{figure}

The results so far are very encouraging, and the prospects are quite good for extending these calculations. The use of the larger operator basis employed in the spectrum calculations, supplemented with multi-particle operators, and including the correction factors from the resonant structure contained in phase shifts, should allow for the determination of multiple excited-level transition form factors up to about $Q^2\approx 3\mbox{ GeV}^2$.

\subsection{Form factors at $Q^2\approx 6\mbox{ GeV}^2$} 

The traditional steps in a lattice form-factor calculation involve choosing suitable creation and annihilation operators with the quantum numbers of interest, and typically where the quark fields are spatially smeared so as to optimize overlap with the state of interest, often the ground state. These smearing parameters are typically chosen to optimize the overlap of a hadron at rest or at low momentum. As the momentum is increased, the overlap of the boosted operator with the desired state in flight becomes small and statistically noisy. One method to achieve high $Q^2$ is to decrease the quark smearing, which has the effect of increasing overlap onto many excited states. By choosing a suitably large basis of smearing, one can then project onto the desired excited state at high(er) momentum.
This technique can extend the range of $Q^2$ in form-factor calculations until lattice discretization effects become dominant. An earlier version of this technique (with smaller basis) was used for a quenched calculation of the Roper transition form factor reaching about $6\mbox{ GeV}^2$~\cite{Lin:2008gv,Lin:2008qv}.
Figure~\ref{fig:Fpi} shows an example from $N_f=2+1$ at 580, 875, 1350~MeV pion masses using extended basis to extract pion form factors with $Q^2$ reaching nearly 7~GeV$^2$~\cite{Lin:2011sa} for the highest-mass ensemble. The extrapolated form factor at the physical pion mass shows reasonable agreement with JLab precision measurements. Future attempts will focus on decreasing the pion masses and exploring $Q^2$-dependence of pion form factors for yet higher $Q^2$.

As before, these form-factor calculations need to be extended to use a larger operator basis of single and multi-particle operators to overlap with the levels within the resonant region of the excited state, say the Roper. These operator constructions are suitable for projecting onto excited states with high momentum, as demonstrated in Ref.~\cite{Thomas:2011rh}. Future work will apply these techniques to form-factor calculations.

\subsection{Form factors at high $Q^2\gg 10\mbox{ GeV}^2$}

 At very high $Q^2$, lattice discretization effects can become quite large. A costly method to control these effects is to go to much smaller lattice spacing, basically $a \sim 1/Q$. An alternative method that was been devised long ago is to use renormalization-group techniques~\cite{Caracciolo:1994ud}, and in particular, step-scaling techniques introduced by the ALPHA collaboration. The step-scaling method was initially applied to compute the QCD running coupling and quark masses. The technique was later extended to handle heavy-quark masses with a relativistic action~\cite{Guagnelli:2002jd,Guazzini:2007ja}. The physical insight is that the heavy-quark mass dependence of ratios of observables is expected to be milder than the observable itself. For form factors, the role of the large heavy-quark mass scale is now played by the large momentum scale $Q$. Basically, the idea is to construct ratios of observables (form factors) such that the overall $Q^2$ dependence is mild, and that suitable products of these ratios, evaluated at different lattice sizes and spacings, can be extrapolated to equivalent results at large volume and fine lattice spacing. The desired form factor is extracted from the ratios.

The technique, only briefly sketched here, is being used now in a USQCD lattice-QCD proposal by D.~Renner (Ref.~\cite{Renner:2011}) to compute the pion form-factor at large $Q^2$, and the technique is briefly discussed in Ref.~\cite{Lin:2011sa}. In principle, the same technique can be used to compute excited-state transition form factors, and although feasibility has yet to be established, it seems worth further investigation.

\subsection{Outlook} 

There has been considerable recent progress in the determination of the highly excited spectrum of QCD using lattice techniques. While at unphysically large pion masses and small lattice volumes, already some qualitative pictures of the spectrum of mesons and baryons is obtained. With the inclusion of multi-hadron operators, the outlook is quite promising for the determination of the excited spectrum of QCD. Anisotropic lattice configurations with several volumes are available now for pion masses down 230~MeV. Thus, it seems quite feasible to discern the resonant structure for at least a few low-lying states of mesons and baryons, of course within some systematic uncertainties, in the two-year timeframe. One of the more open questions is how to properly handle multi-channel decays which becomes more prevalent for higher-lying states. Some theoretical work has already been done using coupled-channel methods, but more work is needed and welcomed.

With the spectrum in hand, it is fairly straightforward to determine electromagnetic transition form factors for the lowest few levels of $N^\ast$, and up to some moderate $Q^2$ of a few GeV$^2$, in the two-year time-frame. Baryon form factors will probably continue to drop purely disconnected terms from the current insertion. Meson transition form factors, namely an exotic to non-exotic meson will be the first target in the short time-frame (less than two years), with the aim to determine photo-couplings. It might well be possible that with the new baryon operator techniques developed, the transition form factors can be extracted to $Q^2\approx 6$~GeV$^2$. Going to an isotropic lattice with a small lattice spacing, it seems feasible to reach higher $Q^2$, say 10~GeV$^2$, and this could be available in less than five years. To reach $Q^2\gg 10$~GeV$^2$ will probably require step-scaling techniques. The high-$Q^2$ limit is of considerable interest since it allows for direct comparisons with perturbative methods.

\newcommand{\lsim}{\mathrel{\rlap{\lower4pt\hbox{\hskip0pt$\sim$}}
\raise1pt\hbox{$<$}}}           
\newcommand{\gsim}{\mathrel{\rlap{\lower4pt\hbox{\hskip0pt$\sim$}}
\raise1pt\hbox{$>$}}}           

\section{Illuminating the Matter of Light-Quark Hadrons \label{DSE}}

\subsection{Heart of the problem}
Quantum chromodynamics (QCD) is the strong-interaction part of the Standard Model of Particle Physics.  Solving this theory presents a fundamental problem that is unique in the history of science.  Never before have we been confronted by a theory whose elementary excitations are not those degrees-of-freedom readily accessible through experiment; i.e., whose elementary excitations are \emph{confined}.  Moreover, there are numerous reasons to believe that QCD generates forces which are so strong that less-than 2\% of a nucleon's mass can be attributed to the so-called current-quark masses that appear in QCD's Lagrangian; viz., forces capable of generating mass \emph{from nothing} (see Sec.\,\ref{dcsb}).  This is the phenomenon known as dynamical chiral symmetry breaking (DCSB).  Elucidating the observable predictions that follow from QCD is basic to drawing the map that explains how the Universe is constructed.

The need to determine the essential nature of light-quark confinement and dynamical chiral symmetry breaking (DCSB), and to understand nucleon structure and spectroscopy in terms of QCD's elementary degrees of freedom, are two of the basic motivations for an upgraded JLab facility.  In addressing these questions one is confronted with the challenge of elucidating the role of quarks and gluons in hadrons and nuclei.   Neither confinement nor DCSB is apparent in QCD's Lagrangian and yet they play the dominant role in determining the observable characteristics of real-world QCD.  The physics of hadronic matter is ruled by \emph{emergent phenomena}, such as these, which can only be elucidated and understood through the use of nonperturbative methods in quantum field theory.  This is both the greatest novelty and the greatest challenge within the Standard Model.  Essentially new ways and means must be found in order to explain precisely via mathematics the observable content of QCD.

Bridging the gap between QCD and the observed properties of hadrons is a key problem in modern science.  The international effort focused on the physics of excited nucleons is at the heart of this program.  It addresses the questions: Which hadron states and resonances are produced by QCD, and how are they constituted?  The $N^\ast$ program therefore stands alongside the search for hybrid and exotic mesons and baryons as an integral part of the search for an understanding of the strongly interacting piece of the Standard Model.

\subsection{Confinement}

Regarding confinement, little is known and much is misapprehended.  It is therefore important to state clearly that the static potential measured in numerical simulations of quenched lattice-QCD is not related in any known way to the question of light-quark confinement.  It is a basic feature of QCD that light-quark creation and annihilation effects are fundamentally nonperturbative; and hence it is impossible in principle to compute a potential between two light quarks \cite{Bali:2005fu,Chang:2009ae}.  Thus, in discussing the physics of light-quarks, linearly rising potentials, flux-tube models, etc., have no connection with nor justification via QCD.

\begin{figure}[t]
\vspace*{9ex}

\hspace*{-1ex}\begin{minipage}[t]{1.0\textwidth}
\begin{minipage}[t]{1.0\linewidth}
\leftline{\includegraphics[clip,width=0.53\linewidth]{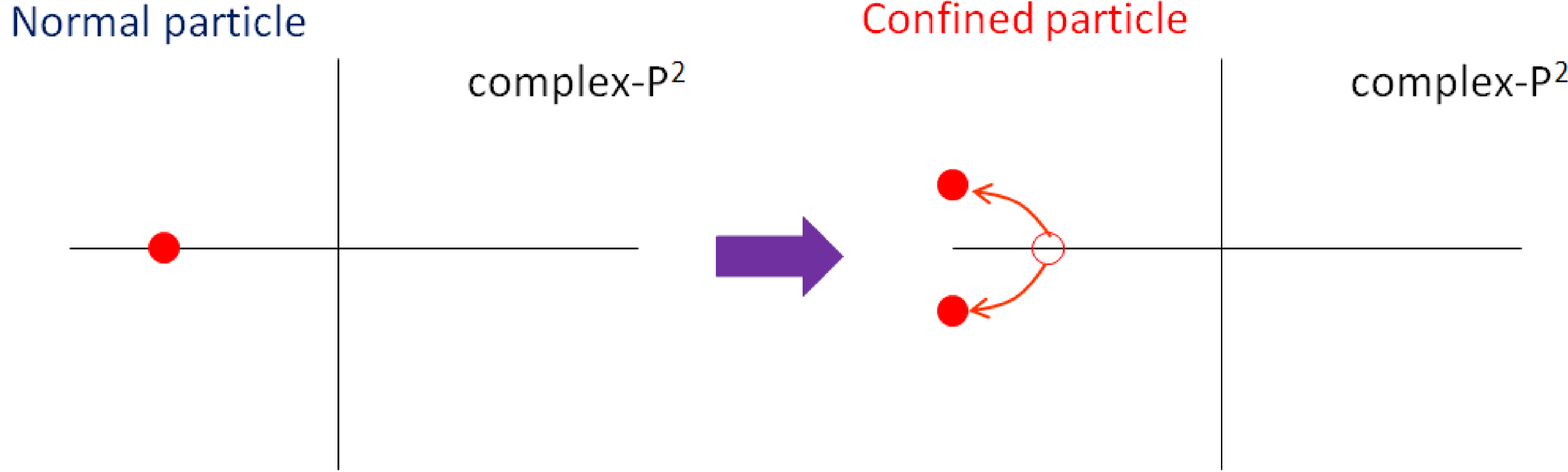}}
\end{minipage}\vspace*{-22ex}

\begin{minipage}[t]{1.0\textwidth}
\rightline{\includegraphics[clip,width=0.45\linewidth]{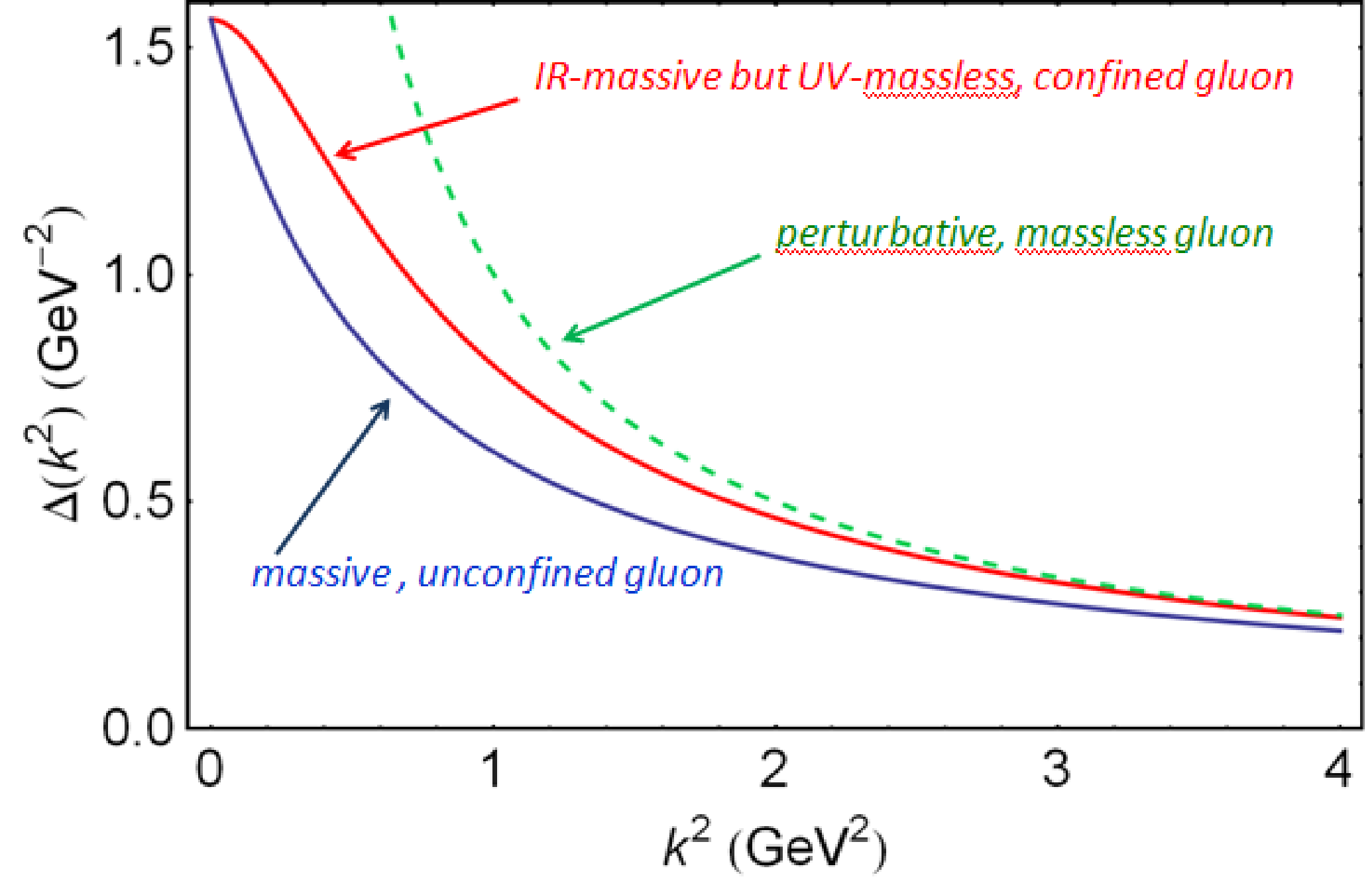}}
\end{minipage}\vspace*{3ex}
\end{minipage}
\caption{\label{fig:confined} \emph{Left panel} -- An observable particle is associated with a pole at timelike-$P^2$, which becomes a branch point if, e.g., the particle is dressed by photons.
\emph{Middle panel} -- When the dressing interaction is confining, the real-axis mass-pole splits, moving into pairs of complex conjugate singularities.  No mass-shell can be associated with a particle whose propagator exhibits such singularity structure. The imaginary part of the smallest magnitude singularity is a mass-scale, $\mu_\sigma$, whose inverse, $d_\sigma=1/\mu_\sigma$ is a measure of the dressed-parton's fragmentation length.
\emph{Right panel} -- $\Delta(k^2)$, the function that describes dressing of a Landau-gauge gluon propagator, plotted for three distinct cases.
A bare gluon is described by $\Delta(k^2) = 1/k^2$ (the dashed line), which is convex on $k^2\in (0,\infty)$.  Such a propagator has a representation in terms of a non-negative spectral density.
In some theories, interactions generate a mass in the transverse part of the gauge-boson propagator, so that $\Delta(k^2) = 1/(k^2+m_g^2)$, which can also be represented in terms of a non-negative spectral density.
In QCD, however, self-interactions generate a momentum-dependent mass for the gluon, which is large at infrared momenta but vanishes in the ultraviolet \protect\cite{Boucaud:2011ug}.  This is illustrated by the curve labeled ``IR-massive but UV-massless.''  With the generation of a mass-\emph{function}, $\Delta(k^2)$ exhibits an inflexion point and hence cannot be expressed in terms of a non-negative spectral density \cite{Roberts:2007ji}.
}
\end{figure}

A perspective on confinement drawn in quantum field theory was laid out in Ref.\,\cite{Krein:1990sf} and exemplified in Sec.~2 of Ref.\,\cite{Roberts:2007ji}.  It draws upon a long list of sources; e.g., Refs.\,\cite{Gribov:1999ui,Munczek:1983dx,Stingl:1983pt,Cahill:1988zi}, and, expressed simply, relates confinement to the analytic properties of QCD's Schwinger functions, which are often called Euclidean-space Green functions or propagators and vertices.  For example, one reads from the reconstruction theorem \cite{Streater:1989vi,GJ81} that the only Schwinger functions which can be associated with expectation values in the Hilbert space of observables; namely, the set of measurable expectation values, are those that satisfy the axiom of reflection positivity.  This is an extremely tight constraint whose full implications have not yet been elucidated.

There is a deep mathematical background to this perspective.  However, for a two-point function; i.e., a propagator, it means that a detectable particle is associated with the propagator only if there exists a non-negative spectral density in terms of which the propagator can be expressed.  No function with an inflexion point can be written in this way.  This is readily illustrated and Fig.\,\ref{fig:confined} serves that purpose.  The simple pole of an observable particle produces a propagator that is a monotonically-decreasing convex function, whereas the evolution depicted in the middle-panel of Fig.\,\ref{fig:confined} is manifest in the propagator as the appearance of an inflexion point at $P^2 > 0$.  To complete the illustration, consider $\Delta(k^2)$, which is the single scalar function that describes the dressing of a Landau-gauge gluon propagator.  Three possibilities are exposed in the right-panel of Fig.\,\ref{fig:confined}.  The inflexion point possessed by $M(p^2)$, visible in Fig.\,\ref{mass_funct_p11}, entails, too, that the dressed-quark is confined.

With the view that confinement is related to the analytic properties of QCD's Schwinger functions, the question of light-quark confinement may be translated into the challenge of charting the infrared behavior of QCD's universal $\beta$-function. (The behavior of the $\beta$-function on the perturbative domain is well known.)  This is a well-posed problem whose solution is a primary goal of hadron physics; e.g., Refs.\,\cite{Qin:2011dd,Brodsky:2010ur,Aguilar:2010gm}.  It is the $\beta$-function that is responsible for the behavior evident in Figs.\,\ref{fig:confined} and  \ref{mass_funct_p11}, and thereby the scale-dependence of the structure and interactions of dressed-gluons and -quarks.  One of the more interesting of contemporary questions is whether it is possible to reconstruct the $\beta$-function, or at least constrain it tightly, given empirical information on the gluon and quark mass functions.

\begin{figure}[t]
\leftline{\includegraphics[clip,width=0.48\textwidth]{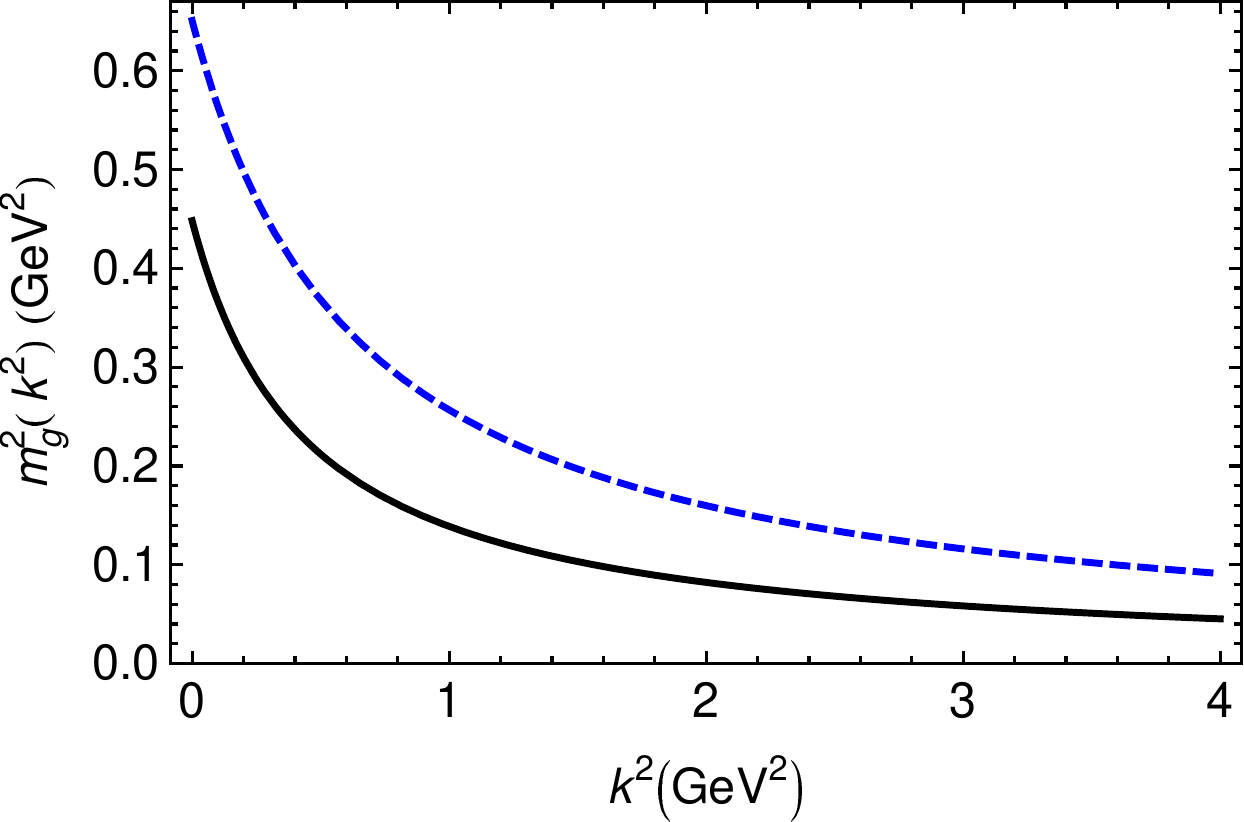}}
\vspace*{-36ex}

\rightline{\includegraphics[clip,width=0.49\textwidth]{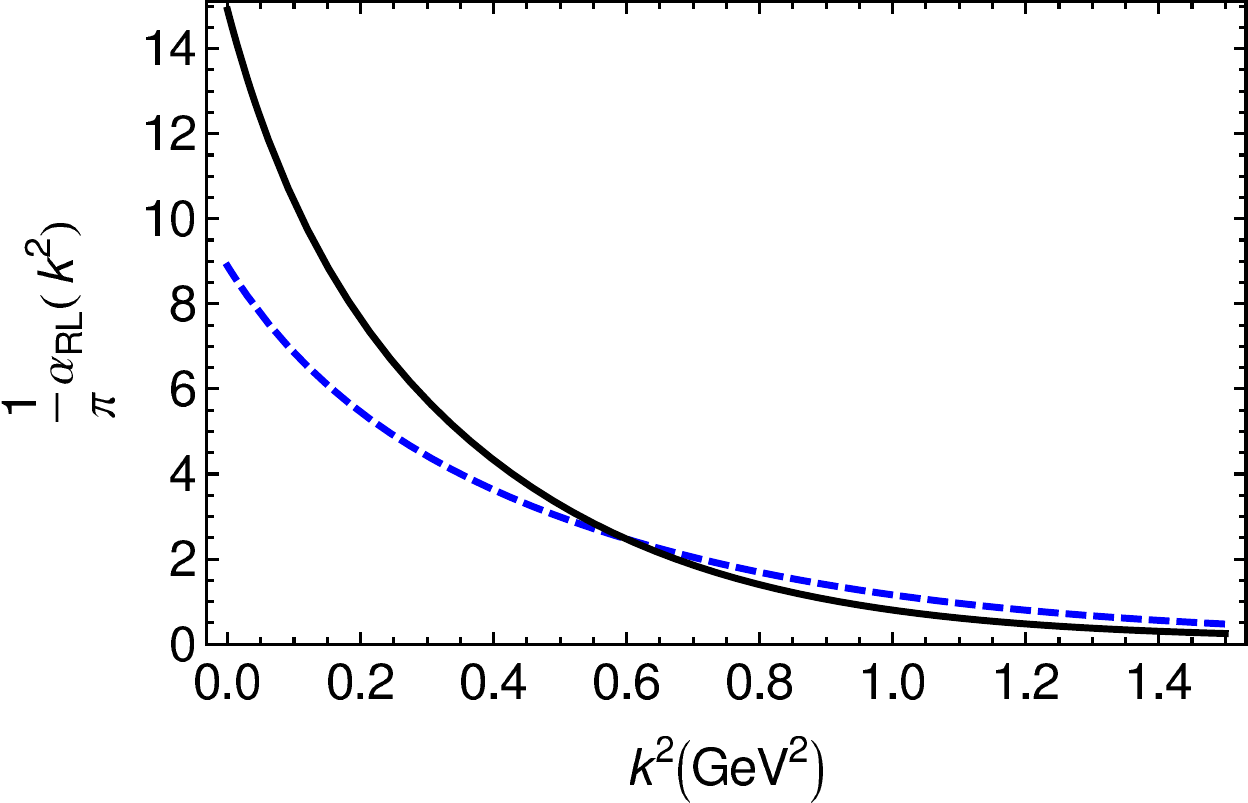}}

\caption{\label{alphaeff}
\emph{Left panel} -- Rainbow-ladder gluon running-mass; and
\emph{right panel} -- rainbow-ladder effective running-coupling, both determined in a DSE analysis of properties of light-quark mesons.  The dashed curves illustrate forms for these quantities that provide the more realistic picture \protect\cite{Qin:2011dd,Qin:2011xq}.  (Figures drawn from Ref.\,\protect\cite{Qin:2011dd}.)
}
\end{figure}

Experiment-theory feedback within the $N^\ast$-programme shows promise for providing the latter \cite{Roberts:2011rr,Gothe:2011up,Aznauryan:2011ub}.  This is illustrated through Fig.\,\ref{alphaeff}, which depicts the running-gluon-mass, analogous to $M(p)$ in  Fig.\,\ref{mass_funct_p11}, and the running-coupling determined by analyzing a range of properties of light-quark ground-state, radially-excited and exotic scalar-, vector- and flavored-pseudoscalar-mesons in the rainbow-ladder truncation, which is leading order in a symmetry-preserving DSE truncation scheme \cite{Bender:1996bb}.  Consonant with modern DSE- and lattice-QCD results \cite{Boucaud:2011ug}, these functions derive from a gluon propagator that is a bounded, regular function of spacelike momenta, which achieves its maximum value on this domain at $k^2=0$ \cite{Aguilar:2010gm,Bowman:2004jm,Aguilar:2009nf}, and a dressed-quark-gluon vertex that does not possess any structure which can qualitatively alter this behavior \cite{Skullerud:2003qu,Bhagwat:2004kj}.  In fact, the dressed-gluon mass drawn here produces a gluon propagator much like the curve labeled ``IR-massive but UV-massless'' in the right-panel of Fig.\,\ref{fig:confined}.

Notably, the value of $M_g=m_g(0)\sim 0.7\,$GeV is typical \cite{Bowman:2004jm,Aguilar:2009nf}; and the infrared value of the coupling, $\alpha_{RL}(M_g^2)/\pi = 2.2$, is interesting because a context is readily provided.
With nonperturbatively-massless gauge bosons, the coupling below which DCSB breaking is impossible via the gap equations in QED and QCD is $\alpha_c/\pi \approx 1/3$ \cite{Roberts:1989mj,Bloch:2002eq,Bashir:1994az}.
In a symmetry-preserving regularization of a vector$\,\times\,$vector contact-interaction used in rainbow-ladder truncation, $\alpha_c/\pi \approx 0.4$; and a description of hadron phenomena requires $\alpha/\pi \approx 1$ \cite{Roberts:2011wy}.
With nonperturbatively massive gluons and quarks, whose masses and couplings run, the infrared strength required to describe hadron phenomena in rainbow-ladder truncation is unsurprisingly a little larger.
Moreover, whilst a direct comparison between $\alpha_{RL}$ and a coupling, $\alpha_{QLat}$, inferred from quenched-lattice results is not sensible, it is nonetheless curious that $\alpha_{QLat}(0)\lsim\alpha_{RL}(0)$ \cite{Aguilar:2010gm}.
It is thus noteworthy that with a more sophisticated, nonperturbative DSE truncation \cite{Chang:2009zb,Chang:2011ei}, some of the infrared strength in the gap equation's kernel is shifted from the gluon propagator into the dressed-quark-gluon vertex.  This cannot materially affect the net infrared strength required to explain observables but does reduce the amount attributed to the effective coupling.  (See, e.g., Ref.\,\cite{Chang:2011ei}, wherein $\alpha(M_g^2) = 0.23\, \pi$ explains important features of the meson spectrum.)

\subsection{Dynamical chiral symmetry breaking}
\label{dcsb}
Whilst the nature of confinement is still debated, Fig.\,\ref{mass_funct_p11} shows that DCSB is a fact.  This figure displays the current-quark of perturbative QCD evolving into a constituent-quark as its momentum becomes smaller.  Indeed, QCD's dressed-quark behaves as a constituent-like-quark or a current-quark, or something in between, depending on the momentum with which its structure is probed.

Dynamical chiral symmetry breaking is the most important mass generating mechanism for visible matter in the Universe.  This may be illustrated through a consideration of the nucleon.  The nucleon's $\sigma$-term is a Poincar\'e- and renormalization-group-invariant measure of the contribution to the nucleon's mass from the fermion mass term in QCD's Lagrangian \cite{Flambaum:2005kc}:
\begin{equation}
\label{sigmaN}
\sigma_N \stackrel{K^2=0}{=} \mbox{\small $\frac{1}{2}$} (m_u+m_d) \langle N(P+K) |  J(K) | N(P) \rangle \approx 0.06 \,m_N,
\end{equation}
where $J(K)$ is the dressed scalar vertex derived from the source $[\bar u(x) u(x) + \bar d(x) d(x)]$
and $m_N$ is the nucleon's mass.  Some have imagined that the non-valence $s$-quarks produce a non-negligible contribution but it is straightforward to estimate \cite{Chang:2009ae,Young:2009zb}
\begin{equation}
\label{sigmaNs}
\sigma_N^s =  0.02 - 0.04\, m_N\,.
\end{equation}
Based on the strength of DCSB for heavier quarks \cite{Flambaum:2005kc}, one can argue that they do not contribute a measurable $\sigma$-term.  It is thus plain that more than 90\% of the nucleon's mass finds its origin in something other than the quarks' current-masses.

\begin{figure}[t] 

\begin{center}
\includegraphics[clip,width=0.48\textwidth]{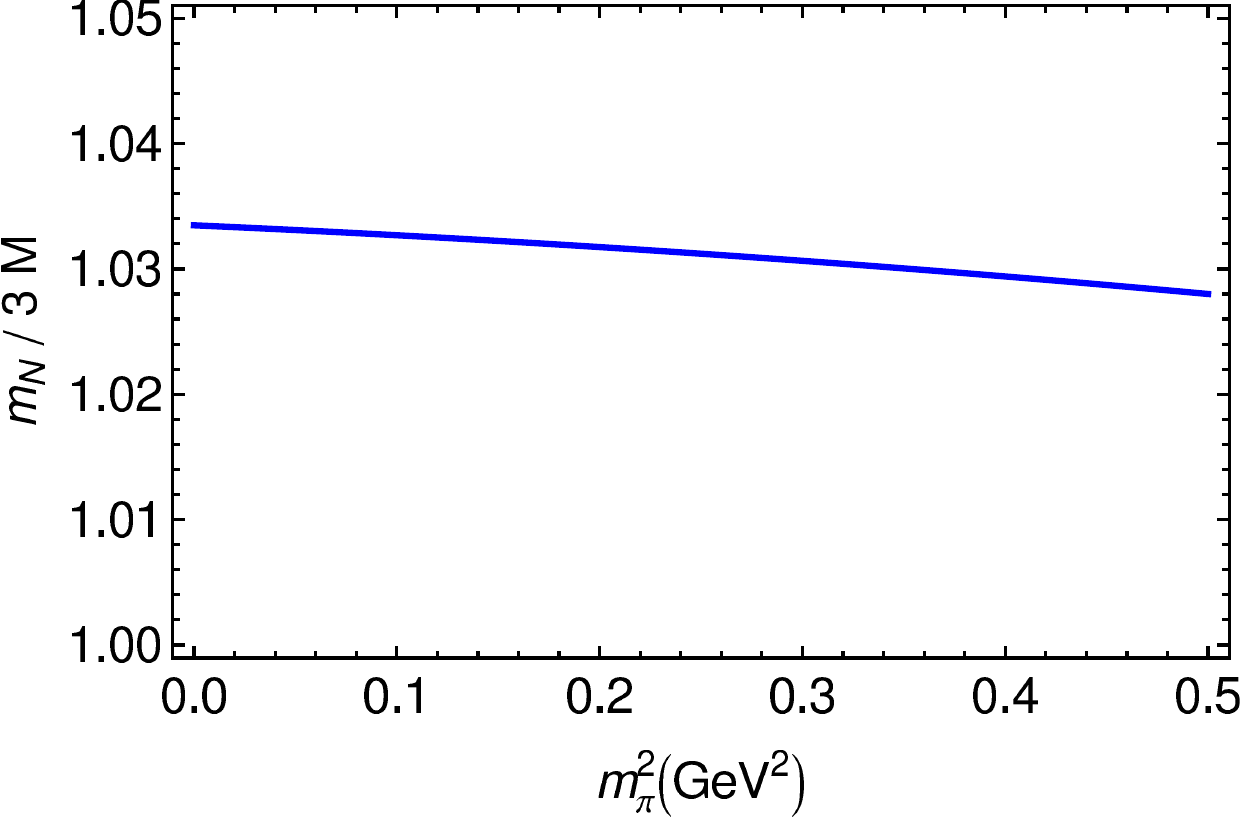}
\end{center}

\caption{\label{nucleonO0}
Evolution with current-quark mass of the ratio $m_N/[3 M]$, which varies by less-than 1\% on the domain depicted.  The calculation is described in Ref.\,\protect\cite{Roberts:2011cf}. NB. The current-quark mass is expressed through the computed value of $m_\pi^2$: $m_\pi^2=0.49\,$GeV$^2$ marks the $s$-quark current-mass.
}

\end{figure}

The source is the physics which produces DCSB.  As we have already mentioned, Fig.\,\ref{mass_funct_p11} shows that even in the chiral limit, when $\sigma_N \equiv 0 \equiv \sigma_N^s$, the massless quark-parton of perturbative QCD appears as a massive dressed-quark to a low-momentum probe, carrying a mass-scale of approximately $(1/3)m_N$.  A similar effect is experienced by the gluon-partons: they are perturbatively massless but are dressed via self-interactions, so that they carry an infrared mass-scale of roughly $(2/3) m_N$, see Fig.\,\ref{alphaeff}.  In such circumstances, even the simplest symmetry-preserving Poincar\'e-covariant computation of the nucleon's mass will produce $m_N^0 \approx 3\,M^0_Q$, where $M^0_Q \approx 0.35\,$GeV is a mass-scale associated with the infrared behavior of the chiral-limit dressed-quark mass-function.  The details of real-world QCD fix the strength of the running coupling at all momentum scales.  That strength can, however, be varied in models; and this is how we know that if the interaction strength is reduced, the nucleon mass tracks directly the reduction in $M^0_Q$  (see Fig.\,\ref{nucleonO0} and Sec.\,\ref{unified}).  Thus, the nucleon's mass is a visible measure of the strength of DCSB in QCD.  These observations are a contemporary statement of the notions first expressed in Ref.\,\cite{Nambu:1961tp}.

It is worth noting in addition that DCSB is an amplifier of explicit chiral symmetry breaking.  This is why the result in Eq.\,(\ref{sigmaN}) is ten-times larger than the ratio $\hat m/m_N$, where $\hat m$ is the renormalization-group-invariant current-mass of the nucleon's valence-quarks.  The result in Eq.\,(\ref{sigmaNs}) is not anomalous: the nucleon contains no valence strangeness.  Following this reasoning, one can view DCSB as being responsible for roughly 98\% of the proton's mass, so that the Higgs mechanism is (almost) irrelevant to light-quark physics.

The behavior illustrated in Figs.\,\ref{fig:confined}--\ref{alphaeff} has a marked influence on hadron elastic form factors.  This is established, e.g., via comparisons between Refs.\,\cite{Maris:2000sk,Bhagwat:2006pu,Cloet:2008re,Eichmann:2008ef,Eichmann:2011vu} and Refs.\,\cite{Roberts:2011wy,GutierrezGuerrero:2010md,Roberts:2010rn}.  Owing to the greater sensitivity of excited states to the long-range part of the interaction in QCD \cite{Qin:2011dd,Qin:2011xq,Holl:2005vu}, we expect this influence to be even larger in the $Q^2$-dependence of nucleon-to-resonance electrocouplings, the extraction of which, via meson electroproduction off protons, is an important part of the current CLAS program and studies planned with the CLAS12 detector \cite{Aznauryan:2011ub,Aznauryan:2011qj,Burkert:2012rh,Gothe:2011up,Gothe:clas12,Aznauryan:2009da}.  In combination with well-constrained QCD-based theory, such data can potentially, therefore, be used to chart the evolution of the mass function on $0.3 \lsim p \lsim 1.2$, which is a domain that bridges the gap between nonperturbative and perturbative QCD.  This can plausibly assist in unfolding the relationship between confinement and DCSB.

In closing this subsection we re-emphasize that the appearance of running masses for gluons and quarks is a quantum field theoretical effect, unrealizable in quantum mechanics.  It entails, moreover, that: quarks are not Dirac particles; and the coupling between quarks and gluons involves structures that cannot be computed in perturbation theory.  Recent progress with the two-body problem in quantum field theory \cite{Chang:2009zb} has enabled these facts to be established \cite{Chang:2010hb}.  One may now plausibly argue that theory is in a position to produce the first reliable symmetry-preserving, Poincar\'e-invariant prediction of the light-quark hadron spectrum \cite{Chang:2011ei}.

\subsection{Mesons and baryons: Unified treatment}
\label{unified}

\begin{figure}[t]
\centerline{\includegraphics[clip,width=0.95\textwidth]{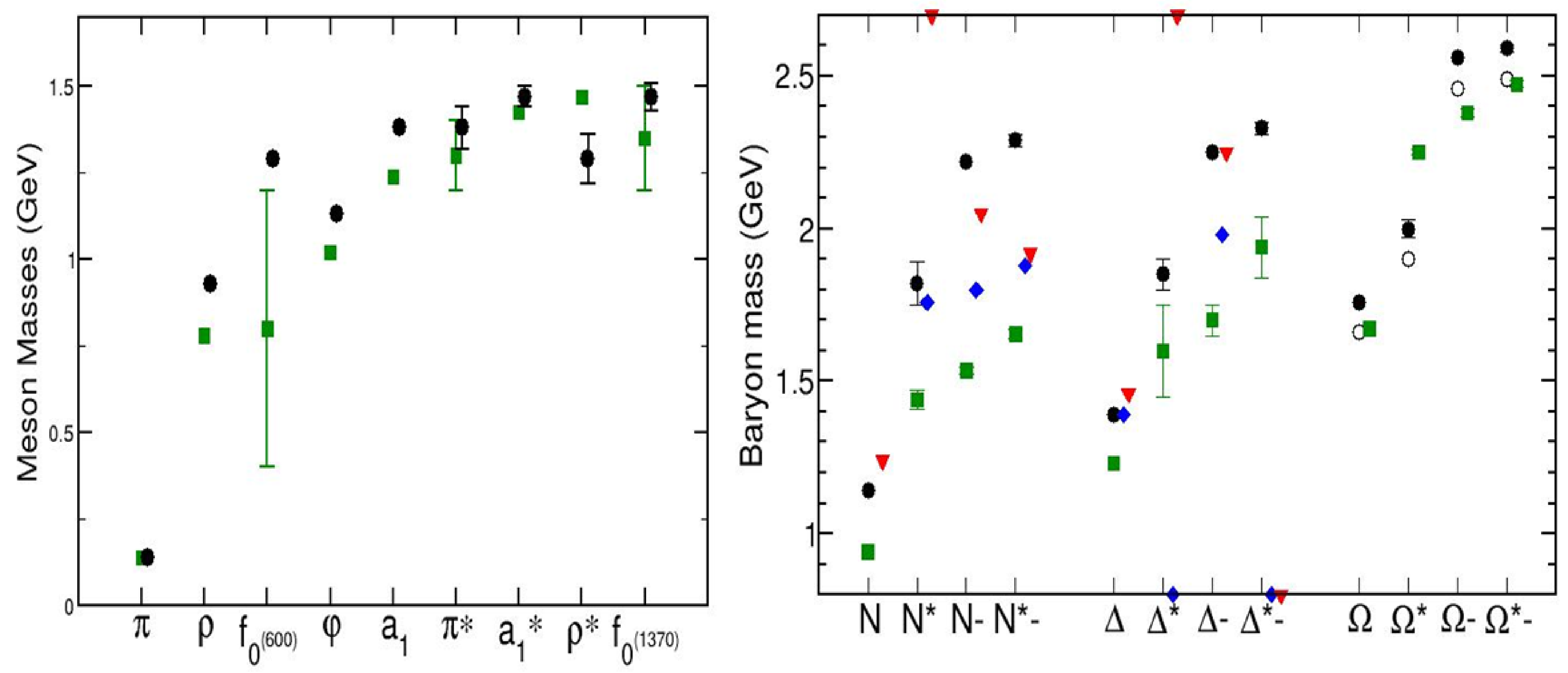}}
\caption{\label{Fig2}
Comparison between DSE-computed hadron masses (\emph{filled circles}) and: bare baryon masses from the Excited Baryon Analysis Center (EBAC), \protect\cite{Suzuki:2009nj} (\emph{filled diamonds}) and J\"ulich,\protect\cite{Gasparyan:2003fp} (\emph{filled triangles}); and experiment \protect\cite{Nakamura:2010zzi}, \emph{filled-squares}.
For the coupled-channels models a symbol at the lower extremity indicates that no associated state is found in the analysis, whilst a symbol at the upper extremity indicates that the analysis reports a dynamically-generated resonance with no corresponding bare-baryon state.
In connection with $\Omega$-baryons the \emph{open-circles} represent a shift downward in the computed results by $100\,$MeV.  This is an estimate of the effect produced by pseudoscalar-meson loop corrections in $\Delta$-like systems at a $s$-quark current-mass.
}
\end{figure}

Owing to the importance of DCSB, it is only within a symmetry-preserving, Poincar\'e-invariant framework that full capitalization on the results of the $N^\ast$-program is possible.  One must be able to correlate the properties of meson and baryon ground- and excited-states within a single, symmetry-preserving framework, where symmetry-preserving includes the consequence that all relevant Ward-Takahashi identities are satisfied.  This is not to say that constituent-quark-like models are worthless.  As will be seen in this article, they are of continuing value because there is nothing better that is yet providing a bigger picture.  Nevertheless, such models have no connection with quantum field theory and therefore not with QCD; and they are not ``symmetry-preserving'' and hence cannot veraciously connect meson and baryon properties.

An alternative is being pursued within quantum field theory via the Faddeev equation.  This analogue of the Bethe-Salpeter equation sums all possible interactions that can occur between three dressed-quarks.  A tractable equation \cite{Cahill:1988dx} is founded on the observation that an interaction which describes color-singlet mesons also generates nonpointlike quark-quark (diquark) correlations in the color-antitriplet channel \cite{Cahill:1987qr}.  The dominant correlations for ground state octet and decuplet baryons are scalar ($0^+$) and axial-vector ($1^+$) diquarks because, e.g., the associated mass-scales are smaller than the baryons' masses and their parity matches that of these baryons.  On the other hand, pseudoscalar ($0^-$) and vector ($1^-$) diquarks dominate in the parity-partners of those ground states \cite{Roberts:2011cf,Chen:2012qr}.  This approach treats mesons and baryons on the same footing and, in particular, enables the impact of DCSB to be expressed in the prediction of baryon properties.

Incorporating lessons learnt from meson studies \cite{Chang:2011vu}, a unified spectrum of $u,d,s$-quark hadrons was obtained using symmetry-preserving regularization of a vector$\,\times\,$vector contact interaction \cite{Roberts:2011cf,Chen:2012qr}.  These studies simultaneously correlate the masses of meson and baryon ground- and excited-states within a single framework; and in comparison with relevant quantities, they produce $\overline{\mbox{rms}}\lsim 15$\%, where $\overline{\mbox{rms}}$ is the root-mean-square-relative-error$/$degree-of freedom.  As indicated by Fig.\,\ref{Fig2}, they uniformly overestimate the PDG values of meson and baryon masses \cite{Nakamura:2010zzi}.  Given that the truncation employed omits deliberately the effects of a meson-cloud in the Faddeev kernel, this is a good outcome, since inclusion of such contributions acts to reduce the computed masses.

Following this line of reasoning, a striking result is agreement between the DSE-computed baryon masses \cite{Roberts:2011cf} and the bare masses employed in modern coupled-channels models of pion-nucleon reactions \cite{Suzuki:2009nj,Gasparyan:2003fp}, see Fig.\,\ref{Fig2}.  The Roper resonance is very interesting.  The DSE studies \cite{Roberts:2011cf,Chen:2012qr} produce a radial excitation of the nucleon at $1.82\pm0.07\,$GeV.  This state is predominantly a radial excitation of the quark-diquark system, with the diquark correlations in their ground state.  Its predicted mass lies precisely at the value determined in the analysis of Ref.\,\cite{Suzuki:2009nj}.  This is significant because for almost 50 years the Roper resonance has defied understanding.

Discovered in 1963/64 \cite{Roper:1964zz}, the Roper appears to be an exact copy of the proton except that its mass is 50\% greater.  The mass was the problem: hitherto it could not be explained by any symmetry-preserving QCD-based tool.  That has now changed.  Combined, see Fig.\,\ref{fig:p11-traj}, Refs.\,\cite{Roberts:2011cf,Suzuki:2009nj,Chen:2012qr} demonstrate that the Roper resonance is indeed the proton's first radial excitation, and that its mass is far lighter than normal for such an excitation because the Roper obscures its dressed-quark-core with a dense cloud of pions and other mesons.  Such feedback between QCD-based theory and reaction models is critical now and for the foreseeable future, especially since analyses of CLAS data on nucleon-resonance electrocouplings suggest strongly that this structure is typical; i.e., most low-lying $N^\ast$-states can best be understood as an internal quark-core dressed additionally by a meson cloud \cite{2012sha}.  This is highlighted further by a comparison between the DSE results and the bare masses obtained in the most complete Argonne-Osaka coupled-channels analysis to date, see Table~\ref{cfANLOsaka}.\footnotemark[2] \footnotetext[2]{With the closing of EBAC at JLab in March 2012, a collaboration between scientists at Argonne National Laboratory and the University of Osaka has accepted the coupled-channels challenge posed by extant and forthcoming CLAS data on the electromagnetic transitions between ground and excited nucleon states.}

\begin{table}[t]
\begin{tabular}{|c|c|c|c|c|c|c|}
\hline
 & P$_{11}$ & S$_{11}$ & S$_{11}$ & P$_{33}$ & P$_{33}$ & D$_{33}$ \\
 \hline
 ANL-Osaka & 1.83 & 2.04 & 2.61 & 1.28 & 2.16 & 2.17 \\
 \hline
 DSE & 1.83 & 2.30 & 2.35 & 1.39 & 1.84 & 2.33 \\
 \hline
 | Rel. Err. | & 0 & 11.3\% & 11.1\% & 7.9\% & 17.4\% & 8.6\% \\
 \hline
\end{tabular}
\caption{Bare masses (GeV) determined in an Argonne-Osaka coupled-channels analysis of single- and double-pion electroproduction reactions compared with DSE results for the mass of each baryon's dressed-quark core.  The notation is as follows: $P_{11}$ corresponds to the $N(1440)$; $S_{11}$ to the $N(1535)$ and the second state in this partial wave; $P_{33}$ to the $\Delta$ and the next state in this partial wave; and $D_{33}$ to the parity partner of the $\Delta$.  The rms-$|\mbox{rel.\,error}|=9.4\pm 5.7$\%.
\label{cfANLOsaka}}
\end{table}

Additional analysis within the framework of Refs.\,\cite{Roberts:2011cf,Chen:2012qr} suggests a fascinating new possibility for the Roper, which is evident in Table.\,\ref{massesN}.  The nucleon ground state is dominated by the scalar diquark, with a significantly smaller but nevertheless important axial-vector diquark component.  This feature persists in solutions obtained with more sophisticated Faddeev equation kernels (see, e.g., Table~2 in Ref.\,\cite{Cloet:2008re}).  From the perspective of the nucleon's parity partner and its radial excitation, the scalar diquark component of the ground-state nucleon actually appears to be unnaturally large.  Expanding the study to include baryons containing one or more $s$-quarks, the picture is confirmed: the ground state $N$, $\Lambda$, $\Sigma$, $\Xi$ are all characterized by $\sim 80$\% scalar diquark content \cite{Chen:2012qr}, whereas their parity partners have a $50-50$ mix of $J=0,1$ diquarks.

One can nevertheless understand the structure of the octet ground-states.  As with so much else in hadron physics, the composition of these flavor octet states is intimately connected with DCSB.  In a two-color version of QCD, scalar diquarks are Goldstone modes \cite{Roberts:1996jx,Bloch:1999vk}.  (This is a long-known result of Pauli-G\"ursey symmetry.)  A ``memory'' of this persists in the three-color theory, for example: in low masses of scalar diquark correlations; and in large values of their canonically normalized Bethe-Salpeter amplitudes and hence strong quark$+$quark$-$diquark coupling within the octet ground-states.  (A qualitatively identical effect explains the large value of the $\pi N$ coupling constant and its analogues involving other pseudoscalar-mesons and octet-baryons.)  There is no commensurate enhancement mechanism associated with the axial-vector diquark correlations.  Therefore the scalar diquark correlations dominate within octet ground-states.

Within the Faddeev equation treatment, the effect on the first radial excitations is dramatic: orthogonality of the ground- and excited-states forces the radial excitations to be constituted almost entirely from axial-vector diquark correlations.  It is critical to check whether this outcome survives with Faddeev equation kernels built from a momentum-dependent interaction.

\begin{table}[t]
\begin{tabular}{|c|c|c|c|c|c|c|c|c|}
\hline
 & N & N(1440) & N(1535) & N(1650) & $\Delta$(1232) & $\Delta$(1600) & $\Delta$(1700) & $\Delta$(1940) \\
 \hline
 0$^{+}$ & 77\% & & & & & & & \\
 \hline
 1$^{+}$ & 23\% & 100\% & & & 100\% & 100\% & & \\
 \hline
 0$^{-}$ & & & 51\% & 43\% & & & & \\
 \hline
 1$^{-}$ & & & 49\% & 57\% & & & 100\% & 100\% \\
 \hline
\end{tabular}
\caption{\label{massesN}
Diquark content of the baryons' dressed-quark cores, computed with a symmetry-preserving regularization of a vector$\,\times\,$vector contact interaction \protect\cite{Roberts:2011ym}.
}

\end{table}

This brings us to another, very significant observation; namely, the match between the DSE-computed level ordering and that of experiment, something which has historically been difficult for models to obtain (see, e.g., the discussion in Ref.\,\cite{Capstick:2000qj}) and is not achieved in contemporary numerical simulations of lattice-regularized QCD (see, e.g., Ref.\,\cite{Edwards:2011jj}).  In particular, the DSE calculations produce a parity-partner for each ground-state that is always more massive than its first radial excitation so that, in the nucleon channel, e.g., the first $J^P=\frac{1}{2}^-$ state lies above the second $J^P=\frac{1}{2}^+$ state.

A veracious expression of DCSB in the meson spectrum is critical to this success.  One might ask why and how?  It is DCSB that both ensures the dressed-quark-cores of pseudoscalar and vector mesons are far lighter than those of their parity partners and produces strong quark$+$antiquark$-$meson couplings, which are expressed in large values for the canonically normalized Bethe-Salpeter amplitudes (Table~3 in Ref.\,\cite{Chen:2012qr}).  The remnants of Pauli-G\"ursey symmetry described previously entail that these features are carried into the diquark sector: as evident in Fig.~3 and Table~5 of Ref.\,\cite{Chen:2012qr} and their comparison with Fig.~2 and Table~3 therein.  The inflated masses but, more importantly, the suppressed values of the Bethe-Salpeter amplitudes for negative-parity diquarks, in comparison with those of positive-parity diquarks, guarantee the computed level ordering: attraction in a given channel diminishes with the square of the Bethe-Salpeter amplitude (see App.\,C in Ref.\,\cite{Chen:2012qr}).  Hence, an approach within which DCSB cannot be realized or a simulation whose parameters are such that the importance of DCSB is suppressed will both necessarily have difficulty reproducing the experimental ordering of levels.

The computation of spectra is an important and necessary prerequisite to the calculation of nucleon transition form factors, the importance of which is difficult to overestimate given the potential of such form factors to assist in charting the long-range behavior of QCD's running coupling.  To place this in context, Refs.\,\cite{Roberts:2011wy,Roberts:2011cf,GutierrezGuerrero:2010md,Roberts:2010rn} explored the sensitivity of a range of hadron properties to the running of the dressed-quark mass-function.  These studies established conclusively that static properties are not a sensitive probe of the behavior in Fig.\,\ref{mass_funct_p11} and Fig.\ref{alphaeff}; viz., regularized via a symmetry-preserving procedure, a vector$\,\times\,$vector contact-interaction predicts masses, magnetic and quadrupole moments, and radii that are practically indistinguishable from results obtained with the most sophisticated QCD-based interactions available currently \cite{Qin:2011dd,Maris:1999nt}.

\subsection{Nucleon to resonance transition form factors}

\begin{figure}[t]

\centerline{\includegraphics[clip,width=0.62\textwidth]{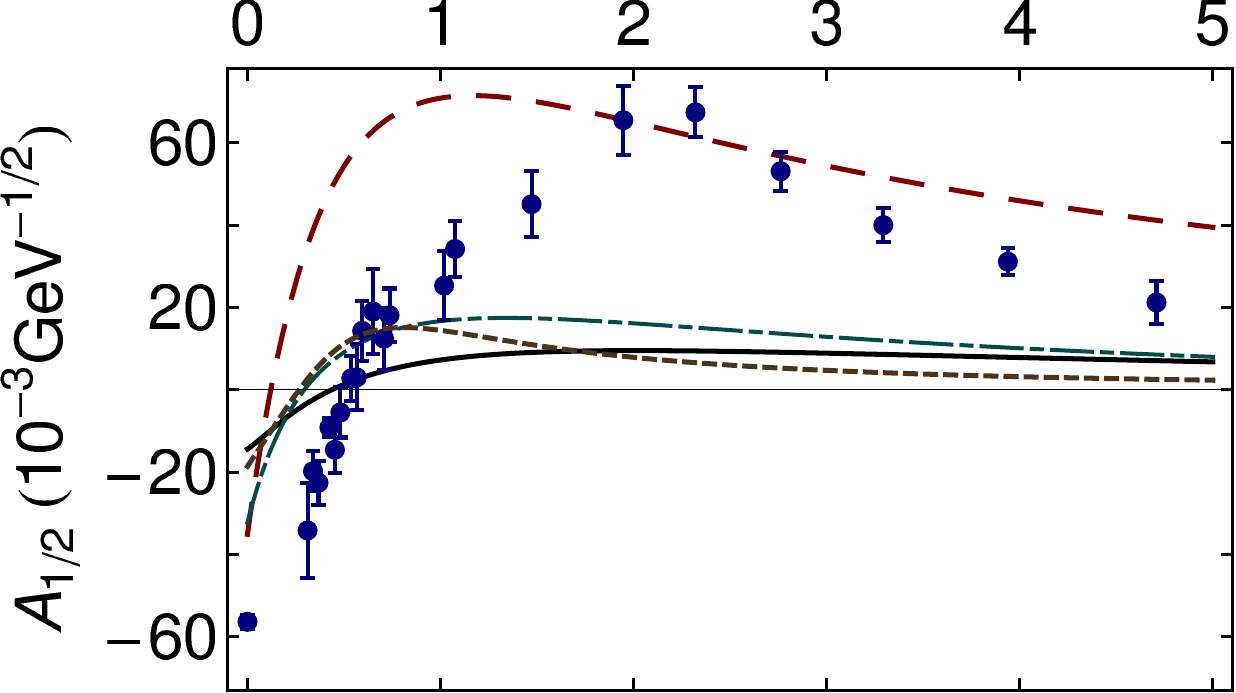}}

\centerline{\rule{1.3ex}{0ex}\includegraphics[clip,width=0.604\textwidth]{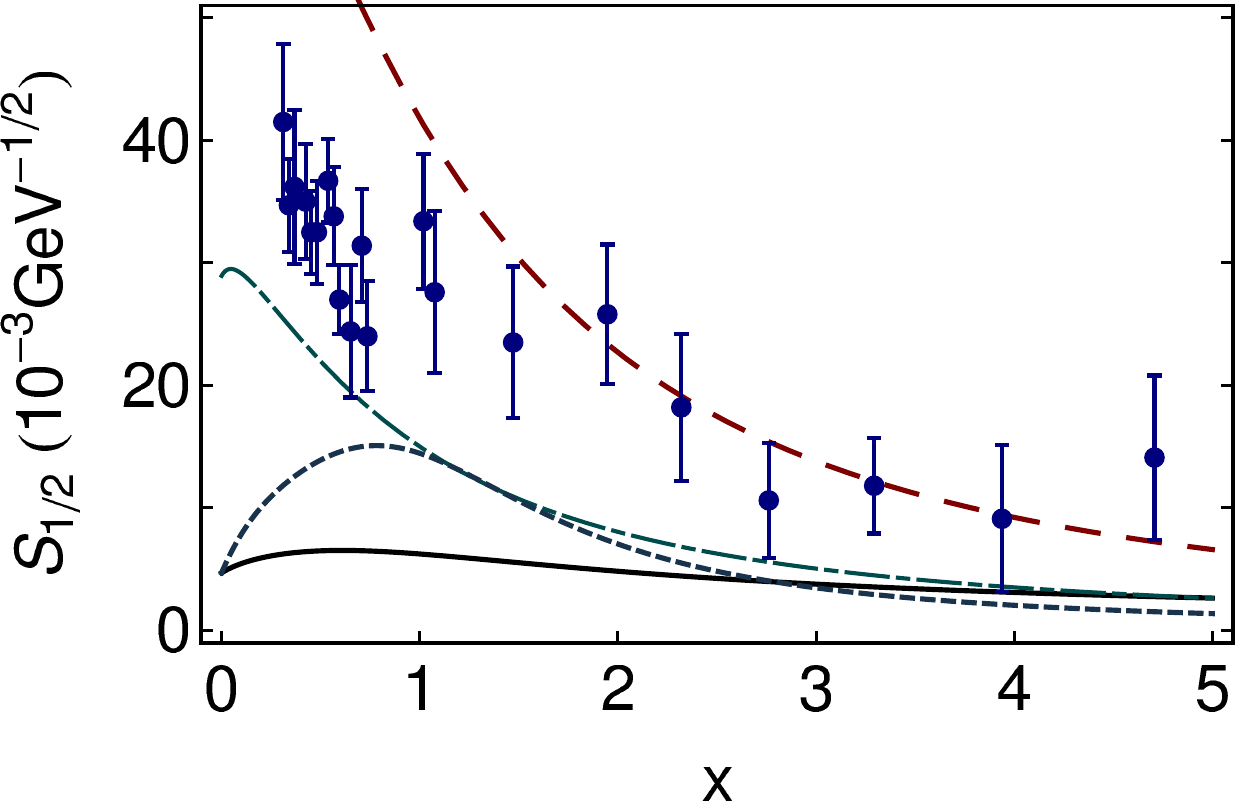}}
\caption{\label{fig:RoperHelicity}
Helicity amplitudes for the $\gamma^\ast p \to P_{11}(1440)$ transition, with $x=Q^2/m_N^2$:
$A_{1/2}$ (upper panel); and $S_{1/2}$ (lower panel).
Solid curves -- DSE computation of Ref.\,\protect\cite{Wilson:2011aa}, obtained using a contact interaction but amended here via an estimate of the impact of the dressed-quark mass in  Fig.\,\protect\ref{mass_funct_p11}, which softens the $x>1$ behavior without affecting $x<1$;
dashed curves -- the light-front constituent quark model results from Ref.\,\protect\cite{Aznauryan:2007ja};
long-dash-dot curves -- the light-front constituent quark model results from Ref.\,\protect\cite{Cardarelli:1996vn};
short-dashed curves -- a smooth fit to the bare form factors inferred in Ref.\,\protect\cite{Suzuki:2010yn,JuliaDiaz:2009ww,LeePrivate:2011};
and data -- Refs.\,\protect\cite{Aznauryan:2009mx,Dugger:2009pn,2012sha}.
}
\end{figure}

The story is completely different, however, with the momentum-dependence of form factors; e.g., in the case of the pion, the difference between the form factor obtained with $M(p)=\,$constant and that derived from $M(p^2)$ in Fig.\,\ref{mass_funct_p11} is dramatically apparent for $Q^2>M^2(p=0)$ \cite{GutierrezGuerrero:2010md}.  The study of diquark form factors in Ref.\,\cite{Roberts:2011wy} has enabled another reference computation to be undertaken; namely, nucleon elastic and nucleon-to-Roper transition form factors \cite{Wilson:2011aa}.  It shows that axial-vector-diquark dominance of the Roper, Table~\ref{massesN}, has a material impact on the nucleon-to-Roper transition form factor.

We choose to illustrate the analysis of Ref.\,\cite{Wilson:2011aa} via Fig.\,\ref{fig:RoperHelicity}.
The figure displays results obtained using a light-front constituent-quark model \cite{Aznauryan:2007ja}, which employed a constituent-quark mass of $0.22\,$GeV and identical momentum-space harmonic oscillator wave functions for both the nucleon and Roper (width$\,=0.38\,$GeV) but with a zero introduced for the Roper, whose location was fixed by an orthogonality condition.
The quark mass is smaller than that which is typical of DCSB in QCD but a more significant difference is the choice of spin-flavour wave functions for the nucleon and Roper.  In Ref.\,\cite{Aznauryan:2007ja} they are simple $SU(6)\times O(3)$ $S$-wave states in the three-quark center-of-mass system, in contrast to the markedly different spin-flavour structure produced by Faddeev equation analyses of these states.

Owing to this, in Fig.\,\ref{fig:RoperHelicity} we also display the light-front quark model results from Ref.\,\cite{Cardarelli:1996vn}.  It is stated therein that large effects accrue from ``configuration mixing;'' i.e., the inclusion of $SU(6)$-breaking terms and high-momentum components in the wave functions of the nucleon and Roper.  In particular, that configuration mixing yields a marked suppression of the calculated helicity amplitudes in comparison with both relativistic and non-relativistic results based on a simple harmonic oscillator \emph{Ansatz} for the baryon wave functions, as used in Ref.\,\cite{Aznauryan:2007ja}.

There is also another difference; namely, Ref.\,\cite{Cardarelli:1996vn} employs Dirac and Pauli form factors to describe the interaction between a photon and a constituent-quark \cite{Cardarelli:1995dc}.
As apparent in Fig.\,2 of Ref.\,\cite{Cardarelli:1996vn}, they also have a noticeable impact, providing roughly half the suppression on $0.5\lsim Q^2/{\rm GeV}^2 \lsim 1.5$.  The same figure also highlights the impact on the form factors of high-momentum tails in the nucleon and Roper wave functions.

In reflecting upon constituent-quark form factors, we note that the interaction between a photon and a dressed-quark in QCD is not simply that of a Dirac fermion \cite{Chang:2010hb,Ball:1980ay,Curtis:1990zs,Alkofer:1993gu,Frank:1994mf,Roberts:1994hh,Maris:1999bh}.  Moreover, the interaction of a dressed-quark with the photon in Ref.\,\cite{Wilson:2011aa} is also modulated by form factors, see Apps.\,A3, C6 therein.  On the other hand, the purely phenomenological form factors in Refs.\,\cite{Cardarelli:1996vn,Cardarelli:1995dc} are inconsistent with a number of constraints that apply to the dressed-quark-photon vertex in quantum field theory; e.g., the dressed-quark's Dirac form factor should approach unity with increasing $Q^2$ and neither its Dirac nor Pauli form factors may possess a zero.  Notwithstanding these observations, the results from Ref.\,\cite{Cardarelli:1996vn} are more similar to the DSE curves than those in Ref.\,\cite{Aznauryan:2007ja}.

In an interesting new development, the study of Ref.\,\cite{Aznauryan:2007ja} has been updated \cite{Aznauryan:2012ec}.  The new version models the impact of a running dressed-quark mass within the light-front formulation of quantum mechanics and yields results that are also closer to those produced by the DSE analysis.

Helicity amplitudes can also be computed using the Argonne-Osaka Collaboration's dynamical coupled-channels framework \cite{Matsuyama:2006rp}.  In this approach, one imagines that a Hamiltonian is defined in terms of bare baryon states and bare meson-baryon couplings; the physical amplitudes are computed by solving coupled-channels equations derived therefrom; and the parameters characterizing the bare states are determined by requiring a good fit to data.
In connection with the $\gamma^\ast p \to P_{11}(1440)$ transition, results are available for both helicity amplitudes \cite{Suzuki:2010yn,JuliaDiaz:2009ww,LeePrivate:2011}.  The associated bare form factors are reproduced in Fig.\,\ref{fig:RoperHelicity}: for $Q^2<1.5\,$GeV$^2$ we depict a smooth interpolation; and for larger $Q^2$ an extrapolation based on perturbative QCD power laws ($A_{\frac{1}{2}}\sim 1/Q^3 \sim S_{\frac{1}{2}}$).


The bare form factors are evidently similar to the results obtained in Ref.\,\cite{Wilson:2011aa} and in Ref.\,\cite{Cardarelli:1996vn}: both in magnitude and $Q^2$-evolution.  Regarding the transverse amplitude, Ref.\,\cite{Suzuki:2010yn} argues that the bare component plays an important role in changing the sign of the real part of the complete amplitude in the vicinity of $Q^2=0$.  In this case the similarity between the bare form factor and the DSE results is perhaps most remarkable -- e.g., the appearance of the zero in $A_{\frac{1}{2}}$, and the $Q^2=0$ magnitude of the amplitude (in units of $10^{-3}\,{\rm GeV}^{-1/2}$)
\begin{equation}
\begin{array}{lcccc}
                    & \mbox{Ref.\,\protect\cite{Aznauryan:2007ja}} &
                    \mbox{Ref.\,\protect\cite{Cardarelli:1996vn}} &
                    \mbox{Ref.\,\protect\cite{Suzuki:2010yn,JuliaDiaz:2009ww,LeePrivate:2011}} &
                                        \mbox{Ref.\,\protect\cite{Wilson:2011aa}} \\
A_{\frac{1}{2}}(0)   & -35.1 & -32.3 & -18.6 & -16.3
\end{array}.
\end{equation}
These similarities strengthen support for an interpretation of the bare-masses, -couplings, etc.,  inferred via coupled-channels analyses, as those quantities comparable with hadron structure calculations that exclude the meson-baryon coupled-channel effects which are determined by multichannel unitarity conditions.

An additional remark is valuable here.  The Argonne-Osaka Collaboration computes electroproduction form factors at the resonance pole in the complex plane and hence they are complex-valued functions.  Whilst this is consistent with the standard theory of scattering \cite{JRTaylor:1972}, it differs markedly from phenomenological approaches that use a Breit-Wigner parametrization of resonant amplitudes in fitting data.  As concerns the $\gamma^\ast p \to P_{11}(1440)$ transition, the real parts of the Argonne-Osaka Collaboration's complete amplitudes are qualitatively similar to the results in Refs.\,\cite{Aznauryan:2009mx,Dugger:2009pn,2012sha} but the Argonne-Osaka Collaboration's amplitudes also have sizeable imaginary parts.  This complicates a direct comparison between theory and extant data.

\subsection{Prospects}
A compelling goal of the international theory effort that works in concert with the $N^\ast$-program is to understand how the interactions between dressed-quarks and -gluons create nucleon ground- and excited-states, and how these interactions emerge from QCD.  This compilation shows no single approach is yet able to provide a unified description of all $N^\ast$ phenomena; and that intelligent reaction theory will long be necessary as a bridge between experiment and QCD-based theory.  Nonetheless, material progress has been made since the release of the White Paper on ``Theory Support for the Excited Baryon Program at the JLab 12-GeV Upgrade'' \cite{Aznauryan:2009da}, in developing strategies, methods and approaches to the physics of nucleon resonances.  Some of that achieved via the Dyson-Schwinger equations is indicated above.  Additional contributions relevant to the $N^\ast$ program are: verification of the accuracy of the diquark truncation of the quark-quark scattering matrix within the Faddeev equation \cite{Eichmann:2011vu}; and a computation of the $\Delta\to \pi N$ transition form factor \cite{Mader:2011zf}.

A continued international effort is necessary if the goal of turning experiment into a probe of the dressed-quark mass function and related quantities is to be achieved.  In our view, precision data on nucleon-resonance transition form factors provides a realistic means by which to constrain empirically the momentum evolution of the dressed-quark mass function and therefrom the infrared behavior of QCD's $\beta$-function; in particular, to locate unambiguously the transition boundary between the constituent- and current-quark domains that is signalled by the sharp drop apparent in  Fig.\,\ref{mass_funct_p11}.  That drop can be related to an inflexion point in QCD's $\beta$-function.  Contemporary theory indicates that this transition boundary lies at $p^2 \sim 0.6\,$GeV$^2$.  Since a probe's input momentum $Q$ is principally shared equally amongst the dressed-quarks in a transition process, then each can be considered as absorbing a momentum fraction $Q/3$.  Thus in order to cover the domain $p^2\in [0.5,1.0]\,$GeV$^2$ one requires $Q^2\in [5,10]\,$GeV$^2$; i.e., the upgraded JLab facility.

In concrete terms, a DSE study of the $N \to N(1535)$ transition is underway, using the contact-interaction, for comparison with data \cite{Aznauryan:2009mx,Denizli:2007tq} and other computations \cite{Braun:2009jy}; and an analysis of the $N\to \Delta$ transition has begun, with the aim of revealing the origin of the unexpectedly rapid $Q^2$-evolution of the magnetic form factor in this process.

At the same time, the Faddeev equation framework of Ref.\,\cite{Cloet:2008re}, is being applied to the $N\to N(1440)$ transition.  The strong momentum dependence of the dressed-quark mass function is an integral part of this framework.  Therefore, in this study it will be possible, e.g., to vary artificially the position of the marked drop in the dressed-quark mass function and thereby identify experimental signatures for its presence and location.  In addition, it will provide a crucial check on the results in Table~\ref{massesN}.  
It is notable that DCSB produces an anomalous electromagnetic moment for the dressed-quark.  This is known to produce a significant modification of the proton's Pauli form factor at $Q^2\lsim 2\,$GeV$^2$ \cite{Chang:2011tx}.  It is also likely to be important for a reliable description of $F_2^\ast$ in the nucleon-to-Roper transition.

The Faddeev equation framework of Ref.\,\cite{Cloet:2008re} involves parametrizations of the dressed-quark propagators that are not directly determined via the gap equation.  An important complement would be to employ the \emph{ab initio} rainbow-ladder truncation approach of Ref.\,\cite{Eichmann:2008ef,Eichmann:2011vu} in the computation of properties of excited-state baryons, especially the Roper resonance.  Even a result for the Roper's mass and its Faddeev amplitude would be useful, given the results in Table~\ref{massesN}.  In order to achieve this, however, technical difficulties must be faced and overcome.  Here there is incipient progress, made possible through the use of generalized spectral representations of propagators and vertices.

In parallel with the program outlined herein, an effort is beginning with the aim of providing the reaction theory necessary to make reliable contact between experiment and predictions based on the dressed-quark core.  While rudimentary estimates can and will be made of the contribution from pseudoscalar meson loops to the dressed-quark core of the nucleon and its excited states, a detailed comparison with experiment will only follow when the DSE-based results are used to constrain the input for dynamical coupled channels calculations.


\newcommand{\dev}{$\chi^2/\textrm{d.o.f}$}
\newcommand{\msb}{$\overline{\text{MS}}$}

\newcommand{\nn}{\nonumber}
\newcommand{\up}{{\uparrow}}
\newcommand{\down}{{\downarrow}}



\section{Light-Cone Sum Rules: A Bridge between Electrocouplings  and Distribution
  Amplitides of Nucleon Resonances \label{Distribution Amplitudes}}

We expect that at photon virtualities from 5 to 10 GeV$^2$ of CLAS12 
the electroproduction cross sections of nuclear resonances will become
amenable to the QCD description in terms of quark partons, whereas the 
description in terms of meson-baryon degrees of freedom becomes
much less suitable than at smaller momentum transfers. 
The major challenge for theory is that quantitative description of 
form factors in this transition region must include nonperturbative  
contributions. In Ref.~\cite{Braun:2009jy} we have suggested to use
a combination of light-cone sum rules (LCSRs) and lattice calculations.
To our opinion this approach presents
a reasonable compromise between theoretical rigour and the necessity to
make phenomenologically relevant predictions. 

\subsection{Light-front wave functions and distribution amplitudes}

The quantum-mechanical picture of a nucleon as a superposition of states 
with different number of partons assumes the infinite momentum frame or
light-cone quantization. Although \emph{a priori} there is no reason to 
expect that the states with, say, 100 partons (quarks and gluons) are suppressed
as compared those with the three valence quarks, the phenomenological success 
of the quark model allows one to hope that only a first few  Fock components are 
really necessary. In hard exclusive reactions which involve a large
momentum transfer to the nucleon, the dominance of valence states is widely expected 
and can be proven, at least  
within QCD perturbation theory~\cite{Lepage:1980fj,Chernyak:1983ej}.  

The most general parametrization of the three-quark sector involves six scalar 
light-front wave functions ~\cite{Ji:2002xn,Ji:2003yj} which correspond to different 
possibilities to couple the quark helicities $\lambda_i$ and orbital angular momentum $L_z$ to produce 
the helicity-$1/2$ nucleon state: $\lambda_1+\lambda_2+\lambda_3 + L_z = 1/2$.
In particular if the quark helicities 
$\lambda_i$ sum up to $1/2$, then zero angular momentum is allowed, $L=0$.   
The corresponding contribution can be written as~\cite{Lepage:1980fj,Chernyak:1983ej,Ji:2002xn}:  
\begin{eqnarray}
\label{def:nucleonWF}
 |N(p)^\uparrow\rangle^{L=0} &=& \frac{\epsilon^{abc}}{\sqrt{6}}
\int\frac{[dx][d^2\vec{k}]}{\sqrt{x_1x_2x_3}}
\,\,\Psi_N(x_i,\vec{k}_i) |u_a^\uparrow(x_1,\vec{k}_1)\rangle
\nonumber\\&&{}\times
\Big[
 \big|u_b^\downarrow(x_2,\vec{k}_2)\rangle |d_c^\uparrow(x_3,\vec{k}_3)\rangle
-\big|d_b^\downarrow(x_2,\vec{k}_2)\rangle |u_c^\uparrow(x_3,\vec{k}_3)\rangle
 \Big].
\end{eqnarray}
Here   
$\Psi_N(x_i,\vec{k}_i)$ is the  light-front wave function that 
depends on the  momentum fractions $x_i$ and transverse momenta $\vec{k}_i$ of the quarks, $|u_a^\uparrow(x_i,\vec{k}_i)\rangle$ is a quark state
with the indicated momenta and color index $a$, and
$\epsilon^{abc}$ is the fully antisymmetric tensor; arrows indicate
helicities.
The integration measure is defined as 
\begin{eqnarray}
\int [dx] &=& \int_0^1  dx_1 dx_2 dx_3\, \delta\big(\sum x_i-1\big)\,,
\nonumber\\
\int [d^2\vec{k}] &=& (16\pi^3)^{-2} \int  d\vec{k}_1 d\vec{k}_2 d\vec{k}_3\, \delta\big(\sum \vec{k}_i\big)\,.
\end{eqnarray}
In hard processes the contribution of $\Psi(x_i,\vec{k}_i)$ is dominant whereas the other existing three-quark 
wave functions give rise to a power-suppressed correction, i.e. a correction of higher twist.  

The light-front description of a nucleon is very attractive for model building.  The calculation of light-front wave functions from QCD can in principle be done using light-front Hamiltonian methods.
A first approximation to the QCD Light-Front equation of motion and  corresponding model solutions for the light-front wavefunctions of mesons and baryons has recently been obtained using light-front holography.
This is discussed in Section \ref{Light-Front Holographic}. In particular there are subtle issues with renormalization and gauge dependence.
An alternative approach has been to describe nucleon structure in terms of \emph{distribution amplitudes}
(DA) corresponding to matrix elements of nonlocal gauge-invariant light-ray operators. 
The classification of DAs goes in twist rather than number of constituents as for the wave functions.
For example the leading-twist-three nucleon (proton) DA is defined by the 
matrix element~\cite{Braun:2000kw}:  
\begin{equation}
\label{varphi-N}
\langle 0 | 
\epsilon^{ijk}\! \left(u^{\up}_i(a_1 n) C \!\!\not\!{n} u^{\down}_j(a_2 n)\right)  
\!\not\!{n} d^{\up}_k(a_3 n) 
|N(p)\rangle 
= - \frac12 f_N\,p \cdot n\! \not\!{n}\, u_N^\up(p)\! \!\int\! [dx] 
\,e^{-i p \cdot n \sum x_i a_i}\, 
\varphi_N(x_i)\,,
\end{equation}
where $q^{\up(\down)} = (1/2) (1 \pm \gamma_5) q$ are quark fields of given 
helicity, $p_\mu$, $p^2=m_N^2$, is the proton momentum, $u_N(p)$ the usual Dirac spinor in 
relativistic normalization, $n_\mu$ an auxiliary light-like vector $n^2=0$ and $C$
the charge-con\-ju\-ga\-tion matrix. 
The Wilson lines that ensure gauge invariance are inserted between the quarks;
they are not shown for brevity.
The normalization constant $f_N$ is  defined in such a way that 
\begin{equation}
\label{norm}
  \int [dx]\, \varphi_N(x_i) =1\,.
\end{equation}
In principle, the complete set of nucleon DAs carries full information on the 
nucleon structure, same as the complete basis of light-front wave functions.
In practice, however, both expansions have to be truncated and usefulness of  
a truncated version, taking into account either a first few Fock states or a few 
lowest twists, depends on the physics application.

Using the wave function in Eq.~(\ref{def:nucleonWF}) to calculate 
the matrix element in Eq.~(\ref{varphi-N}) 
it is easy to show that the DA $\varphi_N(x_i)$ is 
related to the integral of the wave function $\Psi_N(x_i,\vec{k}_i)$ over transverse
momenta, which corresponds to the limit of zero transverse separation between the
quarks in the position space~\cite{Lepage:1980fj}:
\begin{equation}
 f_N(\mu) \, \varphi_N(x_i,\mu) \sim \int\limits_{|\vec{k}|<\mu} [d^2\vec{k}]\, \Psi_N(x_i,\vec{k}_i)\,.
\end{equation}
Thus, the normalization constant $f_N$ can be interpreted as the nucleon 
wave function at the origin (in position space). 

Higher-twist three-quark DAs
are related, in a loose sense, with similar integrals of the wave functions including 
extra powers of the transverse momentum, and with contributions of the other existing 
wave functions which correspond to nonzero quark orbital angular momentum.

As always in a field theory, extraction of the asymptotic 
behavior produces divergences that have to be regulated.  
As the result, the DAs become scheme- and scale-dependent. 
In the calculation of physical observables this dependence is cancelled 
by the corresponding dependence of the coefficient functions.
The DA $\varphi_N(x_i,\mu)$ can be expanded in the set of orthogonal polynomials 
$\mathcal{P}_{nk}(x_i)$
defined as eigenfunctions of the corresponding one-loop evolution equation:
\begin{equation}
   \varphi_N(x_i,\mu) = 120 x_1 x_2 x_3 \sum_{n=0}^\infty\sum_{k=0}^N c^N_{nk}(\mu) \mathcal{P}_{nk}(x_i)\,,
\label{expand-varphi}
\end{equation}
where 
\begin{equation}
   \int [dx]\, x_1 x_2 x_3 \mathcal{P}_{nk}(x_i)\mathcal{P}_{n'k'} = \mathcal{N}_{nk}\delta_{nn'}\delta_{kk'}\,
\end{equation}
and 
\begin{equation}
  c^N_{nk}(\mu) = c^N_{nk}(\mu_0)\left(\frac{\alpha_s(\mu)}{\alpha_s(\mu_0)}\right)^{\gamma_{nk}/\beta_0}.
\end{equation}
Here $\mathcal{N}_{nk}$ are  convention-dependent normalization factors, $\beta_0 = 11-\frac23 n_f$
and $\gamma_{nk}$ the corresponding anomalous dimensions. The double sum in Eq.~(\ref{expand-varphi})
goes over all existing orthogonal polynomials $\mathcal{P}_{nk}(x_i)$, $k=0,\ldots,n$, of degree $n$. 
Explicit expressions for the polynomials $\mathcal{P}_{nk}(x_i)$ for $n=0,1,2$ 
and the corresponding anomalous dimensions can be found in Ref.~\cite{Braun:2008ia}.

 In what follows we will refer to the coefficients $c_{nk}(\mu_0)$ as shape parameters.
The set of these coefficients 
together with the normalization constant $f_N(\mu_0)$ at a reference scale $\mu_0$ specifies the
momentum fraction distribution of valence quarks on the nucleon.
They are nonperturbative quantities that can be related to matrix elements of local gauge-invariant 
three-quark operators (see below).

In the last twenty years there had been mounting evidence that the simple-minded picture of a proton 
with the three valence quarks in an S-wave is insufficient, so that for example the proton spin 
is definitely not constructed from the quark spins alone. If the orbital angular momenta
of quarks and gluons are nonzero, the nucleon is intrinsically deformed.  
The general classification of three-quark light-front wave functions with nonvanishing angular 
momentum has been worked out in Refs.~\cite{Ji:2002xn,Ji:2003yj}. 
In particular the wave functions with $L_z=\pm 1$ play a decisive role in hard processes involving 
a helicity flip, e.g. the Pauli electromagnetic form factor $F_2(Q^2)$ of the proton~\cite{Belitsky:2002kj}.
These wave functions are related, in the limit of small transverse separation,  
to the twist-four nucleon DAs introduced in Ref.~\cite{Braun:2000kw}: 
\begin{eqnarray}
\label{twist-4}
\langle 0 | \epsilon^{ijk}\! 
\left(u^{\up}_i(a_1 n) C\slashed{n} u^{\down}_j(a_2 n)\right)  
\!\slashed{p} d^{\up}_k(a_3 n) |N(p)\rangle 
&=& -\frac14
\,p \cdot n\, \slashed{p}\, u_{N^*}^{\up}(p)\! \!\int\! [dx] 
\,e^{-i p \cdot n \sum x_i a_i}\,
\nonumber\\
&&\times \left[f_{N}\Phi^{N,WW}_4(x_i)+\lambda^N_1\Phi^{N}_4(x_i)\right],   
\nonumber\\
\langle 0 | \epsilon^{ijk}\! 
\left(u^{\up}_i(a_1 n) C \slashed{n}\gamma_{\perp}\slashed{p} u^{\down}_j(a_2 n)\right)  
\gamma^{\perp}\slashed{n} d^{\up}_k(a_3 n) |N(p)\rangle 
&=&
-\frac12
\, p \cdot n\! \not\!{n}\,m_{N} u_{N}^{\up}(p)\! \!\int\! [dx] 
\,e^{-i p \cdot n \sum x_i a_i}\,
\nonumber\\
&&\times \left[f_{N}\Psi^{N,WW}_4(x_i)-\lambda^N_1\Psi^{N}_4(x_i)\right],
\nonumber\\
\langle 0 | \epsilon^{ijk}\! 
\left(u^{\up}_i(a_1 n) C\slashed{p}\,\slashed{n} u^{\up}_j(a_2 n)\right)  
\!\not\!{n} d^{\up}_k(a_3 n) |N(p)\rangle 
&=& \frac{\lambda^N_2}{12}\, p \cdot n\! \not\!{n}\, m_{N} u_{N}^{\up}(p)\! \!\int\! [dx] 
\,e^{-i p \cdot n \sum x_i a_i}
\nonumber\\ &&{}\times\,\Xi^{N}_4(x_i) \,,
\end{eqnarray}
where $\Phi^{N,WW}_4(x_i)$ and $\Psi^{N,WW}_4(x_i)$ are the so-called 
Wandzura-Wilczek contributions, which can be expressed in terms of the 
leading-twist DA $\varphi_N(x_i)$~\cite{Braun:2008ia}. 
The two new constants $\lambda_1^N$ and $\lambda_2^N$
are defined in such a way that the integrals of the ``genuine'' twist-4 
DAs $\Phi_4$, $\Psi_4$, $\Xi_4$ are normalized to unity, similar to Eq.~(\ref{norm}).
They are related to certain normalization integrals 
of the light-front wave functions for the three-quark states with $L_z=\pm 1$, 
see Ref.~\cite{Belitsky:2002kj} for details.

Light-front wave functions and DAs of all baryons, including the nucleon resonances,
can be constructed in a similar manner, taking into account spin and flavor symmetries. They can be constructed for all baryons of
arbitrary spin without any conceptual complications, although it will
become messy. The problem is only that "construct" means basically
that one can enumerate different independent components and find their
symmetries. To calculate them nonperturbatively is becoming increasingly
difficult, however.
This extension is especially simple for the parity doublets of the usual $J^P = \frac12^+$ 
octet since the nonlocal operators entering the definitions of nucleon DAs 
do not have a definite parity. Thus the same operators couple also 
to $N^*(1535)$ and one can define the corresponding leading-twist DA by the same expression 
as for the nucleon: 
\begin{equation}
\label{varphi-Nstar}
\langle 0 | \epsilon^{ijk}\! 
\left(u^{\up}_i(a_1 n) C \!\!\not\!{n} u^{\down}_j(a_2 n)\right)  
\!\not\!{n} d^{\up}_k(a_3 n) |N^*(p)\rangle 
=  \frac12 f_{N^*}\, p \cdot n\! \not\!{n}\, u_{N^*}^{\up}(p)\! 
\!\int\! [dx] \,e^{-i p \cdot n \sum x_i a_i}\, 
\varphi_{N^*}(x_i)\,,
\end{equation}
where, of course, $p^2=m_{N^*}^2$. The constant $f_{N^*}$ has a physical 
meaning of the wave function of $N^*(1535)$ at the origin.  The 
DA $\phi_{N^*}(x_i)$ is normalized to unity (\ref{norm}) and has an expansion
identical to (\ref{expand-varphi}):
\begin{equation}
   \varphi_{N^*}(x_i,\mu) = 120 x_1 x_2 x_3 \sum_{n=0}^\infty\sum_{k=0}^N 
c^{N^*}_{nk}(\mu) \mathcal{P}_{nk}(x_i)\,,
\label{expand-varphi-star}
\end{equation}
albeit with different shape parameters $c^{N^*}_{nk}$.

Similar as for the nucleon, there exist three independent subleading twist-4 
distribution amplitudes for the $N^*(1535)$ resonance: 
$\Phi^{N^*}_4$, $\Psi^{N^*}_4$ and $\Xi^{N^*}_4$. Explicit expressions are given in Ref.~\cite{Braun:2009jy}.

\subsection{Moments of distribution amplitudes from lattice QCD}

The normalization constants $f$, $\lambda_1$, $\lambda_2$ and the shape
parameters $c_{nk}$ are related to matrix elements of local three-quark
operators between vacuum and the baryon state of interest, and can be 
calculated using lattice QCD. Investigations of excited hadrons using 
this method are generally much more difficult compared to the ground
states. On the other hand, the states of opposite parity can be separated
rather reliably as  propagating forwards and backwards in euclidian 
time. For this reason, for the time being we concentrate on the study
of the ground state baryon octet $J^P=\frac{1}{2}^+$, and the
lowest mass octet with negative parity, $J^P=\frac{1}{2}^-$, $N^*(1535)$
being the prime example.   

{}Following the exploratory studies reported in Refs.~\cite{Gockeler:2008xv,Braun:2008ur,Braun:2009jy} 
 QCDSF collaboration is investing significant effort to make such 
calculations 
 fully quantitative. 
The calculation is rather involved and 
requires the following steps: 
(1)~
Find lattice (discretized) operators that transform according to irreducible representations of 
spinorial group $\overline{H(4)}$;
(2)~
Calculate nonperturbative renormalization constants for these operators;
(3)~
Compute matrix elements of these operators on the lattice from suitable correlation functions,
and  
(4)~
Extrapolate $m_\pi\to m_\pi^{\rm phys}$, lattice volume $V\to \infty$ and lattice spacing 
$a\to 0$.

Irreducibly transforming $\overline{H(4)}$ multiplets for three-quark operators have been 
constructed in Ref.~\cite{Kaltenbrunner:2008pb}. Nonperturbative renormalization and 
one-loop scheme conversion factors RI-MOM$\to\overline{\rm MS}$ have been calculated 
in Ref.~\cite{Gockeler:2008we}. A consistent perturbative renormalization scheme for the 
three-quarks operators in dimensional regularization has been 
found~\cite{Kraenkl:2011qb} and the calculation of two-loop conversion factors 
using this scheme is in progress. 
  
The matrix elements of interest are calculated from correlation functions of the form 
$\langle\mathcal{O}_{\alpha\beta\gamma}(x)\bar{\mathcal{N}}(y)_\tau\rangle$, where 
$\mathcal{N}$ is a smeared nucleon interpolator and $\mathcal{O}$ is a local 
three-quark operator with up to two derivatives, and applying the parity 
``projection'' operator $(1/2)(1\pm m\gamma_4/E)$ \cite{Lee:1998cx}. In this way we get
access to the normalization constants, the first and the second moments of the 
distribution amplitudes. Calculation of yet higher moments is considerably more difficult
because one cannot avoid mixing with operators of lower dimension.
 
\begin{figure}
  \includegraphics[height=.30\textheight, clip=true]{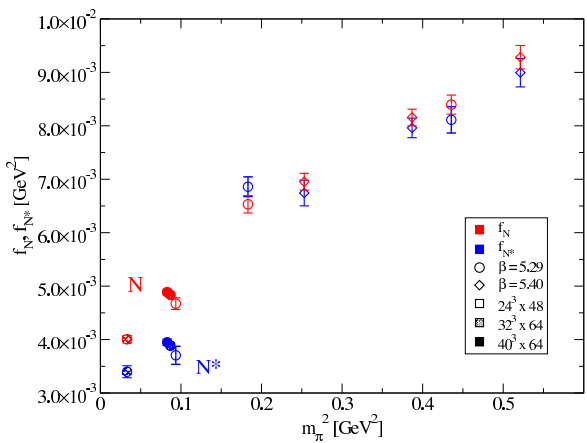}
\caption{Probability amplitude $f_N$, $f_{N^*}$ to find the three valence quarks in the 
nucleon and $N^*(1535)$ at the same space-time point (wave function at the origin).}
\label{fig:1}
\end{figure}

The correlation functions were evaluated using $N_f=2$ dynamic Wilson (clover) 
fermions on several lattices and a range of pion masses $m_\pi\ge 180$~MeV. 
Our preliminary  results for the wave functions the nucleon and $N^*(1535)$ at the origin 
are summarized in Fig.~\ref{fig:1}~\cite{Schiel:2011av}.   
\begin{figure}
  \includegraphics[height=.26\textheight, clip=true]{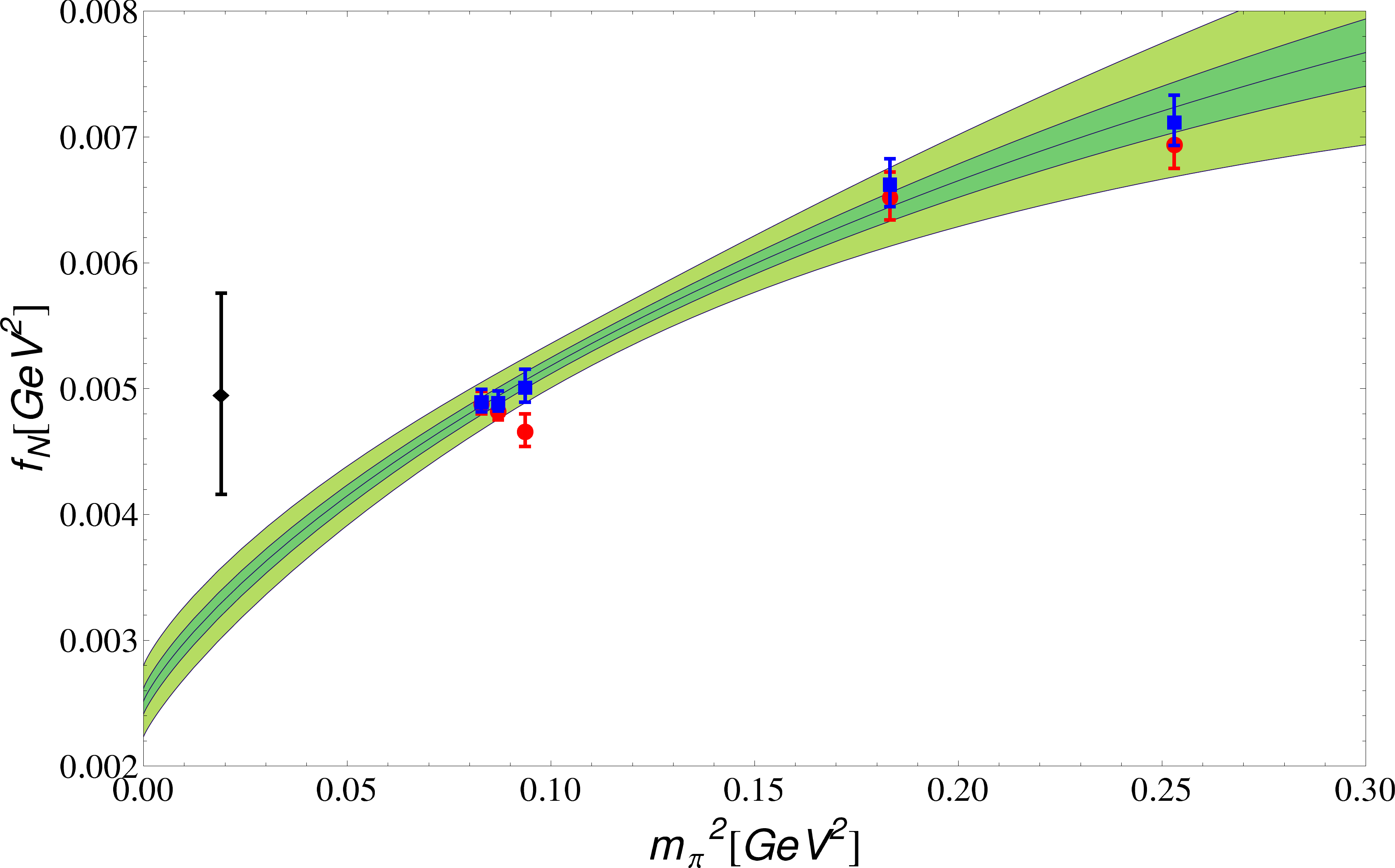}
\caption{The chiral extrapolation of $f_N$ to the physical (light) quark masses.
The red points are lattice data and the blue points are corrected for finite volume effects.   
The green bands are the 1- and 2-$\sigma$ errors, respectively. 
The left-most black ``data point'' at the physical mass shows the recently updated estimate from
QCD sum rule calculations \cite{Gruber:2010bj}.
 }
\label{fig:2}
\end{figure}
The extrapolation of the results for the nucleon to the physical pion mass and infinite 
volume as well as the analysis of the related systematic errors are in progress. 
An example of such an analysis is shown in Fig.~\ref{fig:2}.

This analysis  will be done using one-loop chiral perturbation theory. The necessary expressions 
have been worked out in Ref.~\cite{Wein:2011ix}. Whereas the pion mass dependence of 
nucleon couplings is generally in agreement with expectations, we observe a large difference
(up to a factor of three) in $N^*(1535)$ couplings calculated with heavy and light pions: 
All couplings drop significantly in the transition region where 
the decay $N^*\to N\pi$ opens up. 
This effect can be due to the change in the structure of the wave function, but 
also to contamination of our $N^*(1535)$ results by the contribution 
of the $\pi N$ scattering state, or some other lattice artefact. This is one of
the issues that have to be clarified in future. 

\begin{figure}
  \includegraphics[height=.30\textheight,clip=true]{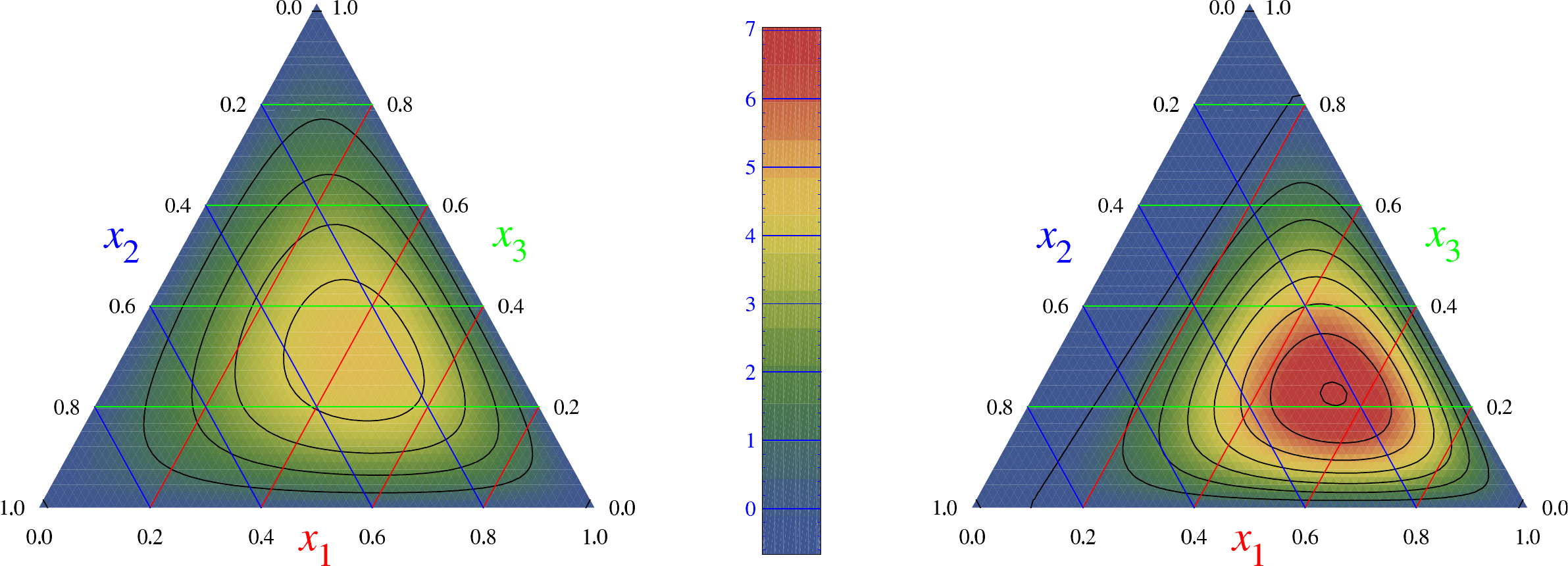}
  \caption{Leading-twist distribution amplitudes of the nucleon (left) and $N^*(1535)$ (right)
           in barycentric coordinates $x_1+x_2+x_3=1$.
           }
\label{fig:3}
\end{figure}

We also find that the wave function of the $N^*(1535)$ resonance is much more asymmetric 
compared to the nucleon: nearly 50\% of the total momentum is carried by the $u$-quark
with the same helicity. 
This shape is illustrated in Fig.~\ref{fig:3}
where the leading-twist distribution amplitudes of the nucleon (left) and $N^*(1535)$ (right)
are shown in barycentric coordinates $x_1+x_2+x_3=1$; $x_i$ are the momentum fractions carried 
by the three valence quarks. 

Our plans for the coming 2-3 years are as follows. The final analysis of the QCDSF lattice data 
using two flavors of dynamic fermions is nearly completed and in future we will go over to $N_f=2+1$ 
studies, i.e. include dynamic strange quarks. The generation of the corresponding gauge configurations
is in progress and first results are expected in one year from now. 
We will continue the studies of the lowest mass states in the $J^P=1/2^+$ and $J^P=1/2^-$ baryon octets. 
In particular the distribution amplitudes of the $\Lambda$ and $\Sigma$ baryons
will be studied for the first time. At a later stage we hope to be able to do similar calculations 
for the $J^P=3/2^\pm$ decuplets.
We are working on the calculation of two-loop conversion factors RI-MOM$\to\overline{\rm MS}$ 
using the renormalization scheme suggested in~\cite{Kraenkl:2011qb} and plan to employ them in the future
studies. Main attention will be payed to the analysis of various sources of systematic uncertainties.
With the recent advances in the algorithms and computer hardware the quark mass and finite volume extrapolations 
of lattice data have become less of a problem, which allows us to concentrate on more subtle issues. 
Our latest simulations for small pion masses make possible, for the first time, to study the
transition region where decays of resonances, e.g. $N^*\to N\pi$, become kinematically allowed.
We have to understand the influence of finite resonance width on the calculation of 
operator matrix elements and to this end plan to consider $\rho$-meson distribution amplitudes
as a simpler example. We will also make detailed studies  of meson (pion) distribution amplutides 
in order to understand better the lattice discretization errors and work out a concrete procedure to 
minimize their effect. 
The full program is expected to last five years and is part of the researh proposal for the Transregional Collaborative Research Center
(SFB Transregio 55 "Hadron physics with Lattice QCD) funded by the
German Research Council (DFG).

\subsection{Light-cone distribution amplitudes and form factors}

The QCD approach to hard reactions is based on the concept of factorization: 
one tries to identify the short distance subprocess which 
is calculable in perturbation theory and take into account the contributions of large distances 
in terms of nonprerturbative parton distributions.  

\begin{figure}[ht]
  \includegraphics[width=.85\textwidth,clip=true]{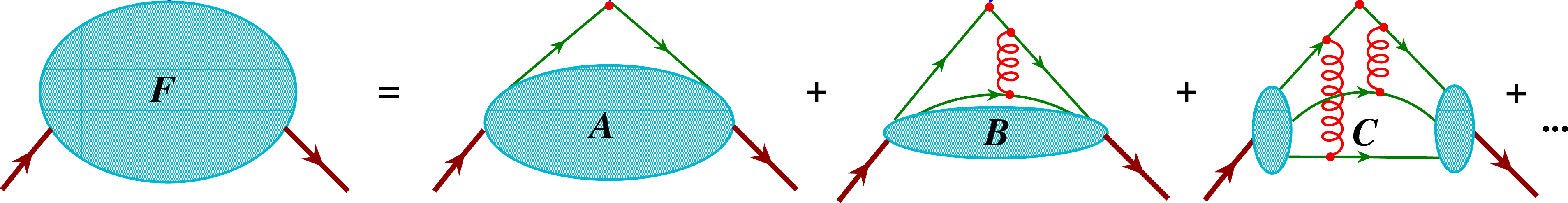}
\caption{\label{figwan}\small
Structure of QCD factorization for baryon form factors.
}
\end{figure}

The problem is that in the case of the baryon form factors
the hard perturbative QCD (pQCD) contribution  is only the third term of the factorization 
expansion. Schematically, one can envisage the expansion of, 
say, the Dirac electromagnetic nucleon form factor
$F_1(Q^2)$ of the form
\begin{eqnarray}
\label{schema}
F_1(Q^2) &~\sim~& A(Q^2)
+ \left ( \frac{\alpha_s(Q^2)}{\pi}\right )  \frac{B(Q^2)}{Q^2} 
+ \left ( \frac{\alpha_s(Q^2)}{\pi}\right ) ^2  \frac{C}{Q^4} + 
\ldots 
\end{eqnarray}
where $C$ is  a constant determined by the nucleon DAs, while $A(Q^2)$ and  $B(Q^2)$
are  form-factor-type functions generated by contributions
of low virtualities, see Fig.~\ref{figwan}.
The soft functions  $A(Q^2)$ and  $B(Q^2)$ are purely nonperturbative and cannot be 
further simplified e.g. factorized in terms of DAs. 
In the light-cone formalism, they are determined by 
overlap integrals of the soft parts of  hadronic wave functions corresponding to large 
transverse separations. 
%
%
Various estimates suggest that $A(Q^2)\lesssim 1/Q^6$, $B(Q^2)\lesssim 1/Q^4$ 
and at very large $Q^2$ they are further  suppressed by 
the Sudakov form factor. To be precise, in higher orders in $\alpha_s(Q)$ there exist double-logarithmic 
contributions $\sim 1/Q^4$ \cite{Duncan:1979hi} that are not 
factorized in the standard manner; however, also they are suppressed by the Sudakov mechanism
\cite{Duncan:1979hi,Kivel:2010ns}. Thus, the third term in (\ref{schema}) is formally 
the leading one at large $Q^2$ to power accuracy.   

The main problem of the pQCD approach~\cite{Lepage:1980fj,Chernyak:1983ej} is a numerical 
suppression of each hard gluon exchange by the $\alpha_s/\pi$ factor which is a standard 
perturbation theory penalty for each extra loop. 
If, say,  $\alpha_s/\pi \sim 0.1$, the pQCD 
contribution to baryon form factors is suppressed by a factor of 
100 compared to the purely soft term.  As the result, 
the onset of the perturbative regime is postponed to very large momentum transfers since 
the factorizable pQCD contribution ${O}(1/Q^4)$ has to win over nonperturbative effects 
that are suppressed by  extra powers of $1/Q^2$, but do not involve small coefficients.  
There is an (almost) overall consensus that ``soft'' contributions play the dominant role at present 
energies. Indeed, it is known for a long time that the use of 
QCD-motivated models for the  wave functions allows one to
obtain, without much effort,  soft contributions comparable in size 
to  experimentally observed values. Also models of generalized parton distributions 
usually are chosen such that the experimental data on form factors 
are described by the soft contributions alone.
A subtle  point for these semi-phenomenological 
approaches is to avoid double counting of hard rescattering 
contributions  ``hidden'' in the model-dependent hadron wave functions
or GPD parametrizations. 

One expects that the rapid development of lattice QCD will allow one to calculate
several benchmark baryon form factors to sufficient precision from first principles. 
Such calculations are necessary and interesting in its own right, but do not add to our 
understanding of how QCD actually ``works'' to transfer the large momentum along the nucleon 
constituents, the quarks and gluons. The main motivation to study ``hard''processes has always 
been to understand hadron properties in terms of 
quark and gluon degrees of freedom; for example, the rationale for the continuing measurements
of the total inclusive cross section in deep inelastic scattering is to extract quark and 
gluon parton distributions. 
Similar, experimental measurements of the electroproduction of nucleon resonances 
at large momentum transfers should eventually allow one to get insight in their structure 
on parton level, in particular momentum fraction distributions of the valence quarks 
and their orbital angular momentum encoded in DAs, 
and this task is obscured by the presence of large ``soft'' contributions 
which have to be subtracted.  

Starting in Ref.~\cite{Braun:2001tj} and in subsequent publications we have been 
developing an approach to hard exclusive processes with baryons
based on light-cone  sum rules (LCSR)~\cite{Balitsky:1989ry,Chernyak:1990ag}.  
This technique is attractive because in LCSRs  ``soft'' contributions to the form 
factors are calculated in terms of the same DAs that enter the pQCD calculation and 
there is no double counting. Thus, the LCSRs provide one with the most 
direct relation of the hadron form factors and distribution amplitudes 
for realistic momentum transfers of the order of $2-10$~GeV$^2$ that is available at present, 
with no other nonperturbative parameters.  It is also sufficiently general and
can be applied to many hard reactions. 

\begin{figure}[ht]
  \includegraphics[width=.35\textwidth,clip=true]{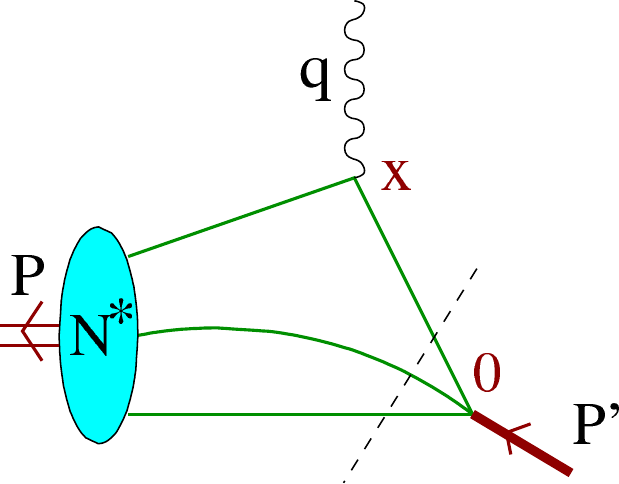}
\caption{\small
Schematic structure of the light-cone sum rule for electroproduction
of nucleon resonances.
}
\label{figsum}
\end{figure}

The basic object of the LCSR approach is the 
correlation function 
\begin{equation}
\int\! dx\, e^{iqx}\langle N^*(P)| T \{ \eta (0) j(x) \} | 0 \rangle 
\end{equation}
 in which $j$  represents the electromagnetic (or weak) probe and $\eta$
is a suitable operator with nucleon quantum numbers.  The nucleon resonance in the final 
state is explicitly represented by its state vector 
 $| N^*(P)\rangle $, see a schematic representation in Fig.~\ref{figsum}.
 When both  the momentum transfer  $q^2=-Q^2$ and 
 the momentum $(P')^2 = (P+q)^2$ flowing in the $\eta$ vertex are large and negative,
 the asymptotic of the correlation function is governed by the light-cone kinematics $x^2\to 0$ and
 can be studied using the operator product expansion (OPE)   
$T \{ \eta(0) j(x) \} \sim \sum C_i(x) {\cal O}_i(0)$ on the 
light-cone $x^2=0$.   The  $x^2$-singularity  of a particular perturbatively calculable
short-distance factor  $C_i(x)$  is determined by the twist of the relevant
composite operator ${\cal O}_i$, whose matrix element $\langle N^*|  {\cal O}_i(0)| 0 \rangle $
is given by an appropriate moment of the $N^*$ DA.
Next, one can represent the answer in form of the dispersion integral in 
$(P')^2$ and define the nucleon contribution
by the cutoff in the quark-antiquark invariant mass, 
the so-called interval of duality $s_0$ (or continuum threshold).
The main role of the interval of duality is that it does not allow large momenta $|k^2| > s_0$ to 
flow through the  $\eta$-vertex; to the lowest order $O(\alpha_s^0)$ one obtains a purely soft    
contribution to the form factor as a sum of terms ordered by twist of the relevant operators and
hence including both the leading- and the higher-twist nucleon DAs. Note that, in difference to the 
hard mechanism, the  contribution of higher-twist DAs is only suppressed by powers of 
the interval of duality $s_0\sim 2$~GeV$^2$ (or by powers of the Borel parameter if one applies 
some standard QCD sum rule machinery), but not by powers of $Q^2$. 
This feature is in agreement with the common wisdom that soft contributions are not 
constrained to small transverse separations.

\begin{figure}[t]
\begin{center}
\includegraphics[width=0.55\textwidth,clip]{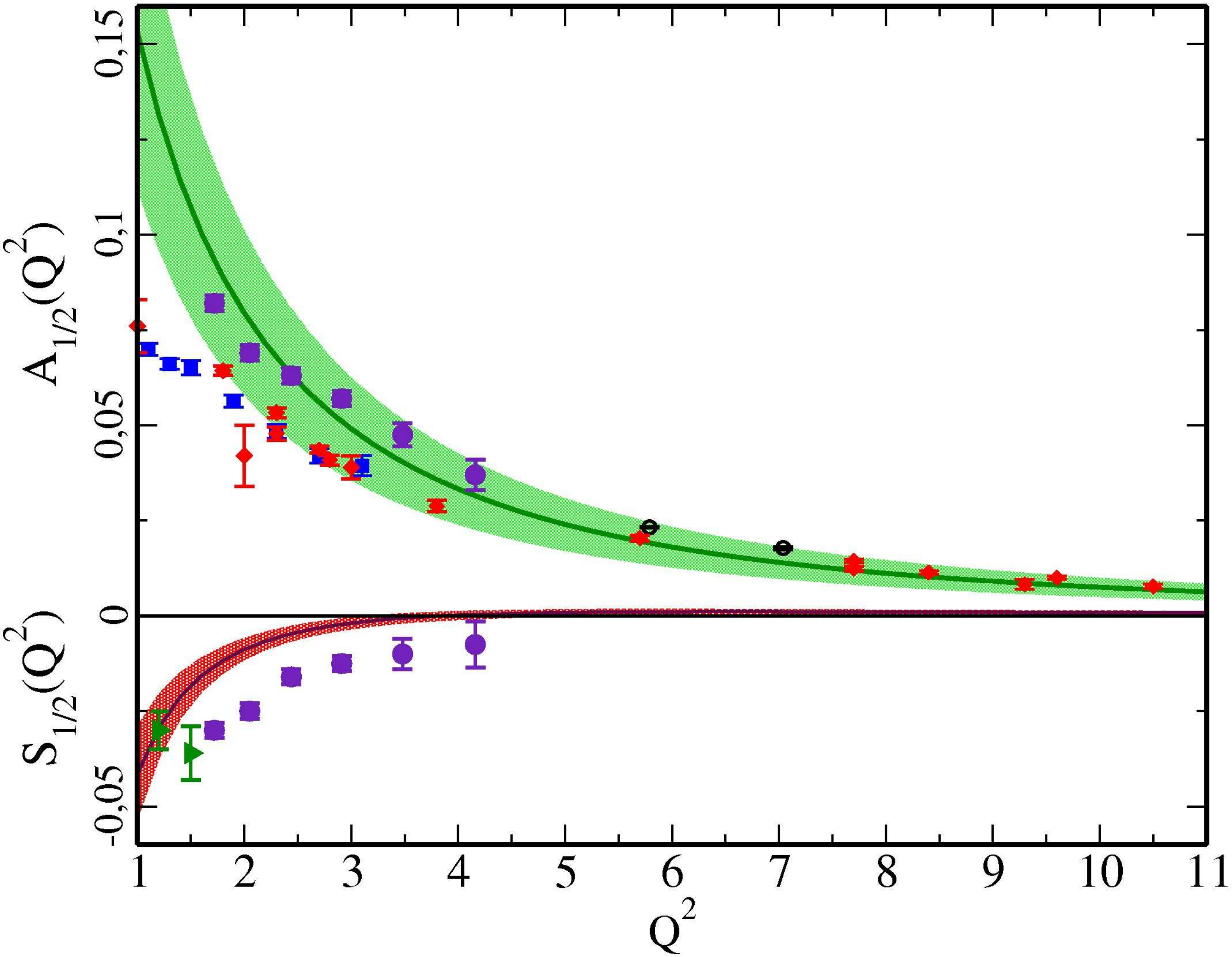}
\end{center}
\caption{The LCSR calculation for the helicity amplitudes
$A_{1/2}(Q^2)$ and $S_{1/2}(Q^2)$ for the electroproduction of the 
$N^*(1535)$ resonance using the lattice results 
for the lowest moments of the  $N^*(1535)$ DAs. 
The curves are obtained using the central values of the lattice 
parameters, and the shaded areas show the corresponding uncertainty.
Figure taken from Ref.~\cite{Braun:2009jy}.}
\label{fig:A12S12}
\end{figure}

We stress that LCSRs are not based on any nonperturbative model of the nucleon 
structure, but rather present a relation between the physical observables (form factors) 
and baryon wave functions at small transverse separation (distribution amplitudes).

\begin{figure}[t]
\begin{center} 
\includegraphics[width=0.45\textwidth,angle=0]{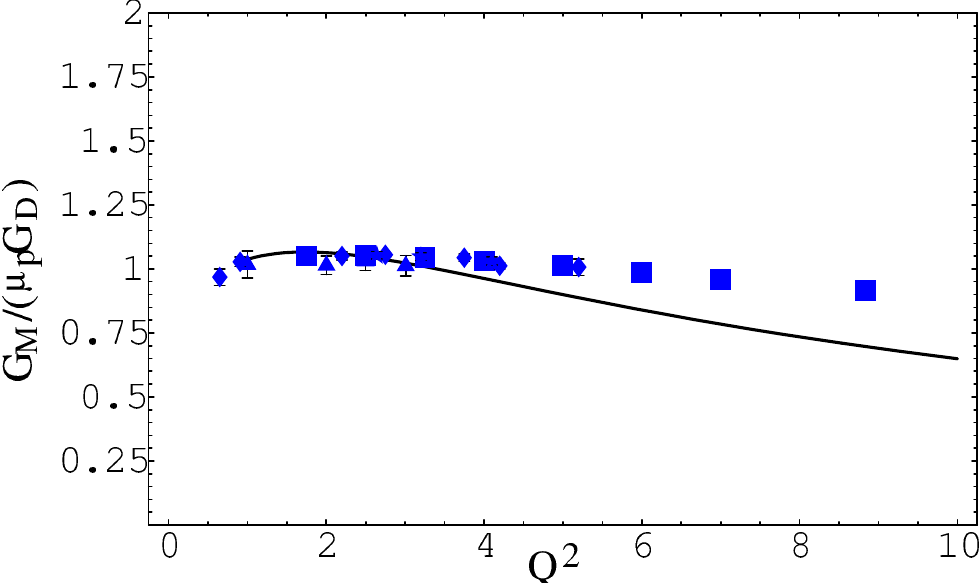}
~~~
\includegraphics[width=0.45\textwidth,angle=0]{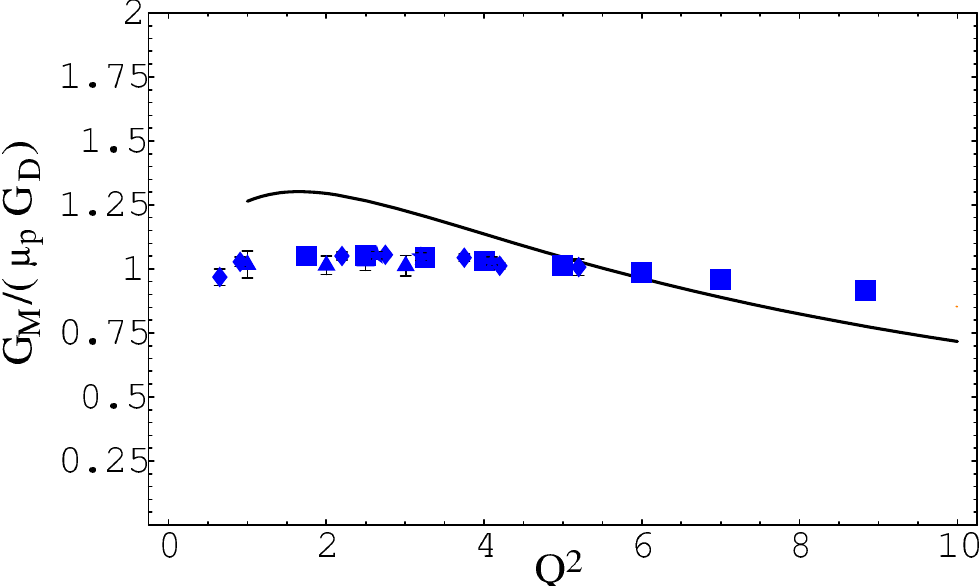}
\end{center}
\caption{LCSR results for the magnetic proton form factor (normalized to
 the dipole formula) for a realistic model of nucleon distribution amplitudes
 \cite{Braun:2006hz}. Left panel: Leading order (LO); right panel: 
next-to-leading order (NLO) for twist-three contributions.
Figure adapted from Ref.~\cite{PassekKumericki:2008sj}.}
\label{fig:Korn1}
\end{figure}

Historically, LCSRs were developed in Refs.~\cite{Balitsky:1989ry,Chernyak:1990ag} 
in an attempt to overcome difficulties of the Shifman-Vainstein-Zakharov QCD sum rule
approach~\cite{Shifman:1978bx} for exclusive processes dominated by the light-cone kinematics.   
In the last 20 years LCSRs have been applied extensively to the 
exclusive $B$-decays 
and remain to be the only nonperturbative technique that allows one to calculate the 
corresponding form factors directly at large recoil. 
In fact the value of the CKM matrix element $V_{ub}$ quoted by the Particle Data Group
as the one extracted from exclusive semileptonic decay 
$B\to\pi\ell\nu_\ell$ is largely based on the recently updated LCSR calculations of the form 
factor $f_+^{B\to\pi}(0)$ \cite{Ball:2004ye,Duplancic:2008ix} (although the  
lattice QCD calculations have become competitive).  
Another important application of LCSRs was for calculation of the 
electromagnetic pion form factor.
More references and further details can be found in the review 
articles \cite{Braun:1997kw,Colangelo:2000dp}.

LCSRs for meson form factors have achieved a certain degree of maturity. One lesson 
is that they are fully consistent with pQCD and factorization theorems.
 In particular the LCSRs also  contain terms  
generating the asymptotic pQCD contributions. 
In the pion case, it was explicitly demonstrated 
that the contribution of hard rescattering is correctly reproduced in the LCSR   
approach as a part of the $O(\alpha_s)$ correction.
It should be noted that  the  diagrams of LCSR that 
contain the ``hard'' pQCD  contributions also possess ``soft'' parts,
i.e., one should perform  a separation  of ``hard'' and ``soft''
terms inside each diagram.  As a result, 
the distinction between ``hard'' and ``soft'' contributions appears to 
be scale- and scheme-dependent.
Most of the LCSRs for meson decays have been derived to the next-to-leading-order
(NLO) accuracy in the strong coupling. The first NLO LCSR calculations were done
in 1997--1998 and since then the NLO accuracy has become standard in this field. 
The size of NLO corrections depends on the form factor in question but typically is of the 
order of 20\%, for the momentum transfers of interest.

\begin{figure}[t]
\begin{center} 
\includegraphics[width=0.45\textwidth,angle=0]{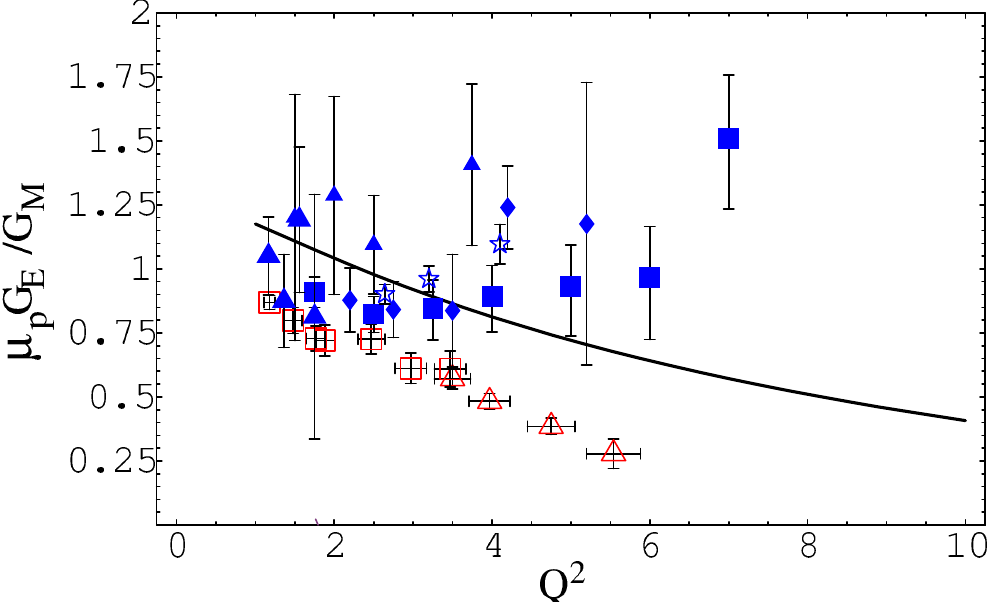}
~~~
\includegraphics[width=0.45\textwidth,angle=0]{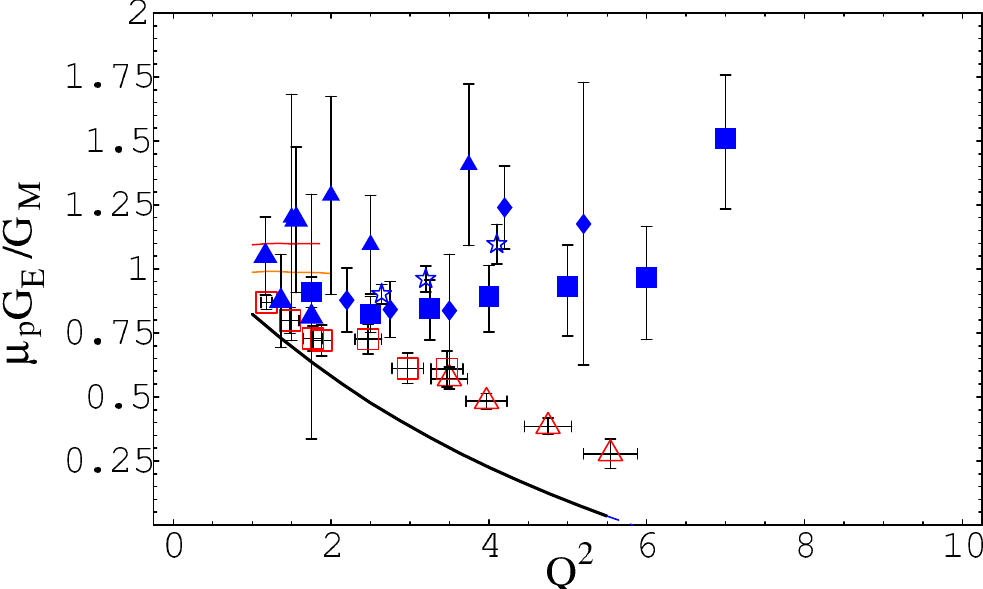}
\end{center}
\caption{LCSR results for the electric to magnetic proton form factor ratio for 
a realistic model of nucleon distribution amplitudes \cite{Braun:2006hz}. Left
 panel: Leading order (LO); right panel: 
next-to-leading order (NLO) for twist-three contributions.
Figure adapted from Ref.~\cite{PassekKumericki:2008sj}.}
\label{fig:Korn2}
\end{figure}

Derivation of LCSRs for exclusive reactions involving baryons is, conceptually,
a straightforward generalization of the LCSRs for mesons. On the other hand,
there are a few new technical issues that had to be resolved, and also
the calculations become much more challenging.
The development so far was mainly to explore the existing possibilities and identify 
potential applications. 
Following the first application to the electromagnetic and axial form factors of the nucleon 
in Refs.~\cite{Braun:2001tj,Braun:2006hz}, LCSRs have been considered for 
the $\gamma^* N\to \Delta$ transition~\cite{Braun:2005be}, heavy baryon decays 
(see \cite{Khodjamirian:2011jp} and references therein)
and various transitions
between baryons in the octet and the decuplet (e.g.~\cite{Aliev:2011uf}). 
In the work  \cite{Braun:2009jy} we have suggested  to use the same approach to 
the study of electroproduction of resonances at large momentum transfers
and in particular $N^*(1535)$.   
Since the structure of sum rules for the nucleon elastic form factors and 
electroproduction of $N^*(1535)$ is very similar, the difference in form factors 
should expose directly the difference in the wave functions, which is of prime interest.
The results for the helicity amplitudes
$A_{1/2}(Q^2)$ and $S_{1/2}(Q^2)$ using the lattice results 
for the lowest moments of the  $N^*(1535)$ DAs appear to be in a good agreement with 
the existing data, see Fig.~\ref{fig:A12S12}.

 All existing LCSRs for baryons are written to the leading order in the strong 
coupling which corresponds, roughly speaking, to the parton model
level description of deep-inelastic scattering. Combined with realistic models of DAs
the existing sum rules yield a reasonable description of the existing data to the 
expected 30-50\% accuracy.
In order to match the accuracy of the future experimental data 
and also of the next generation of lattice results, 
the LCSRs will have to be advanced to include NLO radiative corrections, as it has 
become standard for meson decays.

The first step towards LCSRs to the NLO accuracy was done in 
Ref.~\cite{PassekKumericki:2008sj} where the $\mathcal{O}(\alpha_s)$ 
corrections are calculated for the (leading) twist-three contributions 
to the sum rules for electromagnetic (elastic) nucleon form factors 
derived in \cite{Braun:2001tj,Braun:2006hz}. The results are shown in
Fig.~\ref{fig:Korn1} and Fig.~\ref{fig:Korn2}.

The NLO corrections are large and their effect {\it increases} with 
$Q^2$ which may be counterintuitive. This behavior is, however, expected 
on general grounds because the leading regions for large momentum 
transfers corresponding to the ERBL (Efremov-Radyushkin-Brodsky-Lepage)
collinear factorization appear at the NNLO level only, i.e.
$\mathcal{O}(\alpha^2_s)$. The corrections for the $G_E/G_M$ ratio 
are larger than for the magnetic form factor $G_M$ itself, which is 
again expected since the electric form factor suffers from cancellations 
between chirality-conserving and chirality-violating contributions.

Large NLO corrections can be compensated by the change in the nucleon DA,
similar as it happens with parton distributions --- e.g. the
small-$x$ behavior of the LO and NLO gluon distribution is very different ---
but such an analysis would so far be premature since NLO corrections 
have not been calculated so far for the contributions of twist-four DAs
that take into account the effects of orbital angular momentum. 

In addition, it is necesary to develop a technique for the 
resummation of ``kinematic'' corrections to the sum rules 
that are due to nonvanishing masses of the resonances. The corresponding 
corrections to the total cross section of the deep-inelastic scattering
are known as Wandzura-Wilczek corrections and can be resummed to all 
orders in terms of the Nachtmann variable; we are looking for a generalization
of this method to non-forward kinematics which is also important in
a broader context~\cite{Braun:2011zr}. 

With these improvements, we expect that the LCSR approach can be used to 
constrain light-cone DAs of the nucleon and its resonances from the 
comparison with the electroproduction data. These constraints can 
then be compared with the lattice QCD calculations. 
In order to facilitate this comparison, a work is in progress to derive general 
expressions for the necessary light-cone sum rules to the NLO accuracy. 
The project is to have the LCSRs 
available as a computer code allowing one to calculate elastic electromagnetic 
and axial form factors and also a range of transition form factors involving 
nucleon resonances from a given set of distribution amplitudes.
Although gross features of the wave functions of resonances can definitely be 
extracted from such an analysis, the level of details ``seen'' in sum rule 
calculations will have to be tested on case by case basis. For this reason 
we are also working on similar calculations
for the ``gold-plated'' decays like $\gamma^*\to\pi\gamma$, $\gamma^*\to\eta\gamma$,
see \cite{Agaev:2010aq}, where the theoretical 
uncertainties are expected to be small.

\section{Quark-Hadron Duality and Transition Form Factors}\label{duality}

\subsection{Historical perspective}
\label{sec:hist}

Understanding the structure and interactions of hadrons at intermediate
energies is one of the most challenging outstanding problems in nuclear
physics.  While many hadronic observables can be described in terms of
effective meson and baryon degrees of freedom at low energies,
at energies $\gg$ the nucleon mass $M$ perturbative QCD has been very
successful in describing processes in terms of fundamental quark and
gluon (parton) constituents.

\begin{figure}[ht]
\begin{center}
\includegraphics[width=10cm]{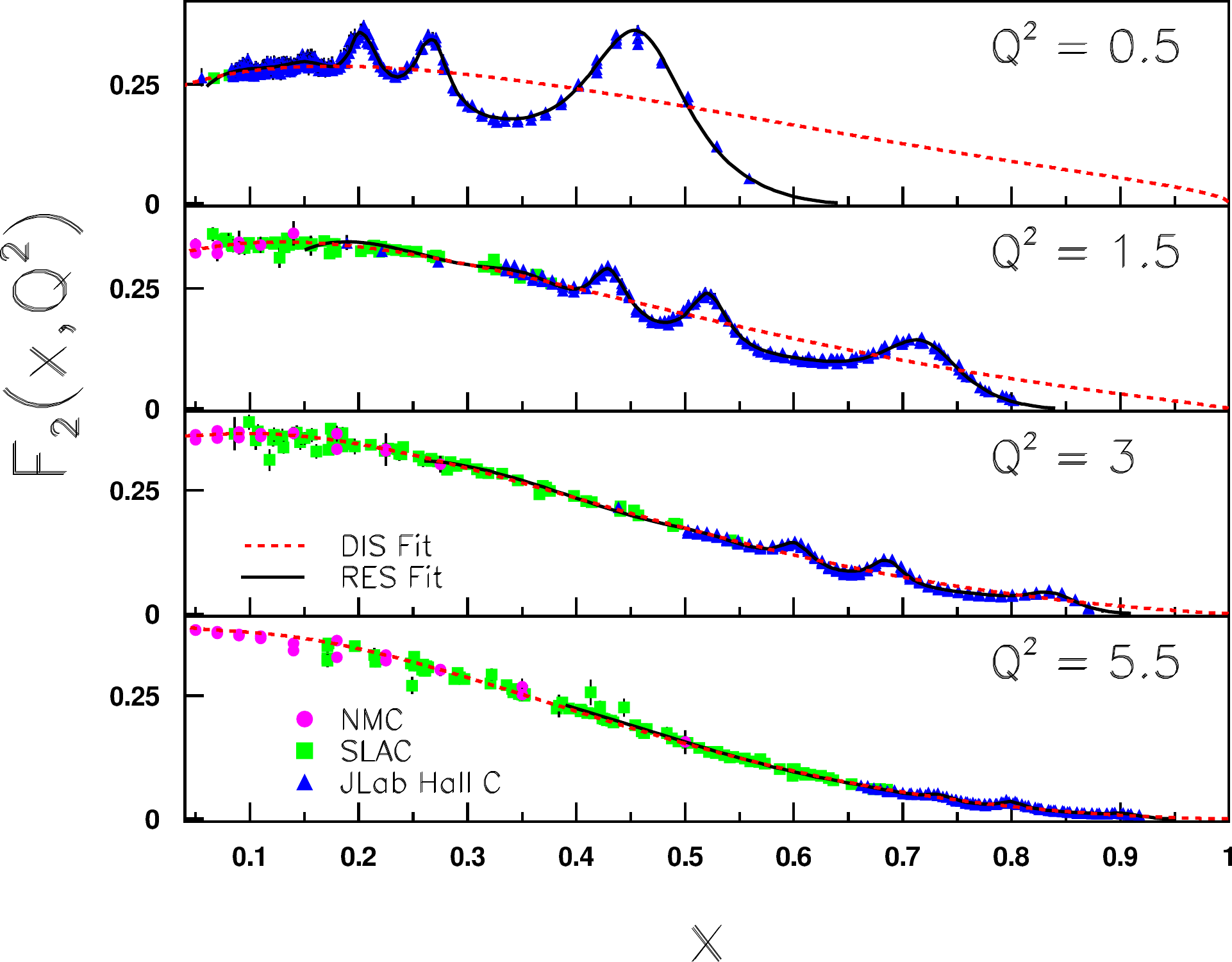}
\caption{\label{fig:f2}
	Proton $F_2$ structure function data from Jefferson Lab Hall~C
	\cite{Malace:2009kw,Liang:2004tj,E00-002}, SLAC \cite{Whitlow:phd,Whitlow:1991uw},
        and NMC \cite{Aglietta:1997pw} at $Q^2 = 0.5, 1.5, 3$ and 5.5~GeV$^2$,
        compared with an empirical fit \cite{Christy:2007ve} to the transverse
	and longitudinal resonance cross sections (solid), and a global
	fit to DIS data (dashed).
	(Figure from Ref.~\cite{Christy:2011cv}.)}
\end{center}
\end{figure}

A connection between the low and high energy realms is realized
through the remarkable phenomenon of quark-hadron duality, where one
often finds dual descriptions of observables in terms of either explicit
partonic degrees of freedom, or as averages over hadronic variables.
In principle, with access to complete sets of either hadronic or partonic
states, the realization of duality would be essentially trivial,
effectively through a simple transformation from one complete set of
basis states to another.
In practice, however, at finite energies one is typically restricted to
a limited set of basis states, so that the experimental observation of
duality raises the question of not {\it why} duality exists, but rather
{\it how} it arises {\it where} it exists, and how we can make use of it.

\begin{figure}[ht]
\begin{center}
\includegraphics[width=12cm]{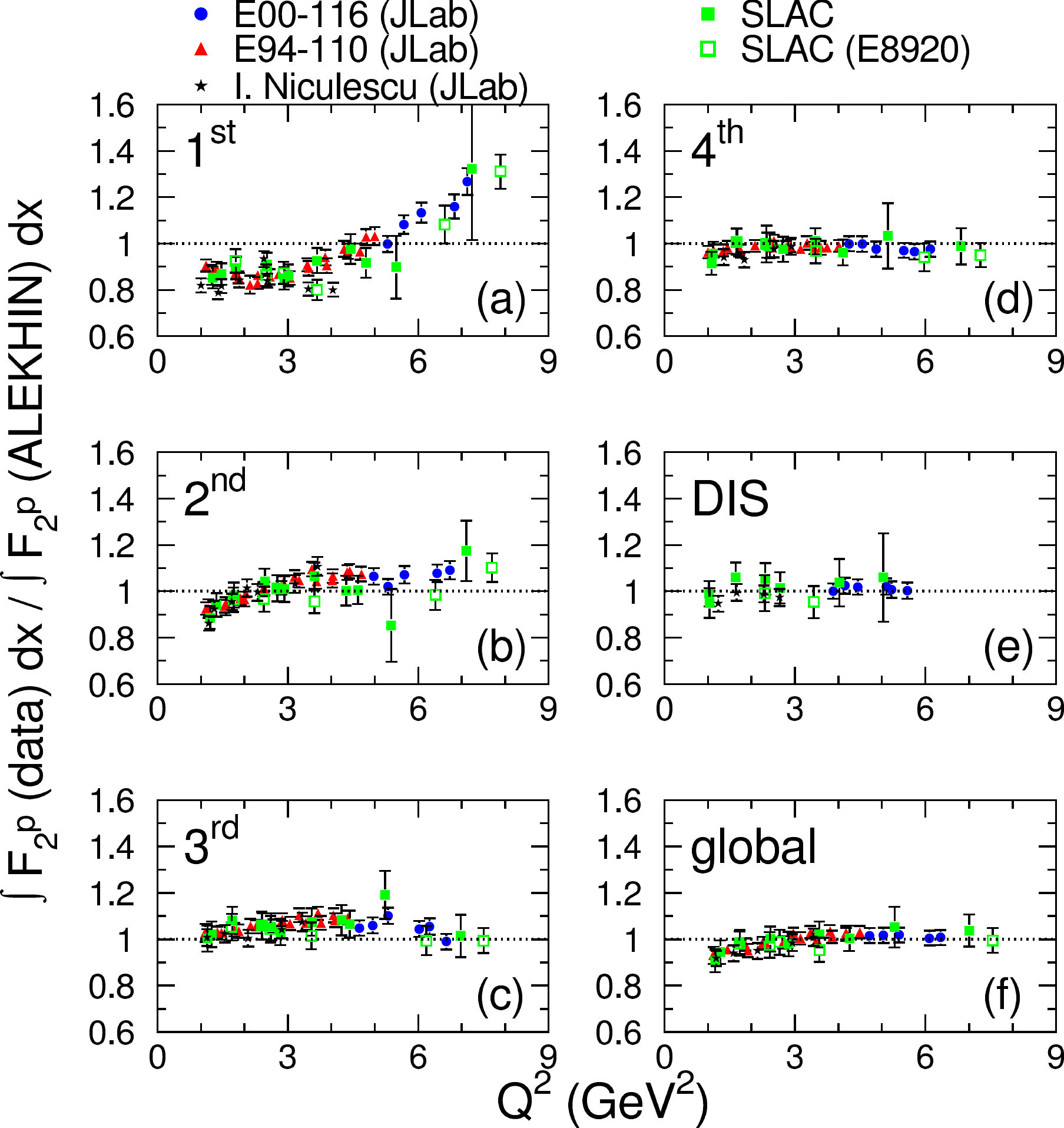}
\caption{\label{fig:mom}
	Ratio of proton $F_2^p$ structure functions integrated over
	specific resonance regions, relative to the global fit of
	parton distributions from Alekhin {\it et al.} \cite{Alekhin:2005gq,Alekhin:2000ch}.
	(Figure from Ref.~\cite{Malace:2009kw}.)}
\end{center}
\end{figure}

Historically, duality in the strong interaction physics represented
the relationship between the description of hadronic scattering
amplitudes in terms of $s$-channel resonances at low energies,
and $t$-channel Regge poles at high energies \cite{Fukugita:1976ab}.
The merger of these dual descriptions at intermediate energies remained
a prized goal of physicists in the decade or so before the advent of QCD.
Progress towards synthesizing the two descriptions was made with the
development of finite energy sum rules (FESRs) \cite{Collins:1977jy,Donnachie:2002en},
\begin{eqnarray}
\int_0^{\nu_{\rm max}} d\nu\ \nu^n\ \Im m\ {\cal A}(\nu,t)
&=& \int_0^{\nu_{\rm max}} d\nu\ \nu^n\
    \Im m\ {\cal A}_{\rm asy}(\nu,t)\ \ \ \ \ \ \ \ \ [\rm FESR]
\label{eq:fesr}
\end{eqnarray}
relating the imaginary part of the amplitude ${\cal A}$ at finite
energy to the asymptotic high energy amplitude ${\cal A}_{\rm asy}$,
where $s$, $t$ and $u$ are the usual Mandelstam variables and
$\nu \equiv (s-u)/4$. The asymptotic amplitude ${\cal A}_{\rm asy}$ is then extrapolated
into the $\nu < \nu_{\rm max}$ region and compared with the
measured amplitude ${\cal A}$ through Eq.~(\ref{eq:fesr}).
The assumption made here is that beyond some maximum energy
$\nu > \nu_{\rm max}$ the scattering amplitude can be represented
by its asymptotic form, calculated within Regge theory.

The FESRs are generalizations of superconvergence relations in Regge
theory relating dispersion integrals over the amplitudes at low
energies to high-energy parameters.  They constitute a powerful
tool allowing one to use experimental information on the low energy
cross sections for the analysis of high energy scattering data.
Conversely, they can be used to connect low energy parameters
(such as resonance widths and couplings) to parameters describing
the behavior of cross sections at high energy.
It was in the context of FESRs, in fact, that the early expressions
of Bloom-Gilman duality were made in the early 1970s \cite{Bloom:1971ye,Bloom:1970xb},
suitably extended to lepton scattering kinematics.

\subsection{Duality in nucleon structure functions}
\label{sec:BG}

One of the most dramatic realizations of duality in nature is in
inclusive electron--nucleon scattering, usually referred to as
``Bloom-Gilman'' duality, where structure functions averaged over
the resonance region are found to be remarkably similar to the leading
twist structure functions describing the deep-inelastic scattering
(DIS) continuum \cite{Bloom:1971ye,Bloom:1970xb,Niculescu:2000tk,Niculescu:2000tj,Melnitchouk:2005zr,Malace:2009kw,Christy:2011cv}.
As Fig.~\ref{fig:f2} illustrates, the resonance data are seen to
oscillate around the scaling curve and slide along it with increasing
$Q^2$.

An intriguing feature of the lepton scattering data is that the duality
appears to be realized not just over the entire resonance region as a
whole, $W \lesssim 2$~GeV, where $W^2 = M^2 + Q^2 (1-x)/x$, but also
in individual resonance regions.
This is illustrated in Fig.~\ref{fig:mom} for the ratios of structure
functions integrated over specific intervals of $W$ at fixed $Q^2$,
with the 1st, 2nd, 3rd and 4th resonance regions defined by
$1.3 \leq W^2 \leq 1.9$~GeV$^2$,
$1.9 \leq W^2 \leq 2.5$~GeV$^2$,
$2.5 \leq W^2 \leq 3.1$~GeV$^2$, and
$3.1 \leq W^2 \leq 3.9$~GeV$^2$, respectively.
The ``DIS'' region in Fig.~\ref{fig:mom} is defined to be
$3.9 \leq W^2 \leq 4.5$~GeV$^2$.
In all cases the duality is realized at the $\lesssim 10-15\%$ level,
suggesting that Bloom-Gilman duality exists {\it locally} as well as
globally.

Understanding the microscopic origin of quark-hadron duality has
proved to be a major challenge in QCD.  Until recently the only
rigorous connection with QCD has been within the operator product
expansion (OPE), in which moments (or $x$-integrals) of structure
functions are expanded as a series in inverse powers of $Q^2$.
The leading, ${\cal O}(1)$ term is given by matrix elements of
(leading twist) quark-gluon bilocal operators associated with free
quark scattering, while the ${\cal O}(1/Q^2)$ and higher terms
correspond to nonperturbative (higher twist) quark-gluon interactions.
In the language of the OPE, duality is then synonymous with the
suppression of higher twist contributions to the moments \cite{DeRujula:1976tz}.

This close relationship between the leading twist cross sections and
the resonance-averaged cross sections suggests that the total higher
twist contributions are small at scales $Q^2 \sim 1$~GeV$^2$.
This implies that, on average, nonperturbative interactions between
quarks and gluons are not dominant at these scales, and that a highly
nontrivial pattern of interferences emerges between the resonances
(and the nonresonant background) to effect the cancellation of the
higher twist contributions.  The physics of parton distributions
and nucleon resonances is therefore intimately connected.
In fact, in the limit of a large number of colors, the spectrum of
hadrons in QCD is one of infinitely narrow resonances \cite{Einhorn:1976uz},
which graphically illustrates the fact that resonances are an
integral part of scaling structure functions.

The phenomenological results raise the question of how can a scaling
structure function be built up entirely from resonances, each of whose
contribution falls rapidly with $Q^2$ \cite{Isgur:2001bt}?  A number of studies
using various nonperturbative models have demonstrated how sums over
resonances can indeed yield a $Q^2$ independent function
(see Ref.~\cite{Melnitchouk:2005zr} for a review).
The key observation is that while the contribution from each individual
resonance diminishes with $Q^2$, with increasing energy new states
become accessible whose contributions compensate in such a way as
to maintain an approximately constant strength overall.  At a more
microscopic level, the critical aspect of realizing the suppression of
the higher twists is that at least one complete set of even and odd   
parity resonances must be summed over for duality to hold \cite{Close:2001ha}.
Explicit demonstration of how this cancellation takes place was made
in the SU(6) quark model and its extensions \cite{Close:2001ha,Close:2003wz,Close:2009yj}.

One of the ultimate goals of duality studies is to determine the
extent to which resonance region data can be used to learn about
leading twist structure functions.  At present, most global analyses
of parton distribution functions impose strong cuts on $Q^2$ and $W^2$
for lepton scattering data in order to exclude the region where higher
twists and other subleading effects are important.
By relaxing the cuts to just inside the traditional resonance region,
$W \gtrsim 1.7$~GeV, the CTEQ-Jefferson Lab (CJ) collaboration could
increase the statistics of the DIS data by a factor $\sim 2$
\cite{Accardi:2009br,Accardi:2011fa}!
Not only were the fits found to be stable with the weaker cuts,
the larger database led to significantly reduced errors,
up to 40-60\% at large $x$, where data are scarce.
Future plans include extending these cuts to even lower values
of $W$, which demands better understanding of the resonance region
and the procedures for systematically averaging over the resonance
structure functions.

The determination of parton distributions at large $x$ is vital not
just for understanding the dynamics of valence quarks in the nucleon
\cite{Melnitchouk:1995fc,Holt:2010vj}, which are currently obscured by nuclear corrections
in deuterium DIS data needed to extract the structure function of the
free neutron.  It is also critical in applications to experiments at
high energy colliders, where uncertainties at large $x$ in the $d$ quark 
distribution in particular feeds down to lower $x$ at higher $Q^2$
\cite{Kuhlmann:1999sf} and can have important consequences for searches for
new particles, such as $W'$ and $Z'$ bosons \cite{Brady:2011hb}.
Thus in an indirect way, better knowledge of the nucleon resonance
region can have a profound impact on physics at the LHC!

\subsection{Duality in inclusive meson production}

\begin{figure}[hbt]
\epsfig{file=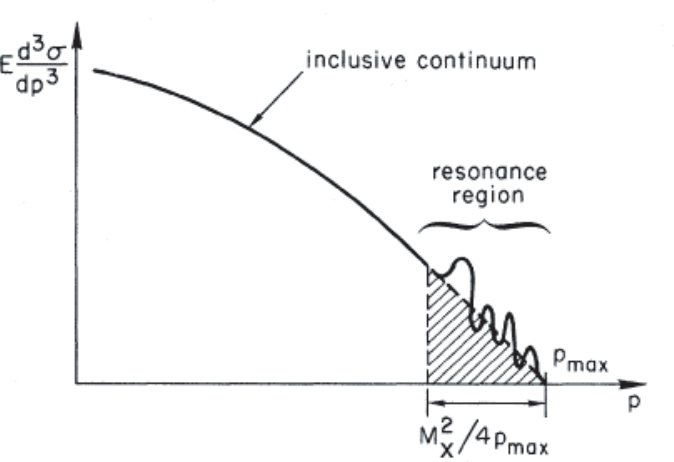,height=6cm}
\caption{\label{fig:corresp}
        Momentum spectrum of produced hadrons in the inclusive hadron
        production reaction $\gamma^* N \to M X$.
        From Ref.~\protect\cite{Bjorken:1973gc}.}
\end{figure}

\begin{figure}[hbt]
\vspace*{0.5cm}
\epsfig{file=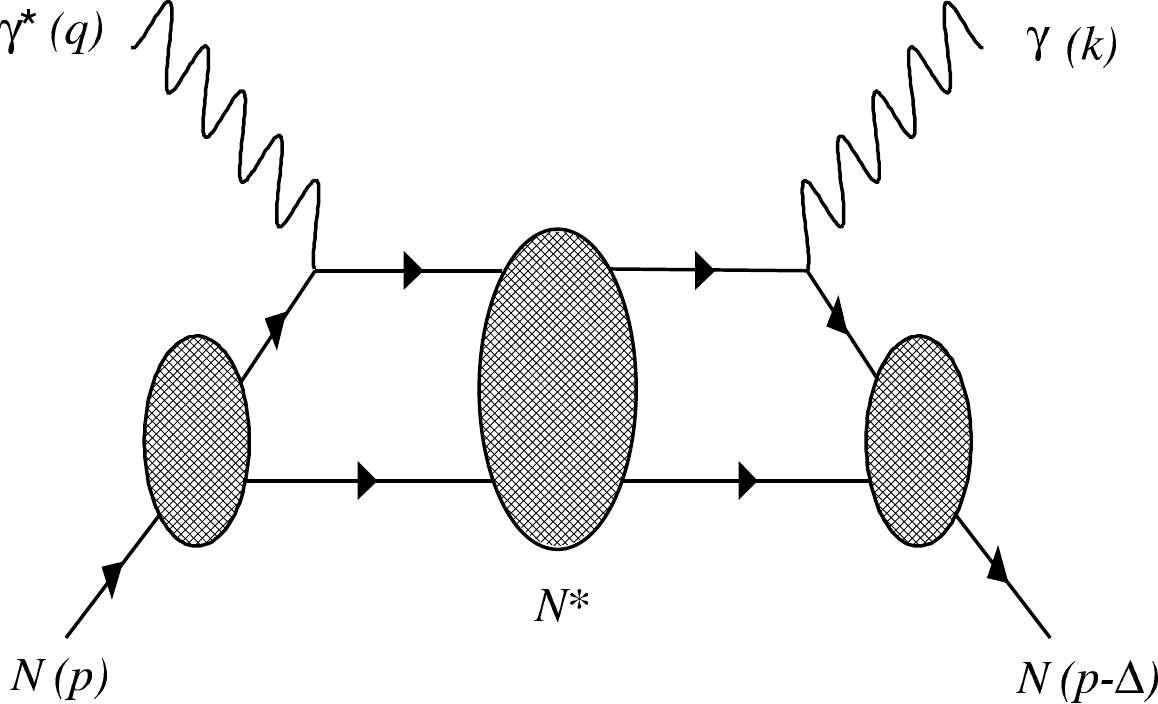,height=5cm}
\caption{\label{fig:dvcs}
	Deeply virtual (nonforward) Compton scattering at the
	hadronic level, with excitation of nucleon resonances $N^*$
	in the intermediate state.
	(Adapted from Ref.~\cite{Close:2002tm}.)}
\end{figure}

Extending the concept of duality to less inclusive reactions,
we can ask whether the semi-inclusive production of mesons 
displays a similar relation between partonic and resonance-based
descriptions.  Such studies have only recently been performed,
for ratios of semi-inclusive $\pi^+$ to $\pi^-$ cross sections
measured at Jefferson Lab as a function of $z = E_{\pi}/\nu$,
where $E_\pi$ is the pion energy and $\nu$ is the energy transfer
to the target \cite{Navasardyan:2006gv}.

The data displayed a smooth behavior in $z$, consistent with earlier
observations at higher energies at CERN \cite{Arneodo:1989ic}, prompting
suggestions that factorization of semi-inclusive cross sections into
scattering and fragmentation sub-processes may hold to relatively
low energies.  Such factorization was found in fact in simple quark
models by explicitly summing over $N^*$ resonances in the $s$-channel
of $\gamma^* N \to \pi N$ scattering \cite{Close:2009yj}.

At the quark level, the (normalized) semi-inclusive cross section
for the production of pions from a nucleon target can be factorized
(at leading order in $\alpha_s$) into a product of a parton
distribution function describing the hard scattering from a parton
in the target, and the probability of the struck parton fragmenting
into a specific hadron,
\begin{equation}
{d\sigma \over dx dz}\
\propto\ \sum_q\ e_q^2\ q^N(x)\ D_q^\pi(z),
\label{eq:Nxz}
\end{equation}
where $e_q$ is the quark charge, and $D_q^\pi$ is the fragmentation
function for quark $q$ to produce a pion with energy fraction $z$.
As pointed out by Close and Isgur \cite{Close:2001ha}, duality between structure
functions represented by (incoherent) parton distributions and by a
(coherent) sum of squares of form factors can be achieved by summing
over neighboring odd and even parity states.
In the SU(6) model this is realized by summing over states in the
$56^+$ ($L=0$, even parity) and $70^-$ ($L=1$, odd parity)
multiplets, with each representation weighted equally.

The pion production cross sections at the hadronic level are
constructed by summing coherently over excited nucleon resonances
($N_1^*$) in the $s$-channel intermediate state and in the final state
($N_2^*$) of $\gamma N \to N_1^* \to \pi N_2^*$, where both $N_1^*$
and $N_2^*$ belong to the $56^+$ and $70^-$ multiplets.
Within this framework, the probabilities of the
$\gamma N \to \pi N_2^*$ transitions can be obtained by summing
over the intermediate states $N_1^*$ spanning the $56^+$ and
$70^-$ multiplets, with the differential cross section
\begin{eqnarray}
{d\sigma \over dx dz}\
&\propto& \sum_{N_2^*}
\left| \sum_{N_1^*}
       F_{\gamma N \to N_1^*}(Q^2,M_1^*)\
       {\cal D}_{N_1^* \to N_2^* \pi}(M_1^*,M_2^*)\
\right|^2.
\end{eqnarray}
Here $F_{\gamma N \to N^*}$ is the $\gamma N \to N^*$ transition
form factor, which depends on the masses of the virtual photon and
excited nucleon ($M_1^*$), and ${\cal D}_{N_1^* \to N_2^* \pi}$ is a
function representing the decay $N_1^* \to \pi N_2^*$, where $M_2^*$
is the invariant mass of the final state $N_2^*$.

Summing over the $N_2^*$ states in the $56^+$ and $70^-$ multiplets,
one finds ratios of unpolarized $\pi^-$ to $\pi^+$ semi-inclusive
cross sections consistent with the parton model results for ratios of
parton distributions satisfying SU(6) symmetry \cite{Close:2001ha,Close:2003wz, Close:2009yj}.
Duality was also found to be realized in more realistic scenarios with
broken SU(6) symmetry, with sums over resonances able to reproduce
parton model semi-inclusive cross section ratios \cite{Close:2009yj}.
The absence of strong resonant enhancement on top of the smooth       
background is indeed one of the notable features of the Jefferson Lab
Hall~C data \cite{Navasardyan:2006gv}, in accord with expectations from duality.

\subsection{Exclusive-inclusive connection}

The general folklore in hadronic physics is that duality works more
effectively for inclusive observables than for exclusive, due to the
presence in the latter of fewer hadronic states over which to average.
For exclusive processes, such as the production of a meson $M$ in
coincidence with and a baryon $B$, $e N \to e M B$, duality may be
more speculative.  Nevertheless, there are correspondence arguments
formulated long ago which relate the exclusive cross sections at
low energy to inclusive production rates at high energy.
The exclusive--inclusive connection dates back to the early dates of
DIS and the discussion of scaling laws in high energy processes.
Bjorken \& Kogut \cite{Bjorken:1973gc} proposed the correspondence relations
by demanding the continuity of the dynamics as one goes from one (known)
region of kinematics to another (which is unknown or poorly known).

For processes such as $\gamma^* N \to M B$, the correspondence
principle relates properties of exclusive (resonant) final states
with inclusive particle spectra for the corresponding reaction
$\gamma^* N \to M X$.  This is illustrated in Fig.~\ref{fig:corresp}
for a typical inclusive momentum spectrum $E d^3\sigma/dp^3$,
where $E$ and $p$ are the energy and momentum of the observed final
state particle $M$.  As $p$ increases, the inclusive continuum
gives way to to the region dominated by resonances.
The correspondence argument postulates that the resonance contribution
to the cross section should be comparable to the continuum contribution
extrapolated from high energy into the resonance region,
\begin{eqnarray}
\int_{p_{\rm min}}^{p_{\rm max}} dp
\left. E\, { d^3\sigma \over dp^3 } \right|_{\rm incl}
&\sim&\ \ \sum_{\rm res}
       \left. E\, { d\sigma \over dp_T^2 } \right|_{\rm excl}\ ,
\label{eq:corresp}
\end{eqnarray}

where the integration region over the inclusive cross section
includes contributions up to a missing mass $M_X$, with
$p_{\rm min} = p_{\rm max} - M_X^2/4 p_{\rm max}$.
The correspondence relation (\ref{eq:corresp}) is another manifestation
of the FESR in Eq.~(\ref{eq:fesr}), in which the cross section in the
resonance region for $p_{\rm min} < p < p_{\rm max}$ is dual to the
high-energy cross section extrapolated down to the same region.

The inclusive cross section $d^3\sigma/dp^3$ is generally a function
of the longitudinal momentum fraction $x$, the transverse momentum
$p_T$, and the invariant mass squared $s$,
\begin{eqnarray}
{ E \over \sigma }
{ d^3\sigma \over dp^3 }
&\equiv& f(x,p_T^2,sQ^2)\ .
\end{eqnarray}
At large $s$ or large $Q^2$ this effectively reduces to a function
of only $x$ and $p_T^2$,
\begin{eqnarray}
f(x,p_T^2,sQ^2) &\to& f(x,p_T^2)\ ,\ \ \ \ s \to \infty\ .
\end{eqnarray}
The continuity relation (\ref{eq:corresp}) implies that there should be
no systematic variation of either side of the equation with external
parameters.

Applications of the exclusive--inclusive correspondence have also
been made to real Compton scattering cross sections from the proton
at large center of mass frame angles \cite{Scott:1974gc}, as well as to
hard exclusive pion photoproduction \cite{Scott:1975eh,Eden:2001ci,Zhao:2003db},
and more recently to deeply virtual Compton scattering,
$e p \to e \gamma p$ \cite{Close:2002tm}.
The latter in particular used a simple model with scalar constituents
confined by a harmonic oscillator potential to show how sums over
intermediate state resonances, Fig.~\ref{fig:dvcs}, lead to destructive
interference between all but the elastic contribution, and the emergence
of scaling behavior for the associated generalized parton distributions
(GPDs).

Future work will build on these exploratory studies, generalizing the
calculations to include spin-1/2 quarks and non-degenerate multiplets,
as well as incorporating nonresonant background within same framework
\cite{Fiore:2003dg,Fiore:2002re}.  Extension to flavor non-diagonal transitions will also
establish a direct link between transition form factors and the GPDs,
with duality providing a crucial link between the hadronic and
partonic descriptions.

\section{Light-Front Holographic  QCD \label{Light-Front Holographic}}

\begin{figure}
  \includegraphics[width=3.0in]{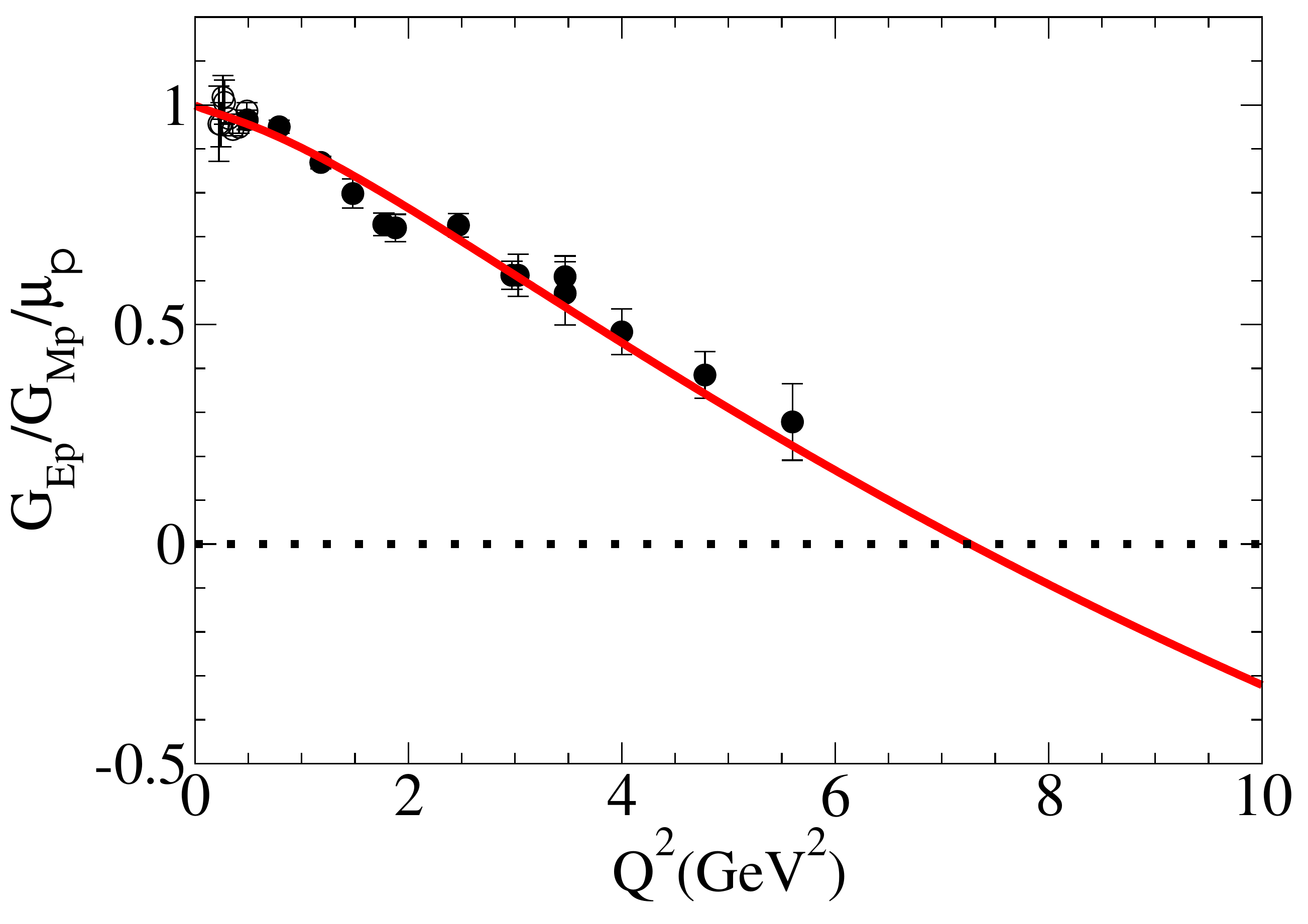}
  \hspace{.1cm}
  \includegraphics[width=3.0in]{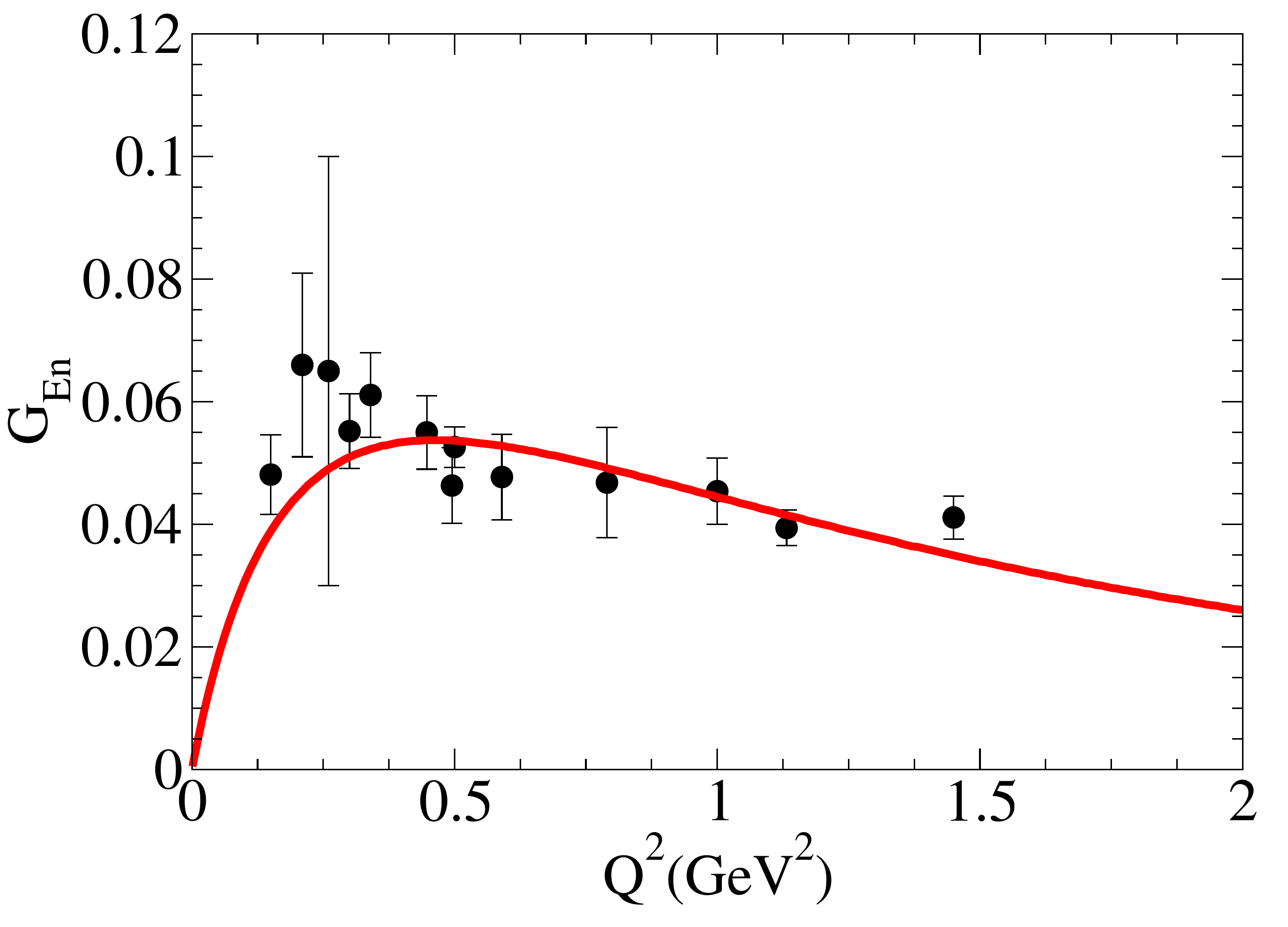}
  \caption{Nucleon form factors.
Model II of Ref.~\cite{Gross:2006fg}.
Left panel: $\mu_p\,G_{Ep}/G_{Mp}$ ratio,
including the Jefferson Lab data.
Right panel: neutron electric form form factor.
}
\label{figNucleon}
\end{figure}

The relation between the hadronic short-distance  constituent quark and gluon particle limit and the long-range confining domain is yet one of the most challenging aspects of particle physics due to the strong coupling nature of Quantum ChromoDynamics, the fundamental theory of the strong interactions.  The central question is how one can compute hadronic properties from first principles; i.e., directly from the QCD Lagrangian. The most successful theoretical approach thus far has been to quantize QCD on discrete lattices in Euclidean space-time.~\cite{Wilson:1974sk} Lattice
numerical results follow from computation of frame-dependent moments of distributions in Euclidean space and dynamical observables in Minkowski space-time, such as the time-like hadronic form factors, are not amenable to Euclidean lattice computations.  The  Dyson-Schwinger methods have led to many important insights, such as the infrared fixed point behavior of the strong coupling constant,~\cite{Cornwall:1981zr} but
 in practice, the analyses are limited to ladder approximation in Landau gauge.
 Baryon spectroscopy and the excitation dynamics of nucleon resonances encoded in the nucleon transition form factors  can provide fundamental insight into the strong-coupling dynamics of QCD. New theoretical  tools are thus  of primary interest for the interpretation of the results expected at the new mass scale and kinematic regions accessible to the JLab 12-GeV Upgrade Project.

The AdS/CFT correspondence between gravity or string theory on a higher-dimensional anti--de Sitter (AdS) space and conformal field theories in physical space-time~\cite{Maldacena:1997re}
 has led to a semi-classical approximation for strongly-coupled QCD, which provides physical insights into its nonperturbative dynamics. The correspondence is holographic in the sense that it determines a duality between  theories in different number of space-time dimensions. This geometric approach leads in fact   to a simple analytical
 and phenomenologically compelling nonperturbative approximation to the full light-front  QCD Hamiltonian -- ``Light-Front Holography".~\cite{deTeramond:2008ht}
 Light-Front Holography  is in fact one of the most remarkable features of 
 the AdS/CFT correspondence.~\cite{Maldacena:1997re}
 The Hamiltonian equation of motion in the light-front (LF) is frame independent and has a structure similar to eigenmode equations in AdS space. This makes a direct connection of QCD with AdS/CFT methods possible.~\cite{deTeramond:2008ht}  Remarkably, the AdS equations
correspond to the kinetic energy terms of  the partons inside a
hadron, whereas the interaction terms build confinement and
correspond to the truncation of AdS space in an effective dual
gravity  approximation.~\cite{deTeramond:2008ht}

One can also study the gauge/gravity duality starting from the bound-state structure of hadrons in QCD quantized in the light-front. The LF Lorentz-invariant Hamiltonian equation for the relativistic bound-state system is
\begin{equation}
P_\mu P^\mu \vert  \psi(P) \rangle =  \left(P^+ P^- \! - \mathbf{P}^2_\perp\right)\vert  \psi(P) \rangle=  M^2 \vert  \psi(P) \rangle,    P^\pm = P^0 \pm P^3, 
\end{equation}
\noindent where the LF time evolution operator $P^-$ is determined canonically from the QCD Lagrangian.~\cite{Brodsky:1997de}
 To a first semi-classical approximation, where quantum loops
and quark masses are not included, this leads to a LF Hamiltonian equation which describes the bound-state dynamics of light hadrons  in terms of an invariant impact variable $\zeta$~\cite{deTeramond:2008ht}
which measures the separation of the partons within the hadron at equal light-front time
$\tau = x^0 + x^3$.~\cite{Dirac:1949cp} 
This allows us to identify the holographic variable $z$ in AdS space with an impact variable $\zeta$.~\cite{deTeramond:2008ht} The resulting Lorentz-invariant Schr\"odinger equation for general spin incorporates color confinement and is systematically improvable.

Light-front   holographic methods  were originally
introduced~\cite{Brodsky:2006uqa, Brodsky:2007hb} by matching the  electromagnetic
current matrix elements in AdS space~\cite{Polchinski:2002jw} with
the corresponding expression  using LF   theory in
physical space time. It was also shown that one obtains  identical
holographic mapping using the matrix elements of the
energy-momentum tensor~\cite{Brodsky:2008pf} by perturbing the AdS
metric around its static solution.~\cite{Abidin:2008ku}

A gravity dual to QCD is not known, but the mechanisms of
confinement can be incorporated in the gauge/gravity
correspondence by modifying the AdS geometry in the  large
infrared (IR) domain  $z \sim 1/\Lambda_{\rm QCD}$, which also
sets the scale of the strong
interactions.~\cite{Polchinski:2001tt}  In this simplified
approach we consider the propagation of hadronic modes in a fixed
effective gravitational background asymptotic to AdS space, which
encodes salient properties of the QCD dual theory, such as the
ultraviolet (UV) conformal limit at the AdS boundary, as
well as modifications of the background geometry in the large $z$
IR region to describe confinement. The modified theory 
generates the point-like hard behavior expected from QCD,~\cite{Brodsky:1973kr,  Matveev:1973ra} instead
of the soft behavior characteristic of extended
objects.~\cite{Polchinski:2001tt}

\subsection{Nucleon form factors}

In the higher dimensional gravity theory, hadronic amplitudes for the  transition $A\to B$ correspond to
the  coupling of an external electromagnetic (EM) field $A^M(x,z)$  propagating in AdS space with a fermionic mode $\Psi_P(x,z)$  given by the left-hand side of the equation below 
 \begin{eqnarray*} \label{FF}
 \int d^4x \, dz \,  \sqrt{g}  \,  \bar\Psi_{B, P'}(x,z)
 \,  e_M^A  \, \Gamma_A \, A_q^M(x,z) \Psi_{A, P}(x,z)  & 
 \sim  \\ 
 (2 \pi)^4 \delta^4 \left( P'  \! - P - q\right) \epsilon_\mu  \langle \psi_B(P'), \sigma'  \vert J^\mu \vert \psi_A(P), \sigma \rangle,
 \end{eqnarray*} 
 where the coordinates of AdS$_5$ are the Minkowski coordinates $x^\mu$ and $z$ labeled $x^M = (x^\mu, z)$,
 with $M, N = 1, \cdots 5$, $g$ is the determinant of the metric tensor and $e^A_M$ is the vielbein with tangent indices
 $A, B = 1, \cdots, 5$.
The expression on the right-hand side  represents the QCD EM transition amplitude in physical space-time. It is the EM matrix element of the quark current  $J^\mu = e_q \bar q \gamma^\mu q$, and represents a local coupling to pointlike constituents.  Can the transition amplitudes be related for arbitrary values of the momentum transfer $q$? How can we recover hard pointlike scattering at large $q$ from the soft collision of extended objects?~\cite{Polchinski:2002jw}
Although the expressions for the transition amplitudes look very different, one can show that a precise mapping of the $J^+$ elements  can be carried out at fixed LF time, providing an exact correspondence between the holographic variable $z$ and the LF impact variable $\zeta$ in ordinary space-time.~\cite{Brodsky:2006uqa}

A particularly interesting model is the ``soft wall'' model of Ref.~\cite{Karch:2006pv},  since it leads to linear Regge trajectories consistent with the light-quark hadron spectroscopy and avoids the ambiguities in the choice of boundary conditions at the infrared wall. In this case the effective potential takes the form of a harmonic oscillator confining potential $\kappa^4 z^2$. For a hadronic state with twist $\tau = N + L$  ($N$ is the number of components and $L$ the internal orbital angular momentum)
the elastic form factor is expressed as a $\tau - 1$ product of poles along the vector meson Regge radial trajectory
($Q^2 = -q^2 >0$)~\cite{Brodsky:2007hb}
\begin{equation} \label{F}
 F(Q^2) =  \frac{1}{{\Big(1 + \frac{Q^2}{M^2_\rho} \Big) }
 \Big(1 + \frac{Q^2}{M^2_{\rho'}}  \Big)  \cdots 
       \Big(1  + \frac{Q^2}{M^2_{\rho^{\tau-2}}} \Big)} ,
\end{equation}
where $M^2_{\rho_n} \to 4 \kappa^2(n + 1/2)$.
For a pion, for example, the lowest Fock state -- the valence state -- is a twist-2 state, and thus the form factor is the well known monopole form.
The remarkable analytical form of Eq. (\ref{F}), expressed in terms of the $\rho$ vector meson mass and its radial excitations, incorporates the correct scaling behavior from the constituent's hard scattering with the photon~\cite{Brodsky:1973kr,  Matveev:1973ra}  and the mass gap from confinement.

\subsection{Computing nucleon form factors in light-front holographic QCD}

As an illustrative example we consider in this section the spin non-flip elastic proton form factor and the 
form factor for the $\gamma^* p \to N(1440) P_{11}$ transition measured recently at JLab. 
In order to compute the separate features of the proton an neutron form factors  one needs to incorporate the spin-flavor structure of the nucleons,  properties  which are absent in the usual models of the gauge/gravity correspondence.
This can be readily included in AdS/QCD by weighting the different Fock-state components  by the charges and spin-projections of the quark constituents; e.g., as given by the $SU(6)$  spin-flavor symmetry.

Using the $SU(6)$ spin-flavor symmetry the expression for the spin-non flip proton form factors for the transition $n,L \to n' L$ is~\cite{Brodsky:2008pg} 
\begin{equation} \label{F1}
{F_1^p}_{n, L \to n', L}(Q^2)  =    R^4 \int \frac{dz}{z^4} \, \Psi_+^{n' \!, \,L}(z) V(Q,z)  \Psi_+^{n, \, L}(z),
\end{equation}
where we have factored out the plane wave dependence of the AdS fields
\begin{equation} \label{Psip}
\Psi_+(z) = \frac{\kappa^{2+L}}{R^2}  \sqrt{\frac{2 n!}{(n+L+1)!}} \,z^{7/2+L} L_n^{L+1}\!\left(\kappa^2 z^2\right) 
e^{-\kappa^2 z^2/2}.
\end{equation}
The bulk-to-boundary propagator $V(Q,z)$  has the integral representation~\cite{Grigoryan:2007my}
\begin{equation} \label{V}
V(Q,z) = \kappa^2 z^2 \int_0^1 \! \frac{dx}{(1-x)^2} \, x^{\frac{Q^2}{4 \kappa^2}} 
e^{-\kappa^2 z^2 x/(1-x)} ,
\end{equation}
with $V(Q = 0, z) = V(Q, z = 0) =1$. The orthonormality of the Laguerre polynomials in (\ref{Psip}) implies that the nucleon form factor at $Q^2 = 0$ is one if $n = n'$ and zero otherwise. Using  (\ref{V}) in (\ref{F1}) we find
\begin{equation} \label{protonF1}
F_1^p(Q^2) =  \frac{1}{{\Big(1 + \frac{Q^2}{M^2_\rho} \Big) }
 \Big(1 + \frac{Q^2}{M^2_{\rho'}}  \Big) },
 \end{equation}
 for the elastic proton Dirac form factor and
 \begin{equation} \label{RoperF1}
 {F_1^p}_{N \to N^*}(Q^2) = \frac{\sqrt{2}}{3} \frac{\frac{Q^2}{M^2_\rho}}{\Big(1 + \frac{Q^2}{M^2_\rho} \Big) 
 \Big(1 + \frac{Q^2}{M^2_{\rho'}}  \Big)
       \Big(1  + \frac{Q^2}{M^2_{\rho^{''}}} \Big)},
\end{equation}
for the EM spin non-flip proton to Roper  transition form factor. The results (\ref{protonF1}) and (\ref{RoperF1}),
 compared with available data in Fig.~\ref{pFFs}, correspond to the valence approximation.   The transition form factor
 (\ref{RoperF1}) is expressed in terms of  the mass of the $\rho$ vector meson and its first two radial excited states, with no additional parameters. The results  in Fig.~\ref{pFFs}  are in good agreement with experimental data. The transition form factor to the  $N(1440) P_{11}$ state shown in Fig.~\ref{pFFs} corresponds to  the first radial excitation of the three-quark  ground state of the nucleon. In fact,  the Roper resonance $N(1440)P_{11}$ and the $N(1710)P_{11}$ are well accounted in the light-front holographic framework as the first and second radial states of the nucleon family, likewise the $\Delta(1600)P_{33}$ corresponds to the first radial excitation of the $\Delta$ family as shown in Fig.~\ref{Baryons}   for the positive-parity light-baryons.~\cite{deTeramond:2009xk}
In the case of massless quarks, the nucleon eigenstates have Fock components with different orbital angular momentum, $L = 0$ and $L = 1$, but with equal probability. In effect, in AdS/QCD the nucleons angular momentum is carried by quark orbital angular momentum since soft gluons do not appear as quanta in the proton.

 Light-front holographic QCD methods have also been used to obtain general parton distributions (GPDs) in
 Ref.~\cite{Vega:2010ns}, and a study of the EM nucleon to $\Delta$ transition form factors and
nucleon to the $S_{11}(1535)$ negative parity nucleon state has been
carried out in the framework of the Sakai and Sugimoto model in Refs.
\cite{Grigoryan:2009pp} and \cite{BallonBayona:2012jy} respectively.  It is certainly worth to extend the simple computations described here and perform a systematic study of the different transition form factors measured at JLab. This study will help to discriminate among models and compare with the new results expected from  the JLab 12-GeV Upgrade Project,  in particular at photon
virtualities $Q^2>5  ~{\rm GeV}^2$, which correspond to the experimental
coverage of the CLAS12 detector.

\newpage

\section{The  $N^{*}$ Electrocoupling Interpretation within the Framework
of Constituent Quark Models \label{qm}}

\begin{figure}
  \includegraphics[width=3.0in]{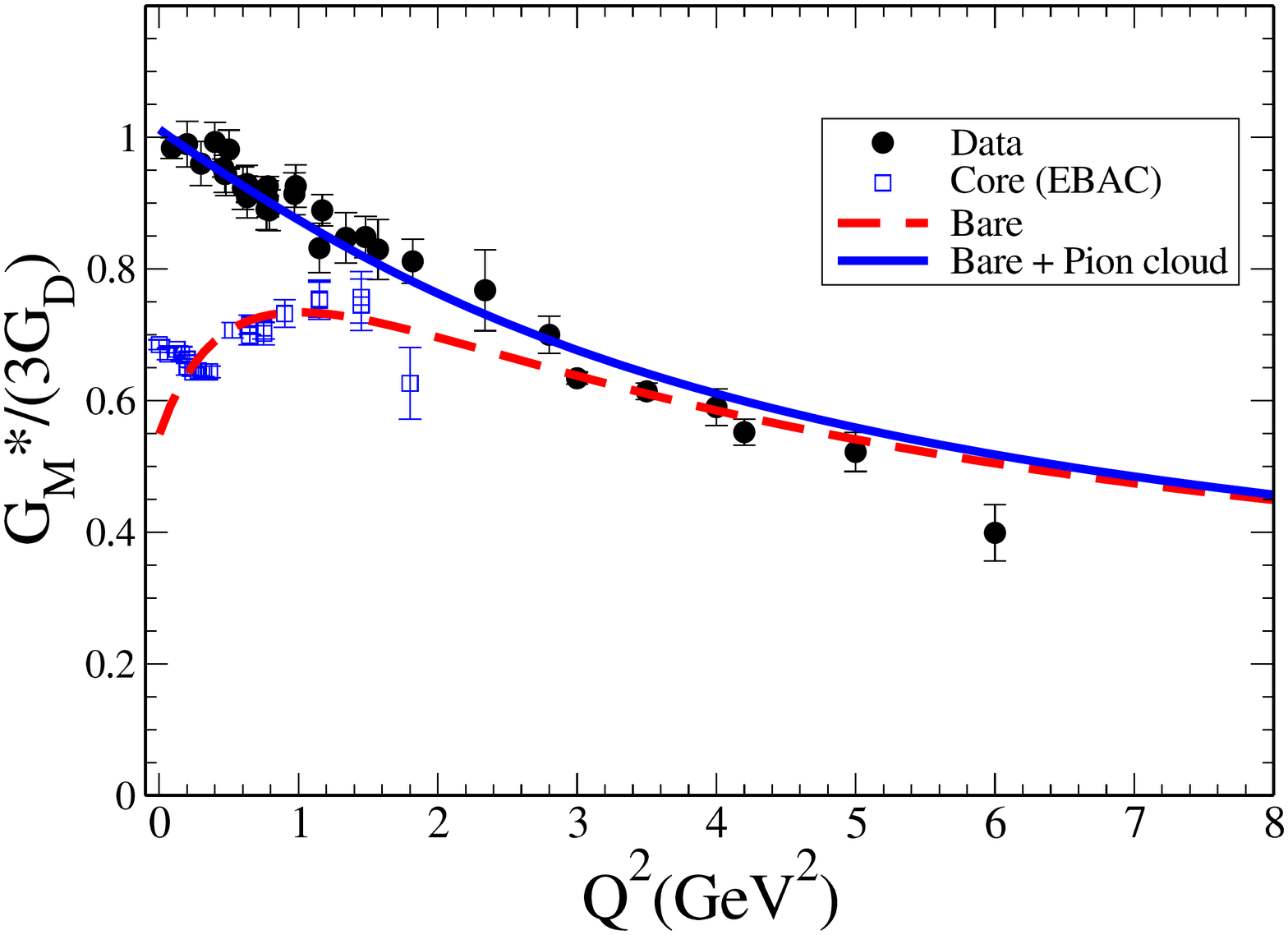}
  \hspace{.1cm}
  \includegraphics[width=3.0in]{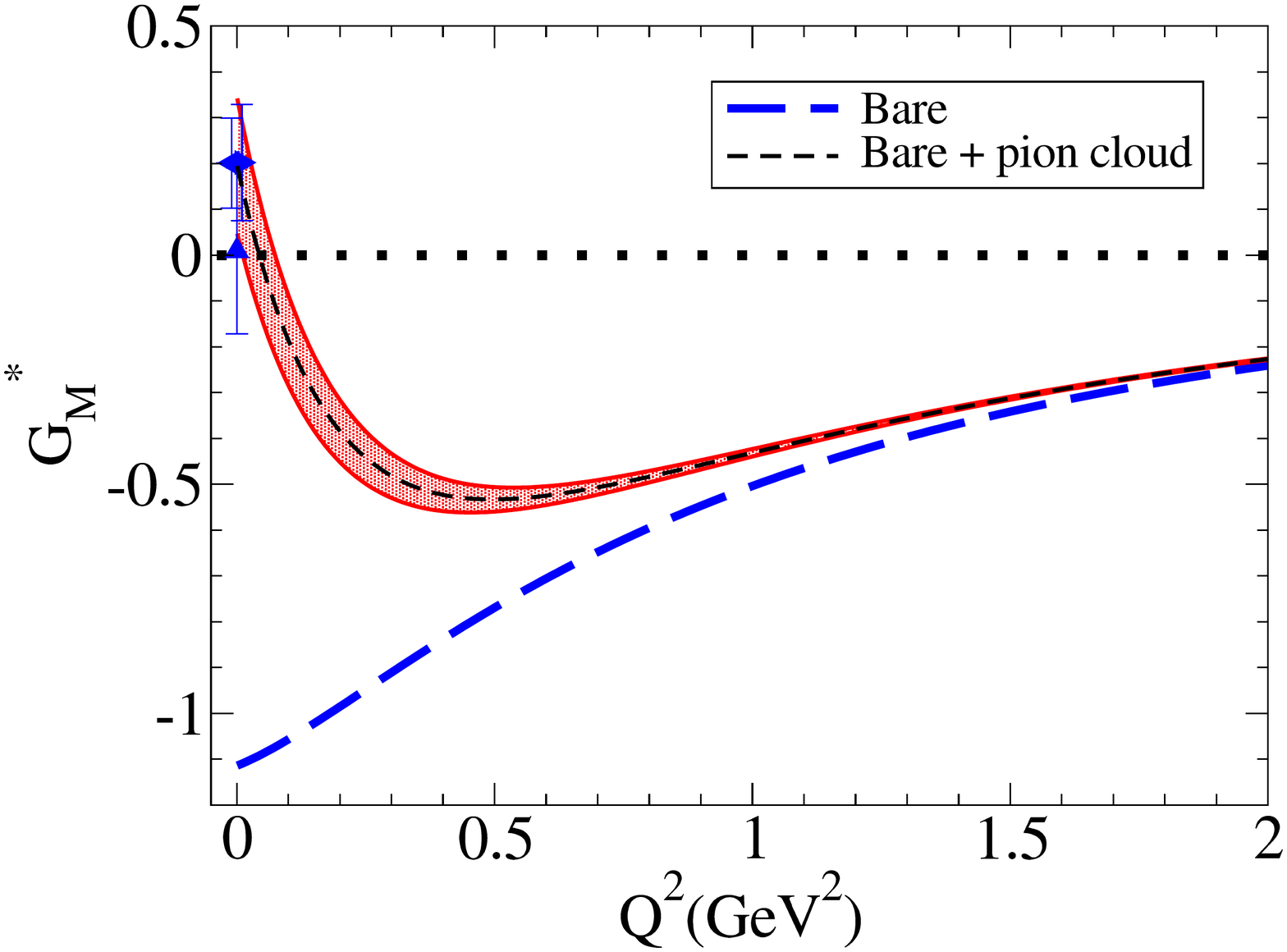} 
  \caption{Nucleon electromagnetic transition for
spin 3/2 resonances.
Left panel: $G_M^\ast/(3 G_D)$
($G_D$ is the nucleon dipole form factor)
for the $\gamma N \to \Delta(1232)$ reaction \cite{Ramalho:2008dp}.
Right panel: $G_M^\ast$ for the
$\gamma N \to \Delta(1600)$ reaction \cite{Ramalho:2010cw}.
In both cases the dashed line gives the valence
quark contribution and the solid line the full result.}
\label{figNDelta}
\end{figure}

\begin{figure}
  \includegraphics[width=3.0in]{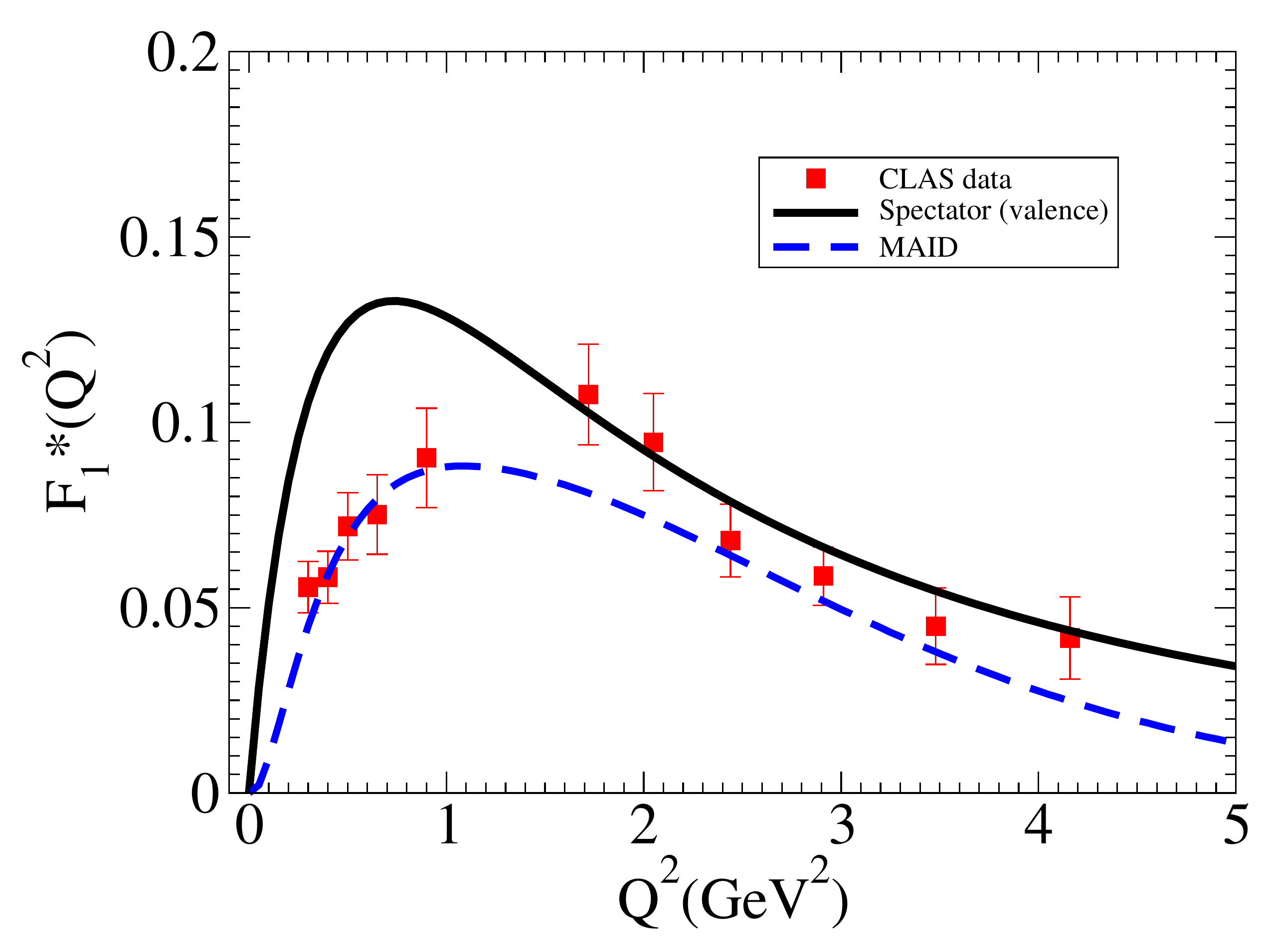}
  \hspace{.1cm}
  \includegraphics[width=3.0in]{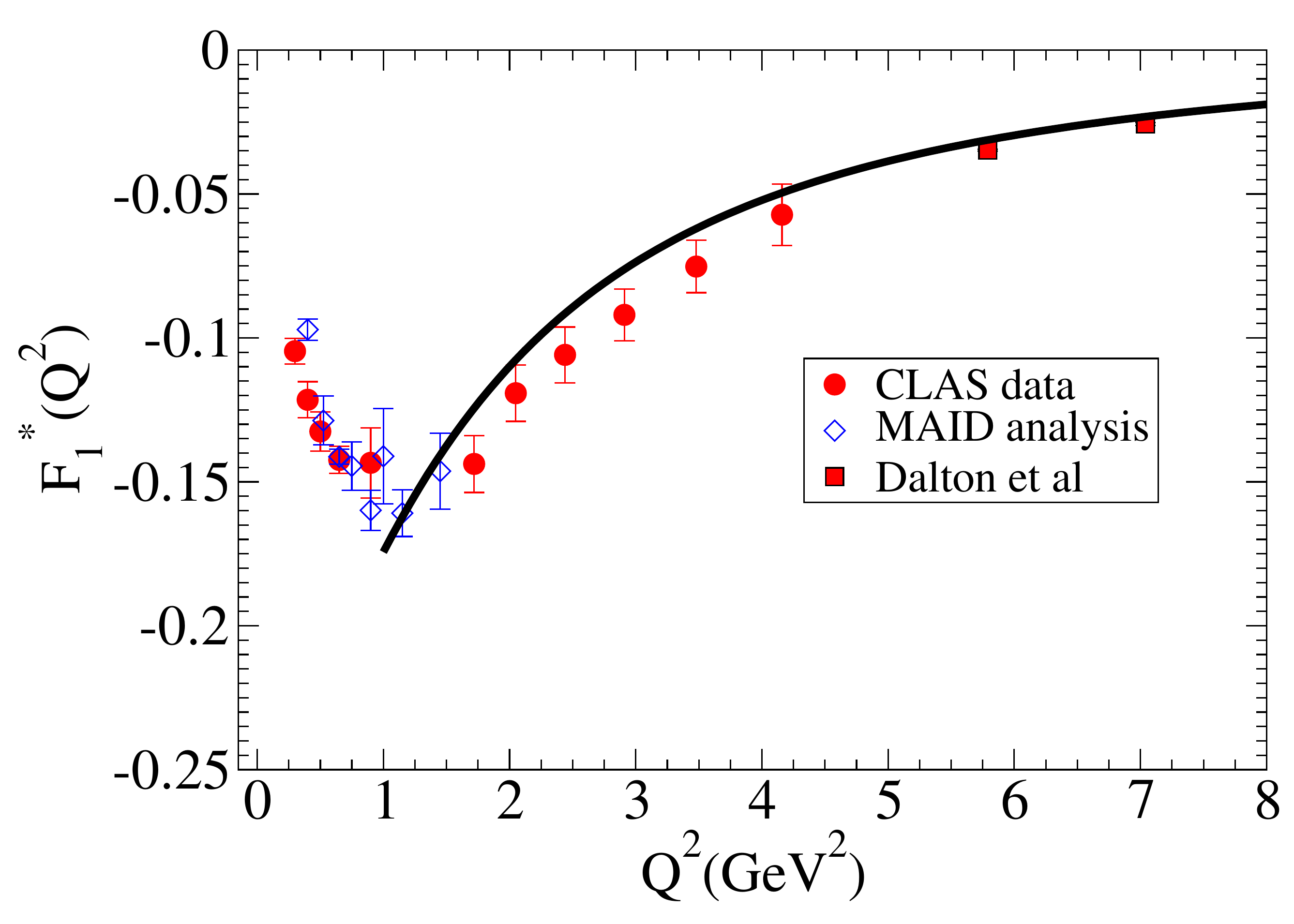}  
  \caption{Dirac-type form factors $F_1^\ast$ for
$\gamma N \to N^ \ast$ transitions.
Left panel:  $\gamma N \to P_{11}(1440)$ reaction \cite{Ramalho:2010js}.
Right panel:  $\gamma N \to S_{11}(1535)$ reaction \cite{Ramalho:2011ae}.
}
\label{figF1}
\end{figure}

\begin{figure}[ht]
\begin{center}
\includegraphics[width=12cm]{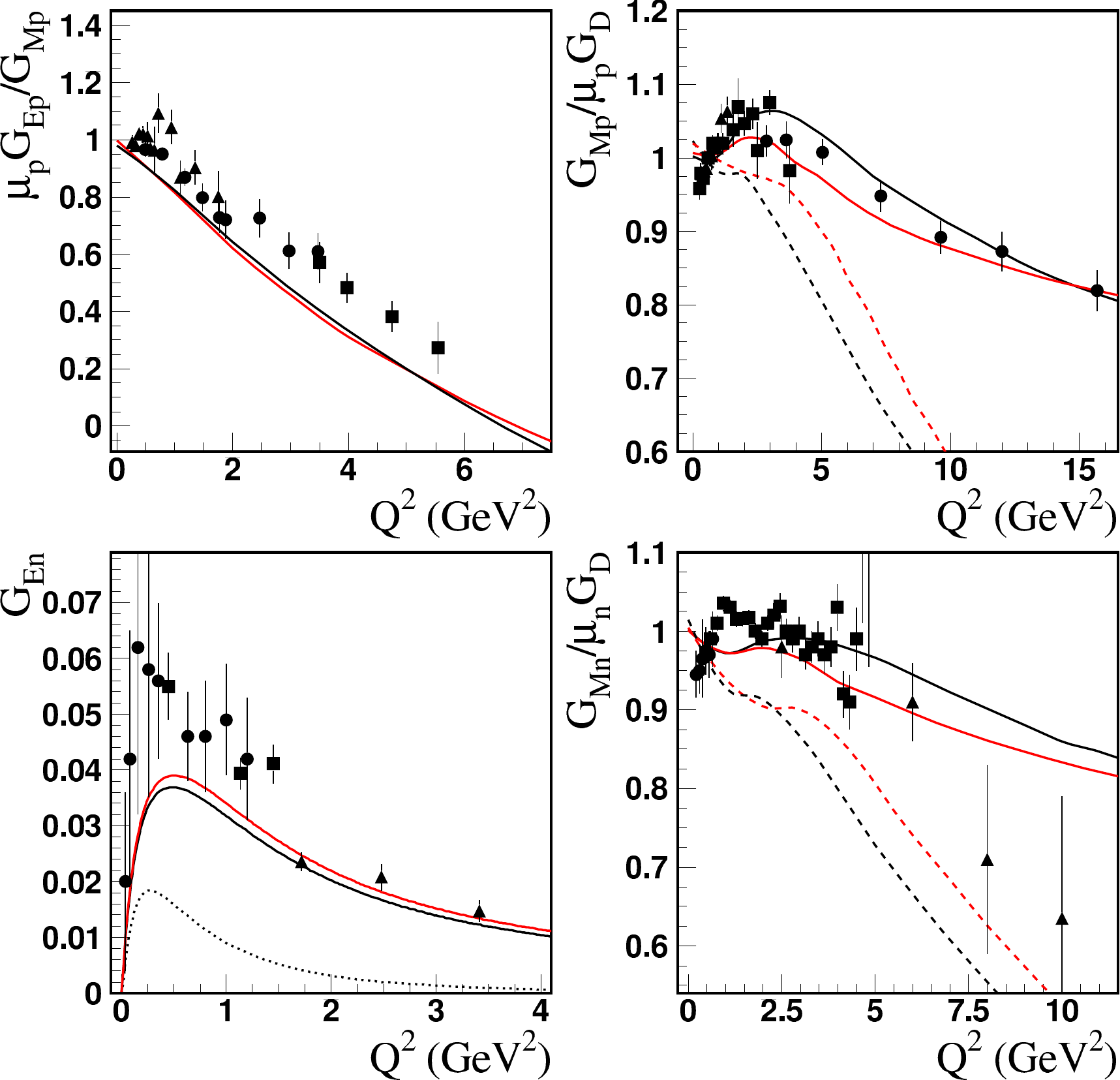}
\end{center}
\caption
{Nucleon electromagnetic form factors.
The solid  curves
correspond to the results obtained
taking into account two contributions to the nucleon
(Eq. \ref{eq:nuc1}): the pion-cloud  \cite{Miller:2002ig}
and the $3q$ core with the running quark masses (\ref{eq:nuc3})
for the wave functions $\Phi_1$ (black curves) and $\Phi_2$ (red curves)
in Eqs. (\ref{eq:sec10}).
The black and red dashed  curves are
the results obtained
for the nucleon taken as a pure $3q$ state
with the parameters (\ref{eq:nuc2}) and
constant quark mass.
Dotted curve for $G_{En}(Q^2)$ is the pion cloud
contribution \cite{Miller:2002ig}.
Data are from Refs.
\cite{Jones:1999rz,Gayou:2001qd,Sill:1992qw,Bartel:1973rf,Madey:2003av,
Riordan:2010id,Anderson:2006jp,Lachniet:2008qf,
Rock:1982gf}. }
\label{fig:fig1}
\end{figure}

\begin{figure}[ht]
\begin{center}
\includegraphics[width=12cm]{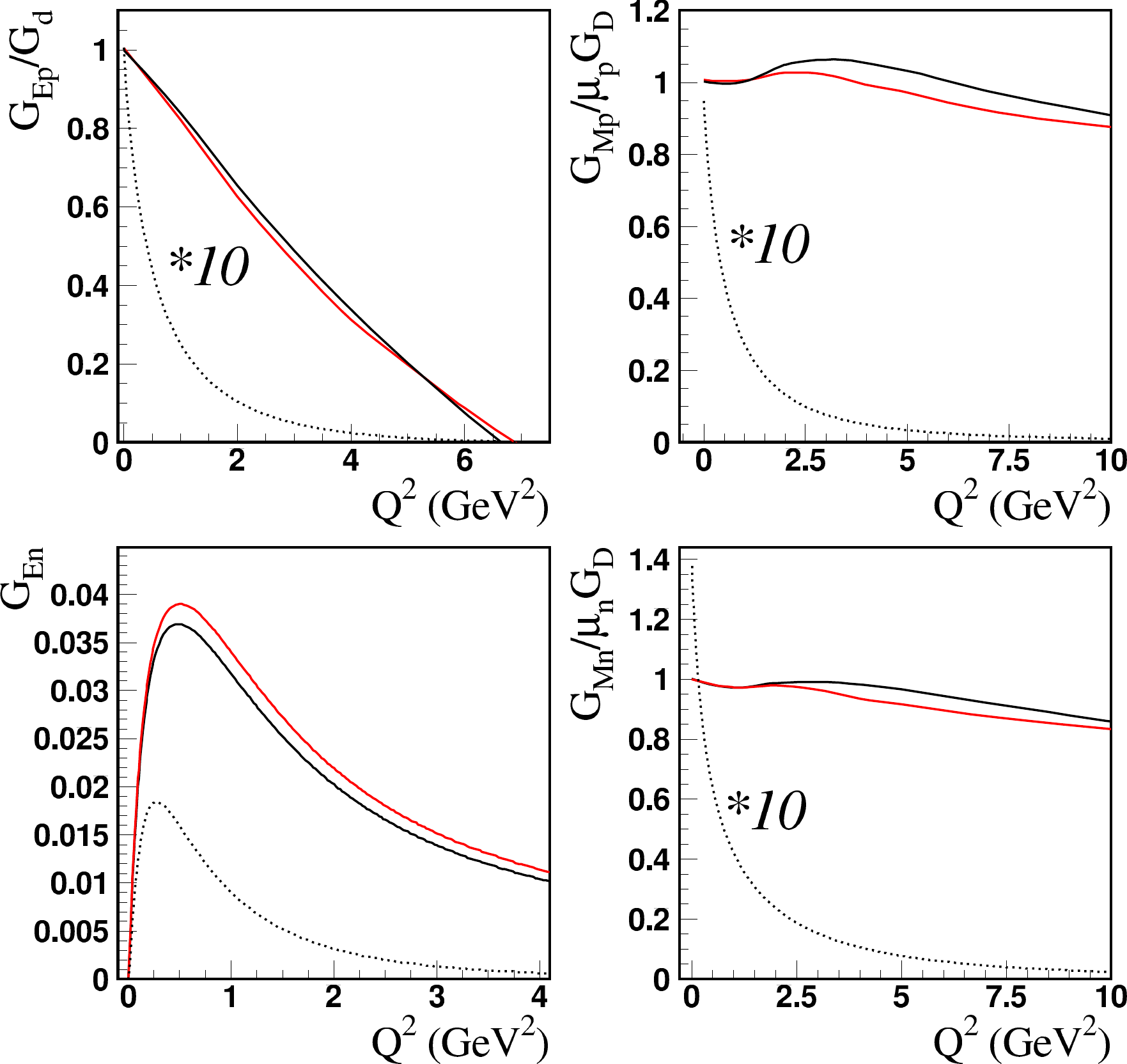}
\end{center}
\caption
{Nucleon electromagnetic form factors.
The legend for
the black and red solid  curves
is as for Fig. \ref{fig:fig1}.
Dotted curves are the pion cloud
contributions \cite{Miller:2002ig}.}
\label{fig:cloud}
\end{figure}

\begin{figure}[ht]
\vspace{1cm}
\begin{center}
\includegraphics[width=12cm]{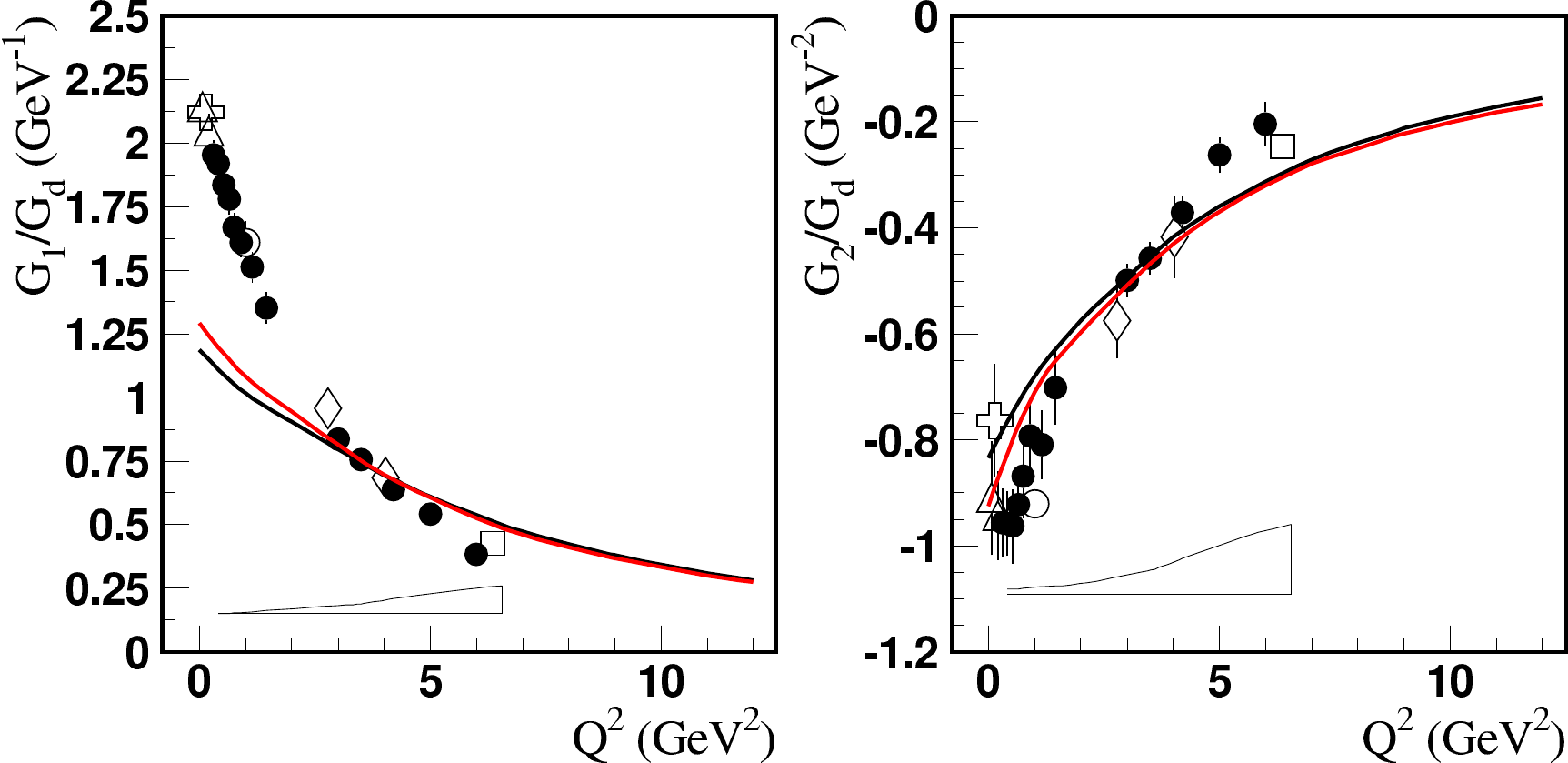}
\end{center}
\caption
{The $\gamma^*p\rightarrow \Delta(1232)$P$_{33}$
transition form factors;
$G_1(Q^2)\sim G_M-G_E$. Weight factors are $c_{N^*}^{(1)}=0.67\pm 0.04$
and $c_{N^*}^{(2)}=0.72\pm 0.04$ for the wave functions 
$\Phi_1$ (black curves) and $\Phi_2$ (red curves)
in Eqs. (\ref{eq:sec10}).
Solid circles correspond to the amplitudes extracted from
the CLAS data by JLab group \cite{Aznauryan:2009mx}, bands represent model
uncertainties of these results.
The results from other experiments are: open triangles
\cite{Stave:2006ea,Sparveris:2006uk,Stave:2008aa};
open crosses
\cite{Mertz:1999hp,Kunz:2003we,Sparveris:2004jn}; open rhombuses
\cite{Frolov:1998pw};
open boxes \cite{Villano:2009sn};
and open circles
\cite {Kelly:2005jj,Kelly:2005jy}.
}
\label{fig:fig2}
\end{figure}

\begin{figure}[ht]
\vspace{1cm}
\begin{center}
\includegraphics[width=12cm]{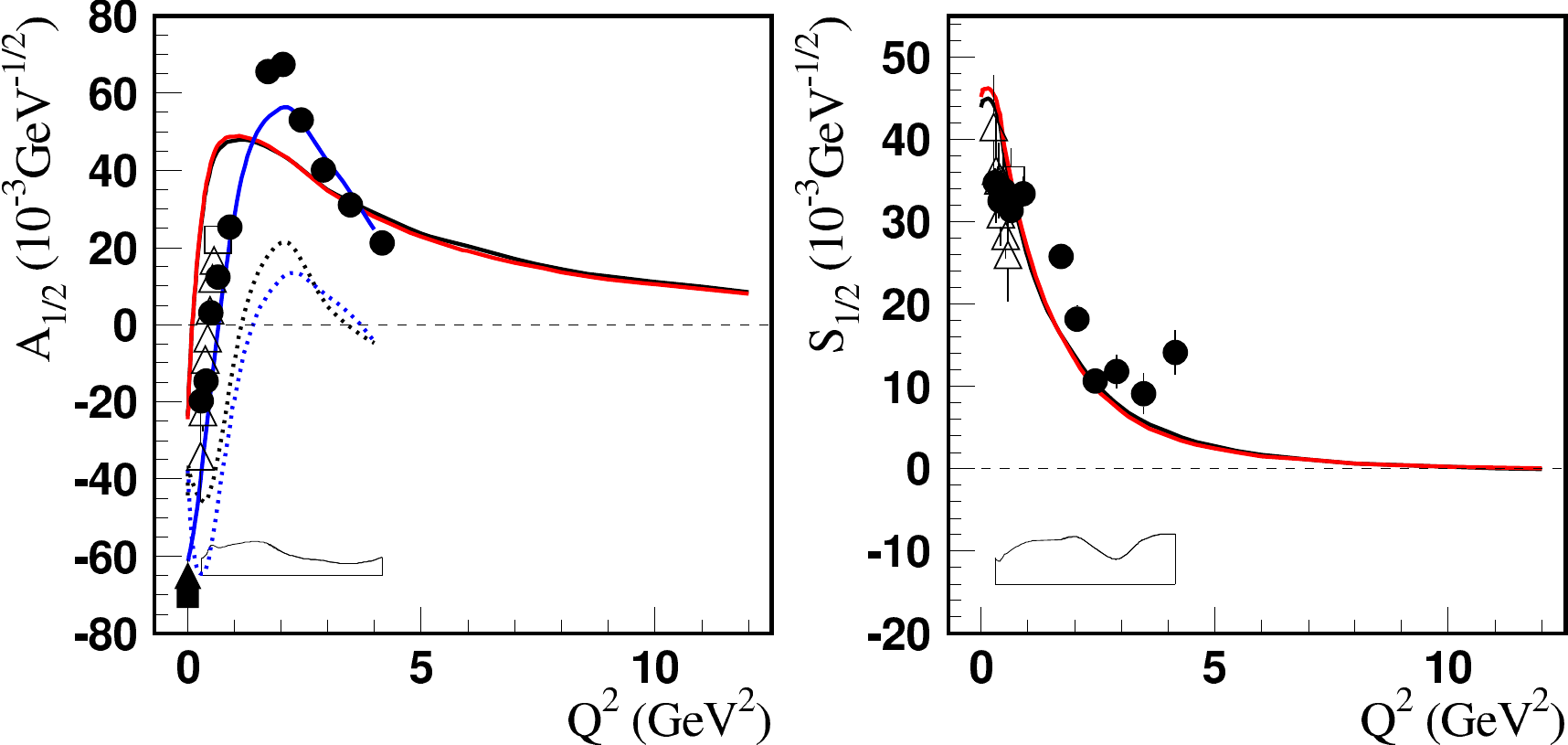}
\end{center}
\caption{
The $\gamma^*p\rightarrow N(1440)$P$_{11}$
transition amplitudes. Blue lines correspond to the MAID
results \cite{Drechsel:1998hk,Kamalov:2000en}. Dotted curves are estimated pion-cloud
contributions.
$c_{N^*}^{(1)}=0.73\pm 0.05$,
$c_{N^*}^{(2)}=0.77\pm 0.05$. 
The open triangles correspond to the amplitudes  extracted from
CLAS 2$\pi$ electroproduction data 
\cite{Mokeev:2008iw}.
Other legend is as for Fig. \ref{fig:fig2}. }
\label{fig:fig3}
\end{figure}

\begin{figure}[ht]
\vspace{1cm}
\begin{center}
\includegraphics[width=12cm]{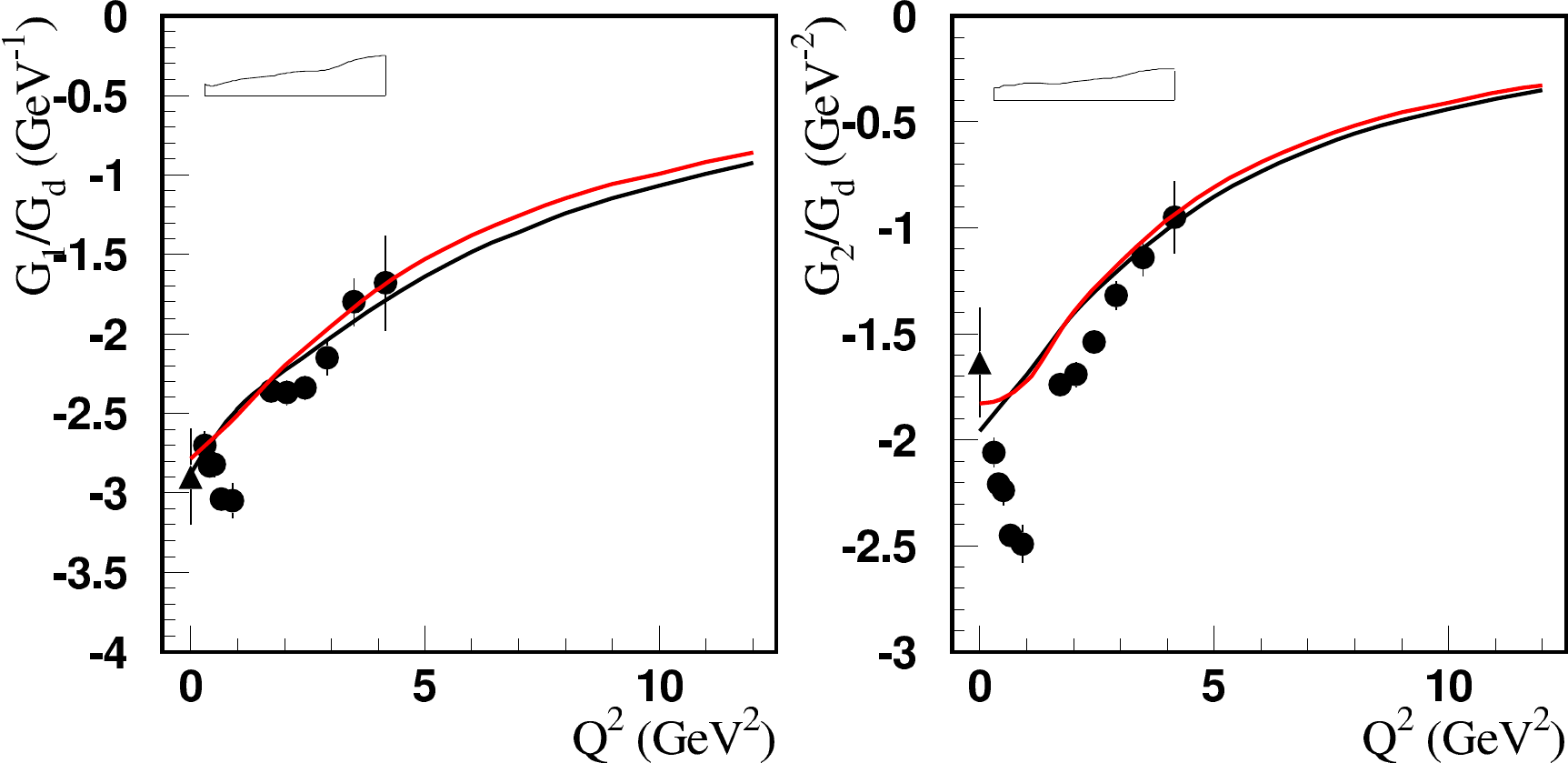}
\end{center}
\caption{
The $\gamma^*p\rightarrow N(1520)$D$_{13}$
transition form factors;
$G_1(Q^2)\sim A_{1/2}-A_{3/2}/\sqrt{3}$.
$c_{N^*}^{(1)}=0.78\pm 0.06$,
$c_{N^*}^{(2)}=0.82\pm 0.06$. 
Other legend is as for Fig. \ref{fig:fig2}. }
\label{fig:fig4}
\end{figure}

\begin{figure}[ht]
\vspace{1cm}
\begin{center}
\includegraphics[width=12cm]{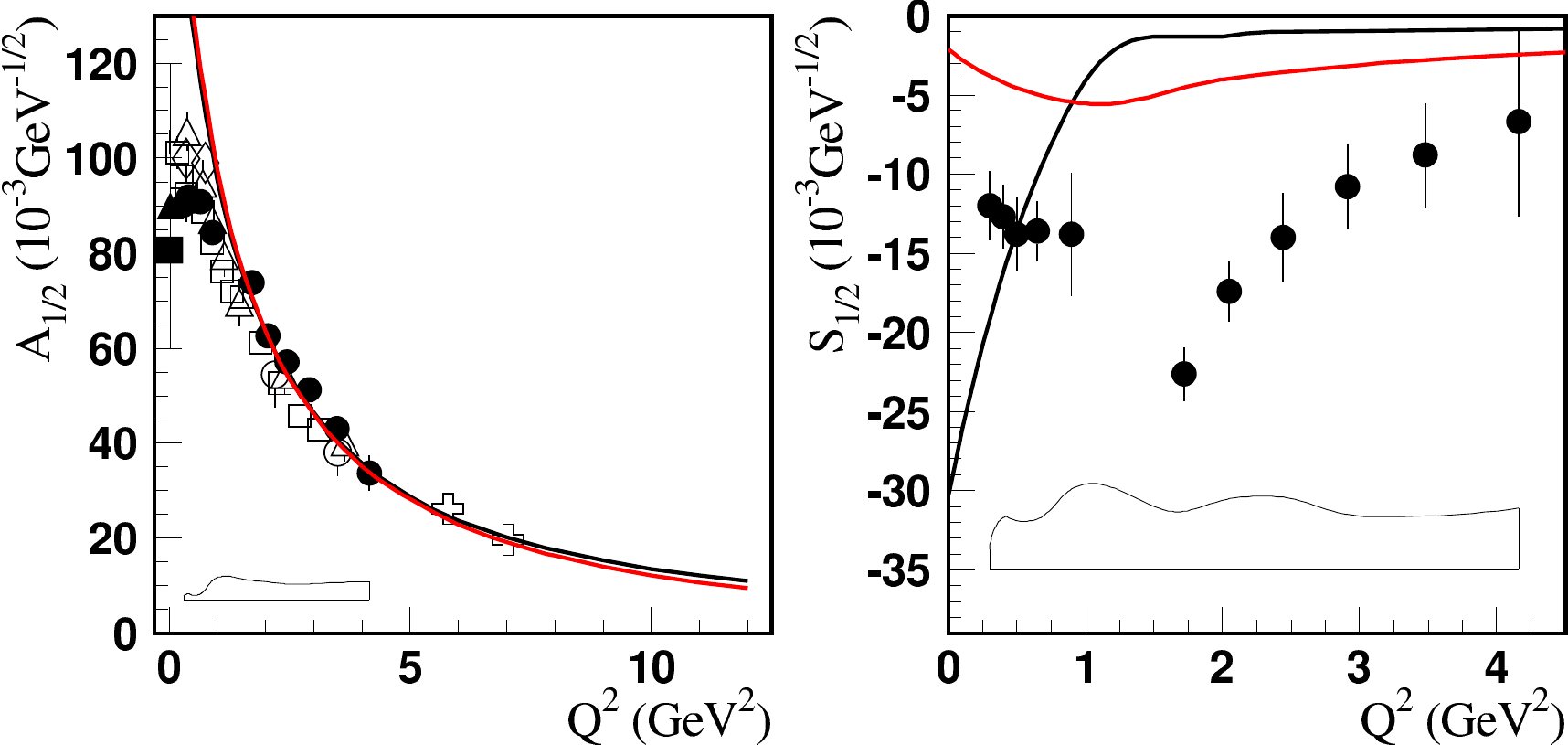}
\end{center}
\caption{
The $\gamma^*p\rightarrow N(1535)$S$_{11}$
transition amplitudes. 
The amplitudes extracted from 
the CLAS and JLab/Hall C data on $ep\rightarrow e\eta p$
are:
the stars \cite{Thompson:2000by},
the open boxes \cite{Denizli:2007tq},
the open circles \cite{Armstrong:1998wg},
the crosses \cite{Dalton:2008aa}, and
the rhombuses \cite{Aznauryan:2004jd,Aznauryan:2009mx}.
$c_{N^*}^{(1)}=0.88\pm 0.03$,
$c_{N^*}^{(2)}=0.94\pm 0.03$. 
Other legend is as for Fig. \ref{fig:fig2}. }
\label{fig:fig5}
\end{figure}

 \begin{figure}[ht]
 
 \includegraphics[width=2.3in]{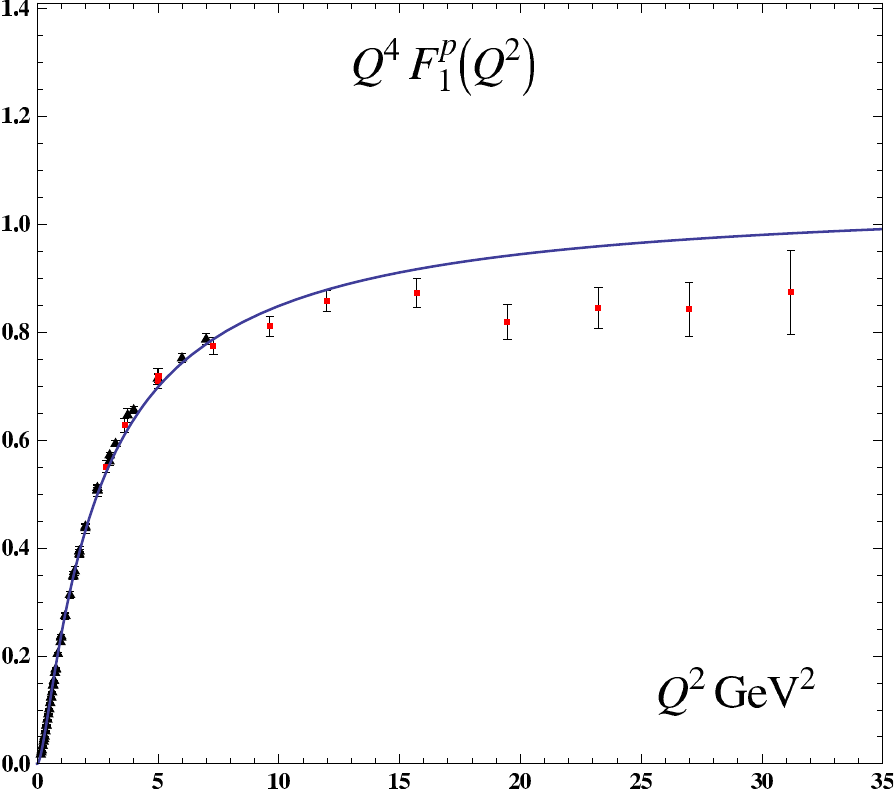}
\includegraphics[width=2.3in]{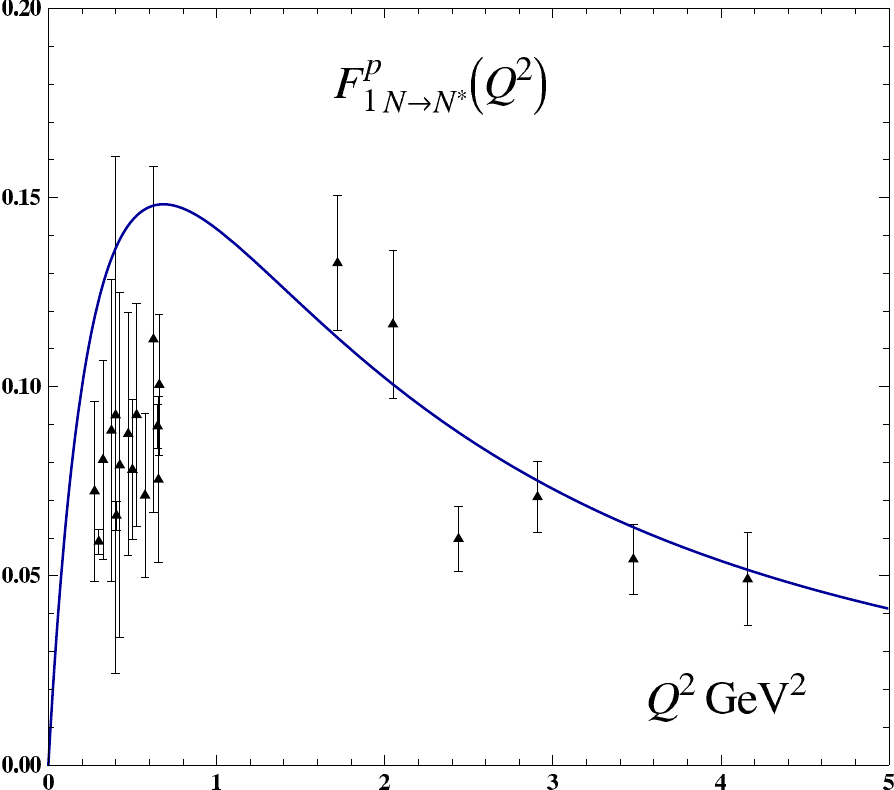}

\caption{Dirac proton form factors in light-front holographic QCD. Left: scaling of proton elastic form factor  $Q^4 F_1^p(Q^2)$. Right: proton transition form factor  ${F_1^p}_{N\to N^*}(Q^2)$ for the $\gamma^* p \to N(1440) P_{11}$  transition. Data compilation  from Diehl~\cite{Diehl:2005wq} (left) and CLAS $\pi$ and $2 \pi$ electroproduction 
data~\cite{Aznauryan:2005tp, Aznauryan:2008pe, Mokeev:2008iw, Aznauryan:2009mx}  (right).}
\label{pFFs}
\end{figure}

 \begin{figure}[ht]
\begin{center}
\includegraphics[width=5in]{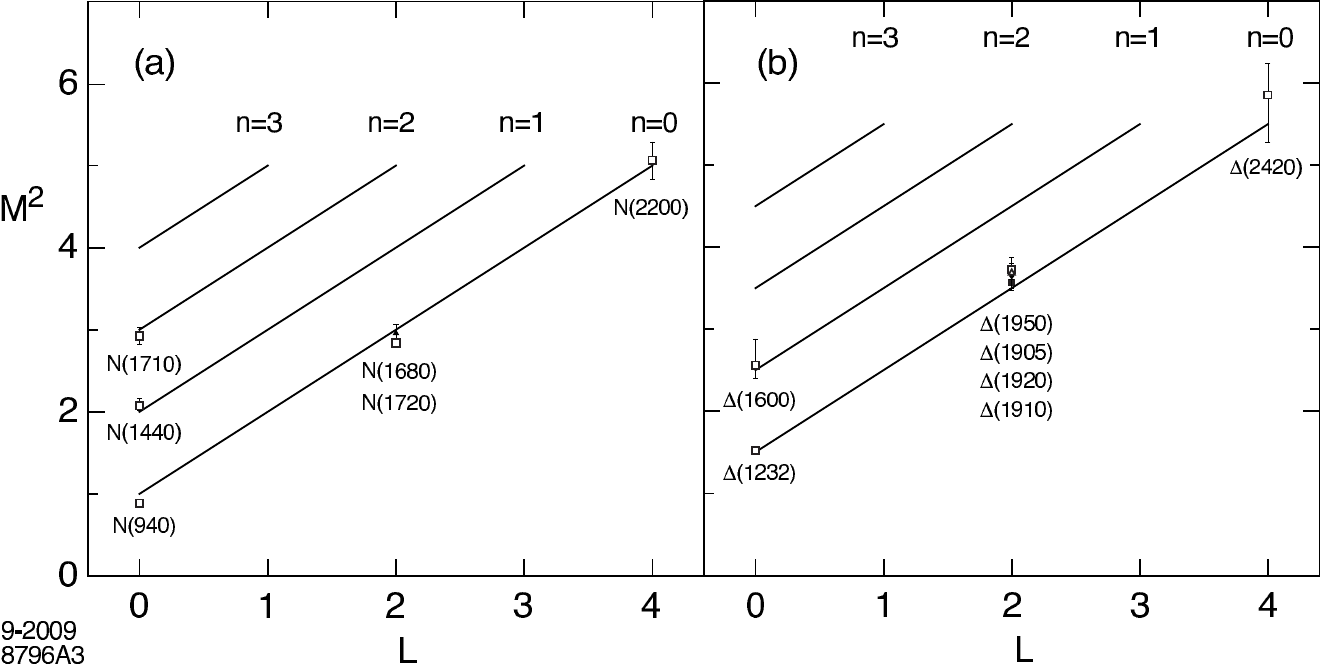}
\caption{Positive parity Regge trajectories for  the  $N$ and $\Delta$ baryon families for $\kappa= 0.5$ GeV.
Only confirmed PDG~\cite{Amsler:2008zzb} states are shown.}
\label{Baryons}
\end{center}
\end{figure}

\subsection{Introduction}
The study of the electromagnetic excitation of the nucleon resonances is
expected to provide a good test for our knowledge concerning the internal
structure of baryons. From a fundamental point of view, the description of the
resonance spectrum and excitation should be performed within a Quantum ChromoDynamics (QCD)
approach, which,  however, does not allow  up to now to extract all the hadron
properties in a systematic way. Therefore, one has to rely on
models, such as the  Constituent Quark Models (CQM).
In CQMs quarks
are considered as effective internal degrees of freedom and can
acquire a mass and a finite size. Phenomenological results of these models provide useful constraints based on experimental data for the development of QCD-based approaches such as LQCD and DSEQCD.  For instance the choice of basis configurations in the recent LQCD studies of $N^{*}$ spectrum \cite{Dudek:2012ag} was motived by quark model results.

In the following we report some results of recent approaches using the constituent quark idea in the framework of various light-front (LF) formulations of the quark wave function (Secs.~\ref{Light-Front Holographic},\ref{sec:CST},\ref{sec:LFquark}) and a discussion on the use of CQM for the interpretation of resonance electrocouplings at high $Q^2$ with particular attention to some future perspectives (Sec.~\ref{sec:cqm2}).

\subsection{Covariant quark-diquark model
for the $N$ and $N^*$ electromagnetic transition
form factors \label{sec:CST}}

The study of hadron structure using the fundamental theory,
Quantum ChromoDynamics, can in practise be done
only in the large $Q^2$ regime or, by means of
lattice simulations, in the unphysical quark masses regime
\cite{Aznauryan:2009da}.
For this reason one has to rely on effective descriptions
either with the degrees of freedom of  QCD (quarks and gluons)
within the Dyson-Schwinger framework \cite{Aznauryan:2009da},
or in terms of the degrees of freedom
observed at low $Q^2$, the meson cloud
and the light baryon core, using a dynamical coupled-channels reaction
(dynamical models or DM) framework
\cite{Burkert:2004sk,Aznauryan:2009da}.
The DSEQCD helps  to understand the transition
between the perturbative regime of QCD
and the low $Q^2$ regime, where the quarks
acquire masses and structure dynamically due to the gluon dressing,
although the meson degrees of freedom are
not included till the moment \cite{Aznauryan:2009da}.
Dynamical models, on the other hand,
help to explain the transition between
the low $Q^2$ picture, in terms of a finite size baryon
and the surrounding meson cloud, and the intermediate
region when $Q^2> 2$ GeV$^2$, where the baryon core effects
become increasingly important \cite{Burkert:2004sk}.
To complete the picture a parameterization
of the structure of the baryon core is required, and
a possibility is to use the
meson-baryon dressing model to extract from the data
the contributions of the core,
that can be interpreted as a 3-valence quark system \cite{JuliaDiaz:2006xt,JuliaDiaz:2009ww}.

Alternative descriptions comprise
effective chiral perturbation theory,
that can be used to interpolate lattice QCD results
but is restricted to the low $Q^2$ regime,
perturbative QCD that works only at very large $Q^2$
with a threshold that is still under discussion,
QCD sum rules 
and constituent quark models
that can include also chiral symmetry and/or
unquenched effects \cite{Aznauryan:2009da}.

CQMs include the gluon and
quark-antiquark polarization in the
quark substructure (that also generates
the constituent quark mass)
with effective inter-quark interactions \cite{Aznauryan:2009da}.
There are different versions according to the
inter-quark interaction potential and
the kinematic considered (nonrelativistic, or relativistic).
Among the relativistic descriptions there are, in particular,
different implementations of relativity
based on the Poincare invariance \cite{Aznauryan:2009da}.

We discuss now with some detail the covariant quark-diquark model, 
also known as the spectator quark model, and present some of its results.
Contrarily to other CQMs, this model 
is not based on a wave equation
determined by some
complex and nonlinear potential. 
For that reason, the model is not used to
predict the baryonic spectrum.
Instead, the wave functions are built
from  the baryon internal symmetries only,
with the shape of the wave functions determined
directly by the experimental data,
or lattice data for some ground state systems \cite{Ramalho:2010vr}.

In the covariant spectator theory (CST) \cite{Gross:1991pm}
the 3-body baryon systems
are described in terms of a vertex
function $\Gamma$
where 2 quarks are on-mass-shell \cite{Stadler:1997iu,Gross:2004nt,Gross:2006fg}.
In this approach confinement ensures that the
vertex $\Gamma$ vanishes when the 3 quarks
are simultaneously on-mass-shell,
and the singularities associated with the propagator of the
off-mass-shell quark are canceled by the vertex $\Gamma$
\cite{Gross:2004nt,Stadler:1997iu}. The baryon state can then be described
by a wave function
\mbox{$\Psi(P,k)= (m_q- \not \! k - i \varepsilon)^{-1} \Gamma(P,k)$},
where $P$ is the baryon momentum,
$m_q$ the quark mass
and $k$ the quark four-momentum \cite{Gross:2004nt,Gross:2006fg}.

The CST formulation is motivated by the fact that in impulse 
approximation only one quark
interacts with the photon, while the two other quarks are spectators.
Therefore, by integrating over the relative momentum of these two quarks,
one can reduce the 3-quark system to a quark-diquark
system,  where the effective diquark has 
an averaged mass $m_D$ \cite{Gross:2004nt,Gross:2006fg}.
In these conditions the baryon is described
by a wave function for the quark-diquark,
with individual states associated with the
internal symmetries (color, flavor, spin, momentum, etc.).
The electromagnetic interaction current is given
in impulse approximation by the coupling of the photon
with the off-mass-shell quark, while the diquark acts as a
spectator on-mass-shell particle \cite{Ramalho:2010vr,Gross:2006fg,Ramalho:2009gk}.

The photon-quark interaction is parameterized
by using the vector meson dominance (VMD) mechanism,
based on a combination of two poles
associated with vector mesons:
a light vector meson (mass $m_v= m_\rho \simeq m_\omega$)
and an effective heavy meson with mass $M_h=2 M$,
where $M$ is the nucleon mass,
which modulates the short range structure \cite{Ramalho:2010vr,Gross:2006fg,Ramalho:2009gk}.
The free parameters of the current were
calibrated for the SU(3) sector
by nucleon electromagnetic form factor data \cite{Gross:2006fg}
and with lattice QCD simulations associated
with the baryon decuplet \cite{Ramalho:2009gk}.
A parameterization based on VMD has the advantage
in the generalization to the lattice QCD regime
\cite{Ramalho:2009gk,Ramalho:2008aa,Ramalho:2009df,Ramalho:2011pp}
and also for the time-like region ($Q^2< 0$) \cite{Ramalho:2012ng}.


The covariant spectator quark model was applied to
the description of the nucleon elastic form factors
using a simple model where the quark-diquark
motion is  taken in the  S-state approximation \cite{Gross:2006fg}.
The nucleon data were used to fix the
quark current as well
as the radial wave function \cite{Gross:2006fg}.
A specific model with no explicit pion cloud effects,
except the effects included in the VMD parameterization
is presented in Fig.~\ref{figNucleon}.
This parameterization, based only on
the valence quark degrees of freedom, was extended successfully
for the nucleon on the lattice regime \cite{Hagler:2007xi}.

The model was also applied to the first nucleon resonance
the $\Delta(1232)$, in particular to the
$\gamma N \to \Delta(1232)$ transition.
Within a minimal model where the $\Delta$ is
described as an S-state of 3-quarks with the total spin
and isospin 3/2, one obtains, for dominant transition form factor
 $G_M^\ast(0) \le 2.07\, {\cal I}\le 2.07$,
where ${\cal I}\le 1$, is the overlap integral between
the nucleon and $\Delta$ radial wave functions
(both are S-states) in the limit $Q^2=0$ \cite{Ramalho:2008ra}.
This simple relation,
which is a consequence of the normalization
of the nucleon and $\Delta$ quark wave functions,
 illustrates the incapability
in describing the  $\gamma N \to \Delta(1232)$,
with quark degrees of freedom only,
since the experimental result is $G_M^\ast(0) \simeq 3$.
The discrepancy,
is common to all constituent quark models,
and is also a manifestation
of the importance of the pion excitation
which contributes with about 30\%-40\%
of the strength of the reaction \cite{Burkert:2004sk,Aznauryan:2009da,JuliaDiaz:2006xt}.
The model can however explain
the quark core contribution in the transition, as
extracted from the data using the EBAC model~\cite{JuliaDiaz:2006xt},
when the meson-baryon cloud is subtracted  \cite{Ramalho:2008ra}.
The comparison of the model with the EBAC estimate
is presented in Fig.~\ref{figNDelta} (left panel, dashed line),
and also with the $G_M^\ast$ data, when meson-baryon cloud is included (solid line).
The model was also extended successfully
to the reaction in the lattice regime \cite{Ramalho:2008aa,Ramalho:2009df}.
The description of the quadrupole
form factors $G_E^\ast$ (electric) and $G_C^\ast$ (Coulomb)
is also possible once small D state components are included
\cite{Ramalho:2008dp,Ramalho:2009df}.
In that case, the lattice QCD data can be well
described by an extension of the model with
an admixture of  D-states less than 1\% \cite{Ramalho:2009df},
but the experimental data are fairly explained
only when the meson-baryon cloud and valence quark
degrees of freedom  are combined \cite{Ramalho:2009df}.
Finally, the model was also applied to the
first radial excitation of the $\Delta(1232)$,
the $\Delta(1600)$ resonance \cite{Ramalho:2010cw}.
In this case no extra parameters are necessary,
and the meson-baryon cloud effects are largely dominant at low $Q^2$.
The results for $G_M^\ast$  are presented in the
right panel of Fig.~\ref{figNDelta}.
In both systems the valence quark effects are dominant for $Q^2> 2$ GeV$^2$.

\begin{figure}[ht!]

\includegraphics[width=3in]{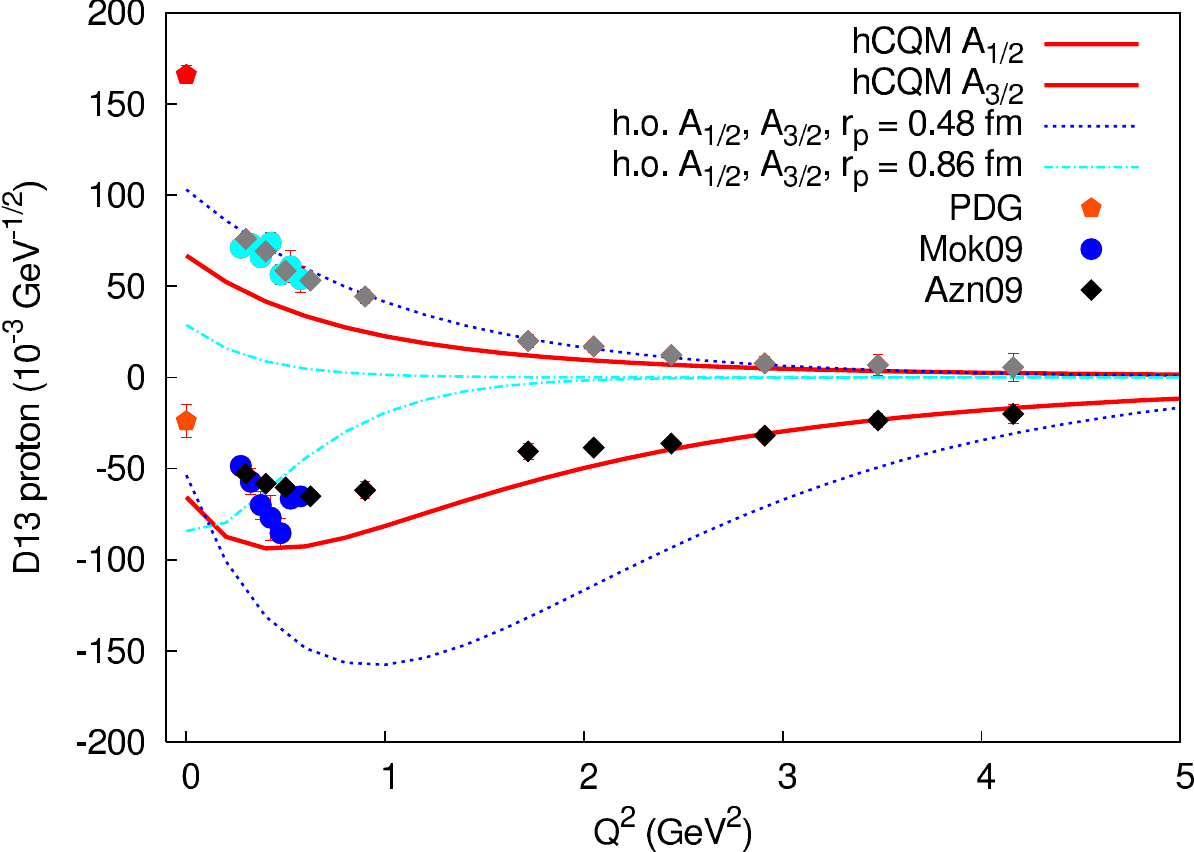} 
\caption{The $D_{13}(1520)$  helicity amplitudes $A_{3/2}$ (upper part) and $A_{1/2}$ (lower part)  predicted by the hCQM (full curves), in comparison with the data of Refs. \cite{Mokeev:2008iw,Aznauryan:2009mx,2012sha} and the PDG values \cite{Nakamura:2010zzi} at the photon point. The h.o.results for two different values of the proton r.m.s. radius ($0.5 fm$ and $0.86 fm$)Å are also shown.}
\label{d13_qm}
\end{figure}

\begin{figure}[ht!]
\includegraphics[width=3in]{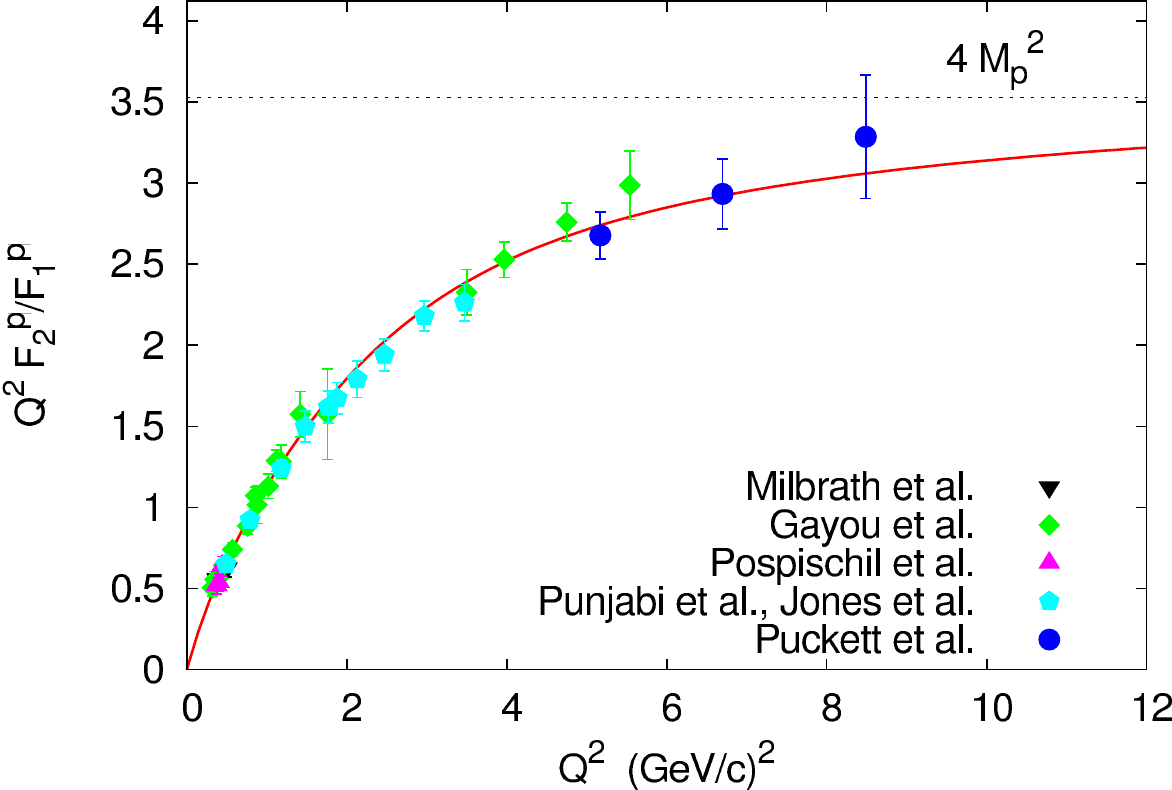}
\caption{ The ratio $F_p$ at high $Q^2$ calculated using the theoretical form factors of Ref. \cite{Sanctis:2007zz}. }
\label{fig:F_p}
\end{figure}

The model was also extended to the spin 1/2 state $N(1440)$ (Roper),
interpreted as the first radial excitation of the nucleon \cite{Ramalho:2010js}.
The $N(1440)$ shared with the nucleon the
spin and isospin structure, differing in the radial
wave function.
Under that assumption we calculated the
transition form factors for the
$\gamma N \to N(1440)$ reaction
based exclusively on the valence quark degrees of freedom \cite{Ramalho:2010js}.
As an example, we present the Dirac-type form factor
$F_1^\ast$ in Fig.~\ref{figF1} (left panel).
The model is also consistent with the lattice data \cite{Ramalho:2010js}.
The covariant spectator quark model
was also applied to the chiral partner of the nucleon
$N(1535)$ (negative parity) under two approximations:
a pointlike diquark and a quark core restricted to spin 1/2 states \cite{Ramalho:2011ae}.
Under these approximations the
 $\gamma N \to N(1535)$ transition form factors were
calculated for the region $Q^2 \gg 0.23$ GeV$^2$ \cite{Ramalho:2011ae}.
The result for $F_1^\ast$ is presented
in Fig.~\ref{figF1} (right panel).
In both reactions the results are consistent with the data
for $Q^2> 1.5$ GeV$^2$
\cite{Ramalho:2010js,Ramalho:2011ae}, except for $F_2^\ast$ for the
reaction with $ N(1535)$.
Our results support the idea that the valence quark dominance
for the intermediate and high $Q^2$ region,
but also the necessity of
the meson excitations 
for the lower $Q^2$ region ($Q^2< 2$ GeV$^2$).
The form factor $F_2^\ast$ for the $\gamma N \to N(1535)$ reaction
is particularly interesting from the perspective
of a quark model, since the data
suggest that $F_2^\ast \approx 0$ for $Q^2>2$ GeV$^2$,
contrarily to the result of the spectator quark model.
These facts suggest that
the valence quark and meson cloud contributions have
opposite signs and cancel in the sum \cite{Ramalho:2011ae,Ramalho:2012im}.
The direct consequence of the result for $F_2^\ast \approx 0$
is the proportionality between the amplitudes
$A_{1/2}$ and $S_{1/2}$ 
for $Q^2>2$ GeV$^2$ \cite{Ramalho:2011fa}.

Other applications of the covariant
spectator quark model are the elastic
electromagnetic
form factors of the baryon octet 
(spin 1/2) \cite{Gross:2009jg,Ramalho:2011pp}, and the
baryon decuplet (spin 3/2)
\cite{Ramalho:2009gk,Ramalho:2009vc,Ramalho:2010xj,Ramalho:2010rr}, as well as
the electromagnetic transition between octet and decuplet
baryons, similarly to the $\gamma N \to \Delta(1232)$ reaction
\cite{Ramalho:2013uza}.
The study of the octet electromagnetic structure
in the nuclear medium is also in progress \cite{Ramalho:2012pu}.


Future work will establish how higher angular momentum states 
in the wave function, namely P and D states,
may contribute to the nucleon form factors.
This work  will be facilitated  by the results in  Ref.~\cite{Gross:2012si,Gross:2012sj}, 
where it was already possible to constrain those terms 
of the wave function by existing deep inelastic scattering data.

Extensions for higher resonances are underway
for $P_{11}(1710)$, $D_{13}(1520)$ and $S_{11}(1650)$.
The last two cases depend on the
inclusion of an isospin 1/2, spin 3/2 core in a state of
the total angular momentum 1/2.
These states are expected to
be the same as that in the part
of the nucleon structure  \cite{Gross:2012si}.

In future developments,
the
quality and quantity of the future lattice QCD studies will be crucial
to constrain the parameterization
of the wave functions, and clarify the effect
of the valence quarks and meson cloud,
following the successful applications
to the lattice QCD regime for the
nucleon \cite{Gross:2012si,Ramalho:2011pp}, $\gamma N \to \Delta(1232)$
transition \cite{Ramalho:2009df} and Roper \cite{Ramalho:2010js}.

In parallel, the comparison with the
estimate of the quark core contributions
performed by the EBAC group  preferentially
for $Q^2>2$ GeV$^2$ \cite{JuliaDiaz:2006xt,JuliaDiaz:2009ww},
will be also very useful in the next two years.
To complement the quark models, the use of dynamical models
and/or effective chiral models \cite{Jido:2007sm}
to estimate the meson cloud effects are also very important. 
This is particularly relevant for
the $\gamma N \to N(1535)$ reaction.
From the experimental side, new accurate measurements
in the low $Q^2$ region as well as the high $Q^2$ region,
as will be measured in the future after the Jefferson Lab 12-GeV upgrade,
will be crucial,
for the purposes of
either to test the present parameterizations
at high $Q^2$, or to calibrate the models
for new calculations at even larger $Q^2$.
The clarifications between the different analysis
of the data such as EBAC,
CLAS, SAID, MAID,
J\"{u}lich and Bonn-Gatchina,
will also have an important role
\cite{Aznauryan:2009da,Arndt:2006bf,Arndt:1985vj,Aznauryan:2009mx,Drechsel:2007if,Doring:2009bi,Anisovich:2009zy}.

\subsection{Nucleon electromagnetic form factors and electroexcitation of low lying nucleon resonances up to $Q^2 = 12$~GeV$^2$ in a light-front relativistic quark model \label{sec:LFquark}}

\subsubsection{Introduction}
In recent decade, with the advent of the
new generation of electron beam facilities,
there is  dramatic progress in the investigation of the
electroexcitation of nucleon resonances
with significant extension of the range of $Q^2$.
The most accurate and complete information has been obtained
for  the electroexcitation amplitudes of
the four lowest excited states,
which have been measured in a range of $Q^2$
up to $8$ and $4.5~$GeV$^2$
for the $\Delta(1232)P_{33}$, $N(1535)S_{11}$ and
$N(1440)P_{11}$, $N(1520)D_{13}$, respectively
(see reviews \cite{Aznauryan:2011qj,Tiator:2011pw} and the recent update \cite{2012sha,Aznauryan:2012ec}.
At relatively small $Q^2$,
nearly massless Goldstone bosons (pions) can produce
significant pion-loop contributions. However
it is expected
that the corresponding hadronic component, including
meson-cloud contributions,
will be losing strength with increasing
$Q^2$. The Jefferson Lab 12-GeV upgrade
will open up a new era in the exploration of excited
nucleons when the ground state and excited nucleon's
quark core will be fully exposed to the electromagnetic probe.

Our goal is to predict $3q$ core contribution
to the electroexcitation amplitudes of the resonances
$\Delta(1232)P_{33}$, 
$N(1440)P_{11}$, $N(1520)D_{13}$, and  $N(1535)S_{11}$.
The approach we use is based on light-front (LF)
dynamics which realizes Poincar\'e invariance and the description
of the vertices $N(N^*)\rightarrow 3q,N\pi$ 
in terms of wave functions.
The corresponding LF relativistic model
for bound states is formulated in Refs. 
\cite{Berestetsky:1977zk,Berestetsky:1976um,Aznaurian:1982yd,Aznaurian:1992ej}.
The parameters of the model for the $3q$ contribution
have been specified via description of the nucleon electromagnetic
form factors in the approach that combines 
$3q$ and pion-cloud contributions.
The pion-loop contributions to nucleon electromagnetic form factors
have been described according to the LF approach of Ref. \cite{Miller:2002ig}.

\subsubsection{Quark core contribution to transition amplitudes \label{sec:qqcore}}
The $3q$ contribution to the $\gamma^* N\rightarrow N(N^*)$
transitions has been evaluated within the approach
of Refs. \cite{Aznaurian:1982yd,Aznaurian:1992ej} where the 
LF relativistic quark model
is formulated in infinite
momentum frame (IMF).
The IMF is chosen in such a way, that the 
initial hadron moves
along the $z$-axis with the momentum $P\rightarrow \infty$,
the virtual photon momentum is
$ k^{\mu}=\left(
\frac {m_{out}^2-m_{in}^2-\mathbf{Q}^2_{\perp}}{4P},
\mathbf{Q}_{\perp}, 
-\frac {m_{out}^2-m_{in}^2-\mathbf{Q}^2_{\perp}}{4P}\right)$,
the final hadron momentum is
$P'=P+k$, and $Q^2\equiv -k^2=\mathbf{Q}_{\perp}^2$; $m_{in}$ and $m_{out}$
are masses of the initial and final hadrons, respectively. 
The matrix elements of the electromagnetic current
are related to the $3q$-wave functions in the following way:
\begin{equation}
 \frac{1}{2P}<N(N^*),S'_z|J_{em}^{0,3}|N,S_z>|_
{P\rightarrow\infty}
=e\Sigma_i\int \Psi'^+ Q_i\Psi d\Gamma,
\label{eq:sec1}
\end{equation}
where $S_z$ and $S'_z$ are the projections of the hadron
spins on the $z$-direction, $Q_i$ $(i=a,b,c)$
are the charges of the quarks in units of $e$, $e^2/4\pi=\alpha$,
$\Psi$ and $\Psi'$ are wave functions
in the vertices $N(N^*)\rightarrow 3q$, and
$d\Gamma$ is the phase space volume:
\begin{equation}
d\Gamma=(2\pi)^{-6}\frac
{d\mathbf{q}_{b\perp}d\mathbf{q}_{c\perp}dx_b dx_c}
{4x_ax_bx_c}.
\label{eq:sec2}
\end{equation}

The quark momenta in the initial
and final hadrons are parameterized via:
\begin{eqnarray}
&&\mathbf{p}_i=x_i\mathbf{P}+\mathbf{q}_{i\perp},~
~~\mathbf{p}'_i=x_i\mathbf{P}'+\mathbf{q}'_{i\perp},
\label{eq:sec3}\\
&&\mathbf{P}\mathbf{q}_{i\perp}=\mathbf{P}'\mathbf{q}'_{i\perp}=0,~~~
\Sigma\mathbf{q}_{i\perp}=\Sigma\mathbf{q}'_{i\perp}=0,
~~~\mathbf{q}'_{i\perp}=\mathbf{q}_{i\perp}-y_i\mathbf{Q}_{\perp},
\label{eq:sec4}\\
&&\Sigma x_i=1,~~~y_a=x_a-1,~~y_b=x_b,~~y_c=x_c.
\label{eq:sec5}
\end{eqnarray}
Here we have supposed that quark $a$ is an active quark.

The wave function $\Psi$ is related
to the wave function in the c.m.s. of the 
system of three quarks 
through Melosh matrices \cite{Melosh:1974cu}:
\begin{equation}
\Psi=U^+(p_a)U^+(p_b)U^+(p_c)\Psi_{fss}
\Phi(\mathbf{q}_{a},\mathbf{q}_{b},\mathbf{q}_{c}),
\label{eq:sec6}
\end{equation}
where we have separated the flavor-spin-space
part of the wave function $\Psi_{fss}$ in the c.m.s.
of the quarks
and its spatial part $\Phi(\mathbf{q}_{a},\mathbf{q}_{b},\mathbf{q}_{c})$.
The Melosh matrices are defined by
\begin{equation}
U(p_i)=\frac{m_q+M_0x_i+i\epsilon _{lm}\sigma_l q_{im}}
{\sqrt{(m_q+M_0x_i)^2+\mathbf{q}_{i\perp}^2}},
\label{eq:sec7}
\end{equation}
where $m_q$ is the quark mass.
The flavor-spin-space
parts of the wave functions 
are constructed according to commonly used rules \cite{Capstick:1994ne,Koniuk:1979vy}.
To construct these parts
we need also the $z$-components
of quark momenta in the c.m.s. of quarks. They are defined
by:
\begin{equation}
q_{iz}=\frac{1}{2}\left(x_iM_0-\frac{m_q^2+\mathbf{q}_{i\perp}^2}
{x_iM_0}\right),
~~
q'_{iz}=\frac{1}{2}\left(x_iM'_0-\frac{m_q^2+\mathbf{q'}_{i\perp}^2}
{x_iM'_0}\right),
\label{eq:sec8}
\end{equation}
where $M_0$ and $M'_0$ are invariant masses
of the systems of initial and final quarks:
\begin{equation}
M_0^2=
\Sigma\frac{\mathbf{q}_{i\perp}^2+m_q^2}{x_i},~~~~
{M'_0}^2=
\Sigma\frac{\mathbf{q'}_{i\perp}^2+m_q^2}{x_i}.
\label{eq:sec9}
\end{equation}

To study sensitivity to the form of the quark wave function,
we employ two widely used forms of the spatial parts of wave functions:

\begin{equation} 
\Phi_1\sim exp(-M_0^2/6\alpha_1^2),~~~~~
\Phi_2\sim  
exp\left[-({\bf{q}}_1^2+{\bf{q}}_2^2+{\bf{q}}_3^2)/2\alpha_2^2\right],
\label{eq:sec10}
\end{equation}
used, respectively, in Refs. 
\cite{Berestetsky:1977zk,Berestetsky:1976um,Aznaurian:1982yd,Aznaurian:1992ej}
and \cite{Capstick:1986bm}. 
 
\subsubsection{Nucleon \label{sec:nucleon}}

The nucleon 
electromagnetic form factors were described by
combining the
$3q$-core and pion-cloud contributions
to the nucleon wave function.
With the pion loops evaluated according to
Ref. \cite{Miller:2002ig},   
the nucleon wave function has the form: 
\begin{equation}
|N>=0.95|3q>+0.313|N\pi>,
\label{eq:nuc1}
\end{equation}
where the portions of different contributions 
were found from the condition the charge
of the proton be unity: $F_{1p}(Q^2=0)=1$.
The value of the quark mass at $Q^2=0$
has been taken equal to $m_q(0)=0.22~$GeV
from the description of baryon and meson masses 
in the relativized quark model \cite{Godfrey:1985xj,Capstick:1986bm}.
Therefore, the only unknown parameters in the description
of the $3q$ contribution to nucleon formfactors
were the quantities $\alpha_1$ and $\alpha_2$
in Eqs. (\ref{eq:sec10}).
These parameters were found equal to 
\begin{equation}
\alpha_1=0.37~{\rm GeV},~~~\alpha_2=0.405~{\rm GeV}
\label{eq:nuc2}
\end{equation}
from the description of the magnetic moments at $Q^2=0$
(see Fig. \ref{fig:fig1}).
The parameters (\ref{eq:nuc2}) give very close magnitudes for
the mean values of invariant masses and momenta
of quarks at $Q^2=0$:
$<M^2_0>\approx 1.35~$GeV$^2$ and 
$<{\mathbf{q}}_{i}^2>\approx 0.1~$GeV$^2,~i=a,b.c$.
  
The constant value of the quark mass 
gives rapidly decreasing  
form factors $G_{Ep}(Q^2)$, $G_{Mp}(Q^2)$, and $G_{Mn}(Q^2)$
(see Fig. \ref{fig:fig1}). The wave functions (\ref{eq:sec10}) increase
as $m_q$ decreases. Therefore, to describe the
experimental data we have assumed the $Q^2$-dependent
quark mass that decreases with increasing $Q^2$:

\begin{equation}
m_q^{(1)}(Q^2)=\frac{0.22{\rm \; GeV}}{1+Q^2/60{\rm \;  GeV^{2}}},~~~
m_q^{(2)}(Q^2)=\frac{0.22{\rm \;  GeV}}{1+Q^2/10{\rm \;  GeV^{2}}}
\label{eq:nuc3}
\end{equation}

for the wave functions $\Phi_1$ and $\Phi_2$, respectively. Momentum dependent quark mass allowed us to obtain good description of the nucleon electromagnetic 
form factors up to $Q^2=16~$GeV$^2$.
From Fig. \ref{fig:cloud} it can be seen
that at $Q^2>2~$GeV$^2$,
these form factors
are dominated by the $3q$-core contribution.                                     
 
\subsubsection{Nucleon resonances $\Delta(1232)P_{33}$, 
$N(1440)P_{11}$, $N(1520)D_{13}$, 
and $N(1535)S_{11}$ \label{sec:nucleonres}}

No calculations are available that allow for the separation
of the $3q$ and $N\pi$ (or nucleon-meson) contributions
to nucleon resonances. Therefore, the
weights $c^*~(c^*<1)$ of the $3q$ contributions to the resonances: 
$|N^*>=c^*|3q>+...$, are unknown.
We determine these weights by fitting to experimental
amplitudes at $Q^2=2-3~$GeV$^2$,
assuming that at these $Q^2$ the transition amplitudes
are dominated by the $3q$-core contribution, as is the case
for the nucleon. Then we predict the transition amplitudes
at higher $Q^2$ (see Figs. \ref{fig:fig2}-\ref{fig:fig5}).

As it is shown in Refs. \cite{Aznauryan:85,Keister:1993mg},
there are difficulties in the utilization of the LF
approaches \cite{Berestetsky:1976um,Berestetsky:1977zk,Aznaurian:1982qc,Aznaurian:1993np,Capstick:1994ne}
for the hadrons with spins $J\geq 1$. These difficulties
can be avoided if Eq. (\ref{eq:sec1}) is used to
calculate only those matrix elements that
correspond to $S'_z=J$ \cite{Aznauryan:85}.
This restricts the number of transition form factors
that can be calculated for the resonances  
 $\Delta(1232)P_{33}$ and $N(1520)D_{13}$,
and only two transition form factors
can be investigated for these resonances: $G_1(Q^2)$
and $G_2(Q^2)$ (the definitions can be found in review 
\cite{Aznauryan:2011qj}).
For these resonances we can not present
the results for the transition helicity amplitudes.
The results for the resonances with $J=\frac{1}{2}$:
$N(1440)P_{11}$ and $N(1535)S_{11}$,
are presented in terms of the transition helicity amplitudes. 
 
\subsubsection{Discussion}

The important feature of the obtained predictions
for the resonances is the fact that at $Q^2> 2-3~$GeV$^2$
both investigated amplitudes for each resonance 
are described well by the
$3q$ contribution by fitting the only parameter, that is
the weight of this contribution to the resonance.
These predictions need to be checked at higher $Q^2$.

The results for the resonances allow us also 
to make conclusions on the size and form
of expected pion-cloud and/or meson-baryon
contributions to the amplitudes. According to
our predictions for the $3q$ contributions,
one can expect that pion-cloud contributions to the
form factor $G_2(Q^2)$ for the $\Delta(1232)P_{33}$,
to $S_{1/2}$ amplitude for the $N(1440)P_{11}$,
and to the form factor $G_1(Q^2)$ for the $N(1520)D_{13}$
are small.
Large contributions are expected to the
longitudinal amplitude for the $N(1535)S_{11}$
and to the form factor $G_2(Q^2)$ for the $N(1520)D_{13}$.
The expected pion-cloud contributions to the
form factor $G_1(Q^2)$ for the $\Delta(1232)P_{33}$
and to $A_{1/2}$ amplitude for the $N(1535)S_{11}$
have $Q^2$ behavior similar to that
in the nucleon formfactors $G_{Mp(n)}(Q^2)$. 
In Fig. \ref{fig:fig3} by dotted curves we show
estimated pion-cloud
contribution to 
$A_{1/2}$ amplitude for the Roper resonance.
It can be seen that non-trivial $Q^2$-dependence
of this contribution can be expected.

The remarkable feature that follow from the
description of the nucleon electromagnetic formfactors
in our approach is
the decreasing quark mass with increasing $Q^2$.
This
is in qualitative agreement with the QCD lattice calculations
and with Dyson-Schwinger equations
\cite{Bowman:2005vx,Bhagwat:2003vw,Bhagwat:2006tu} where
the running quark mass is generated dynamically.
The mechanism that generates the running quark mass
can generate also the quark 
anomalous magnetic moments and
form factors. This should be incorporated
in model calculations. Introducing quark form factors
will cause a faster $Q^2$ fall-off of electromagnetic
form factors in quark models. This will force
$m_q(Q^2)$ to drop faster with $Q^2$ to describe
the data.

\subsection{Constituent Quark Models and the interpretation of the nucleon form factors \label{sec:cqm2}}

Various Constituent Quark Models (CQM) have been proposed in the past
decades after the pioneering work of Isgur and Karl (IK) \cite{Isgur:1978wd}. Among
them
let us quote the relativized Capstick-Isgur model (CI)
\cite{Capstick:1986bm}, the algebraic approach (BIL) \cite{Bijker:1994yr}, the hypercentral CQM
(hCQM) \cite{Ferraris:1995ui}, the chiral Goldstone Boson Exchange model ($\chi$CQM)
\cite{Glozman:1995fu} and the Bonn instanton model (BN) \cite{Loring:2001kv,Loring:2001ky}. They are all able
to fairly reproduce the baryon spectrum, which is the first test to be performed
before applying any model to the description of other baryon properties.
The models, although different, have a simple
general structure, since, according to the prescription provided by the early Lattice QCD calculations \cite{DeRujula:1975ge},
 the three-quark interaction $V_{3q}$  is split into a
spin-flavor independent
part $V_{inv}$, which is $SU(6)$-invariant and contains the confinement
interaction, and a $SU(6)$-dependent part $V_{sf}$, which contains spin
and eventually flavor dependent interactions

\begin{equation} \label{v3q}
V_{3q}÷=÷V_{inv}÷+÷V_{sf}
\end{equation}

\noindent In Tab. \ref{cqm}, a summary of the main features of various Constituent Quark Models is reported.

After having checked that these models provide a reasonable
description of the baryon spectrum, they have been applied to the calculation of many baryon properties, 
including electrocouplings. One should however not
forget that in many cases
the calculations referred to as  CQM calculations are actually performed
using a simple h.o. wave function for the internal quark motion either in
the non relativistic (HO) or relativistic (relHO) framework. 
The former (HO) applies to the calculations of Refs.   \cite{Copley:1969qn} and  \cite{Koniuk:1979vy}, while the latter (relHO) is valid 
for Ref. \cite{Capstick:1994ne}.
The relativized model (CI) of Ref. \cite{Capstick:1986bm} is used for a systematic calculation of the transition amplitudes in Ref. \cite{Capstick:1992uc}  and, within a light-front approach in Refs. \cite{Cardarelli:1995ug} and Ref. \cite{Cardarelli:1996vn} for the transitions to the $\Delta$ and Roper resonances respectively.
In the algebraic approach \cite{Bijker:1994yr}, a particular form of the charge distribution along the string is assumed and used 
for both the elastic and transition form factors; the elastic form factors are fairly well reproduced, but there are problems with the transition amplitudes, specially at low $Q^2$. There is no helicity amplitude calculation with the GBE model, whereas the BN model has been also used for the helicity amplitudes \cite{Merten:2002nz}, with particular attention to the strange baryons \cite{VanCauteren:2005sm}.
Finally, the hCQM has produced predictions for the transverse excitation of the negative parity resonances \cite{Aiello:1998xq} and also for the main resonances, both for the longitudinal and transverse excitation \cite{PhysRevC.86.065202}.

In some recent approaches the CQ idea is used to derive relations between the various electromagnetic form factors, relations which, after having fitted one selected quantity, say the elastic proton form factor (Sec.~\ref{sec:qqcore})  or the helicity amplitude at intermediate $Q^2$ (Sec.~\ref{sec:nucleon}), are used to predict the other quantities of interest. A remarkable prediction of both the proton elastic form factor and the proton transition to the Roper resonance is provided by the light-front holographic approach (Sec.~\ref{sec:nucleonres}).

The works briefly illustrated above have shown that the three-quark idea is able to fairly reproduce a large variety of observables, in particular the helicity amplitudes at medium $Q^2$, however, a detailed comparison with data shows that, besides the fundamental valence quarks, other issues are or presumably will be of relevant importance for the interpretation of the transition amplitudes. These issues are: relativity, meson cloud and quark-antiquark pair effects, and quark form factors.

\begin{table}
\centering
\caption[]{Illustration of the features of various CQMs}
\vspace{15pt}
\label{photo}
\begin{tabular}{|c|c|c|c|c|c|}
\hline
$CQM$ & Kin. Energy & $V_{inv}$ & $V_{sf}$ &  Ref. \\
\hline
 Isgur-Karl  & non rel. & h.o. + shift & OGE &  \cite{Isgur:1978wd}\\
 \hline
Capstick-Isgur & rel & string +coul-like & OGE &  \cite{Capstick:1986bm} \\
\hline
$U(7)  B.I.L.$ & rel $M^2$ & vibr + L &  G\"{u}rsey-Rad & \cite{Bijker:1994yr}\\
\hline
Hypercentral G.S. & non rel/ rel & $O(6)$: lin + hyp.coul & OGE & \cite{Ferraris:1995ui} \\
\hline
Glozman-Riska & rel & h.o. / linear & GBE & \cite{Glozman:1995fu} \\
\hline
Bonn & rel & linear + $3$ body & instanton & \cite{Loring:2001kv} \\

\hline
\end{tabular}
\label{cqm}
\end{table}

A consistent relativistic treatment is certainly important for the description of the elastic nucleon form factors. In fact, in the non-relativistic hCQM \cite{Ferraris:1995ui}, the proton radius compatible with the spectrum is too low, about $0.5 fm$, and the resulting form factors \cite{DeSanctis:1998ck} are higher than data. However, the introduction of the Lorentz boosts improves the description of the elastic form factors  \cite{DeSanctis:1998ck} and determines a ratio $\mu_p G_e^p/G_M^p$ lower than 1 \cite{Sanctis:2000eg}. Using a relativistic formulation of the hCQM in the Point Form approach, in which again the unknown parameters are fitted to the spectrum, the predicted elastic nucleon form factors are nicely close to data \cite{Sanctis:2007zz}. Furthermore, if one introduces quark form factors, an accurate description of data is achieved \cite{Sanctis:2007zz}. Since such form factors are fitted, this means that they contain, in an uncontrolled manner, all the missing contributions.

Applying hCQM to the excitation of higher resonances demonstrated that the inclusion of relativity is less crucial, since the Lorentz boosts affect only slightly the helicity amplitudes \cite{DeSanctis:1998tu}. A quite different situation occurs for the excitation to the $\Delta$, which is a spin-isospin excitation of the nucleon and as such it shares with the nucleon the spatial structure. In this case relativity is certainly important, however it does not seem to be sufficient even within LF approaches. In fact,  the good results of the Rome group  \cite{Cardarelli:1995ug} are obtained introducing quark form factors, while in Sec.~\ref{sec:qqcore} the quark wave function fitted to the elastic nucleon form factor leads to a lack of strength at low $Q^2$ in the $\Delta$ excitation. In Sec.~\ref{sec:nucleon} a pion cloud term is present from the beginning in the nucleon form factor, nevertheless the transition to the $\Delta$ is too low at low $Q^2$.

Of course, the future data at high $Q^2$ will force, at least for consistency reasons, to use a relativistic approach also for the other resonances.

At medium-low $Q^2$ the behavior of the helicity amplitudes is often described quite well, also in  a non relativistic approach \cite{PhysRevC.86.065202}. An example is provided by Fig. \ref{d13_qm}, where the hCQM results are compared with the more recent JLab data.  In Fig.  \ref{d13_qm} there are also the h.o. results, which do not seem to be able to reproduce the data. The good agreement achieved by the hCQM has a dynamical origin. Let us remind that in hCQM the $SU(6)$-invariant part of the quark potential of Eq. (\ref{v3q}) is 
 
\begin{equation}
V_{inv}^{hCQM} ÷=÷-\frac{\tau}{x}÷+÷\alpha x
\end{equation}

($x=\sqrt{\rho^2+\lambda^2}$ is the hyperradius) however the main responsible of the medium-high $Q^2$ behavior of the helicity amplitudes is the hypercoulomb interaction $÷-\frac{\tau}{x}$. In fact, in the analytical version of hCQM presented in Ref. \cite{Santopinto:1997jz}, it is shown that the helicity amplitudes provided by the $÷-\frac{\tau}{x}$ term are quite similar to the ones calculated with the full hCQM.

The main problem with the description provided by CQM (non relativistic or relativistic) is the lack of strength at low $Q^2$, which is attributed, with general consensus, to the missing meson cloud or quark-antiquark pair effects \cite{Aiello:1998xq,JuliaDiaz:2009ww,Krewald:2012zz,Tiator:2011pw}. In fact, it has been shown within a dynamical model \cite{JuliaDiaz:2009ww,Krewald:2012zz,Tiator:2011pw} that the meson-cloud contributions are relevant at low $Q^2$ and tend to compensate the lack of strength of unquenched three-quark models \cite{Tiator:2003uu}.

To conclude, a fully relativistic and unquenched hCQM is not yet available and work is now in progress in this direction, but certainly it will be a valuable tool for the interpretation of the helicity amplitudes at high $Q^2$.

However, also taking into account the one pion contribution there seems to be some problem. In Sec.~\ref{sec:qqcore}, the quark wave function is chosen in order to reproduce the proton form factor, in this way all possible extra contributions (meson cloud, quark form factors,....) are implicitly included, but the description of the $N-\Delta$ transition needs an extra pion term. On the other hand, in  Sec.~\ref{sec:nucleon} it is shown that the pion term explicitly included in the fit to the proton is not sufficient for the description of the $N-\Delta$ transition. In fact, the inclusion of a pion cloud term, either fitted or calculated (as e.g. in Ref. \cite{Chen:2006bza}) seems to be too restrictive, since it is equivalent to only one quark-antiquark configuration. If one wants to include consistently all quark-antiquark effects, one has to proceed to unquenching the CQM, as it has been done in \cite{Geiger:1991qe}. Such an unquenching is achieved by summing over all quark loops, that is over all intermediate meson-baryon states; the sum is in particular necessary in order to preserve the OZI rule. 

This unquenching has been recently performed also for the baryon sector \cite{Bijker:2007zz}. The  state for a baryon A is written as 

\begin{equation} 
|\Psi_A> = \mathit{N} \left[ |A> + \sum_{BClj} \int d\vec{k}~ |BC \vec{k}lJ> \frac{<BC\vec{k}lJ | T^\dagger  |A>}{M_A-E_B-E_C} \right] \label{ttt}
\end{equation}

\noindent where B (C) is any intermediate baryon (meson), $E_B (E_C)$ are  the corresponding energies, $M_A$ is the baryon mass, $T^{\dag}$ is the $^3P_0$ pair creation operator and $\vec{k}$, $\vec{l}$ and $\vec{J}$ are the relative momentum, the orbital and total angular momentum, respectively. Such unquenched model, with the inclusion of the quark-antiquark pair creation mechanism,  will allow to build up a consistent   description of all the baryon properties (spectrum, form factors, ...). There are already some applications \cite{Bijker:2007zz,Santopinto:2010zza,Bijker:2012zza}, in particular it has been checked that, thanks to the summation over all the intermediate states described in Eq. (\ref{ttt}), the good reproduction of the baryon magnetic moments provided by the CQM is preserved after renormalization  \cite{Bijker:2007zz}. Using an interaction containing the quark-antiquark production the resonances acquire a finite width, at variance with what happens in all CQMs, allowing a consistent description of both electromagnetic and strong vertices.

The structure of the state in Eq. (\ref{ttt}) is more general than the one containing a single pion contribution. The influence of the quark-antiquark cloud will be certainly important at low $Q^2$, but one can also expect that the multiquark components, which are  mixed with the standard 3q states as in Eq. (\ref{ttt}), may have a quite different behavior \cite{an:2006zf,An:2008xk,JuliaDiaz:2006av} in the medium-high $Q^2$ region, leading therefore to some new and interesting behavior also at short distances. Actually, there are some clues that this may really happen. First,  it has been shown in \cite{Aznauryan:2009mx} that the quantity $Q^3 A^p_{1/2}$ seems to become flat in the range around $4 ~GeV^2$ (see Fig.~\ref{scaling}), while the CQM calculations do not show any structure.

A second important issue is the ratio $R_{p}=\mu_{p} G_{e}^{p}/G_{M}^{p}$ between the proton form factors. A convenient way of understanding its behavior is to consider the ratio $Q^{2} F_{2}^{p} / F_{1}^{P}$, which is expected to saturate at high $Q^{2}$ \cite{Brodsky:1974vy,Lepage:1979za}, while it should pass through the value $4 M_{p}^{2} / \kappa_{p}$ in correspondence of a zero for $R_{p}$ \cite{Santopinto:2010zz}. The predictions of the hCQM \cite{Sanctis:2007zz} are compared with the JLab data \cite{Jones:1999rz,Gayou:2001qd,Gayou:2001qt,Milbrath:1997de,Punjabi:2005wq,Puckett:2010ac} in Fig. \ref{fig:F_p}. For a  pure three-quark state, even in presence of quark form factors as in \cite{Sanctis:2007zz} the occurrence of  a zero seems to be difficult, while an interference between three- and  multi-quark configurations may be a possible candidate for the generation of a dip in the electric form factors \cite{Santopinto:2010zz}.

It is interesting to note that the Interacting Quark Diquark model introduced in Ref.  
(\cite{Santopinto:2004hw}) and its relativistic reformulation  (\cite{Ferretti:2011zz}), both of which do not exhibit  
missing states in the non strange sector under $ 2 GeV^2 $, give rise to a ratio  
$R=\mu_p {G^p}_E/{G^p}_M$  that goes through a zero at around $8 GeV^2$ after the 
introduction of quark form factors, as calculated in Ref. (\cite{DeSanctis:2011zz}).

Once the quark-antiquark pair creation effects have been included consistently in the CQM,   it will be possible to disentangle the quark forms factors from the other dynamic mechanisms. The presence of structures with a finite dimension has been  shown in a recent analysis of deep inelastic electron-proton scattering \cite{Petronzio:2003bw}.

\section{Conclusions and Outlook  \label{Conclusion}}

Studying $\gamma_{v}NN^{*}$ electrocouplings gives insight into the
relevant degrees of freedom of the baryon structure and its evolution
with distance. The CLAS 6~GeV program has already provided information
on the transition in the $N^{*}$ structure from a super\-position of
meson-baryon and quark degrees of freedom to a quark-core dominance.
The approved experiment, Nucleon Resonance Studies with CLAS12
\cite{Gothe:clas12}, will run within the first year following the
commissioning of the CLAS12 detector in Hall B of Jefferson Lab.
This experiment seeks to extract all prominent excited state
electro\-couplings by analyzing all the major exclusive meson
electroproduction channels, including $\pi^+ n$, $\pi^0 p$,
and $\pi^+\pi^-p$, in the almost unmeasured $Q^{2}$ region
of 5 to 12~GeV$^2$, and, as in the 6~GeV program, we expect
high-quality electro\-coupling results for all prominent $N^{*}$s.
As delineated in this paper, there are plans to develop advanced
reaction models to reliably extract the $\gamma_{v}pN^{*}$
electrocouplings in this unexplored range of photon virtualities.
These models will explicitly take into account the contributions
from quark-gluon degrees of freedom, and will be extended in scope
to incorporate the $\eta p$ and $KY$ exclusive channels.
The evaluation of the resonance electrocouplings can also ensue
from the less constrained semi-inclusive meson electroproduction
data, as soon as reliable modelling of non-resonant contributions
to $\pi$s and $\eta$s in semi-inclusive electroproduction becomes
available.

The 12~GeV $N^*$ program, with the highest $Q^{2}$ reach worldwide,
will allow direct access to quarks decoupled from the meson-baryon
cloud and to the evolution of quark and gluon interactions with
distance that are ultimately responsible for the formation of the
excited states. These nucleon resonances have larger transverse
sizes in comparison with the ground nucleon states. Data on the
electrocouplings of excited proton states will enable exploration of
the nonperturbative interactions at larger transverse separations
between quarks, which are especially sensitive to the nonperturbative
contributions to baryon structure. Systematic studies of
$\gamma_{v}NN^*$ electrocouplings at high photon virtualities are
key in accessing the very essence of the strong interaction in the 
nonperturbative regime; that is, the region where the active degrees
of freedom in hadron structure, along with their interactions,
become very different from those of current quarks, gluons,
and their interactions, as described in the perturbative expansion
of the QCD Lagrangian. This is the unique feature of the strong
interaction evolution with distance that shapes the structure of
hadronic matter.

The data expected from the CLAS12 $N^*$ experiment \cite{Gothe:clas12}
will make possible the study of the kinematic regime of momenta
$0.5 < p < 1.1$~GeV running over the dressed-quark propagator,
where $p = \sqrt{Q^{2}}/3$. This kinematic region spans the
transition from the almost-completely dressed constituent quarks to
the almost-completely undressed current quarks. The $\gamma_{v}pN^*$
electrocouplings will be sensitive to the transition from the
confinement regime of strongly bound dressed quarks and gluons at 
small momenta, $p < 0.5$~GeV, to the pQCD regime for $p > 2$~GeV,
where almost-undressed and weakly-interacting current quarks and
gluons gradually emerge as the relevant degrees of freedom in
resonance structure with increasing $Q^2$. The momentum dependence
of the dressed-quark mass should affect all dressed-quark propagators
and therefore the $Q^2$ evolution of $\gamma_{v}NN^*$
electro\-couplings. Virtual photon interactions with the
dressed-quark electromagnetic current should thus be sensitive
to the dynamical structure of dressed quarks, including spin
and helicity-flip dependent parts of the dressed quark electromagnetic 
current that is generated nonperturbatively through DCSB.
The dressed-quark dynamical structure and its mass function should be practically independent of the excited state quantum numbers.
Therefore, the combined physics analyses of the data on
electrocouplings of several prominent $N^*$ states will
considerably improve our knowledge of the momentum dependence
of dressed-quark masses and their dynamical structure.

The theoretical interpretation of experimental results on resonance 
electrocouplings will be a critical component of the successful
realization of the CLAS12 $N^*$ program.  QCD-based approaches,
as well as more model-dependent frameworks, will be important for
examining the capability of QCD as the fundamental theory of strong
interactions in describing the full complexity of nonperturbative
processes, which generate all $N^*$ states from quarks and gluons
through the strong interaction.

Lattice QCD results on electrocouplings of prominent $N^*$ states
can be directly compared with their values determined from
experimental data analyses, offering a unique way to explore
the capability of this method in describing the nonperturbative
QCD interactions responsible for the generation of $N^*$ states
with different quantum numbers. LQCD methods will be further
developed to provide reliable evaluations of $\gamma_{v}NN^*$
electrocouplings employing a realistic basis of projection
operators, approaching the physical pion mass, and carried out
in boxes that fully include the physics of the pion cloud.
Available analyses from CLAS experimental results on
$\gamma_{v}pN^*$ electrocouplings \cite{Aznauryan:2009mx,2012sha}
allow one to develop and cross check LQCD simulations in the
regime of small and intermediate photon virtualities,
$Q^2 < 5$~GeV$^2$. The extension of LQCD calculations to
higher $Q^2$ (up to $Q^2=12$~GeV$^2$) represents an important
additional challenge.

The Dyson-Schwinger equation of QCD allows for mapping out the
momentum dependence of the dressed-quark running mass, structure,
and nonperturbative interaction from the data on excited nucleon
electro\-couplings, such for the $Q^2$ evolution of $P_{11}(1440)$
state. Since DSEQCD can currently only describe the quark-core
contribution, these predictions will be compared with the
experimental data at large photon virtualities, $Q^2 > 5$~GeV$^2$,
where the quark degrees of freedom are expected to dominate
resonance structure. Analyses of the future data on
$\gamma_{v}NN^*$ electrocouplings expected from CLAS12 within
this approach will provide the first pieces of information on
the generation of the dominant part of $N^*$ masses through 
dynamical chiral symmetry breaking, and will allow one to explore
how confinement in baryons emerges from the pQCD regime.
Experiments, coupled with theoretical studies, therefore lay
out an ambitious program to map out the momentum-dependent
dressed-quark mass function and thereby seek evidence for how
dynamical chiral symmetry breaking generates more than 98\% of
baryon masses and to explore the nature of confinement in baryons.

In conjunction with LQCD and DSEQCD, the light-front sum rule
approach opens up prospects for constraining quark distribution
amplitudes for various $N^*$ states from the data on
$\gamma_{v}pN^*$ electrocouplings, thereby offering access to
the partonic degrees of freedom in excited nucleons for the
first time. This approach can be employed and tested in analyses
of available CLAS data on $\gamma_{v}NN^*$ electrocouplings for
$Q^2$ from 2 to 5~GeV$^2$. After that, it is straightforward to use
this approach for accessing the partonic degrees of freedom in the
$N^*$ structure at high photon virtualities up to 12~GeV$^2$ in the
analyses of future results from CLAS12. These studies will provide
important information for an extension of generalized parton
distributions to the transition GPDs from ground to excited
nucleon states.

Theory and experiment go hand in hand and all these studies of the
structure of excited nucleons will further refine the quark models of the 
$N^*$ structure. 
The AdS/QCD light-front holographic approach is particularly intriguing
for nucleon and nucleon resonance physics. This nonperturbative model
predicts both hadron dynamics as well as spectroscopy, accounting for
elastic and transition form factors and the overall characteristics of
the complete nucleon resonance spectrum, based on one mass scale
parameter $\kappa^2$.
The AdS/QCD light-front holographic model is confining, satisfies
chiral invariance, and agrees with quark counting rules at high $Q^2$.
It also predicts nonzero quark orbital angular momentum and the
frame-independent light-front wavefunctions underlying many hadron
observables. The form of the color-confining potential is uniquely
determined by general arguments based on chiral symmetry and conformal
invariance~\cite{Brodsky:2013ar}. The formalism  is extended to hadrons of general
spin in Ref.~\cite{deTeramond:2013it}
Despite the shortcomings of quark models in not being
directly related (according to our currently understanding) to the
QCD Lagrangian, they offer very useful phenomenological tools for
analyses of the experimental results on the structure of all excited
nucleon states predicted by these models. We expect that analyses
of resonance electrocouplings within the framework of quark models
will provide important information for a QCD-based theory of hadron
structure. As an example, the first encouraging attempt to observe
the manifestation of the momentum dependence of the dressed-quark
mass and structure was undertaken in interpreting the low-lying
resonance electrocouplings within the framework of the light-front
quark model \cite{Aznauryan:2012ec}.
Finally, from a more phenomenological perspective, a better
understanding of quark-hadron duality in exclusive reactions holds
the prospect of utilizing information on the $N^*$ resonance
structure to constrain the behavior of exclusive processes at
higher energies, in kinematic regions difficult to access
experimentally. Conversely, the details of the quark-hadron
transition can be tested by comparing the future CLAS12 $N^*$
data with results of quark-gluon calculations extrapolated into
the low-$W$ region.

Understanding the underlying structure of the proton and neutron
is of primary importance to the 12~GeV upgrade for Jefferson Lab.
Detailed information on the structure of the ground nucleon states
in terms of different partonic structure functions, including
access to the structure in 3-dimensions expected from results on
generalized parton distributions and transverse momentum dependent
distribution functions is the driving component of the physics
program. Nonetheless, all these studies together still mark only
the first step in the exploration of nucleon structure as they
are limited to studying the ground state of the nucleon.
Nonperturbative QCD interactions generate both the ground and
excited nucleon states, thereby manifesting themselves differently
in the case of excited states with their different quantum numbers.
Comprehensive information on the ground state nucleon structure must therefore be extended with the results on transition
$\gamma_{v}NN^*$ electrocouplings. This will provide deeper insights
into the essence of the nonperturbative strong interactions
responsible for the formation of ground and excited nucleon
states from quarks and gluons, and their emergence from QCD.
These studies will allow us to finally answer the central and
most challenging questions in present-day hadron physics:
(a) how more than 98\% of hadron masses are generated
nonperturbatively;
(b) how quark and gluon confinement and dynamical chiral
symmetry breaking emerge from QCD; and
(c) how the nonperturbative strong interaction generates
the ground and excited nucleon states with various quantum
numbers from quarks and gluons.

\section{Acknowledgments}
The authors thank  B.~Juli\'a-D\'iaz, H. Kamano,
 A.~Matsuyama, S.~X.~Nakamura, T.~Sato,
 and N.~Suzuki
for their collaborations at EBAC, and
 would also like to thank  M.~Pennington and A.~W.~Thomas for their strong support 
and many constructive discussions.
We also acknowledge valuable discussions with S.-x.~Qin, H.\,L.\,L.~Roberts and P.\,C.~Tandy, and thank the University of South Carolina for their support of the most recent EmNN* 2012 Workshop.
This work is supported by the U.S. National Science Foundation under Grants  NSF-PHY-0856010, NSF-PHY-0903991, and NSF-PHY-1206082, and U.S. Department of Energy, Office of Nuclear Physics Division,
under Contract No. DE-AC02-76SF00515, DE-AC02-06CH11357 and DE-AC05-06OR23177 under which Jefferson Science Associates operates the Jefferson Lab, European Union under the HadronPhysics3 Grant No.~283286, Funda\c{c}\~ao para a Ci\^encia e a Tecnologia under Grant No. SFRH/BPD/26886/2006 and PTDC/FIS/113940/2009,  {\em Programa de Cooperaci\'on Bilateral M\'exico-Estados Unidos} CONACyT Project 46614-F, Coordinaci\'on de la Investigaci\'on Cient\'{i}fica (CIC) Project No. 4.10, Forschungszentrum J\"ulich GmbH, the University of Adelaide and the Australian Research Council through Grant No.~FL0992247, and Funda\c{c}\~ao de Amparo \`a Pesquisa do Estado de S\~ao Paulo, Grant No.~2009/51296-1 and 2010/05772-3. This research used resources of the National Energy Research Scientific Computing Center, 
which is supported by the Office of Science of the U.S. Department of Energy 
under Contract No. DE-AC02-05CH11231, resources provided on ``Fusion,'' 
a 320-node computing cluster operated by the Laboratory Computing Resource Center 
at Argonne National Laboratory, and resources of Barcelona Sucpercomputing Center (BSC/CNS).

\bibliography{bibref}


\end{document}